\documentclass[10pt]{article}
\pdfoutput=1
\usepackage{jheparxiv_blue}%jheparxiv_blue
\usepackage[utf8]{inputenc}
\usepackage{amsmath}
\usepackage{amsfonts}
\usepackage{amssymb}
\usepackage{latexsym}
\usepackage{mathrsfs}
\usepackage[usenames,dvipsnames,svgnames,x11names]{xcolor}
\usepackage{graphicx}
\usepackage{shuffle}
\usepackage{subcaption}
\usepackage{slashed}
\usepackage{diagbox}
\usepackage{arydshln}
\usepackage{enumerate}
\usepackage{silence}
\usepackage[all]{xy}
\usepackage{tikz}
\usepackage[compat=1.1.0]{tikz-feynman}
\usepackage[colorinlistoftodos]{todonotes}

\usepackage{etoolbox}
\appto\appendix{\addtocontents{toc}{\protect\setcounter{tocdepth}{1}}}

\usepackage[colorlinks=true]{hyperref}
\usepackage{cleveref}

\usetikzlibrary{arrows.meta,decorations,shapes}
\tikzset{
  mid arrow/.style={postaction={decorate,decoration={
        markings,
        mark=at position .5 with {\arrow[#1]{Stealth[length=2.5mm]}}
      }}},
}
\tikzfeynmanset{
  fermion1/.style={
    /tikz/postaction={
      /tikz/decoration={
        markings,
        mark=at position 0.55 with {
          \node[
            transform shape,
            xshift=-0.5mm,
            fill,
            dart tail angle=100,
            inner sep=1.0pt,
            draw=none,
            dart
          ] { };
        },
      },
      /tikz/decorate=true,
    },
  },
  fermion2/.style={
    /tikz/postaction={
      /tikz/decoration={
        markings,
        mark=at position 0.55 with {
          \arrow{>[length=6pt, width=5pt]};
        },
      },
      /tikz/decorate=true,
    },
  },
    anti_fermion2/.style={
    /tikz/postaction={
      /tikz/decoration={
        markings,
        mark=at position 0.55 with {
          \arrow{<[length=6pt, width=5pt]};
        },
      },
      /tikz/decorate=true,
    },
  },
}
    \definecolor{light-gray}{gray}{0.95}
    \definecolor{light-grayII}{gray}{0.85}
    \definecolor{firebrick}{rgb}{0.7, 0.13, 0.13}
    \definecolor{lincolngreen}{rgb}{0.11, 0.35, 0.02}

\DeclareFontFamily{U}{mathx}{\hyphenchar\font45}
\DeclareFontShape{U}{mathx}{m}{n}{
      <5> <6> <7> <8> <9> <10>
      <10.95> <12> <14.4> <17.28> <20.74> <24.88>
      mathx10
      }{}
\DeclareSymbolFont{mathx}{U}{mathx}{m}{n}
\DeclareFontSubstitution{U}{mathx}{m}{n}
\DeclareMathAccent{\widecheck}{0}{mathx}{"71}

\newcommand{\ang}[1]{\left\langle #1\right\rangle}
\newcommand{\squ}[1]{\left[ #1\right]}

\newcommand{\cA}{\mathcal{A}}

\newcommand{\cN}{\mathcal{N}}

\newcommand{\C}{\mathbb{C}}

\newcommand{\lambdat}{\tilde{\lambda}}

\newcommand{\diracd}[2]{\delta^{#1}\left(#2\right)}

\newcommand{\cI}{\mathcal{I}}

\newcommand{\tr}{\mathrm{tr}}

\newcommand{\pf}{\text{Pf} \,}

\newcommand{\la}{\left\langle}          %%%
\newcommand{\ra}{\right\rangle}     %%%
\newcommand{\sL}{{\scalebox{0.6}{$L$}}}
\newcommand{\sR}{{\scalebox{0.6}{$R$}}}
\newcommand{\ssL}{{\scalebox{0.5}{$L$}}}
\newcommand{\ssR}{{\scalebox{0.5}{$R$}}}
\newcommand{\sA}{{\scalebox{0.6}{$A$}}}
\newcommand{\sB}{{\scalebox{0.6}{$B$}}}
\newcommand{\sD}{{\scalebox{0.6}{$D$}}}
\newcommand{\sE}{{\scalebox{0.6}{$E$}}}
\newcommand{\sN}{{\scalebox{0.6}{$N$}}}
\newcommand{\sP}{{\scalebox{0.6}{$P$}}}
\newcommand{\sM}{{\scalebox{0.6}{$M$}}}

\newcommand{\DFL}{{\scalebox{0.6}{DFL}}}
\newcommand{\CHY}{{\scalebox{0.6}{$\mathrm{CHY}$}}}  %%%
\newcommand{\lin}{{\scalebox{0.7}{lin}}}
\newcommand{\Def}{{\scalebox{0.7}{$\ell^2$-def}}}
\newcommand{\tell}{\ell_0}
\newcommand{\mI}{\mathcal{I}}
\newcommand{\one}{\hat{1}}
\newcommand{\four}{\hat{4}}

\newcommand{\six}{\hat{6}}
\newcommand{\num} {\mathfrak{n}}
\newtheorem{lemma}{Lemma}[section]

\newcommand{\eq}[1]{\begin{equation}#1\end{equation}}
\newcommand{\eqs}[1]{\begin{equation}\begin{split}#1\end{split}\end{equation}}

\newcommand\Tstrut{\rule{0pt}{2.6ex}}         % = `top' strut
\subheader{\hfill \begin{tabular}{r} \texttt{QMUL-PH-20-15} \end{tabular}}
 
\title{\Large Propagators, BCFW Recursion \\ and New Scattering Equations at One Loop}
\author[a]{Joseph A. Farrow,}
\author[b]{Yvonne Geyer,}
\author[a]{Arthur E. Lipstein,}
\author[c]{Ricardo Monteiro}
\author[c]{\& Ricardo Stark-Much{\~a}o}

\affiliation[a]{Department of Mathematical Sciences, Durham University, Durham, DH1 3LE, United Kingdom}
\affiliation[b]{Department of Physics, Faculty of Science, Chulalongkorn University\\
Thanon Phayathai, Pathumwan, Bangkok 10330, Thailand}
\affiliation[c]{Centre for Research in String Theory, School of Physics and Astronomy \\
        Queen Mary University of London, E1 4NS, United Kingdom}
        
\emailAdd{joseph.a.farrow@durham.ac.uk}
\emailAdd{yjgeyer@gmail.com}
\emailAdd{arthur.lipstein@durham.ac.uk}
\emailAdd{ricardo.monteiro@qmul.ac.uk}
\emailAdd{r.j.stark-muchao@qmul.ac.uk}

\abstract{
We investigate how loop-level propagators arise from tree level via a forward-limit procedure in two modern approaches to scattering amplitudes, namely the BCFW recursion relations and the scattering equations formalism. In the first part of the paper, we revisit the BCFW construction of one-loop integrands in momentum space, using a convenient parametrisation of the $D$-dimensional loop momentum. We work out explicit examples with and without supersymmetry, and discuss the non-planar case in both gauge theory and gravity. In the second part of the paper, we study an alternative approach to one-loop integrands, where these are written as worldsheet formulas based on new one-loop scattering equations. These equations, which are inspired by BCFW, lead to standard Feynman-type propagators, instead of the `linear'-type loop-level propagators that first arose from the formalism of ambitwistor strings. We exploit the analogies between the two approaches, and present a proof of an all-multiplicity worldsheet formula using the BCFW recursion.
}

\begin{document}

\maketitle

%\listoftodos

%%%%%%%%%%%%%%%%%%%%%%%%%%%%%%
%%%%%%%%%%%%%%%%%%%%%%%%%%%%%%

\section{Introduction}\label{sec:intro}

Two important modern advances in understanding perturbative scattering amplitudes in quantum field theory have been the BCFW recursion relations and the formalism of the scattering equations. These were first developed at tree level in \cite{Britto:2005fq,Britto:2004ap,Bedford:2005yy,Cachazo:2005ca} and \cite{Cachazo:2013gna,Cachazo:2013hca,Cachazo:2013iea,Roiban:2004yf,Witten:2003nn,Mason:2013sva}, respectively. Both formalisms exploit `on-shell' constructions in some way, providing alternatives to the traditional Feynman diagram expansion, which involves virtual particles. Both have been fruitful for studying theories of massless particles, most importantly gauge theory and gravity, and have brought numerous insights. They have also been extended to loop level, in various approaches; see, e.g., \cite{Bern:2005hs,Brandhuber:2005kd,ArkaniHamed:2008gz,ArkaniHamed:2009dn,ArkaniHamed:2010kv,Boels:2010nw,Boels:2011mn,Arkani-Hamed:2016byb,Bianchi:2018peu,Edison:2019ovj} for the BCFW recursion relations and \cite{Adamo:2013tsa,Geyer:2015bja,He:2015yua,Geyer:2015jch,Cachazo:2015aol,Adamo:2015hoa,Feng:2016nrf,Geyer:2016wjx,Geyer:2018xwu,Geyer:2019hnn,Feng:2019xiq,Wen:2020qrj,Edison:2020uzf} for the scattering equations.

The connection between these two formalisms has been studied at tree level \cite{Spradlin:2009qr,Dolan:2009wf,Nandan:2009cc,ArkaniHamed:2009dg,Bullimore:2009cb,Farrow:2017eol}. In fact, the BCFW recursion has been used to prove tree-level worldsheet formulas based on the scattering equations, e.g., as in \cite{Cachazo:2012pz, Dolan:2013isa, Albonico:2020mge}. More generally, the connection has inspired novel geometric interpretations of scattering amplitudes, in terms of objects such as the amplituhedron \cite{Arkani-Hamed:2013jha} and the associahedron \cite{Arkani-Hamed:2017mur}. At loop level, however, this connection is yet to be fully explored, because the best known worldsheet formulas for loop integrands lead to an unorthodox representation, with a different factorisation structure than the usual BCFW approach \cite{Geyer:2015bja,Baadsgaard:2015twa}.

In this paper, we will take a close look at how local expressions for one-loop integrands of scattering amplitudes  arise in each of the formalisms. In both cases, loop-level propagators appear from tree level via a forward-limit procedure, reminiscent of Feynman's tree theorem \cite{Feynman:1963ax,Feynman:1972mt,Feynman:2000fh,CaronHuot:2010zt}, but the details of this procedure differ significantly between the two formalisms. Starting from this initial motivation, we will: {\it (i)} present a viewpoint on one-loop BCFW recursion that is particularly suited to $D$-dimensional momentum space, describing the cancellation of spurious poles in simple examples with and without supersymmetry, and also in non-planar theories; and {\it (ii)} present a new set of one-loop scattering equations that yields standard Feynman-type propagators, as opposed to more unorthodox representations of the loop integrand considered previously. This new worldsheet approach is inspired by the BCFW story, but does not exhibit spurious poles in the same way. Just as this connection  proved fruitful at tree level in the past, we hope that our work will foster new ideas for geometric formulations at loop level.

We will now give a taster of the two approaches, using as a toy example the construction of the following expression,
\begin{equation}
\label{eq:ngon}
\frac1{\ell^2}\cdot \frac1{(\ell+k_1)^2(\ell+k_1+k_2)^2\cdots(\ell-k_n)^2}\,.
\end{equation}
This represents the propagator structure of a single $n$-gon scalar diagram. In this paper, we take all external momenta $k_i$ to be null, and to satisfy momentum conservation in the convention that all external particles are incoming. The factor $1/\ell^2$ plays a passive role in both formalisms at one loop, so we detached it for clarity.

\phantom{.} 

\noindent {\bf BCFW recursion} \;\; Let us start with the approach of the BCFW recursion relations. A crucial part of an $n$-point one-loop integrand in this approach is the forward limit of an $(n+2)$-point tree-level amplitude, where we have the external momenta $k_i$ and the back-to-back null momenta $\pm\ell_0$, which we will soon relate to the loop momentum $\ell$. For our $n$-gon example, the analogue of such a forward limit is
\begin{equation}
\label{eq:ngon0}
\frac{1}{\ell^2} \cdot \frac1{(\ell_0+k_1)^2(\ell_0+k_1+k_2)^2\cdots(\ell_0-k_n)^2} \,.
\end{equation}
The BCFW recursion relations are derived from a residue argument, whose starting point is a deformation of the kinematics by a `BCFW shift'. Consider the following shift
\begin{equation}
k_1^\mu \to k_1^\mu+\alpha\, q^\mu\,,\qquad k_n^\mu \to k_n^\mu-\alpha\, q^\mu\,,
\qquad \text{with} \quad k_1\cdot q=k_n\cdot q=q^2=0\,,
\end{equation}
which preserves the null conditions and the conservation of momentum of the deformed external kinematics.
Then, the desired expression \eqref{eq:ngon} is recovered if the relation between $\ell_0$ and the loop momentum $\ell$ is
\begin{equation}
\ell^\mu=\ell_0^\mu + \alpha\, q^\mu\,.
\end{equation}

It is only in special cases, however, that this is the full story. Generically, together with terms of the $n$-gon type \eqref{eq:ngon}, other contributions appear, such as
\begin{equation}
\frac{1}{\ell^2} \cdot  \frac1{(\ell+k_1)^2(\ell+k_1+k_2)^2\cdots(\ell-k_{n-1}-k_n)^2}\cdot \frac1{(k_{n-1}+k_n)^2} \,,
\end{equation}
which corresponds to an $(n-1)$-gon with a massive corner associated to particles $n-1$ and $n$. The analogue of the forward limit is, in this case,
\begin{equation}
\frac{1}{\ell^2} \cdot \frac1{(\ell_0+k_1)^2(\ell_0+k_1+k_2)^2\cdots(\ell_0-k_{n-1}-k_n)^2}\cdot \frac1{(k_{n-1}+k_n)^2} \,.
\end{equation}
If we apply the shift described above, then the propagators involving $\ell_0$ will work out nicely, but not the massive corner propagator, which, due to the shift of $k_n$, will acquire a spurious pole at $\alpha=\frac{k_{n-1}\cdot k_n}{k_{n-1}\cdot q}$\,.
Indeed, the loop-level BCFW recursion relation generically includes tree-level-type factorisation terms, and not only a forward-limit contribution. The combination of these is free of spurious poles and yields the correct loop integrand.

In this paper, we will see in detail how these cancellations work out if we think of the BCFW shift as involving also the loop momentum:
\begin{equation}
k_1^\mu \to k_1^\mu+z\, q^\mu\,,\qquad k_n^\mu \to k_n^\mu-z\, q^\mu\,,
\qquad 
\ell^\mu=\ell_0^\mu + \alpha\, q^\mu\, \to \, \ell_0^\mu + (\alpha-z)\, q^\mu \,.
\end{equation}
Then, in the BCFW residue argument, the forward-limit term corresponds to the residue at $z=\alpha$, and the factorisation terms correspond to residues such as, in our example, at $z=\frac{k_{n-1}\cdot k_n}{k_{n-1}\cdot q}$\,. 

Our treatment of loop-level recursion is implicitly equivalent to a previous approach developed in \cite{Boels:2010nw,Boels:2011mn,Boels:2016jmi}, but we clarify several important aspects such as the mechanism for spurious pole cancellation in momentum space, which is crucial in order to go beyond the simplest example, namely the four-point amplitude in $\mathcal{N}=4$ super-Yang-Mills (SYM). Our approach also clarifies the relation to the all-loop recursion for planar $\mathcal{N}=4$ SYM in momentum twistor space developed in \cite{ArkaniHamed:2010kv,Arkani-Hamed:2016byb}, putting the one-loop realisation of the latter into a larger context beyond maximal supersymmetry, four dimensions, and the planar limit. Moreover, our integrand recursion can be used to rederive, without the need for boundary terms, a previously known recursion for the integrated all-plus amplitudes in pure YM \cite{Bern:1993qk,Mahlon:1993si,He:2014bga}. %Whereas integrated all-plus amplitudes do not vanish when the BCFW shift is taken to infinity, we do not encounter this issue at the level of the integrand.
Finally, as we will describe below, our formulation of loop-level recursion suggests a natural modification of the scattering equations which gives rise to Feynman propagators, and we use the recursion to prove a new worldsheet formula for MHV amplitudes in planar $\mathcal{N}=4$ SYM. Additional considerations, based on the colour structure and on the double copy \cite{Bern:2008qj,Bern:2010ue}, allow us to establish analogous formulas for the non-planar case and for $\mathcal{N}=8$ supergravity. 

\phantom{.} 

\noindent {\bf Scattering equations} \;\; In the approach of the scattering equations, the loop integrand is written as a worldsheet formula times an overall $1/\ell^2$ factor \cite{Geyer:2015bja}. The worldsheet formula is a residue integral on the moduli space of spheres with $n+2$ marked points $\{\sigma_i,\sigma_{\pm}\}$, of which $n$ are associated to external particles and 2 are associated to loop-momentum insertions $\pm\ell$. As opposed to the BCFW recursion approach, the full loop integrand  -- apart from the $1/\ell^2$ factor -- can be understood as a type of forward limit \cite{He:2015yua,Geyer:2015jch,Cachazo:2015aol,He:2016mzd,He:2017spx,Geyer:2017ela}. There is, however,
a price to pay. The corresponding loop-integrand representation has some drawbacks: there are more terms (e.g., $n$ terms for an $n$-gon) and the propagator structure is non-standard. Apart from the overall $1/\ell^2$ factor, the loop propagators are of the type $1/((\ell+K)^2-\ell^2)$, instead of the usual $1/(\ell+K)^2$. For example, one of the $n$ terms associated to the $n$-gon in \eqref{eq:ngon} is
\begin{equation}
\label{eq:ngonpartial}
\frac1{\ell^2}\cdot \frac1{\big(2\ell\cdot k_1\big)\big(2\ell\cdot (k_1+k_2)+(k_1+k_2)^2\big)\cdots\big(-2\ell\cdot k_n\big)}\,.
\end{equation}
This kind of representation can only be matched to a standard one up to shifts in the loop momentum, of the type $\ell\to\ell+K$, and there is a different shift for each of the $n$ terms in the $n$-gon example. It is straightforward to start with a standard representation and obtain this unorthodox representation, but unfortunately the converse statement is not true.

This begs the question of whether there is an alternative story, where the propagator structure is directly the standard one, i.e., that of \eqref{eq:ngon}. The difficulty lies with the fact that the scattering equations only deal with kinematics where the momenta are either null or effectively null. For instance, the loop momentum may enter in a way where it can be thought of as being null in higher-dimensions, if $\ell^2$ does not appear in the worldsheet formula, only $\ell\cdot k_i$\,; this is the case in \eqref{eq:ngonpartial}, excluding the overall $1/\ell^2$ factor. There is previous work in this direction \cite{Gomez:2017lhy,Gomez:2017cpe,Ahmadiniaz:2018nvr}, based on the idea of considering the moduli space of spheres with $n+2+2$ marked points, where two pairs of null momenta $\pm \ell_1$ and $\pm \ell_2$ can be used to write the (non-null) loop momentum, e.g., $\ell=\ell_1+\ell_2$. To this date, however, this ingenious idea of a `double-forward limit' at one loop has only been successfully applied to certain scalar formulas. As we discuss in appendix~\ref{sec:DFL}, a recent proposal along these lines for Yang-Mills theory \cite{Agerskov:2019ryp} does not give the correct loop integrand. (A revised version of that work is in preparation.)

Another relevant result is the use of on-shell diagrams to derive four-dimensional worldsheet formulas for $\mathcal{N}=4$ SYM and $\mathcal{N}=8$ supergravity at four points, by two of the present authors \cite{Farrow:2017eol}. These formulas, which we will review in an appendix, exhibit standard (quadratic) propagators. However, the formulas cannot be straightforwardly extended to higher multiplicity. The reason is that the four-point on-shell diagrams take the form of a BCFW forward limit. At higher points, there are also tree-level-type factorisation contributions, but there is no obvious candidate for a single worldsheet formula which encodes both types of contributions. Nevertheless, the BCFW construction, with its special treatment of adjacent particles $n$ and $1$, provides the inspiration for our new worldsheet story.

Our proposal is to introduce a conventional $\ell^2$ dependence via a deformation of the previously studied one-loop scattering equations. The latter take the form \,$\mathcal{E}_a =0$\, for $a=1,2,\cdots,n,+,-$\,, where
\begin{equation}
\mathcal{E}_i :=\sum_{j\neq i} \frac{2k_i\cdot k_j}{\sigma_{ij}} + 2\ell\cdot k_i\, \frac{\sigma_{+-}}{\sigma_{i+}\sigma_{i-}}\,,
\qquad 
\mathcal{E}_\pm :=\pm \sum_{j} \frac{2\ell\cdot k_j}{\sigma_{\pm j}} \,.
\end{equation}
Of these $n+2$ equations, only $n-1$ are independent, due to the SL$(2,\mathbb{C})$ symmetry. 
This symmetry, which is at the heart of the formalism, is guaranteed by momentum conservation,
\begin{equation}
\label{eq:momcons}
\sum_{j\neq i} {2k_i\cdot k_j} =0= \sum_j 2\ell\cdot k_j  \,.
\end{equation}
The basic observation is that these conditions  can easily be preserved under deformations. 
We introduce the $\ell^2$ deformation by taking a new set of scattering equations, \,$\mathcal{E}_a^{\Def} =0$\,, where we define
\begin{equation}
\mathcal{E}_a^{\Def} := \mathcal{E}_a\big(\;2\ell\cdot k_1 \mapsto \ell^2+2\ell\cdot k_1\,,\; 2\ell\cdot k_n \mapsto -\ell^2+2\ell\cdot k_n\;\big)  \,.
\end{equation}
This deformation preserves the conditions \eqref{eq:momcons}, and therefore the SL$(2,\mathbb{C})$ symmetry.\footnote{The deformation can actually be thought of as $\ell$, $k_1$, $k_n$ being the $D$-dimensional components of null $(D+2)$-dimensional (complex) momenta $\tilde\ell$, $\tilde k_1$, $\tilde k_n$, such that $2\tilde\ell\cdot\tilde k_1=2\ell\cdot k_1+\ell^2$, etc.} Crucially for us, it takes \eqref{eq:ngonpartial} into \eqref{eq:ngon}, which is our original goal. While this happens in a different way from the BCFW case, there is a close analogy. It is important, though, to consider the impact of this deformation on the moduli-space integrands, and not just on the measure associated to the scattering equations. Moduli-space integrands valid with the `old' one-loop scattering equations are not necessarily valid with the `new' one-loop scattering equations, as we shall discuss.

\phantom{.} 

Our treatments of the one-loop BCFW recursion and scattering equations apply most  naturally to planar loop integrands (indeed, we have made a choice in the examples above to have the loop momentum $\ell$ running between particles $n$ and $1$), but we will discuss the possibilities and limitations of these approaches when applied to non-planar gauge theory and gravity.

\phantom{.} 

This paper is organised as follows. In section \ref{sec:BCFW}, we derive a BCFW recursion relation for planar one-loop integrands of generic  massless quantum field theories, and, in sections \ref{sec:BCFW_MHV} and \ref{allplusexampl}, we demonstrate how it works for MHV amplitudes in $\mathcal{N}=4$ SYM and all-plus amplitudes in pure YM, respectively. In section \ref{sec:previous}, we describe how this relates to previous treatments of one-loop BCFW and, in section \ref{nonplanar}, we describe different ways to extend it to non-planar amplitudes. In section \ref{sec:wsquad}, we propose new one-loop scattering equations, which give rise to loop integrands with standard Feynman propagators, and we prove the resulting worldsheet formula for MHV amplitudes in planar $\mathcal{N}=4$ SYM using BCFW recursion in section \ref{sec:BCFW_WS}. In section~\ref{sec:wsquadnp}, we extend that MHV formula to non-planar $\mathcal{N}=4$ SYM and $\mathcal{N}=8$ supergravity. Finally, we present our conclusions and future directions in section \ref{sec:conclusion}. There are also a number of Appendices. Appendices \ref{app:6ptMHV} and \ref{app:allplus} describe technical details for various examples of one-loop recursion in $\mathcal{N}=4$ SYM and pure YM, respectively.  Appendix \ref{wsonshell} describes how Feynman propagators arise from a worldsheet formula for the one-loop four-point superamplitude in $\mathcal{N}=4$ SYM previously deduced from on-shell diagrams in \cite{Farrow:2017eol}, and explains why it is difficult to extend it to higher points. Appendix \ref{sec:DFL} reviews alternative worldsheet formulas based on taking a double-forward limit and describes why they encounter difficulties. Appendix \ref{app:BCFW} fills in technical details of the proof presented in section \ref{sec:BCFW_WS}.

\section{BCFW recursion at one loop}\label{sec:BCFW}

In this section, we apply the BCFW residue argument to obtain a recursion relation for planar one-loop integrands.  %At the end of the section, we discuss boundary terms in the recursion. 
Our treatment of the one-loop recursion is implicitly equivalent to previous approaches, but provides a new perspective and clarifies details like the cancellation of spurious poles in momentum space, and the relation to the recursion in momentum-twistor space. We will discuss the relation to previous approaches in section~\ref{sec:previous}, and the extension to the non-planar case in section \ref{nonplanar}. Moreover, we will use our recursion to motivate and prove a new worldsheet formula for loop integrands with standard Feynman propagators in sections \ref{sec:wsquad} and \ref{sec:BCFW_WS}.

\subsection{Residue argument: planar case} \label{contourplanar} 
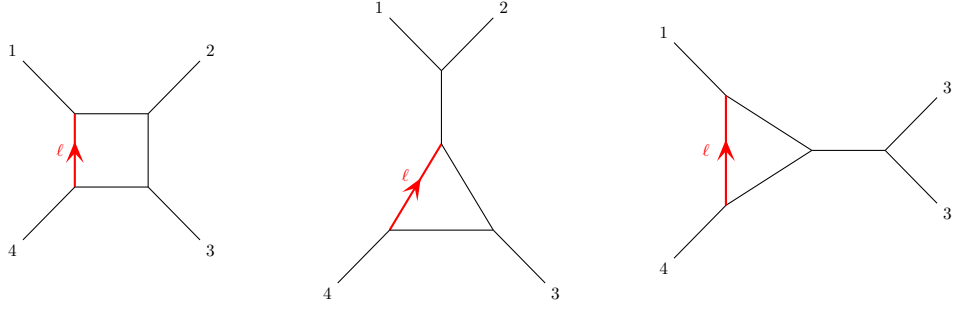
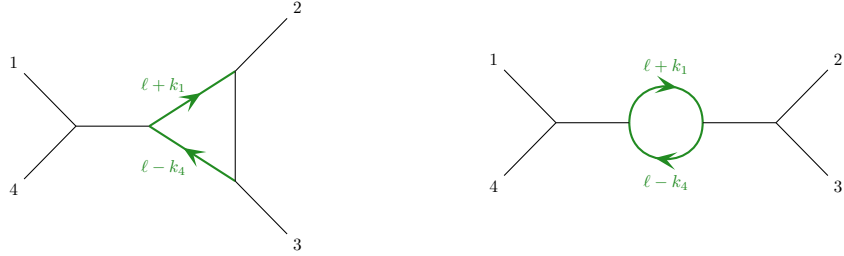
\begin{figure}[ht]
\begin{subfigure}[b]{\textwidth}
\begin{center}
   \begin{tikzpicture}[scale=0.65,transform shape, baseline=(current bounding box.center)]
    \begin{feynman}
      \vertex (a);
      \vertex [below left = of a] (i1) {\(4\)};
      \vertex [      right=of a ] (b);
      \vertex [      above=of b ] (c);
      \vertex [      left =of c ] (d);
      \vertex [      below=of d ] (a);
      \vertex [below right=of b ] (i2) {\(3\)};
      \vertex [above right=of c ] (f2) {\(2\)};
      \vertex [above  left=of d ] (e) {\(1\)};
      \diagram* {
        (d) --  (c) --  (b) -- (a) -- [fermion1, edge label=\(\ell\;\), red, thick] (d),
        (b) --  (i2),
        (a) -- (i1),
        (c) --  (f2),
        (d) --  (e),
      };
    \end{feynman}
  \end{tikzpicture}   \hspace{30pt}
  \begin{tikzpicture}[scale=0.65,transform shape, baseline=(current bounding box.center)]
    \begin{feynman}
      \vertex (center);
      \vertex [above = 20pt of center] (c);
      \vertex [below left=of center ] (b);
      \vertex [below  right = of center] (a);
       \vertex [above =of c ] (d);
      \vertex [below right= of a] (i1) {\(3\)};
      \vertex [below left=of b ] (i2) {\(4\)};
      \vertex [above right=of d ] (f1) {\(2\)};
      \vertex [above  left=of d ] (f2) {\(1\)};
      \diagram* {
        (b) -- [fermion1, edge label=\(\ell\), red, thick] (c) -- (a),
        (a) -- (b),
        (c) --  (d),
        (a) -- (i1),
        (b) -- (i2),
        (f1) -- (d) -- (f2),
      };
    \end{feynman}
  \end{tikzpicture}\hspace{30pt}
  \begin{tikzpicture}[scale=0.65,transform shape, baseline=(current bounding box.center)]
    \begin{feynman}
      \vertex (center);
      \vertex [right = 50pt of center] (c);
      \vertex [above=32pt of center ] (b);
      \vertex [below = 32pt of center] (a);
       \vertex [right =of c ] (d);
      \vertex [below left= of a] (i1) {\(4\)};
      \vertex [above left=of b ] (i2) {\(1\)};
      \vertex [above right=of d ] (f1) {\(3\)};
      \vertex [below right=of d ] (f2) {\(3\)};
      \diagram* {
        (b) --  (c) -- (a),
        (a) -- [fermion1, edge label=\(\ell\), red, thick, inner sep=9pt] (b),
        (c) --  (d),
        (a) -- (i1),
        (b) -- (i2),
        (f1) -- (d) -- (f2),
      };
    \end{feynman}
  \end{tikzpicture}
  \end{center}
  \caption{Diagrams without a massive 1-$n$ corner}
  \end{subfigure}\\

   \begin{subfigure}[b]{\textwidth}
   \begin{center}
     \begin{tikzpicture}[scale=0.65,transform shape, baseline=(current bounding box.center)]
    \begin{feynman}
      \vertex (center);
      \vertex [left = 50pt of center] (c);
      \vertex [below =32 pt of center ] (b);
      \vertex [above  = 32pt  of center] (a);
       \vertex [left =of c ] (d);
      \vertex [above right= of a] (i1) {\(2\)};
      \vertex [below right=of b ] (i2) {\(3\)};
      \vertex [above left=of d ] (f1) {\(1\)};
      \vertex [below left=of d ] (f2) {\(4\)};
      \diagram* {
        (b) -- [fermion1, ForestGreen, thick, edge label=\(\ell-k_4\)] (c) -- [fermion1, ForestGreen, thick, edge label=\(\ell+k_1\)](a),
        (a) -- (b),
        (c) --  (d),
        (a) -- (i1),
        (b) -- (i2),
        (f1) -- (d) -- (f2),
      };
    \end{feynman}
  \end{tikzpicture}
    \hspace{60pt}
   \begin{tikzpicture}[scale=0.65,transform shape,baseline=(current bounding box.center)]
    \begin{feynman}
      \vertex (c);
      \vertex [right = of c] (b);
      \vertex [right  = of b] (a);
       \vertex [left =of c ] (d);
      \vertex [above right= of a] (i1) {\(2\)};
      \vertex [below right=of a] (i2) {\(3\)};
      \vertex [above left=of d ] (f1) {\(1\)};
      \vertex [below left=of d ] (f2) {\(4\)};
      \diagram* {
        (b) -- [fermion1, ForestGreen, thick,  half left, looseness=1.7, edge label=\(\ell-k_4\),  inner sep=9pt] (c) -- [fermion1, ForestGreen, thick, half left,  looseness=1.7, edge label=\(\ell+k_1\),  inner sep=7pt](b) -- (a),
        (c) --  (d),
        (a) -- (i1),
        (a) -- (i2),
        (f1) -- (d) -- (f2),
      };
    \end{feynman}
  \end{tikzpicture}
    \end{center}
      \caption{Diagrams with a massive 1-$n$ corner}
    \end{subfigure}
  \caption{The above four-particle diagrams illustrate the placement of the loop momentum $\ell$ between $k_1$ and $k_{n=4}$ for planar diagrams with ordering $(1234)$. Note that when $1$ and $n$ occur in a  `massive corner', there is no propagator of the form $1/\ell^{2}$.}
  \label{fig:placement_loop4pt}
  \end{figure}

Let us start from a standard Feynman-diagram-like representation of the loop amplitude. Decomposing the loop momentum into null vectors $\ell_0$ and $q$,
\begin{equation}
\ell=\ell_0+\alpha q\,,
\end{equation}
where $q$ will be associated to the BCFW shift, we have
\begin{equation}
 \cA^{(1)}_n=\int d^\sD\ell\,\;\mathfrak{I}_n^{(1)}(\ell)=\int d^{\sD}\ell_{0}\,\delta(\ell_{0}^{2})\; d\alpha\;2\ell_0\cdot q\,\;\mathfrak{I}_n^{(1)}(\ell_0+\alpha q)\,.
\end{equation}
Since we are considering the planar case in this section, we can choose to represent $\mathfrak{I}_n^{(1)}$ such that $\ell$ lies between $k_1$ and $k_n$ for all the diagrams;\footnote{In our convention, the external momenta are incoming, and the loop momentum runs from $n$ to 1, such that all loop propagators are of the form \,$1/(\ell+\sum_{i=1}^j k_i)^2$\,.} see Figure \ref{fig:placement_loop4pt}.

Consider the following BCFW-type shift of the loop integrand:
\begin{subequations}\label{eq:shift}
 \begin{align}
  &\hat k_1=k_1+zq\,, \\
  &\hat k_n=k_n-zq\,, \\
  &\hat \alpha = \alpha-z\,,\;\;\;\; \textrm{i.e.,}\;\;\;\; \hat\ell=\ell-zq=\ell_0+(\alpha-z) q \,,
 \end{align}
\end{subequations}
where \,$q^2=q\cdot k_1=q\cdot k_n=0$\,, so that \,$\hat k_1^2=\hat k_n^2=0$\,. The amplitude is as usual the residue at $z=0$,
\begin{equation}
 \cA^{(1)}_n=\int d^\sD\ell
 \,\;\oint_{z=0}\frac{dz}{2\pi i\,z} \;{\mathfrak{I}}_n^{(1)}(z)\,.
\end{equation}
Then, wrapping the contour around, we find poles at
\begin{enumerate}
\item the tree-level-like factorisation channels for which $\hat K_I^2=(\sum_{i\in I}\hat k_i)^2=0$\,, at values\, $z=z_I$\,,
\item the single-cut contribution for which $\hat \ell^2=0$ , at $z=\alpha$\,,
\item potentially at $z=\infty$\,.
\end{enumerate}
The tree-level-like poles occur for
\begin{equation}
 \hat K_I^2 = K_I^2+2z_I\,K_I\cdot q =0\qquad \Leftrightarrow \qquad z_I=-\frac{K_I^2}{2K_I\cdot q}\,,
\end{equation}
with either $1\in I$ and $n\notin I$, or with $1\notin I$ and $n\in I$; since we are in the planar case, the set $I$ is consecutive in the colour ordering. The corresponding residues are
\begin{equation}
 -\oint_{z=z_I}\frac{dz}{2\pi i\,z} \,{\mathfrak{I}}_n^{(1)}(z) =\frac{1}{K_I^2}\sum_{\text{states}_I}\cA^{(0)}_{n_I+1}(z_I)\; {\mathfrak{I}}_{n-n_I+1}^{(1)}(z_I)\,,
\end{equation}
where the colour ordering is implicit.
Notice that, in each contribution of this type, the loop momentum in the sub-amplitude ${\mathfrak{I}}_{n-n_I+1}^{(1)}$ is shifted, i.e., it is $\ell_0+(\alpha-z_I)q$, whereas the loop measure is not shifted.

\begin{figure}[ht]
\begin{center}
  \begin{tikzpicture}[scale=0.65,transform shape][small, baseline=(current bounding box.center)]
    \begin{feynman}
      \vertex (center);
      \vertex [above = 20pt of center] (c);
      \vertex [below left=of center ] (b);
      \vertex [below  right = of center] (a);
       \vertex [above =of c ] (d);
      \vertex [below right= of a] (i1) {\(2\)};
      \vertex [below left=of b ] (i2) {\(3\)};
      \vertex [above right=of d ] (f1) {\(1\)};
      \vertex [above  left=of d ] (f2) {\(4\)};
      \diagram* {
        (b) -- [fermion1, ForestGreen, thick, edge label=\(\ell-k_4\)] (c) -- [fermion1, ForestGreen, thick, edge label=\(\ell+k_1\)](a),
        (a) -- (b),
        (c) --  (d),
        (a) -- (i1),
        (b) -- (i2),
        (f1) -- (d) -- (f2),
      };
    \end{feynman}
  \end{tikzpicture}
  \hspace{30pt}
\begin{tikzpicture}[scale=0.68,transform shape][small, baseline=(current bounding box.center)]
    \begin{feynman}
      \vertex (a);
      \vertex [below left = of a] (i1) {\(4\)};
      \vertex [      right=of a ] (b);
      \vertex [      above=of b ] (c);
      \vertex [      left =of c ] (d);
      \vertex [      below=of d ] (a);
      \vertex [below right=of b ] (i2) {\(3\)};
      \vertex [above right=of c ] (f2) {\(2\)};
      \vertex [above  left=of d ] (e);
      \vertex [left=of e ] (f1) {\(5\)};
      \vertex [above=of e ] (f3) {\(1\)};
      \diagram* {
        (d) -- [fermion1, ForestGreen, thick, edge label=\(\ell+k_1\)] (c) --  (b) -- (a) -- [fermion1, ForestGreen, thick, edge label=\(\ell-k_5\;\)] (d),
        (b) --  (i2),
        (a) -- (i1),
        (c) --  (f2),
        (d) --  (e),
        (e) -- (f1),
        (e) -- (f3),
      };
    \end{feynman}
  \end{tikzpicture}
  \caption{Examples of diagrams for four and five particles that do not contribute to the single cut. This is due to the BCFW shift \eqref{eq:shift}, which leaves all loop propagators in these diagrams invariant, e.g. $\hat \ell+\hat k_1=\ell+k_1$.}
  \label{fig:massive-1-n-corner}
  \end{center}
  \end{figure}
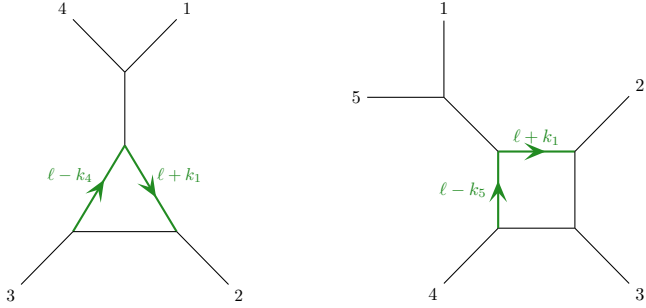

The single-cut term occurs only for $\hat \ell^2=0$ in the planar case, because the shift in $\hat k_1$ cancels the shift in $\hat \ell$ for any propagator other than $1/\hat\ell^2$ due to our chosen placement of the loop momentum (between $n$ and 1). We can write the terms in $\mathfrak{I}_n^{(1)}(z)$ which have a simple pole at $\hat \ell^2=0$ as
\begin{equation}
 \frac{1}{\hat\ell^2}\;{\mathscr{I}}_n^{(1)}(z)
 = \frac{\alpha}{\alpha-z}\, \frac{1}{\ell^2}\;{\mathscr{I}}_n^{(1)}(z)\,.
\end{equation}
Then
\begin{equation}
 -\oint_{z=\alpha}\frac{dz}{2\pi i\,z} \;\mathfrak{I}_n^{(1)}(z) =
 \oint_{z=\alpha}\frac{dz}{2\pi i\,z} \; \frac{\alpha}{z-\alpha}\, \frac{1}{\ell^2}\;{\mathscr{I}}_n^{(1)}(z) = \frac{1}{\ell^2}\;{\mathscr{I}}_n^{(1)}(\alpha)\,.
\end{equation}
Notice that we have $z=\alpha$, and therefore $\hat\ell=\ell_0$ is on-shell. Therefore, ${\mathscr{I}}_n^{(1)}(\alpha)$ is a single cut, and can be interpreted as a forward limit,
\begin{equation}
{\mathscr{I}}_n^{(1)}(\alpha)=\sum_{\text{states}_{0}}\cA^{(0)}_{n+2}\left(\ell_0,\,k_1+\alpha q,\,k_2,\dots, \,k_{n-1}, k_n-\alpha q,\,-\ell_0\right)
=:\sum_{\text{states}_{0}}\cA^{(0)}_{n+2}(\alpha)\,,
\label{eq:FLI}
\end{equation}
which includes a sum over the states with momentum $\ell_0$ running in the cut. Generically, the forward limit of an $(n+2)$-point tree-level amplitude is divergent, and it requires regularisation. In the examples that we will consider, there is no such divergence. It would be interesting, however, to understand this regularisation in explicit examples, which to our knowledge has not been achieved in the literature.\footnote{One possible procedure is that, in a Feynman-diagram-like representation of the $(n+2)$-point tree-level amplitude, we drop the terms that correspond to tadpoles and external bubbles in the forward limit. Such a procedure is in principle not gauge invariant, but there is no reason to expect gauge invariance from a generic loop integrand, before loop integration.}
Finally, notice that not all diagrams with loop propagators contribute to the single cut; see Figure \ref{fig:massive-1-n-corner}.

There may, of course, be a residue at $z=\infty$: a boundary term
\begin{equation}
\label{eq:boundaryterm}
 -\oint_{z=\infty}\frac{dz}{2\pi i\,z} \;\mathfrak{I}_n^{(1)}(z) =
{{\mathcal B}_{n}}\,.
\end{equation}
We will discuss below that this term is absent in Yang-Mills theory for an appropriate choice of the shift vector $q$.

Finally, we can write down the recursion relation for the loop integrand as
\begin{align}
\boxed{
\;\;\mathfrak{I}_n^{(1)} \;\;=\; \sum_{I,\text{\,states}_I}\cA^{(0)}_{n_I+1}(z_I)\;\frac{1}{K_I^2}\; {\mathfrak{I}}_{n-n_I+1}^{(1)}(z_I) \;\;+\;\;
\frac{1}{\ell^2}\sum_{\text{states}_{0}}\cA^{(0)}_{n+2}(\alpha)
\;\;+\;\; {\mathcal B}_n \,.\;}
\label{eq:planarrecursion}
\end{align}

To conclude, we may ask what have we gained from our choice of BCFW shift \eqref{eq:shift}, as opposed to shifting only the external particles as in previous work. The shift in $\alpha$ makes the structure of the recursion more transparent, in our view, and the connection to the on-shell diagrams formalism becomes more manifest, as we will see later.

\subsection{Boundary terms} \label{boundary} 

In this section, we briefly recall the arguments of \cite{ArkaniHamed:2008yf} and \cite{Boels:2010nw} for the absence of boundary terms in Yang-Mills theory at tree level and at loop level, respectively. We also mention some one-loop explicit checks of our own.

\paragraph{Tree-level amplitude.} 
As argued in \cite{ArkaniHamed:2008yf}, amplitudes in the large-$z$ limit have a physical interpretation as a hard particle scattering off a soft background. Using the background field method, we can thus analyse them by expanding the Yang-Mills Lagrangian $\mathcal{L}[\mathscr{A}=A+a]$ to second order in the `hard' field $a_\mu$, where we perturb the gauge field  $\mathscr{A}_\mu =A_\mu+a_\mu$ around a soft background $A_\mu$. The soft and hard fields have individual gauge symmetries. Using Lorentz gauge for the hard field $a$, the quadratic gauge-fixed Lagrangian for this field can be written as
\begin{equation}\label{eq:eff-action}
 \mathcal{L}=-\frac{1}{4}D_\nu a_\mu\,D^\nu a^\mu+\frac{i}{2}\mathrm{tr}\left([a_\mu,a_\nu]F^{\mu\nu}\right)\,,
\end{equation}
where the covariant derivative $D=D[A]$ and the field strength $F=F[A]$ are defined with respect to the background field $A$.\footnote{As usual, terms linear in the hard field $a$ vanish by the equations of motion for the background field $A$.} To manifest the large-$z$-limit, it is convenient to work with the soft fields in $q$-lightcone gauge, also known in the context of BCFW as Arkani-Hamed-Kaplan (AHK) gauge, $ q_\mu A^\mu=0\,,$
whereas the hard field $a$ is in Lorentz gauge as mentioned previously.
From the Feynman rules and from symmetry arguments, it can be shown from power-counting in $z$ that the scattering amplitude takes the following form,
\begin{equation}\label{eq:ampl_z>>1}
 \mathcal{A}(z) = \hat \epsilon_1^\mu\hat\epsilon_n^\nu\,\mathcal{A}_{\mu\nu}=\hat \epsilon_1^\mu\hat\epsilon_n^\nu\,\left(z\,\eta_{\mu\nu}f_1\left(z^{-1}\right)+f_{2,[\mu\nu]}\left(z^{-1}\right)+O\big(z^{-1}\big)\right)\,.
\end{equation}
Here, $\hat\epsilon_{1,n}$ denote the shifted polarisation vectors of  particles 1 and $n$, 
the $f_i$ are polynomials in $z^{-1}$ with non-trivial constant term, and $f_{2,[\mu\nu]}$ is antisymmetric in its indices. At tree level, $f_1$ and $f_2$ can be given explicitly in terms of the one particle off-shell current $J_\mu$ of the soft fields \cite{Boels:2010nw}, but we can show that the boundary terms vanish (for a `good' choice of $q$) using only the general form of $\mathcal{A}_{\mu\nu}(z)$ in \eqref{eq:ampl_z>>1} and gauge invariance of the amplitude, such that
\begin{equation}\label{eq:gauge-inv_z>>1}
  \hat k_1^\mu\hat\epsilon_n^\nu\,\mathcal{A}_{\mu\nu}= \left(k_1 + zq\right)^\mu \hat\epsilon_n^\nu\,\mathcal{A}_{\mu\nu}=0\,.
\end{equation}
To see this, let us consider the form of the shifted polarisation vectors  $\hat\epsilon_{1,n}$. Firstly, we define a notion of helicity in $D$-dimensional spacetime for particles 1 and $n$ by taking into account the four-dimensional subspace Span$(k_1,k_n,q,\bar q)$.\footnote{Recall that $k_1\cdot q=k_n\cdot q=q^2=0$. In this context, $q$ cannot be real, so $\bar q\neq q$, and also  $k_1\cdot \bar q=k_n\cdot \bar q=\bar q^2=0$.}
It is convenient to define the bases of (unshifted) $D-2$ polarisation vectors for these particles as $\{\epsilon_i^+,\epsilon_i^-,\epsilon_i^{T_a}\}$ with
\begin{equation}
 \epsilon_1^-=\epsilon_n^+=q\,,\qquad \epsilon_1^+=\epsilon_n^-=\bar q\,, \qquad \epsilon_1^{T_a}=\epsilon_n^{T_a}=:\epsilon^{T_a}\,,
\end{equation}
where the transverse polarisations $\epsilon^{T_a}$, with $\,a=1,\dots, D-4\,$, are orthogonal to the four-dimensional subspace Span$(k_1,k_n,q,\bar q)$.
Under a BCFW shift, it is then easily verified that the shifted bases are given by 
\begin{equation}
 \hat\epsilon_1^-=\hat \epsilon_n^+=q\,,\qquad \hat\epsilon_1^+=\bar q-z\,\frac{q\cdot \bar q}{k_1\cdot k_n}k_n\,,\qquad \hat\epsilon_n^-=\bar q+z\,\frac{q\cdot \bar q}{k_1\cdot k_n}k_1\,, \qquad \hat\epsilon^{T_a}=\epsilon^{T_a}\,.
\end{equation}
We can now see the power of the gauge-invariance condition \eqref{eq:gauge-inv_z>>1}: if we choose the shift vector $q$ to align with the polarisation of particle 1 such that $q=\epsilon_1$ (i.e., particle 1 has `minus' helicity, $\epsilon_1^-= \hat\epsilon_1^-=q$), then  \eqref{eq:gauge-inv_z>>1} tells us that
\begin{equation}
  \mathcal{A}(z)=\hat \epsilon_1^\mu\hat\epsilon_n^\nu\,\mathcal{A}_{\mu\nu}= q^\mu\hat\epsilon_n^\nu\,\mathcal{A}_{\mu\nu}=-\frac{1}{z}k_1^\mu\hat\epsilon_n^\nu\,\mathcal{A}_{\mu\nu}\,.
  \label{eq:eq:ampl_zqpol1}
\end{equation}
Let us define
\begin{equation}
 \mathcal{A}^{-+}:=\hat \epsilon_{1,-}^\mu\hat\epsilon_{n,+}^\nu\,\mathcal{A}_{\mu\nu}\,,\qquad \mathcal{A}^{--}:=\hat \epsilon_{1,-}^\mu\hat\epsilon_{n,-}^\nu\,\mathcal{A}_{\mu\nu}\,,\qquad \mathcal{A}^{-T_a}:=\hat \epsilon_{1,-}^\mu\hat\epsilon_{n,T_a}^\nu\,\mathcal{A}_{\mu\nu}\,.
\end{equation}
Starting from $\mathcal{A}_{\mu\nu}$ as given in \eqref{eq:ampl_z>>1}, and choosing $q=\epsilon_1$, which leads to \eqref{eq:eq:ampl_zqpol1}, one can easily show that
\begin{equation}
 \mathcal{A}^{-+}=O\big(z^{-1}\big)\,,\qquad \mathcal{A}^{--}=O\big(z^{-1}\big)\,,\qquad \mathcal{A}^{-T_a}=O\big(z^{-1}\big)\,,
\end{equation}
by using $k_1\cdot q=k_1\cdot \bar q=k_1\cdot \epsilon^{T_a}=0$, $k_1^2=0$, and the anti-symmetry of $f_2$.

This means that if we choose $q=\epsilon_1$,  the boundary term $\mathcal{B}_n=\lim_{z\rightarrow\infty}\mathcal{A}(z)=0$ vanishes  for all amplitudes $\mathcal{A}(z)$ -- independently of the polarisation of particle $n$. Other choices for $q$ lead to worse-behaved amplitudes, summarised in table~\ref{table:large-z}.

\begin{table}[ht]
\begin{center}
 \begin{tabular}{c|ccc}
  \backslashbox{$\epsilon_1$}{$\epsilon_n$}& $-$ & $+$ & T\\\hline\Tstrut
  $-$ & $z^{-1}$ & $z^{-1}$& $z^{-1}$\\
  $+$ & $z^{3}$ & $z^{-1}$ & $z$\\
  T1 & $z$ & $z^{-1}$ & $z$\\
   T2 & $z$ & $z^{-1}$ & $z^0$\\
 \end{tabular}
\caption{Large-$z$ behaviour of the amplitude for different polarisations, with $q$ arbitrary. Clearly, the boundary terms vanish if we align $q=\epsilon_1^-$, or $q=\epsilon_n^+$. The cases T1 and T2 are distinguished by $\epsilon_1^T\cdot \epsilon_n^T\neq 0$ (T1), versus  $\epsilon_1^T\cdot \epsilon_n^T= 0$ (T2). Figure taken from \cite{ArkaniHamed:2008yf}.}
\label{table:large-z}
\end{center}
\end{table}

\paragraph{One loop integrand.} 
In \cite{Boels:2010nw}, it was argued that the reasoning leading to table~\ref{table:large-z} at tree level extends to loop integrands at any loop order (not just the planar part),  because they do not rely on any on-shell constraints for the `soft lines'. Using our conventions for the loop momentum and the shift \eqref{eq:shift}, the `hard line' follows the shortest possible path through the diagram, and fixes the $z$-dependence of all propagators.\footnote{Our shift conventions differ from \cite{Boels:2010nw}, where the loop momentum is unshifted, but this does not impact the following discussion. See section~\ref{sec:unshiftedloop} for a more detailed comparison.}
The loop integrand is therefore of the form 
\begin{equation}\label{eq:int_large-z}
\mathfrak{I}(z) = \hat \epsilon_1^\mu\hat\epsilon_n^\nu\,\mathfrak{I}_{\mu\nu}=\hat \epsilon_1^\mu\hat\epsilon_n^\nu\,\left(z\,\eta_{\mu\nu}f_1\left(z^{-1}\right)+f_{2,[\mu\nu]}\left(z^{-1}\right)+O\big(z^{-1}\big)\right)\,.
\end{equation}
The tree-level argument leading to table~\ref{table:large-z} relied on gauge invariance. This is more subtle for the loop integrand, because it is generically not strictly gauge invariant (e.g., non-supersymmetric Yang-Mills), only up to terms that vanish after loop integration in dimensional regularisation. That is,
\begin{equation}\label{eq:gauge-inv_loop_z>>1}
  \hat k_1^\mu\hat\epsilon_n^\nu\,\mathfrak{I}_{\mu\nu}^{\scalebox{0.6}{$(1)$}}= \left(k_1 + zq\right)^\mu \hat\epsilon_n^\nu\,\mathfrak{I}_{\mu\nu}^{\scalebox{0.6}{$(1)$}}=\mathcal{G}(z)\,,
\end{equation}
where $\mathcal{G}(z)$ vanishes after loop integration (i.e., it has no $D$-dimensional unitarity cuts).
With the same tree-level choice of aligning the BCFW vector $q$ with the polarisation of particle 1, we find
\begin{equation}
 \mathfrak{I}^{\scalebox{0.6}{$(1)$}}=\hat \epsilon_1^\mu\hat\epsilon_n^\nu\,\mathfrak{I}_{\mu\nu}^{\scalebox{0.6}{$(1)$}}= q^\mu\hat\epsilon_n^\nu\,\mathfrak{I}_{\mu\nu}^{\scalebox{0.6}{$(1)$}}=-\frac{1}{z}k_1^\mu\hat\epsilon_n^\nu\,\mathfrak{I}_{\mu\nu}^{\scalebox{0.6}{$(1)$}}+ \frac{1}{z}\mathcal{G}(z)\,.
\end{equation}
By the same argument as at tree level, the first term falls off as $O(z^{-1})$, and so does not give a boundary term in the BCFW recursion relation. The second term, on the other hand, could contribute, because an explicit counting gives $\mathcal{G}(z)=O(z^2)$ for some polarisations. Note, however, that  $\mathcal{G}(z)$ has to integrate to zero for any value of $z$, and thus
\begin{equation}
 \int d^D\ell\; \mathrm{Res}_{\infty} \frac{1}{z^2}\mathcal{G}(z)=0\,,
\end{equation}
Boundary terms originating from the lack of gauge invariance of the loop integrand can thus be safely dropped from the recursion relation in Yang-Mills theory. 

We will see that, in many examples of interest such as MHV amplitudes in maximal super Yang-Mills and all-plus amplitudes in pure Yang-Mills, the boundary terms can be explicitly checked to be absent from the loop integrand, and there is no need to identify terms that integrate to zero. We discuss this in detail in  sections \ref{sec:bdy_MHV} and \ref{sec:bdy_all+} for MHV amplitudes in maximal super Yang-Mills  and for all-plus amplitudes in pure Yang-Mills, respectively.

\paragraph{Numerical checks.} We have numerically verified the `good' large-$z$ falloff of the pure Yang-Mills four-particle one-loop integrand in four dimensions by using the explicit representation in \cite{Bern:2013yya}, which satisfies the property of colour-kinematics duality \cite{Bern:2008qj,Bern:2010ue}. In particular, we confirmed that all loop integrands for which
$$
\epsilon_1=\epsilon_1^-=\frac{|1\rangle [\eta| }{ [1 \eta] }\,,\qquad 
\epsilon_4=\epsilon_4^+=\frac{|\eta\rangle [4| }{ \langle4 \eta\rangle }
$$
vanish as $O\big(z^{-1}\big)$ for large $z$, if the BCFW shift vector is $q=\lambda_1 \tilde\lambda_4$, so that $\epsilon_1\cdot q=\epsilon_4\cdot q=0$. In fact, this statement holds for any ordering of the planar amplitudes, even if particles 1 and 4 are not adjacent and next to $\ell$, and therefore applies directly to the colour-dressed amplitude, e.g., via \eqref{eq:DDMYM}. We will also see more explicitly later, in section~\ref{sec:bdy_all+}, that the equal-helicities case (all-plus in that example) also admits a `good' BCFW shift.

In \cite{Boels:2010nw}, it was also suggested that analogous statements hold for gravity. We will study the gravity case in section~\ref{nonplanar}, but take this opportunity to discuss the boundary contributions. Considering the gravity theory obtained from pure Yang-Mills via the `double copy', which corresponds to the universal bosonic sector of supergravity (graviton, dilaton and B-field), we constructed the four-point one-loop integrand numerically using \eqref{eq:DDMgrav}. By performing a BCFW shift of particles 1 and 4, as discussed above for Yang-Mills, we could obtain at best a `bad' behaviour $O\big(z^{0}\big)$ for large $z$. However, this does not exclude the possibility that there is better behaviour up to terms that integrate to zero. Although this is not how BCFW is usually considered, we also checked the application of different BCFW shifts to each term in \eqref{eq:DDMgrav}, and in that case there are choices with `good' behaviour $O\big(z^{-1}\big)$ for the individual terms. We will discuss this alternative perspective on non-planar BCFW in section~\ref{nonplanar}, but we will consider there only maximal supergravity. In that case, it is clear that there is a choice of shift without boundary terms, for the same reason as for maximal super-Yang-Mills, which we discuss in the following section.

%%%%%%%%%%%%%%%%%%%%%%%%%%%%%%
%%%%%%%%%%%%%%%%%%%%%%%%%%%%%%

\section{BCFW examples in maximal super-Yang-Mills: MHV}
\label{sec:BCFW_MHV}

In this section, we will provide some examples of the one-loop BCFW recursion, using the shift \eqref{eq:shift}. We will focus here on the MHV sector of planar $\mathcal{N}=4$ SYM for simplicity. We start by reviewing the spinor-helicity formulation of the recursion, which is purely four-dimensional. Then we will see how the recursion works explicitly up to six points but with a $d$-dimensional loop momentum. We will make use of an explicit Feynman-type representation of the loop integrand that possesses convenient properties.

\subsection{\texorpdfstring{$n$}{n}-point MHV recursion} \label{mhvrecursion}

\begin{figure}
\centering
      \begin{center}
      \begin{tikzpicture}[scale=0.75]
 \draw [fill, light-grayII] (-0.5,0) circle [radius=1];
 \draw [thick] (-0.5,0) circle [radius=1];
 \node at (-0.5,0) {\scalebox{0.9}{$\mathcal{A}_{n,k}^{\scalebox{0.6}{$(L)$}}$}};
 \draw [dotted,thick,domain=115:155] plot ({-0.5+1.5*  cos(\x)}, {1.5 * sin(\x)});
  \draw [dotted,thick,domain=205:245] plot ({-0.5+1.5*  cos(\x)}, {1.5 * sin(\x)});
  %\draw [dotted,thick,domain=295:335] plot ({1.5*  cos(\x)}, {1.5 * sin(\x)});
   \draw [dotted,thick,domain=25:65] plot ({-0.5+1.5*  cos(\x)}, {1.5 * sin(\x)});
  \draw [domain=0.966:1.932] plot ({-0.5+\x},{ tan(15)*\x});
  \draw [domain=-0.966:-1.932] plot ({-0.5+\x},{ tan(15)*\x});
  \draw [domain=0.966:1.932] plot ({-0.5+\x},{ -tan(15)*\x});
  \draw [domain=-0.966:-1.932] plot ({-0.5+\x},{ -tan(15)*\x});
  \draw [domain=0.259:0.518] plot ({-0.5+\x},{ tan(75)*\x});
  \draw [domain=-0.259:-0.518] plot ({-0.5+\x},{ tan(75)*\x});
  \draw [domain=0.259:0.518] plot ({-0.5+\x},{ -tan(75)*\x});
  \draw [domain=-0.259:-0.518] plot ({-0.5+\x},{- tan(75)*\x});
  \draw [domain=0.707:1.414] plot ({-0.5+\x},{ -tan(45)*\x});
 \node at (0.1,-2.2) {\scalebox{0.9}{$1$}};
 \node at (-1.1,-2.2) {\scalebox{0.9}{$2$}};
 \node at (1.1,-1.6) {\scalebox{0.9}{$n$}};
 \node at (2,-0.557) {\scalebox{0.9}{$n-1$}};
  \node at (2,0.55) {\scalebox{0.9}{$n-2$}};
 \node at (3.5,0) {$=$};
 \draw [fill, light-grayII] (6.5,0) circle [radius=1];
 \draw [thick] (6.5,0) circle [radius=1];
 \node at (6.5,0) {\scalebox{0.9}{$\mathcal{A}_{n-1,k}^{\scalebox{0.6}{$(L)$}}$}};
 \draw [dotted,thick,domain=115:155] plot ({6.5+1.5*  cos(\x)}, {1.5 * sin(\x)});
 % \draw [dotted,thick,domain=205:245] plot ({6.5+1.5*  cos(\x)}, {1.5 * sin(\x)}); %OLD
 %  \draw [dotted,thick,domain=25:65] plot ({6.5+1.5*  cos(\x)}, {1.5 * sin(\x)});%OLD
  %\draw [domain=0.966:1.932] plot ({6.5+\x},{ tan(15)*\x}); %OLD
  \draw [domain=-0.966:-1.932] plot ({6.5+\x},{ tan(15)*\x});
  \draw [domain=0.966:1.932] plot ({6.5+\x},{ -tan(15)*\x});
  \draw [domain=-0.966:-1.932] plot ({6.5+\x},{ -tan(15)*\x});
  \draw [domain=0.259:0.518] plot ({6.5+\x},{ tan(75)*\x});
 % \draw [domain=-0.259:-0.518] plot ({6.5+\x},{ tan(75)*\x}); %OLD
  \draw [domain=0.259:0.518] plot ({6.5+\x},{ -tan(75)*\x});
  \draw [domain=-0.259:-0.518] plot ({6.5+\x},{- tan(75)*\x});
     \draw [domain=0.707:1.414] plot ({6.5+\x},{ tan(45)*\x}); %NEW
   \draw [domain=-0.707:-1.414] plot ({6.5+\x},{ tan(45)*\x});%NEW
%
 %\node at (5.9,-2.2) {\scalebox{0.9}{$2$}}; %OLD
% \node at (9,0.55) {\scalebox{0.9}{$n-2$}}; %OLD
 \node at (4.9,-1.6) {\scalebox{0.9}{$2$}}; %NEW
 \node at (8.5,1.6) {\scalebox{0.9}{$n-2$}}; %NEW
 \draw [thick] (7.06,-2.08) circle [radius=.15];
 \draw [thick] (8.59,-0.58) circle [radius=.15];
\draw [fill] (8.59,-2.08) circle [radius=.15];
\draw (8.59,-2.08)  --  (7.21,-2.08);
\draw (8.59,-2.08) --  (8.59,-0.73);
 \draw [domain=-0.04:-0.3] plot ({7.06+\x},{ -2.08+tan(75)*\x});
 \draw [domain=0.15:1.25] plot ({8.59+\x},{ -0.58+tan(15)*\x});
 \draw [domain=0:1] plot ({8.59+\x},{ -2.08-tan(45)*\x});
  \node at  (10.4,-0.25) {\scalebox{0.9}{$n-1$}};
  \node at  (9.8,-3.25) {\scalebox{0.9}{$n$}};
  \node at  (6.7,-3.45) {\scalebox{0.9}{$1$}};
  %\node at (7.25,-1.5) {\scalebox{0.75}{$= \hat1$}};
 \node at (6.6,-1.5) {\scalebox{0.75}{$\widecheck 1$}};
 \node at (8.9,-1.35) {\scalebox{0.75}{$\widecheck n$}};
 \node at (8.05,-0.02) {\scalebox{0.75}{$n\widecheck{-}1$}};
  \node at (12,0) {$+$};
   \draw [fill, light-grayII] (15,0) circle [radius=1];
 \draw [thick] (15,0) circle [radius=1];
 \node at (15,0) {\scalebox{0.9}{$\mathcal{A}_{n+2,k+1}^{\scalebox{0.6}{$(L-1)$}}$}};
 \draw [dotted,thick,domain=115:155] plot ({15+1.5*  cos(\x)}, {1.5 * sin(\x)});
  \draw [dotted,thick,domain=205:245] plot ({15+1.5*  cos(\x)}, {1.5 * sin(\x)});
   \draw [dotted,thick,domain=25:65] plot ({15+1.5*  cos(\x)}, {1.5 * sin(\x)});
  \draw [domain=0.966:1.932] plot ({15+\x},{ tan(15)*\x});
  \draw [domain=-0.966:-1.932] plot ({15+\x},{ tan(15)*\x});
  \draw [domain=0.966:2.515] plot ({15+\x},{ -tan(15)*\x});
  \draw [domain=-0.966:-1.932] plot ({15+\x},{ -tan(15)*\x});
  \draw [domain=0.259:0.518] plot ({15+\x},{ tan(75)*\x});
  \draw [domain=-0.259:-0.518] plot ({15+\x},{ tan(75)*\x});
  \draw [domain=0.259:0.647] plot ({15+\x},{ -tan(75)*\x});
  \draw [domain=-0.259:-0.518] plot ({15+\x},{- tan(75)*\x});
 \draw [thick] (15.7,-2.59) circle [radius=.15];
 \draw [fill] (17.6,-0.72) circle [radius=.15];
  \node at (14.4,-2.2) {\scalebox{0.9}{$2$}};
 \node at (17.5,0.55) {\scalebox{0.9}{$n-1$}};
  \draw [domain=-0.04:-0.3] plot ({15.7+\x},{ -2.59+tan(75)*\x});
 \draw [domain=0.15:1.25] plot ({17.6+\x},{ -0.72+tan(15)*\x});
 \draw [domain=0.1:1.9] plot ({15.7+\x},{ -2.59+tan(44)*\x});
  \node at  (19.15,-0.35) {\scalebox{0.9}{$n$}};
  \node at  (15.35,-4.) {\scalebox{0.9}{$1$}};
  \draw [thick] (15.84,-0.5) to [out=320, in=45] (16.3,-1.3) to [out=225, in=320] (15.56,-0.8);
  \node at (16.9,-1.9) {\scalebox{0.9}{$\alpha$}};
\end{tikzpicture}
\end{center}
    \caption{Recursion for MHV amplitudes using on-shell diagrams, which is equivalent to \eqref{eq:MHVrecursion}. We will say more about on-shell diagrams in section~\ref{momtwistrecursion}. The first term in the recursion is an $L$-loop $(n-1)$-particle MHV amplitude with a soft factor, the second term corresponds to an $(n+2)$-particle $(L-1)$-loop NMHV amplitude in the forward limit. In the figure, the black and white vertices represent 3-particle MHV and $\overline{\text{MHV}}$ amplitudes, and the edges signify integrals over on-shell states.} 
    \label{recursion}
\end{figure}
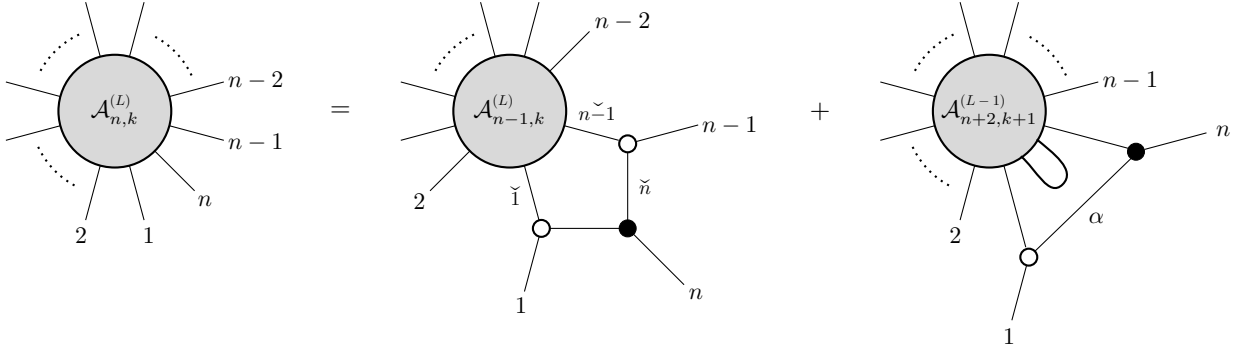

The four-dimensional loop-level BCFW recursion in planar $\mathcal{N}=4$ SYM can be elegantly formulated in terms of on-shell diagrams \cite{ArkaniHamed:2012nw}.\footnote{On-shell diagrams for $\mathcal{N}<4$ SYM were developed in \cite{Benincasa:2015zna}.} In Figure~\ref{recursion}, we illustrate the recursion for MHV amplitudes.
At $n$ points and $L$ loops, there are two contributions: an $L$-loop $(n-1)$-point MHV amplitude with a soft factor, and an $(L-1)$-loop $(n+2)$-point NMHV amplitude with a forward limit.  We can use the spinor-helicity formalism to write the known expression for the superintegrand,\footnote{We use the supersymmetric version of the four-dimensional helicity scheme \cite{Bern:2002zk}, whereby the sum over on-shell loop states is four-dimensional, even though the loop momentum $\ell_0$ is on-shell in $D$ dimensions. In the superamplitude, this sum is represented by the integral over the Grassmann variables; see, e.g., \cite{Elvang:2013cua}.}
\begin{equation}
\label{eq:MHVrecursion}
{\mathfrak{I}}_{\text{MHV}}^{(1)}\!\left(1,2,...,n\right)=\frac{\left\langle n-1,1\right\rangle }{\left\langle n-1,n\right\rangle \left\langle n,1\right\rangle }\;{\mathfrak{I}}_{\text{MHV}}^{(1)}\!\left(\check{1},2,...,\check{n-1}\right)+\frac1{\ell^2}\int d^4\eta_{0}\,\mathcal{A}_{\text{NMHV}}^{(0)}\!\left(\ell_{0},\hat{1},2,...,\hat{n},-\ell_{0}\right) \,,
\end{equation}
where
\begin{equation}
\ell=\ell_{0}+\alpha\, q = \ell_{0}+\alpha\,\lambda_{1}\tilde{\lambda}_{n} \,, \qquad 
\frac{d^{D}\ell}{\ell^{2}}=d^{D}\ell_{0}\delta\left(\ell_{0}^{2}\right)\frac{d\alpha}{\alpha} \,.
\end{equation}
Unlike much of the past literature, we allow the loop momentum to be $D$-dimensional: $\ell_0$ is null in $D$ dimensions. Much of the previous literature is restricted to $D=4$, where
\begin{equation}
\ell_0
=\lambda_{0}\tilde{\lambda}_{0}
\,, 
\qquad\qquad 
\frac{d^{4}\ell}{\ell^{2}}=\frac{d^{2}\lambda_{0}d^{2}\tilde{\lambda}_{0}}{GL(1)}\frac{d\alpha}{\alpha}  \,,
\end{equation}
but we will keep $\ell$ in $D$ dimensions with a view to dimensional regularisation.

Let us focus on the forward-limit term first. As usual, the sum over states is enforced by the fermionic integration, and we can represent the back-to-back particles as
\begin{equation}
k_0=\ell_0\,, \qquad 
k_{n+1}=-\ell_0\,, \qquad 
\eta_0=\eta_{n+1}\,.
\end{equation}
Moreover, we have the super-shift corresponding to $z=\alpha$\,,
\begin{subequations}
\begin{align}
& \hat{k}_1=k_1+\alpha q=\hat{\lambda}_{1}\hat{\tilde{\lambda}}_{1}=\lambda_{1}(  \tilde{\lambda}_{1}+\alpha\,\tilde{\lambda}_{n})\,,  &&  \hat{\eta}_{1}=\eta_{1}+\alpha\,\eta_{n}\,,\\
& \hat{k}_n=k_n-\alpha q=\hat{\lambda}_{n}\hat{\tilde{\lambda}}_{n}=(\lambda_{n}-\alpha\,\lambda_{1})\tilde{\lambda}_{n}\,, &&  \hat{\eta}_{n}=\eta_{n}\,.
\end{align}
\end{subequations}
Loop propagators beyond the overall $1/\ell^2$ in the forward-limit term arise from tree-level propagators as 
\begin{equation}
\frac1{(\ell_0+\hat{k}_1+\cdots)^2}=\frac1{(\ell+k_1+\cdots)^2}\,.
\end{equation}

For the factorisation term, we have
\begin{equation}
z=z_n:= \frac{k_{n-1}\cdot k_{n}}{k_{n-1}\cdot q}
=\frac{\langle n-1,n \rangle }{\langle n-1,1 \rangle}
\,,
\end{equation}
such that \,$k_{n-1}+(k_n-z_n\,q)$\, is on-shell.
The soft factor is simply the result of the tree-level three-point subamplitude and the BCFW propagator, from the point of view of the general recursion \eqref{eq:planarrecursion}.\footnote{This term is often called the `inverse soft limit', because it turns an $(n-1)$-point integrand into (part of) an $n$-point integrand. The associated soft factor is sometimes called `inverse soft factor' for this reason.}  As for the kinematics of the $(n-1)$-point one-loop subamplitude, we will denote the BCFW-shifted quantities $\kappa$ as $\check{\kappa}$ to distinguish them from the forward-limit shift $\hat{\kappa}$, so that we have
\begin{equation}
 \check{k}_{1} :=k_1+z_n\,q= \check{\lambda}_{1} \check{\tilde{\lambda}}_{1} \,,\qquad\qquad 
 \check{k}_{n-1} :={k}_{n-1} + (k_n-z_n\,q) =:\check{\lambda}_{n-1}\check{\tilde{\lambda}}_{n-1} \,,
\end{equation}
with
\begin{subequations} \label{eq:checkdef}
\begin{align}
\check{\lambda}_{1}&={\lambda}_{1} &  \check{\tilde{\lambda}}_{1} &= 
\lambda_{1}+z_n\,{\tilde{\lambda}}_{n}\,, &  \check{\eta}_{1}&=\eta_{1}+z_n\,\eta_{n} \,,\\
\check{\lambda}_{n-1}&=\lambda_{n-1} & \check{\tilde{\lambda}}_{n-1} &= \tilde{\lambda}_{n-1}+\frac{\left\langle n,1\right\rangle }{\left\langle n-1,1\right\rangle }\,\tilde{\lambda}_{n}\,, & \check{\eta}_{n-1}&=\eta_{1}+\frac{\left\langle n,1\right\rangle }{\left\langle n-1,1\right\rangle }\,\eta_{n} \,.
\end{align}
\end{subequations}
The super-shift ensures momentum and super-momentum conservation, i.e.,
\begin{equation}
\sum_{i=1}^{n-1} \lambda_i\check{\tilde\lambda}_i=\sum_{i=1}^n \lambda_i{\tilde\lambda}_i\,,  \qquad
\sum_{i=1}^{n-1} \lambda_i \,\check{\eta}_i=\sum_{i=1}^n \lambda_i \,\eta_i\,.
\end{equation}
The loop momentum in the tree factorisation term is shifted as prescribed in \eqref{eq:shift},
\begin{equation}
\label{eq:checkloopMHV}
\check{\ell}= \ell_0+\left(\alpha - z_n \right) q
=\ell_0+ \left(\alpha -z_n\right) \lambda_1 \tilde{\lambda}_n \,.
\end{equation}
Therefore,
\begin{equation}
\label{eq:measurecheck}
\frac{d^{D}\check{\ell}}{\check{\ell}^{2}}
\;=\;
 \frac{d^{D}{\ell}}{{\ell}^{2}} \,\frac{\alpha}{\alpha-z_n} \,.
\end{equation}
We see that there is a spurious pole at $\alpha=z_n$ in this measure. As we will see in explicit examples, this pole will play the crucial role of cancelling an analogous pole in the forward-limit term.

The decomposition of the loop momentum is re-interpreted at each step in the recursion in order to combine the various terms (factorisation and forward limit) under the same loop-integrand measure. Notice that, in the previous step of the recursion, the integrand of $\mathcal{A}_{n-1,k}^{(1)}$ is constructed with
\begin{equation}
\label{eq:checkloopMHVprime}
\check{\ell} =\ell_0' + \alpha'\, q'
=\ell_0' + \alpha'\, {\lambda}_1 \check{\tilde{\lambda}}_{n-1} \,,
\end{equation}
for some $\ell_0',\,\alpha'$\,. The relation of these to  \eqref{eq:checkloopMHV} can be written explicitly as
\begin{equation}
\label{eq:translateloop}
\ell_0'=\ell_0 + (\alpha-z_n) \left(q-\frac{\ell_0\cdot q}{\ell_0\cdot q'}\, q'\right) \,,
\qquad
\alpha'= \frac{\ell_0\cdot q}{\ell_0\cdot q'} \,(\alpha-z_n)\,.
\end{equation}
If we want to restrict to $D=4$, with $\,\ell_0'=\lambda_0'\tilde{\lambda}_0'$\,, this can be simplified to
\[
\lambda_{0}^{\prime}=\lambda_{0}+(\alpha-z_n)\,\frac{\left[n\,\check{n-1}\right]}{\left[0\,\check{n-1}\right]}\,\lambda_{1}\,,\qquad
\tilde{\lambda}_{0}^{\prime}=\tilde{\lambda}_{0}\,,
\qquad
\alpha^{\prime}=\frac{\left[0\,n\right]}{\left[0\,\check{n-1}\right]}\,(\alpha-z_n).
\]

%%%%%%%%%%%%%%%%%%%%%%%%%
\subsection{Absence of boundary terms for MHV}\label{sec:bdy_MHV}

The MHV recursion relation has no boundary terms. In fact, it is straightforward to verify their absence if we make use of a known form of the $n$-point loop integrand, derived from the field-theory limit of string theory \cite{He:2015wgf}. It is expressed in terms of trivalent diagrams satisfying the colour-kinematics duality \cite{Bern:2008qj,Bern:2010ue}. Here, we will consider the single trace contribution, i.e., the planar case, where we place the loop momentum between particles $n$ and 1 as above.

The algorithm for constructing the MHV loop integrand, with $d$-dimensional loop momentum, was described in \cite{He:2015wgf}. It is sufficient for our purposes to consider the low-point examples. Defining
\begin{align}
\mathfrak{I}^{(1), \text{MHV}}(1,2,\ldots,n;\ell) &\equiv \frac{\delta^8(Q)}{\prod_{i=2}^n \langle 1 i\rangle^2} \ \mathcal{I}_{1,2,\ldots,n} (\ell) \ ,
\label{subamp}
\end{align}
where $\,Q=\sum_i \lambda_i\,\eta_i\,$ is the super-momentum,
we have at four and five points
\begin{align}
\mathcal{I}_{1,2,3,4}(\ell)&=\frac{    X_{2,4} X_{2,3}}{\ell^2 (\ell+k_1)^2 (\ell+k_{12})^2 (\ell+k_{123})^2}
\label{4ptex} \\
\mathcal{I}_{1,2,3,4,5}(\ell)&=\frac{ X_{2,4}X_{2,3}X_{\ell,5}+X_{2,5}X_{2,3}X_{2+3,4}+X_{3,5}X_{\ell,2}X_{2+3,4} }{\ell^2 (\ell+k_1)^2 (\ell+k_{12})^2 (\ell+k_{123})^2 (\ell+k_{1234})^2}+ \frac{ X_{2,3} \ X_{2+3,4} X_{2+3,5}  }{s_{23} \ell^2 (\ell+k_1)^2  (\ell+k_{123})^2 (\ell+k_{1234})^2} \notag \\
 & + \frac{ X_{3,4} \  X_{2,3+4} X_{2,5}  }{s_{34}  \ell^2 (\ell+k_1)^2  (\ell+k_{12})^2 (\ell+k_{1234})^2}   + \frac{ X_{4,5} \ X_{2,3} X_{2,4+5}  }{s_{45}  \ell^2 (\ell+k_1)^2  (\ell+k_{12})^2 (\ell+k_{123})^2} 
  \ .
\label{5ptex}
\end{align}
We use the notation $k_{12\cdots}=k_1+k_2+\cdots$, and
\begin{equation}
X_{A,B} \equiv \langle 1| K_A  K_B | 1 \rangle = -X_{B,A} \ , \qquad
\text{with} \;\; X_{A,B+C}=X_{A,B}+X_{A,C} \,,
\label{defXX}
\end{equation}
where the momenta momenta $K_A$ and $K_B$ are possibly off shell.\footnote{This object is really a rescaled spinor bracket, and satisfies the Shouten identity, \,$X_{A,B}X_{C,D}+X_{B,C}X_{A,D}+X_{C,A}X_{B,D}=0$\,, which underlies the colour-kinematics duality of this representation of the loop integrand.}
For on-shell momenta,
\begin{equation}
X_{i,j} =\langle1i\rangle[ij]\langle j1\rangle \,.
\end{equation}
Particle 1 is chosen to provide a reference spinor $\lambda_1=| 1 \rangle$ in the definition of $X_{A,B}$. Considering the propagator structure, the terms in \eqref{4ptex} and \eqref{5ptex} correspond to trivalent diagrams where particle 1 attaches directly to the loop, i.e., it is never in a massive corner of the loop; this is related to the fact that $\,X_{1,A}=0\,$. These statements hold also at higher points; see, e.g., the six-point case in appendix~\ref{app:6ptMHV}.

The choice of reference spinor $|1\rangle$ is very convenient when we consider the BCFW shift \eqref{eq:shift} such that $q=|1\rangle[ n|$. In \eqref{subamp}, the pre-factor is invariant, because $|1\rangle$ is unshifted, and $|n\rangle$ is shifted along $|1\rangle$. The super-momentum $Q$ is invariant by construction of the super-shift. Conveniently, the objects $X_{A,B}$ are also invariant because $q|1\rangle=0$\,, so only the propagators can be affected by the shift. Since particle 1 is attached directly to the loop, all the terms contain the propagator $1/\ell^2$, whose shift leads to suppression for large $z$, as $\hat{\ell}^2=\ell^2(\alpha-z)/\alpha$\,. The remaining loop propagators are unshifted, since $\hat{\ell}+\hat{k}_1=\ell+k_1$\,. Finally, terms with Mandelstam poles $1/s_{\cdots{n}}$ corresponding to massive corners of the loop that include particle $n$ are also shifted, and are further suppressed. Overall, it is clear that there is suppression at least as $1/z$ for large $z$, so that no contribution arises from \eqref{eq:boundaryterm}. These statements hold for arbitrary multiplicity.

\subsection{MHV, \texorpdfstring{$n=4$}{n=4}}
We will now look at some low-particle examples of the recursion in more detail. At four points, the expression  \eqref{4ptex} matches the well-know answer (up to a sign due to conventions), since some spinor-helicity algebra yields
\begin{equation}
\frac{\delta^8(Q)}{\prod_{i=2}^4 \langle 1 i\rangle^2} \; X_{2,4} X_{2,3}=-\delta^8(Q) \frac{[12][34]}{\langle 12\rangle \langle 34\rangle} \,.
\end{equation}
The same expression is obtained from the recursion relation \eqref{eq:MHVrecursion} because $\,\mathcal{A}_{3}^{(1)}=0$\,, and
\begin{align}
\int d^4\eta_{0}\,\mathcal{A}_{n+2}^{(0),\text{NMHV}}\!\left(\ell_{0},\hat{1},2,3,\hat{4},-\ell_{0}\right)
&= -\delta^8(\hat{Q}) \frac{[\hat{1}2][34]}{\langle 12\rangle \langle 3\hat{4}\rangle} \,
\frac1{(\ell_0+ \hat{k}_1)^2(\ell_0+\hat{k}_1+k_2)^2(\ell_0- \hat{k}_4)^2} \nonumber \\
&= -\delta^8({Q}) \frac{[{1}2][34]}{\langle 12\rangle \langle 3{4}\rangle} \,
\frac1{(\ell+ {k}_1)^2(\ell+{k}_1+k_2)^2(\ell- {k}_4)^2} \,.
\end{align}
The first equality is a well-known result, which can be obtained in a variety of ways -- for instance, directly from on-shell diagrams, as in \cite{Arkani-Hamed:2016byb}, or via a worldsheet formula, as in \cite{Farrow:2017eol}, which is reviewed in appendix~\ref{wsonshell}.

\subsection{MHV, \texorpdfstring{$n=5$}{n=5}} \label{5ptmhv}

Beyond four points, both terms in \eqref{eq:MHVrecursion}, namely the factorisation term and the forward-limit term, contribute to the recursion. Let us consider first the forward-limit term. By itself, the forward-limit term almost reproduces \eqref{5ptex}, with the pre-factor \eqref{subamp}.\footnote{The forward-limit term can be computed in different ways. One way is to import results from worldsheet formulas, by adapting the results in \cite{Geyer:2015bja} or in our section~\ref{sec:BCFW_WS}.}
In the first instance, the only difference is that it reproduces this with the substitutions $1\mapsto\hat{1}$, $5\mapsto\hat{5}$, and $\ell\mapsto\ell_0$. We can notice, however, that the pre-factor and the $X_{A,B}$ quantities are invariant for these substitutions, and the loop propagators involving external momenta satisfy $\,\ell_0+k_{\hat{1}\ldots}=\ell+k_{{1}\ldots}$\,. Therefore, the only difference with respect to \eqref{5ptex} is that instead of the Mandelstam propagator in the last term, \,$1/s_{45}$\,, we have
\begin{equation}
\frac1{s_{4\hat5}} = \frac1{\langle 4 \hat 5 \rangle [54]} = \frac1{s_{45}-\alpha \langle 4 1 \rangle [54]}
= \frac1{s_{45}\left(1-\alpha\frac{\langle 41 \rangle}{\langle  45 \rangle}\right)} \,.
\label{eq:spurious5pt}
\end{equation}
The spurious pole in the forward-limit term is now explicit.

The factorisation term in the recursion relation reads, including the loop integration measure,
\begin{align}
& \frac{\langle 41 \rangle}{\langle 45 \rangle\langle 51 \rangle}
\int {d^{D}\check{\ell}} \;
\; \mathfrak{I}^{(1), \text{MHV}}(\check{1},2,3,\check{4};\check{\ell}) \;%= \nonumber \\
= \; \frac{\langle 41 \rangle}{\langle 45 \rangle\langle 51 \rangle}
\int \frac{d^{D}\check{\ell}}{\check{\ell}^{2}} 
\; \frac{\delta^8( Q)}{\prod_{i=2}^4 \langle 1 i\rangle^2}\; \frac{    X_{2,\check 4} X_{2,3}}{ (\check{\ell}+k_{\check 1})^2 (\check{\ell}+k_{{\check 1}2})^2 (\check{\ell}+k_{{\check1}23})^2} \,.
\label{eq:5pt4ptstep1}
\end{align}
We used the previous step in the recursion, that is \eqref{4ptex}, while noticing that the $\lor$-shift \eqref{eq:checkdef} does not change the four-point pre-factor, because $\check{Q}=Q$ and the $|i\rangle^2$ are $\lor$-unshifted. The $\lor$-shift can affect the objects $X_{A,B}$, and we have
$$
X_{2,\check 4} = \langle 12 \rangle [2\check4] \langle 41 \rangle = \langle 12 \rangle [24] \langle 41 \rangle+\langle 12 \rangle [25] \langle 51 \rangle = X_{2, 4+5} \,.
$$
Now,
$$
\frac{\langle 41 \rangle}{\langle 45 \rangle\langle 51 \rangle}\; \frac1{\prod_{i=2}^4 \langle 1 i\rangle^2} = 
\frac{\langle 41 \rangle \langle 51 \rangle}{\langle 45 \rangle}\; \frac1{\prod_{i=2}^5 \langle 1 i\rangle^2} =
\frac{\langle 14 \rangle [45] \langle 51 \rangle}{s_{45}}\; \frac1{\prod_{i=2}^5 \langle 1 i\rangle^2} =
\frac{X_{4,5}}{s_{45}}\; \frac1{\prod_{i=2}^5 \langle 1 i\rangle^2} ,
$$
so that \eqref{eq:5pt4ptstep1} yields
$$
\int \frac{d^{D}\check{\ell}}{\check{\ell}^{2}} 
\; 
\frac{\delta^8( Q)}{\prod_{i=2}^5 \langle 1 i\rangle^2}\; \frac{ X_{4,5} X_{2,3} X_{2,4+5}}{s_{45}\;(\check{\ell}+k_{\check 1})^2 (\check{\ell}+k_{{\check 1}2})^2 (\check{\ell}+k_{{\check1}23})^2}
=
\int \frac{d^{D}\check{\ell}}{\check{\ell}^{2}} 
\;  \frac{ N^{\text{box}}_{[4,5]123}   }{s_{45}\; (\check{\ell}+k_{\check 1})^2 (\check{\ell}+k_{{\check 1}2})^2 (\check{\ell}+k_{{\check1}23})^2} .
$$
The last step is to deal with the loop momentum. From \eqref{eq:checkloopMHV}, we have
\[
\check{\ell}= \ell_0+\left(\alpha - \frac{\left\langle 45\right\rangle }{\left\langle 41\right\rangle }\right) q\,,
 \qquad\qquad \text{whereas} \qquad\qquad
 {\ell}= \ell_0+\alpha \, q\,.
\]
Hence,
\begin{equation}
\frac{d^{D}\check{\ell}}{\check{\ell}^{2}}
\;=\;
 \frac{d^{D}{\ell}}{{\ell}^{2}} \,\frac{\alpha}{\alpha-\frac{\left\langle 45\right\rangle }{\left\langle 41\right\rangle }} \,,
 \qquad\qquad \text{and} \qquad\qquad  \check{\ell}+k_{{\check1}\dots} = {\ell}+k_{{1}\dots} \,.
\end{equation}
We see that the spurious pole in the measure is the same as in \eqref{eq:spurious5pt}

Combining the problematic term in the forward-limit part and the tree factorisation term, we have
\begin{align}
& \frac1{1-\alpha\frac{\langle 41 \rangle}{\langle  45 \rangle}}\,\frac{ N^{\text{box}}_{[4,5]123}   }{s_{45}\,  \ell^2 (\ell+k_1)^2  (\ell+k_{12})^2 (\ell+k_{123})^2} +\frac{\alpha}{\alpha-\frac{\langle 45 \rangle}{\langle  41 \rangle}}\,
\frac{ N^{\text{box}}_{[4,5]123}   }{s_{45}\,  \ell^2 (\ell+k_1)^2  (\ell+k_{12})^2 (\ell+k_{123})^2} =\nonumber \\
& = \frac{ N^{\text{box}}_{[4,5]123}   }{s_{45}\,  \ell^2 (\ell+k_1)^2  (\ell+k_{12})^2 (\ell+k_{123})^2} \,,
\label{eq:5ptdif}
\end{align}
where the spurious pole disappeared. This concludes the verification that the correct five-point loop integrand is reproduced.

\subsection{MHV, \texorpdfstring{$n=6$}{n=6}}

The higher-point cases are analogous to $n=5$. Let us comment briefly on the six-point case, for which the full loop integrand is written in appendix~\ref{app:6ptMHV}. As at five-points, the full integrand is almost reproduced by the forward-limit contribution. The difference is only in the terms with spurious poles of the type $\,1/s_{\ldots \hat{n}}\,$, which for $\,n=6\,$ are $\,1/s_{5\hat{6}}\,$ and $\,1/s_{45\hat{6}}\,$. Let us consider these terms:
\begin{gather}
\frac{d^D\ell}{\ell^2 }\; \frac{\delta^{(8)}(Q)}{\prod_{i=2}^6 \langle 1i \rangle^2} \left[ \frac{X_{5,6}(X_{2,4}X_{2,3}X_{\ell,5+6} + X_{2,5+6}X_{2,3}X_{2+3,4} + X_{3,5+6}X_{\ell,2}X_{2+3,4})}{s_{5\six}(\ell + k_{{1}})^2(\ell + k_{{1}2})^2 (\ell + k_{{1}23})^2 (\ell + k_{{1}234})^2 (\ell + k_{{1}2345})^2} \right. \nonumber \\[10pt]
\left. 
+ \frac{X_{2,3}X_{5,6}X_{2+3,4}X_{2+3,5+6}}{s_{23}s_{5\six} (\ell + k_{{1}})^2(\ell + k_{{1}23})^2 (\ell + k_{{1}234})^2}
+ \frac{X_{3,4}X_{5,6}X_{2,3+4}X_{2,5+6}}{s_{34}s_{5\six} (\ell + k_{{1}})^2(\ell + k_{{1}2})^2 (\ell + k_{{1}234})^2} \right. \label{eq:6ptterms} \\[10pt] 
\left.
+ \frac{X_{4,5}X_{4+5,6}X_{2,3}X_{2,4+5+6}}{s_{45} s_{45\six} (\ell + k_{{1}})^2(\ell + k_{{1}2})^2 (\ell + k_{{1}23})^2}
+ \frac{X_{6,5}X_{6+5,4}X_{2,3}X_{2,4+5+6}}{s_{5\six} s_{45\six} (\ell + k_{{1}})^2(\ell + k_{{1}2})^2 (\ell + k_{{1}23})^2} \right]. \nonumber
\end{gather}
Here, we have already made use of $\,\hat{Q}=Q\,$, $\,X_{\ell_0,A}=X_{\ell,A}$\,, and \,$\ell_0 + k_{\hat{1}\cdots}=\ell + k_{{1}\cdots}$\,. The five terms correspond to a massive pentagon, two two-mass boxes, and two one-mass boxes, in order of appearance. The only issue is really the presence of $\,1/s_{5\hat{6}}\,$ and $\,1/s_{45\hat{6}}\,$, in the place of $\,1/s_{5{6}}\,$ and $\,1/s_{45{6}}\,$. 

The factorisation contribution provides the correction that cancels the spurious poles. Recalling the $\lor$-shift \eqref{eq:checkdef}, we can write this contribution as
\begin{align}
\frac{d^D\check\ell}{\check\ell^2 }\;&
\frac{\delta^{(8)}(Q)}{\prod_{i=2}^{6} \langle 1i \rangle^2} \;
\frac{X_{5,6}}{s_{56}} 
\left[ \frac{X_{2,4}X_{2,3}X_{\ell, \check{5}} + X_{2,\check{5}}X_{2,3}X_{2+3,4} + X_{3,5}X_{\ell,2}X_{2+3,4}}{(\ell + k_{{1}})^2(\ell + k_{{1} 2})^2(\ell + k_{{1} 23})^2 (\ell + k_{{1} 234})^2} \right.
 \nonumber \\[5pt]
&\left. + \frac{X_{2,3}X_{2+3,4}X_{2+3,\check{5}}}{s_{23} (\ell + k_{{1}})^2 (\ell + k_{{1} 23})^2 (\ell + k_{{1} 234})^2} + \frac{X_{3,4}X_{2,3+4}X_{2,\check{5}} }{s_{34} (\ell + k_{{1}})^2 (\ell + k_{{1} 2})^2 (\ell + k_{{1} 234})^2} \right.
 \\[5pt]
&\left. + \frac{X_{4,\check{5}}X_{2,3}X_{2,4+\check{5}}}{s_{4\check{5}}  (\ell + k_{{1}})^2(\ell + k_{{1} 2})^2 (\ell + k_{{1} 23})^2} \right] \nonumber \,,
\end{align} 
where we have already made use of $\,\check{Q}=Q\,$, $\,X_{\check\ell,A}=X_{\ell,A}$\,, and \,$\check\ell + k_{\check{1}\cdots}=\ell + k_{{1}\cdots}$\,, as well as the relation
\begin{equation*}
\frac{\langle 51 \rangle}{\langle 56 \rangle \langle 61 \rangle}\; \frac{1}{\prod_{i=2}^{5} \langle 1i \rangle^2} = \frac{1}{\prod_{i=2}^{6} \langle 1i \rangle^2}\;\frac{X_{5,6}}{s_{56}}  \,.
\end{equation*}
We can also use
\begin{equation*}
X_{A,\check{5}} = \langle 1 | K_A | \check{5} ] \langle 51 \rangle = \langle 1| K_A |5] \langle 51 \rangle + \langle 1| K_A |6] \langle 61 \rangle = X_{A,5+6}\,,
\end{equation*}
for any momentum $K_A$, leading to
\begin{align}
\frac{d^D\check\ell}{\check\ell^2 }\;&
\frac{\delta^{(8)}(Q)}{\prod_{i=2}^{6} \langle 1i \rangle^2} 
\left[ \frac{X_{5,6}(X_{2,4}X_{2,3}X_{\ell, 5+6} + X_{2,5+6}X_{2,3}X_{2+3,4} + X_{3,5}X_{\ell,2}X_{2+3,4})}{s_{56}(\ell + k_{{1}})^2(\ell + k_{{1} 2})^2(\ell + k_{{1} 23})^2 (\ell + k_{{1} 234})^2} \right.
 \nonumber \\[5pt]
&\left. + \frac{X_{2,3}X_{5,6}X_{2+3,4}X_{2+3,5+6}}{s_{23}s_{56} (\ell + k_{{1}})^2 (\ell + k_{{1} 23})^2 (\ell + k_{{1} 234})^2} + \frac{X_{3,4}X_{5,6}X_{2,3+4}X_{2,5+6} }{s_{34} s_{56}(\ell + k_{{1}})^2 (\ell + k_{{1} 2})^2 (\ell + k_{{1} 234})^2} \right.
\label{eq:6pttermst}
 \\[5pt]
&\left. + \frac{X_{4,\check{5}}X_{2,3}X_{2,4+5+6}X_{5,6}}{s_{4\check{5}} s_{56} (\ell + k_{{1}})^2(\ell + k_{{1} 2})^2 (\ell + k_{{1} 23})^2} \right] \nonumber \,.
\end{align} 
Taking into account that 
\begin{equation}
\frac{d^{D}\check{\ell}}{\check{\ell}^{2}}
\;=\;
 \frac{d^{D}{\ell}}{{\ell}^{2}} \,\frac{\alpha}{\alpha-\frac{\left\langle 56\right\rangle }{\left\langle 51\right\rangle }} \,,
\end{equation}
we can see that the first three terms in \eqref{eq:6pttermst} take the precise form that corrects the first three terms in \eqref{eq:6ptterms}, in a manner that mirrors the $\,n=5$\, case. Although it requires a calculation to check it, the last term in \eqref{eq:6pttermst} does an analogous job for the last two terms in \eqref{eq:6ptterms}; one can readily see they have the same loop propagators.

%%%%%%%%%%%%%%%%%%%%%%%%%%%%
%%%%%%%%%%%%%%%%%%%%%%%%%%%%
\section{BCFW examples in pure Yang-Mills: all-plus} \label{allplusexampl}

\subsection{\texorpdfstring{$n$}{n}-point all-plus recursion} 

In pure Yang-Mills theory, all-plus helicity amplitudes at one loop take a very simple form \cite{Bern:1993qk,Mahlon:1993fe}. Known forms of the loop integrand are also relatively simple, and are in fact related to the maximally-supersymmetric MHV integrands discussed above \cite{Bern:1996ja}. Moreover, the forward limit of tree-level amplitudes can be performed in such a way that it does not require regularisation. All-plus amplitudes provide, therefore, the simplest non-supersymmetric example for the loop-level recursion relation.

We will follow two approaches, a simpler one to be presented shortly, and a more complicated one, which we discuss in appendix~\ref{allplusg}. The simpler approach uses the fact that, from the supersymmetric Ward identities, the one-loop all-plus amplitudes coincide with amplitudes where a minimally-coupled massive scalar runs in the loop \cite{Bern:1996ja}. The more complicated approach is to explicitly consider gluons running in the loop.

As can be anticipated from the `dimension-shifting formula' that relates all-plus amplitudes in pure Yang-Mills and MHV amplitudes in maximal super-Yang-Mills \cite{Bern:1996ja}, the recursion \eqref{eq:MHVrecursion} for the latter has an analogue for the former: 
\begin{equation}
{\mathfrak{I}}^{(1)}\!\left(1^{+},2^{+},...,n^{+}\right)=\frac{\left\langle n-1,1\right\rangle }{\left\langle n-1,n\right\rangle \left\langle n,1\right\rangle }\;{\mathfrak{I}}^{(1)}\!\left(\check{1}^{+},2^{+},...,\check{n-1}^{+}\right)+\frac{2}{\ell^2}
\,\mathcal{A}^{(0)}(\ell_{0},\hat{1}^{+},2^{+},...,\hat{n}^{+},-\ell_{0}) \,.
\label{allplusr}
\end{equation}
The BCFW shifts are the same as for \eqref{eq:MHVrecursion}, and we repeat them here for convenience:
\begin{align} \label{eq:genshifts}
&\hat{k}_1=k_1+\alpha q=\hat{\lambda}_{1}\hat{\tilde{\lambda}}_{1}=\lambda_{1}(  \tilde{\lambda}_{1}+\alpha\,\tilde{\lambda}_{n})\,,\nonumber \\
&\hat{k}_n=k_n-\alpha q=\hat{\lambda}_{n}\hat{\tilde{\lambda}}_{n}=(\lambda_{n}-\alpha\,\lambda_{1})\tilde{\lambda}_{n}\,,
\end{align}
and
\begin{align}
& \check{k}_{1} =k_1+z_n\,q= \check{\lambda}_{1} \check{\tilde{\lambda}}_{1}  = {\lambda}_{1}(\lambda_{1}+z_n\,{\tilde{\lambda}}_{n}) \,,\nonumber \\
& \check{k}_{n-1} ={k}_{n-1} + (k_n-z_n\,q) =\check{\lambda}_{n-1}\check{\tilde{\lambda}}_{n-1} = {\lambda}_{n-1}
\left(\tilde{\lambda}_{n-1}+\frac{\left\langle n,1\right\rangle }{\left\langle n-1,1\right\rangle }\,\tilde{\lambda}_{n}\right) \,,
\end{align}
where
\begin{equation}
z_n= \frac{k_{n-1}\cdot k_{n}}{k_{n-1}\cdot q}
=\frac{\langle n-1,n \rangle }{\langle n-1,1 \rangle}
\,.
\end{equation}
The first term in \eqref{allplusr} is the inverse soft limit.
It corresponds to the factorisation term, where the soft factor is a combination of a three-point amplitude and the BCFW propagator.
The second term is the forward limit of a higher-point tree-level
amplitude, including a sum over the two states of the complex scalar,\footnote{We use the four-dimensional helicity scheme \cite{Bern:1991aq}, whereby the on-shell loop states are four-dimensional, even though the loop momentum is on-shell in $D$ dimensions.} which gives us the factor of 2.

The tree-level amplitude's external states are $n$ gluons and two scalars with back-to-back momenta. These scalars can be thought of as either massless with momentum $\pm \ell_0$ in $D=4-2\epsilon$ dimensions, or as having mass $\mu^2$ and momentum $\ell_0^{(4D)}$ in four dimensions, with
\[
{\ell_0^{(4D)}}^2=\mu^2\,.
\]
An explicit formula for the tree-level amplitudes is \cite{Badger:2005zh,Forde:2005ue}:\footnote{For ease of notation, we use the more recent convention where the momenta inside $[i|\cdots|n]$ are not written as `slashed', as opposed to the original reference \cite{Forde:2005ue}. Moreover, we employ the standard convention under which \,$\ell_0|i]:={\ell_0^{(4D)}}|i]$\,, and \,$[i|K^2|j]:=K^2\,[ij]$\,, so that $\,[i|\ell_0^2|j]=0$\,, whereas $\,[i|\ell_0\ell_0|j]=[i|\ell_0^{(4D)}\ell_0^{(4D)}|j]=\mu^2\,[ij]$\,. It is useful to recall that, inside a bracketed expression, $\,K\,K'+K'\,K\mapsto 2K_{(4D)}\cdot K'_{(4D)}$\,, as in the following example: $\,[1|ij+ji|n]=s_{ij}\,[1n]\,$.} 
\begin{equation}
\mathcal{A}^{(0)}(\ell_{0},{1}^{+},2^{+},...,{n}^{+},-\ell_{0})=\frac{\mu^{2}}{\left\langle 1,2\right\rangle ...\left\langle n-1,n\right\rangle }\frac{\left[1\right|\Pi_{i=2}^{n-1}\left(L_i^2-k_i\, L_{i-1}\right)\left|n\right]}{\Pi_{j=1}^{n-1}L_j^2}
\,,
\label{treescal}
\end{equation}
where we denote
\[
L_{j} = \ell_0+\sum_{i=1}^{j}k_{i}\,.
\]
In the $n=2$ case, we define the tree-level amplitude to vanish. In the $n=3$ case, momentum conservation can be used to show that $\,[1|L_2^2-k_2\, L_1|3]=0$\,. So $n=4$ gives the first non-trivial amplitude.

In terms of the splitting of the loop momentum into a four-dimensional part $\ell_0^{(4D)}+\alpha\,q$, with $q=|1\rangle [n|$, and a $(D-4)$-dimensional part with inner-product $-\mu^2$, we can write the integration measure as
\[
d^D\ell= d^4\ell^{(4D)} d^{-2\epsilon}\mu = d^D\ell_0\,\delta\!\left(\ell_0^2\right) d\alpha\, 2 \ell_0\cdot q =
d^4\ell_0^{(4D)} d^{-2\epsilon}\mu\, \delta\!\left( {\ell_0^{(4D)}}^2\!-\!\mu^2 \right) d\alpha\,2 \ell_0^{(4D)}\cdot q\,  .
\]

%%%%%%%%%%%%%%
\subsection{Absence of boundary terms for all-plus integrand}\label{sec:bdy_all+}

Using Feynman rules in light-cone gauge, the all-plus one-loop integrand can be shown to admit a representation satisfying the BCJ colour-kinematics duality \cite{Boels:2013bi,Monteiro:2011pc}. This means that there are only trivalent diagrams, whose kinematic numerators can be determined from those of $n$-gon diagrams. For the all-plus integrand, the $n$-gon numerators are
\begin{equation}
N(1^+, \, \cdots, n^+) = 2 \,(-1)^n \prod_{i=1}^{n} \frac{1}{\langle \eta i \rangle^2} \;X_{\ell +k_1+ \cdots +k_{i-1}, k_i} \,,
\label{eq:allplusnums}
\end{equation}
where we define
\begin{equation}
X_{A,B} \equiv \langle \eta| K_A  K_B | \eta \rangle = -X_{B,A} \ , \qquad
\text{with} \;\; X_{A,B+C}=X_{A,B}+X_{A,C} \,.
\label{defXXeta}
\end{equation}
This coincides with the previous definition \eqref{defXXeta} for \,$|\eta\rangle=|1\rangle$\,; when dealing with the all-plus case, however, we take a generic reference spinor $|\eta\rangle$, because a gauge choice $|\eta\rangle\to|1\rangle$ is a somewhat messy singular limit that is unnecessary for our purposes. From the $n$-gon numerators \eqref{eq:allplusnums}, we obtain the numerators of other diagrams, $p$-gons with $p<n$, via the Jacobi identity. For example, an $(n-1)$-gon with a massive corner containing (adjacent) momenta $k_i$, $k_j$ has numerator
\begin{equation*}
N(\, \cdots, [i, j],\, \cdots) := N(\, \cdots, i, j,\, \cdots) - N(\, \cdots, j,i ,\, \cdots).
\end{equation*}
Indeed, in the four-point example this is how the numerators for the triangles and the bubbles are defined.

Now, under the BCFW shift \eqref{eq:genshifts},
it is easy to see that $n$-gon numerators \eqref{eq:allplusnums} go as $\mathcal{O}(1)$ for large $z$, for any number of external legs $n$. This is because the leading behaviour $\sim z^2$ from the product of $X$'s is cancelled by the prefactor. On the one hand, because \,$\hat\ell+{\hat{k}}_1 = \ell+k_1\,$\,, only the first and the last $X$ variables get shifted, namely
\begin{align}
& X_{\hat\ell,\hat{k}_1} \sim z\,(-X_{q,k_1}+X_{\ell,q}) = z\,X_{\ell+k_1,q} \,,
\nonumber \\
& X_{\hat\ell+\hat{k}_1+k_2+\cdots+k_{n-1},\hat{k}_n}=X_{\ell-k_n,\hat{k}_n} \sim  -z\,X_{\ell-k_n,q}
 \,. \nonumber
\end{align}
On the other hand, under the shift with $q = |1\rangle[n|$, we have $|\hat n\rangle=| n\rangle-z\, |1\rangle$, and therefore $\langle \eta \hat{n} \rangle^2 \sim z^2\, \langle \eta 1\rangle^2$, and \eqref{eq:allplusnums} scales as $\mathcal{O}(1)$. Finally, the $n$-gon propagator structure is such that all but one propagators are unchanged, because \,$\hat\ell+{\hat{k}}_1 = \ell+k_1\,$\,; the propagator\, $1/\hat{\ell}^2\sim 1/(-2z\,\ell\cdot q)$\, then ensures that this contribution to the loop integrand vanishes for large $z$. 

The numerators of the other diagrams follow via Jacobi relations. The corresponding pieces of the loop integrand can be checked to vanish for large $z$, individually. One may be concerned with diagrams where the shifted external particles appear in massive corners. For example, the $(n-1)$-gon with massive corner involving particles 1 and 2 has the numerator $N([1,2], \, \cdots, n)$, which has behaviour $~z$ for large $z$. This is because now three of the $X$ variables are affected by the shift, instead of two for the $n$-gon. However, there will also be a shifted propagator $1/(\hat{k}_1 + k_2)^2$, which compensates that extra factor of $z$ in the numerator. So overall that diagram's contribution vanishes for large $z$. 

A dangerous looking $(n-1)$-gon is the one with numerator $\,N([n,1],2, \, \cdots,n-1; \ell-{k}_n)$\,, where we explicitly wrote the loop momentum to remind the reader that $\ell$ lies between $n$ and $1$. This diagram is such that no propagator is changed by the shift. However, the $X$ variables are all unchanged, and the shift of the prefactor $1/\prod_i\langle \eta i \rangle^2$ ensures that the contribution from this diagram vanishes for $z\gg1$.

The reasoning above can be applied to any diagram that contributes to the all-plus loop integrand, so there is no boundary term in the BCFW recursion.

\subsection{All-plus, \texorpdfstring{$n=4$}{n=4}} \label{sec:allplusn=4}

Now let us check some examples. As mentioned previously, we will use as building blocks the tree-level amplitudes of $n$ gluons and two back-to-back scalars. In appendix~\ref{allplusg}, we discuss the more standard but involved approach of using tree-level amplitudes of $(n+2)$-gluons.

Since the two- and three-point amplitudes vanish, at four-points we have only the forward-limit contribution to \eqref{allplusr},
\begin{equation}
{\mathfrak{I}}^{(1)}\!\left(1^{+},2^{+},3^{+},4^{+}\right)=\frac{2}{\ell^2}\;\frac{\mu^{2}}{\left\langle 12\right\rangle \left\langle 23\right\rangle \left\langle 3\hat{4}\right\rangle }\frac{\left[\hat{1}\right|\left(\hat{L}_{2}^2-k_{2}\hat{L}_{1}\right)\left(\hat{L}_{3}^2-k_{3}\hat{L}_{2}\right)\left|4\right]}{\hat{L}_{1}^2\,\hat{L}_{2}^2\,\hat{L}_{3}^2}. \label{eq:4ptallplusFL}
\end{equation}
Since this is evaluated for $z=\alpha$\,, we have $\,\hat{L}_i=\ell+\sum_{j=1}^{i}k_j\,$, and the loop propagators are fully reproduced.
Moreover, after some algebra, the numerator can be simplified to
\[
\left[\hat{1}\right|\left(\hat{L}_{2}^2-k_{2}\hat{L}_{1}\right)\left(\hat{L}_{3}^2-k_{3}\hat{L}_{2}\right)\left|4\right]=- \mu^{2}\left[\hat{1}2\right]\left\langle 23\right\rangle \left[34\right]
\,,
\]
and we obtain the known result

\begin{equation}
\mathcal{A}^{(1)}\!\left(1^{+},2^{+},3^{+},4^{+}\right)=
-2\,
\frac{\left[12\right]\left[34\right]}{\left\langle 12\right\rangle \left\langle 34\right\rangle }
\int d^D\ell \;
\frac{\mu^{4}}{\ell^{2}\left(\ell+k_{1}\right)^2\left(\ell+k_{12}\right)^2\left(\ell+k_{123}\right)^2} \,,
\end{equation}
where $\left[\hat{1}2\right]/\left\langle 3\hat{4}\right\rangle =\left[12\right]/\left\langle 34\right\rangle $
by momentum conservation.

\subsection{All-plus, \texorpdfstring{$n=5$}{n=5}}

The five-point case is the first for which the factorisation term in \eqref{allplusr} contributes. Let us start with this term,
\begin{equation*}
\frac{\langle 41\rangle }{\langle 45\rangle  \langle 51\rangle }\;
{\mathfrak{I}}^{(1)}\!\left(\check{1}^{+},2^{+},3^{+},\check{4}^{+}\right)
=
-2\,\frac{\langle 41\rangle }{\langle 45\rangle  \langle 51\rangle }\;
\frac{\left[\check{1}2\right]\left[3\check{4}\right]}{\left\langle 12\right\rangle \left\langle 34\right\rangle }\;
\frac{\alpha}{\alpha-\frac{\langle45\rangle}{\langle41\rangle}}\;
\frac{\mu^{4}}{{\ell}^{2}\left(\ell+k_{1}\right)^2\left({\ell}+k_{12}\right)^2\left(\ell+k_{123}\right)^2}
\,,
\end{equation*}
where we have used the identity $\,\check{\ell}+\check{k}_1=\ell+k_1\,$, with \,$\check{\ell}= \ell- z_5\, q$\,, and made the change to a loop measure $d^D\ell$ through 
\begin{equation}
\frac{d^{D}\check{\ell}}{\check{\ell}^{2}}
\;=\;
 \frac{d^{D}{\ell}}{{\ell}^{2}} \,\frac{\alpha}{\alpha-z_5} \,, \qquad
 z_5=\frac{\langle45\rangle}{\langle41\rangle}\,.
\end{equation}
These steps are familiar from the supersymmetric MHV example above. With the usual Parke-Taylor factor 
\begin{equation*}
\text{PT}_5=\frac{1}{ \langle12\rangle\langle23\rangle\langle34\rangle\langle45\rangle\langle51\rangle}\,,
\end{equation*}
as well as the shifts \eqref{eq:checkdef}, the factorisation term becomes
\begin{equation}
\label{eq:5ptallplusISF}
\frac{\langle 41\rangle }{\langle 45\rangle  \langle 51\rangle }\; {\mathfrak{I}}^{(1)}\!\left(\check{1}^{+},2^{+},3^{+},\check{4}^{+}\right) = 2\,s_{23}\,\text{PT}_5\,\frac{\langle12\rangle[23]\langle34\rangle}{\langle14\rangle} \; \frac{\alpha}{\alpha-\frac{\langle45\rangle}{\langle41\rangle}}\; \frac{\mu^{4}}{D_0 D_1 D_2 D_3 D_4}
\,,
\end{equation}
where, for brevity, we have denoted the loop propagator factors using
\[
D_{i} = \Big(\ell+\sum_{j=1}^{i}k_{j}\Big)^2\,. 
\]

For the forward-limit term, we start from the tree-level amplitude \eqref{treescal}, which reads at five points
\begin{align*}
\mathcal{A}^{(0)}(\ell_0,1^+,2^+,3^+,4^+,5^+,-\ell_0) &= {\mu^2} \frac{\text{PT}_5}{L_1^2\, L_2^2\, L_3^2\, L_4^2} \; [1|(L_2^2-k_2\, L_1)(L_3^2-k_3\, L_2)(L_4^2-k_4\, L_3)|5]\langle 51\rangle \\ 
&= {\mu ^4}\frac{\text{PT}_5}{L_1^2 \, L_2^2 \, L_3^2 \, L_4^2}\left(L_2^2 s_{45} s_{51} + L_4^2 s_{12} s_{23}-[12]\langle 23\rangle [34]\langle 45\rangle \left[5|\ell_0|1\right\rangle \right)
\,.
\end{align*}
The algebraic manipulations used above, mostly employing the Dirac algebra to shuffle the $L_i$ past the external momenta $k_j$, are explained in appendix \ref{sec:slashedvecs}, and allow us to write the forward-limit term in a form which is well-suited to applying the BCFW shifts to legs 5 and 1. Upon applying the shifts, the propagator factors $L_i^2$ turn into $D_i$, and the tree-level amplitude formally becomes the single-cut contribution $\mathscr{I}_5^{(1)}(\alpha)$ in the recursion, as in \eqref{eq:FLI}. Accommodating for the shift in the Parke-Taylor factor and noting that ${s}_{\hat5\hat1} = s_{51}$, we see that the effect of the shift on particle 5 cancels in the first and final terms but not in the second,
\begin{equation*}
\mathscr{I}_5^{(1)}(\alpha) = 2\,
{\mu^4}\frac{\text{PT}_5}{D_1 D_2 D_3 D_4}\left(D_2 s_{45} s_{51}+D_4 s_{23} \langle 21\rangle  \frac{\langle 45\rangle }{\left\langle 4 \hat{5}\right\rangle }\left[\hat{1} 2\right]-\left[\hat{1} 2\right]\langle 23\rangle [34]\langle 45\rangle \left[5|\ell|1\right\rangle \right)
\,,
\end{equation*}
where we have used that $[5|\ell_0|1\rangle = [5|\ell|1\rangle$. For the final term, using that $\alpha[5|\ell|1\rangle = D_0$, we find 
\begin{equation}
\left[\hat{1} 2\right]\langle 23\rangle [34]\langle 45\rangle \left[5|\ell|1\right\rangle =[12]\langle 23\rangle [34]\langle 45\rangle \left[5|\ell|1\right\rangle + D_0[52] \langle 23\rangle  [34] \langle 45\rangle
\,,
\end{equation}
so that the forward-limit contribution is equivalent to
\begin{equation}
\begin{aligned}
\label{eq:5ptallplusFL}
\frac1{\ell^2}\,\mathscr{I}_5^{(1)}(\alpha) = 2\, {\mu^4}\frac{\text{PT}_5}{D_0 D_1 D_2 D_3 D_4}\bigg(&-D_0[52] \langle 45\rangle  \langle 23\rangle [34] + D_2 s_{45} s_{51} \\
&+ D_4\frac{\langle 45\rangle }{\left\langle 4 \hat{5}\right\rangle } [\hat{1} 2]\langle 21\rangle s_{23} - [12]\langle 23\rangle [34]\langle 45\rangle \left[5|\ell|1\right\rangle \bigg).
\end{aligned}
\end{equation}

The full amplitude is the combination the two contributions \eqref{eq:5ptallplusISF} and \eqref{eq:5ptallplusFL}. By noting that
\begin{equation*}
\frac{[23]\langle 34 \rangle}{\langle 14 \rangle} \frac{\alpha}{\alpha - \frac{\langle 45 \rangle}{\langle 41 \rangle}} + \frac{\langle 45 \rangle [\hat{1}2]}{\langle \hat{4}5 \rangle} = [12] \,,
\end{equation*}
we find that the spurious poles cancel, and we arrive at
\begin{equation}
\label{eq:ym5ptallplus1}
2\,\text{PT}_5\int \frac{d^D\ell \, \mu ^4}{D_0 D_1 D_2 D_3 D_4}\bigg(-D_0[52] \langle 45\rangle  \langle 23\rangle [34]+D_2 s_{45} s_{51}+D_4 s_{12} s_{23}-[12]\langle 23\rangle [34]\langle 45\rangle \left[5|\ell|1\right\rangle \bigg)
\,.
\end{equation}
This expression is indeed equal to the full amplitude as required, though to see this explicitly some additional manipulations must be performed. To connect with a known form of the result, we would like to write an integrand without a term linear in $\ell$ in the pentagon numerator. To do this, one may split this term into parity-plus and parity-minus pieces,
\begin{equation*}
[12]\langle 23\rangle [34]\langle 45\rangle [5|\ell|1\rangle =\frac{1}{2} \text{Tr} \left(\slashed{k}_1 \slashed{k}_2 \slashed{k}_3\slashed{k}_4\slashed{k}_5\slashed{\ell}\right)-\frac{1}{2} \text{Tr} \left(\gamma _5 \slashed{k}_1 \slashed{k}_2 \slashed{k}_3\slashed{k}_4\slashed{k}_5\slashed{\ell}\right)
\,.
\end{equation*}
We then have two non-trivial identities for these traces, which we prove in appendix~\ref{sec:slashedvecs},
\begin{align}
\text{Tr}\left(\slashed{k}_1 \slashed{k}_2 \slashed{k}_3\slashed{k}_4\slashed{k}_5\slashed{\ell}\right)&=
-D_0\left(s_{23} s_{34}+\text{Tr}\left(\slashed{k}_5\slashed{k}_2 \slashed{k}_3\slashed{k}_4\right)\right)-D_1 s_{34} s_{45}+D_2 s_{45} s_{51}-D_3 s_{51} s_{12}+D_4 s_{12} s_{23}
\,,  \nonumber \\
\label{eq:parityminustrace}
\text{Tr}\left(\gamma_5 \slashed{k}_1 \slashed{k}_2 \slashed{k}_3\slashed{k}_4\slashed{k}_5\slashed{\ell}\right)
&=
8 i \mu ^2 \varepsilon \left(k_1 k_2 k_3 k_4\right) - D_0 \text{Tr}\left(\gamma_5 \slashed{k}_5 \slashed{k}_2 \slashed{k}_3\slashed{k}_4\right)- 4i(D_0 \varepsilon \left(k_2 k_3 k_4\left(\ell+k_1\right)\right)\\[3pt] 
&\quad +D_1 \varepsilon \left(k_3 k_4 k_5 \ell\right) +D_2 \varepsilon \left(k_4 k_5 k_1 \ell\right)+D_3 \varepsilon \left(k_5 k_1 k_2 \ell\right)+D_4 \varepsilon \left(k_1 k_2 k_3 \ell \right))
\,, \nonumber
\end{align}
where $\varepsilon(\cdot \cdot \cdot \, \cdot)$ denotes contraction with $\varepsilon_{\mu \nu \rho \sigma}$. Note that, in the parity-minus trace, we have added terms quadratic in $\ell$ which add up to zero using Cramer's rule for the linear dependence of five vectors in four dimensions. Hence \eqref{eq:ym5ptallplus1} becomes
\eqs{
&\mathrm{PT}_5\int\frac{d^D\ell \,\mu^4}{D_0D_1D_2D_3D_4}\Big(D_0 \left( s_{23} s_{34}-2[52]\langle 23\rangle [34] \langle 45\rangle  +\text{Tr}\left(\slashed{k}_5\slashed{k}_2 \slashed{k}_3\slashed{k}_4\right)-\text{Tr} \left(\gamma _5 \slashed{k}_5\slashed{k}_2 \slashed{k}_3\slashed{k}_4\right)\right)
\\&\qquad\qquad\qquad\qquad\qquad\qquad  +D_1 s_{34} s_{45}+D_2 s_{45} s_{51}+D_3 s_{12} s_{51}+D_4 s_{12} s_{23}+8 i \mu ^2 \varepsilon\left(k_1 k_2 k_3 k_4\right)\Big)
\\-&4i\,\mathrm{PT}_5\int\frac{d^D\ell \,\mu^4}{D_0D_1D_2D_3D_4} \Big(D_0\varepsilon\left((k_2k_3k_4(\ell+k_1)\right) + D_1\varepsilon(k_3k_4k_5\ell) 
\\& \qquad\qquad\qquad\qquad\qquad\qquad\qquad+ D_2\varepsilon(k_4k_5k_1\ell) + D_3\varepsilon(k_5k_1k_2\ell)+ D_4\varepsilon(k_1k_2k_3\ell)\Big)\,, \notag
}
which we have chosen to split into two integrals. The second integral can be shown to vanish, by adapting an argument in \cite{NigelGlover:2008ur}. To see this, look for example at the final term, with numerator $ D_4\varepsilon(k_1k_2k_3\ell)$. This integral has the propagators of a scalar box depending on the three momenta $k_1$, $k_2$ and $k_3$, with a vector $\ell$ in the numerator. This means that, after integration, the vector-valued integral must lie in the subspace spanned by $k_1$, $k_2$ and $k_3$. The integral is contracted with $\varepsilon(k_1k_2k_3\cdot)$, and hence we can see that it must be zero. The same argument follows for all of the terms in the second integral above. For the first term, it is necessary to transform $\ell \to \ell - k_1$ before applying this argument. Returning to the first integral, we can use that $2[25]\langle23\rangle[34]\langle45\rangle = \mathrm{Tr}\left(\slashed{k}_5\slashed{k}_2 \slashed{k}_3\slashed{k}_4\right) - \mathrm{Tr}\left(\gamma _5 \slashed{k}_5\slashed{k}_2 \slashed{k}_3\slashed{k}_4\right)$. We then arrive at the final form,
\begin{align}
\mathcal{A}^{(1)}(1^+,2^+,3^+,4^+,5^+) = \mathrm{PT}_5\int\frac{d^D\ell \,\mu^4}{D_0D_1D_2D_3D_4}
\Big(&D_0s_{23}s_{34} + D_1s_{34}s_{45} + D_2s_{45}s_{51} 
\\ + &D_3s_{51}s_{12} + D_4s_{12}s_{23}+ 8i\mu^2\varepsilon(k_1k_2k_3k_4)\Big) \,, \nonumber
\end{align}
which matches the five-point amplitude as written in eq. (15) of \cite{Bern:1996ja}.

To conclude, we have explicitly constructed the five-point one-loop all-plus Yang-Mills amplitude using BCFW recursion. Equalities in this section have been checked numerically, often using code from \cite{Farrow:2018cqi}.

%%%%%%%%%%%%%%%%%%%%%%%%%%%%%%%%%
%%%%%%%%%%%%%%%%%%%%%%%%%%%%%%%%%
\section{Relation to previous BCFW literature}
\label{sec:previous}

In this section, we will compare the version of the one-loop BCFW recursion described above with ones appearing in the literature, namely the recursion in momentum twistor space for planar $\mathcal{N}=4$ SYM, the momentum space recursion with unshifted loop momentum, and the recursion for all-plus amplitudes (i.e., after loop integration) in pure Yang-Mills theory.

\subsection{Momentum twistor recursion for \texorpdfstring{$\mathcal{N}=4$}{N=4} SYM} \label{momtwistrecursion}

In this subsection, we will relate the recursion in ordinary momentum space described above to the one proposed for
planar $\mathcal{N}=4$ SYM in \cite{ArkaniHamed:2010kv}. The latter
was defined in terms of variables known as momentum twistors, which
are 4-component objects which transform in the fundamental representation
of the dual conformal group \cite{Hodges:2009hk}. In more detail, the momentum twistor for particle $i$
can be written as $Z_{i}=\left(\lambda_{i},\mu_{i}\right)$, where
$\mu_{i}=x_{i}\cdot\lambda_{i}$ and $x_{i}$ are region momentum
coordinates defined by 

\[
x_{i+1}-x_{i}=k_{i}=\lambda_{i}\tilde{\lambda}_{i}.
\]
Note that a point in region momentum space corresponds to a pair of
momentum twistors via the formula
\begin{equation}
x_{i}=\frac{\lambda_{i}\mu_{i-1}-\lambda_{i-1}\mu_{i}}{\left\langle i-1i\right\rangle }.\label{pointtoline}
\end{equation}
For a more detailed review of momentum twistors, see for example  \cite{Elvang:2013cua} or section
2.1 of \cite{Lipstein:2012vs}. 

The BCFW recursion in \cite{ArkaniHamed:2010kv} was based on the following shift
in momentum twistor space:

\begin{equation}
\hat{Z}_{n}=Z_{n}+wZ_{n-1}.\label{momshif}
\end{equation}
In terms of region momenta, this correponds to the shift \cite{Elvang:2013cua}
\begin{equation}
\hat{x}_{1}=x_{1}+zq\,,\qquad\qquad \text{with }\; z=\frac{\left\langle n-1n\right\rangle w}{\left\langle n-11\right\rangle w+\left\langle n1\right\rangle }\,,\label{ztow}
\end{equation}
and $q=\lambda_{1}\tilde{\lambda}_{n}$,
which can be proven using \ref{pointtoline} and the relation 
\[
\tilde{\lambda}_{i}=\frac{\left\langle i-1i+1\right\rangle \mu_{i}+\left\langle i+1i\right\rangle \mu_{i-1}+\left\langle ii-1\right\rangle \mu_{i+1}}{\left\langle i-1i\right\rangle \left\langle ii+1\right\rangle }.
\]
This implies a standard BCFW shift of external momenta 
\[
\hat{k}_{n}=\hat{x}_{1}-x_{n}=k_{n}-zq\,,\qquad\qquad 
\hat{k}_{1}=x_{2}-\hat{x}_{1}=k_{1}+zq\,.
\]
Moreover at one loop, there is one internal region momentum $x_{0}$, so
we may define the loop momentum to be $\ell=x_{0}-x_{1}$. We will shortly describe a concrete  example of the recursion which will make it clear why this is a natural definition. It follows
that the loop momentum is also shifted,
\[
\hat{\ell}=x_{0}-\hat{x}_{1}=\ell-zq.
\]
Note that this is the same shift we consider in this paper, as stated in \eqref{eq:shift}. In the momentum twistor formalism, however, all momenta are necessarily four-dimensional, including the loop momentum.

\begin{figure}[ht]
\begin{center}
       \begin{tikzpicture}[scale=0.75]
\node at (2.5,0.05) {$\displaystyle \mathcal{A}_n^{\scalebox{0.6}{$(1)$}}$};
 \node at (3.5,0) {$=$};
 \draw [fill, light-grayII] (6.5,0) circle [radius=1];
 \draw [thick] (6.5,0) circle [radius=1];
 \node at (6.5,0) {\scalebox{0.9}{$\mathcal{A}_{n-1}^{\scalebox{0.6}{$(1)$}}$}};
 \draw [dotted,thick,domain=115:155] plot ({6.5+1.5*  cos(\x)}, {1.5 * sin(\x)});
  \draw [domain=-0.966:-1.932] plot ({6.5+\x},{ tan(15)*\x});
  \draw [domain=0.966:2] plot ({6.5+\x},{ -tan(15)*\x});
  \draw [domain=-0.966:-1.932] plot ({6.5+\x},{ -tan(15)*\x});
  \draw [domain=0.259:0.518] plot ({6.5+\x},{ tan(75)*\x});
  \draw [domain=0.259:0.58] plot ({6.5+\x},{ -tan(75)*\x});
  \draw [domain=-0.259:-0.518] plot ({6.5+\x},{- tan(75)*\x});
   \draw [domain=0.707:1.414] plot ({6.5+\x},{ tan(45)*\x});
   \draw [domain=-0.707:-1.414] plot ({6.5+\x},{ tan(45)*\x});
 \node at (4.9,-1.6) {\scalebox{0.9}{$2$}};
 \node at (8.5,1.6) {\scalebox{0.9}{$n-2$}};
\draw (8.59,-2.08)  --  (7.06,-2.08);
\draw (8.59,-2.08) --  (8.59,-0.58);
 \draw [domain=-0:-0.3] plot ({7.06+\x},{ -2.08+tan(75)*\x});
 \draw [domain=0:1.25] plot ({8.59+\x},{ -0.58+tan(15)*\x});
 \draw [domain=0:1] plot ({8.59+\x},{ -2.08-tan(45)*\x});
 \draw [fill, white] (7.06,-2.08) circle [radius=.08];
 \draw [fill, white] (8.59,-0.58) circle [radius=.08];
  \draw [thick] (7.06,-2.08) circle [radius=.08];
 \draw [thick] (8.59,-0.58) circle [radius=.08];
\draw [fill] (8.59,-2.08) circle [radius=.08];
  \node at  (10.4,-0.25) {\scalebox{0.9}{$n-1$}};
  \node at  (9.8,-3.25) {\scalebox{0.9}{$n$}};
  \node at  (6.7,-3.45) {\scalebox{0.9}{$1$}};
 \node[blue] at (6.2,-1.6) {\scalebox{0.9}{$x_2$}};
 \node[blue] at (9.6,-1.4) {\scalebox{0.9}{$x_n$}};
 \node[blue] at (8.5,0.3) {\scalebox{0.9}{$x_{n-1}$}};
  \node[blue] at (8,-2.8) {\scalebox{0.9}{$x_{1}$}};
  \node at (12,0) {$+$};
  \node at (13.5,0) { $\displaystyle \sum_{a=2}^{n-2}$};
   \draw [fill, light-grayII] (16,0.5) circle [radius=0.5];
 \draw [thick] (16,0.5) circle [radius=0.5];
 \draw [domain=-0.354:-1] plot ({16+\x},{0.5- tan(45)*\x});
  \draw [domain=-0.354:-1] plot ({16+\x},{0.5+ tan(45)*\x});
 \draw [dotted,thick,domain=150:210] plot ({16+1*  cos(\x)}, {0.5+1 * sin(\x)});
 % pointing right
  \draw [domain=0.354:1] plot ({16+\x},{0.5- tan(45)*\x});
  \draw [domain=0.354:1] plot ({16+\x},{0.5+ tan(45)*\x});
  \draw [domain=0:0.6] plot ({17+\x},{1.5- tan(45)*\x});
  \draw [domain=0:0.6] plot ({17+\x},{-0.5+ tan(45)*\x});
  \draw (17,1.5) -- (17,-0.5);
  \draw (17.6,0.9) -- (17.6,0.1) -- (18.4,0.1) -- (18.4,0.9) -- (17.6,0.9);
  \draw (18.4,0.9) -- (19,1.5);
  \draw (18.4,0.1) -- (19,-0.5);
  \draw (17,1.5) -- (19,1.5);
  \draw (19,1.5) -- (19,-0.5);
  \draw [domain=0:0.646] plot ({19+\x},{1.5- tan(45)*\x});
  \draw [domain=0:0.646] plot ({19+\x},{-0.5+ tan(45)*\x});
  \draw [fill, white] (17,-0.5) circle [radius=.08];
  \draw [thick] (17,-0.5) circle [radius=.08];
  \draw [fill] (17,1.5) circle [radius=.08];
  \draw [fill, white] (19,1.5) circle [radius=.08];
  \draw [thick] (19,1.5) circle [radius=.08];
  \draw [fill] (19,-0.5) circle [radius=.08];
  \draw [fill, white] (17.6,0.9) circle [radius=.08];
  \draw [thick] (17.6,0.9) circle [radius=.08];
  \draw [fill] (17.6,0.1) circle [radius=.08];
  \draw [fill, white] (18.4,0.1) circle [radius=.08];
  \draw [thick] (18.4,0.1) circle [radius=.08];
  \draw [fill] (18.4,0.9) circle [radius=.08];
  \draw [fill, light-grayII] (20,0.5) circle [radius=0.5];
 \draw [thick] (20,0.5) circle [radius=0.5];
  \draw [domain=0.354:1] plot ({20+\x},{0.5- tan(45)*\x});
  \draw [domain=0.354:1] plot ({20+\x},{0.5+ tan(45)*\x});
  \draw [dotted,thick,domain=330:360] plot ({20+1*  cos(\x)}, {0.5+1 * sin(\x)});
  \draw [dotted,thick,domain=0:30] plot ({20+1*  cos(\x)}, {0.5+1 * sin(\x)});
  \draw (16,0) -- (16,-2.08) -- (20,-2.08) -- (20,0);
  \draw [domain=0:-1] plot ({16+\x},{-2.08+ tan(45)*\x});
  \draw [domain=0:1] plot ({20+\x},{-2.08- tan(45)*\x});
  \draw [fill, white] (16,-2.08) circle [radius=.08];
  \draw [thick] (16,-2.08) circle [radius=.08];
  \draw [fill] (20,-2.08) circle [radius=.08];
  \node at  (21.2,-3.25) {\scalebox{0.9}{$n$}};
  \node at  (14.8,-3.25) {\scalebox{0.9}{$1$}};
  \node at  (14.8,-0.7) {\scalebox{0.9}{$2$}};
  \node at  (14.8,1.7) {\scalebox{0.9}{$a$}};
  \node at  (21.2,-0.7) {\scalebox{0.9}{$n-1$}};
  \node at  (21.2,1.7) {\scalebox{0.9}{$a+1$}};
  \node[blue] at (18,-1) {\scalebox{0.9}{$\widehat{x}_1$}};
  \node[blue] at (18,0.45) {\scalebox{0.9}{$x_{0}$}};
  \node[blue] at (18,-3) {\scalebox{0.9}{$x_{1}$}};
  \node[blue] at (18,2.2) {\scalebox{0.9}{$x_{a+1}$}};
  \node[blue] at  (20.7,-1.8) {\scalebox{0.9}{$x_n$}};
  \node[blue] at  (15.3,-1.8) {\scalebox{0.9}{$x_2$}};
\end{tikzpicture}
    \caption{On-shell diagram recursion relation for one-loop MHV amplitudes, where the first term corresponds to tree-level factorisation, and the second term corresponds to a forward limit.} 
    \label{recursionfig}
    \end{center}
\end{figure}
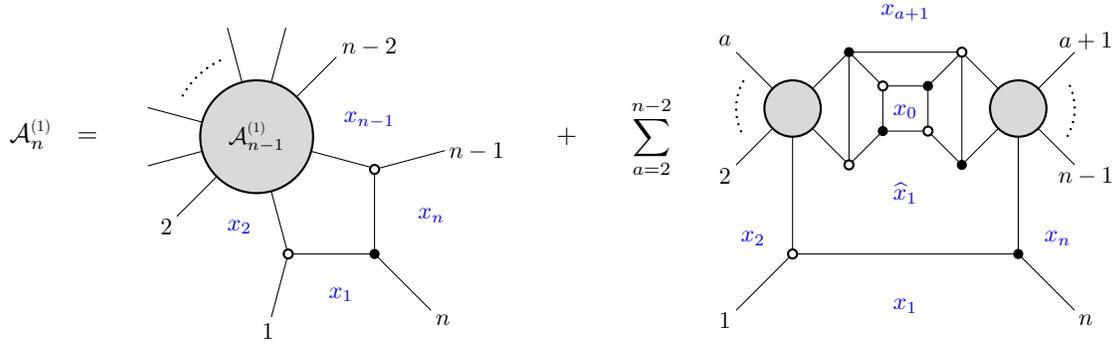

Just as the recursion described in previous sections, the recursion in \cite{ArkaniHamed:2010kv} involves a forward
limit and tree-level factorisations. This can be diagrammatically represented
using on-shell diagrams \cite{ArkaniHamed:2012nw}, as depicted  in Figure~\ref{recursionfig} for one-loop MHV amplitudes. 
In this figure, black and white vertices represent 3-point MHV and $\overline{\text{MHV}}$ amplitudes, respectively, and the 4-point vertices denote mergers of 3-point vertices; the edges represent integrals over on-shell states. The grey blobs represent MHV regions, which are set to unity in momentum twistor space.
The first term describes a tree-level factorisation, while the remaining
terms describe the forward limit of a tree-level $(n+2)$-point amplitude written as a sum over $(n-3)$ factorisation channels. We also indicate the region momenta. Note that $x_0$ is not displayed in the first term, but we will make it explicit for the case $n=5$ below. Looking at the forward limit terms, it is clear why $x_0-x_1$ is a natural definition for the loop momentum. We see that $x_1$ is separated from $x_0$ by a pair of edges whose momenta combine into an off-shell propagator momentum; in our conventions, $x_0-\hat x_1=\ell_0$ and $\hat x_1- x_1=\alpha\,q$, so $x_0-x_1=\ell_0+\alpha\,q=\ell$. Although $x_{0}-x_{a+1}$ can also be thought of as a propagator momentum, this would give a different loop momentum for each diagram; but $x_{0}-x_{2}$ and $x_{0}-x_{n}$ are also natural choices (and indeed the first case is related to the story in the next subsection). Each on-shell diagram corresponds to a so-called Kermit diagram, so that the integrand is given by
\[
\mathcal{A}_{n}^{(1)}=\sum_{1<i<j<n}K(i,j),
\]
where
\begin{equation}
K(i,j)=\frac{\left(\left\langle AB1i\right\rangle \left\langle i+11jj+1\right\rangle +\left\langle ABi+11\right\rangle \left\langle i1jj+1\right\rangle \right)^{2}}{\left\langle AB1i\right\rangle \left\langle ABii+1\right\rangle \left\langle ABjj+1\right\rangle \left\langle ABj+11\right\rangle \left\langle ABi+11\right\rangle \left\langle AB1j\right\rangle },\label{kermit}
\end{equation}
$\left\langle ijkl\right\rangle =\epsilon_{IJKL}Z_{i}^{I}Z_{j}^{J}Z_{k}^{K}Z_{l}^{L}$,
and the region momentum $x_{0}$ is associated to the line in momentum twistor space defined by the pair of twistors
$Z_{A}$ and $Z_{B}$ .

\begin{figure}[ht]
\begin{center}
         \begin{tikzpicture}[scale=0.65]
    \draw [domain=0:0.707] plot ({17-\x},{1.5+tan(45)*\x});
  \draw [domain=0:0.707] plot ({19+\x},{1.5+ tan(45)*\x});
  \draw [domain=0:0.6] plot ({17+\x},{1.5- tan(45)*\x});
  \draw [domain=0:0.6] plot ({17+\x},{-0.5+ tan(45)*\x});
  \draw (17,1.5) -- (17,-0.5);
  \draw (17.6,0.9) -- (17.6,0.1) -- (18.4,0.1) -- (18.4,0.9) -- (17.6,0.9);
  \draw (18.4,0.9) -- (19,1.5);
  \draw (18.4,0.1) -- (19,-0.5);
  \draw (17,1.5) -- (19,1.5)  -- (19,-0.5) -- (17,-0.5);
  \draw [domain=0:-0.5] plot ({17+\x},{-0.5+tan(72)*\x});
  \draw [domain=0:0.5] plot ({19+\x},{-0.5- tan(72)*\x});
  \draw (16.5, -2.039) -- (18,-3) --(19.5,-2.039);
  \draw (18,-3) -- (18,-4);
  \draw (16.5, -2.039) -- (15.6,-2.6);
  \draw (19.5,-2.039) -- (20.4,-2.6);
  \draw [fill, white] (17,-0.5) circle [radius=.08];
  \draw [thick] (17,-0.5) circle [radius=.08];
  \draw [fill] (17,1.5) circle [radius=.08];
  \draw [fill, white] (19,1.5) circle [radius=.08];
  \draw [thick] (19,1.5) circle [radius=.08];
  \draw [fill] (19,-0.5) circle [radius=.08];
  \draw [fill, white] (17.6,0.9) circle [radius=.08];
  \draw [thick] (17.6,0.9) circle [radius=.08];
  \draw [fill] (17.6,0.1) circle [radius=.08];
  \draw [fill, white] (18.4,0.1) circle [radius=.08];
  \draw [thick] (18.4,0.1) circle [radius=.08];
  \draw [fill] (18.4,0.9) circle [radius=.08];
  \draw [fill, white] (16.5, -2.039) circle [radius=.08];
  \draw [thick] (16.5, -2.039) circle [radius=.08];
  \draw [fill, white] (19.5,-2.039) circle [radius=.08];
  \draw [thick] (19.5,-2.039) circle [radius=.08];
  \draw [fill] (18,-3) circle [radius=.08];
  \node[blue] at (18,-1.7) {\scalebox{0.9}{$\widehat{x}_1$}};
  \node[blue] at (18,0.45) {\scalebox{0.9}{$x_{0}$}};
  \node[blue] at (16.2,-0.2) {\scalebox{0.9}{$x_{2}$}};
  \node[blue] at (19.8,-0.2) {\scalebox{0.9}{$x_{4}$}};
  \node[blue] at (18,2.2) {\scalebox{0.9}{$x_{3}$}};
  \node[blue] at (16.8,-3.2) {\scalebox{0.9}{$x_{1}$}};
  \node[blue] at (19.2,-3.2) {\scalebox{0.9}{$x_{5}$}};
  \node at (18,-4.35) {\scalebox{0.8}{$5$}};
  \node at (15.4,-2.7) {\scalebox{0.8}{$1$}};
  \node at (20.6,-2.7) {\scalebox{0.8}{$4$}};
  \node at (16.1,2.35) {\scalebox{0.8}{$2$}};
  \node at (19.9,2.35) {\scalebox{0.8}{$3$}};
  \draw [thick, decoration={brace,  mirror}, decorate] (15.5,-5.3) -- (20.5,-5.3);
  \node at (18,-6) {$=K(2,3)$};
  \node at (21.5, -0.5) {$+$};
  \draw [domain=0:0.707] plot ({24-\x},{1.5+tan(45)*\x});
  \draw [domain=0:0.6] plot ({24+\x},{1.5- tan(45)*\x});
  \draw [domain=0:0.6] plot ({24+\x},{-0.5+ tan(45)*\x});
  \draw (24,1.5) -- (24,-0.5);
  \draw (24.6,0.9) -- (24.6,0.1) -- (25.4,0.1) -- (25.4,0.9) -- (24.6,0.9);
  \draw (25.4,0.9) -- (26,1.5);
  \draw (25.4,0.1) -- (26,-0.5);
  \draw (24,1.5) -- (26,1.5)  -- (26,-0.5);
  \draw [domain=0:0.5] plot ({26+tan(72)*\x},{-0.5-\x});
  \draw [domain=0:0.5] plot ({26+tan(72)*\x},{1.5+\x});
  \draw (27.539,-1) -- (28.5,0.5) -- (27.539,2);
  \draw (28.5,0.5) -- (29.5,0.5);
  \draw (27.539,2) -- (28.1,2.9);
  \draw (27.539,-1) -- (27.539,-2.5) -- (24,-2.5) -- (24,-0.5);
  \draw [domain=0:-0.707] plot ({24+\x},{-2.5+ tan(45)*\x});
  \draw [domain=0:0.707] plot ({27.539+\x},{-2.5-tan(45)*\x});
  \draw [fill, white] (24,-0.5) circle [radius=.08];
  \draw [thick] (24,-0.5) circle [radius=.08];
  \draw [fill] (24,1.5) circle [radius=.08];
  \draw [fill, white] (26,1.5) circle [radius=.08];
  \draw [thick] (26,1.5) circle [radius=.08];
  \draw [fill] (26,-0.5) circle [radius=.08];
  \draw [fill, white] (24.6,0.9) circle [radius=.08];
  \draw [thick] (24.6,0.9) circle [radius=.08];
  \draw [fill] (24.6,0.1) circle [radius=.08];
  \draw [fill, white] (25.4,0.1) circle [radius=.08];
  \draw [thick] (25.4,0.1) circle [radius=.08];
  \draw [fill] (25.4,0.9) circle [radius=.08];
  \draw [fill, white] (27.539,2) circle [radius=.08];
  \draw [thick] (27.539,2) circle [radius=.08];
  \draw [fill, white] (27.539,-1) circle [radius=.08];
  \draw [thick] (27.539,-1) circle [radius=.08];
  \draw [fill] (28.5,0.5) circle [radius=.08];
  \draw [fill, white] (24,-2.5) circle [radius=.08];
  \draw [thick] (24,-2.5) circle [radius=.08];
  \draw [fill] (27.539,-2.5) circle [radius=.08];
  \node[blue] at (25,0.45) {\scalebox{0.9}{$x_{0}$}};
  \node[blue] at (25.77,-1.5) {\scalebox{0.9}{$\widehat{x}_1$}};
  \node[blue] at (25.77,2.2) {\scalebox{0.9}{$x_3$}};
  \node[blue] at (25.77,-3.4) {\scalebox{0.9}{$x_1$}};
  \node[blue] at (23.2,-0.2) {\scalebox{0.9}{$x_{2}$}};
  \node[blue] at (28.6,1.55) {\scalebox{0.9}{$x_{4}$}};
  \node[blue] at (28.5,-1.2) {\scalebox{0.9}{$x_{5}$}};
  \node at (23.1,2.35) {\scalebox{0.8}{$2$}};
  \node at (23.1,-3.3) {\scalebox{0.8}{$1$}};
  \node at (28.2,3.1) {\scalebox{0.8}{$3$}};
  \node at (28.44,-3.4) {\scalebox{0.8}{$5$}};
  \node at (29.75,0.5) {\scalebox{0.8}{$4$}};
  \draw [thick, decoration={brace,  mirror}, decorate] (23,-5.3) -- (29.5,-5.3);
  \node at (26.25,-6) {$=K(2,4)$};
  \node at (31, -0.5) {$+$};
  \draw [domain=0:0.707] plot ({38+\x},{1.5+tan(45)*\x});
  \draw [domain=0:-0.6] plot ({38+\x},{1.5+tan(45)*\x});
  \draw [domain=0:0.6] plot ({38-\x},{-0.5+tan(45)*\x});
  \draw (38,1.5) -- (38,-0.5);
  \draw (37.4,0.9) -- (37.4,0.1) -- (36.6,0.1) -- (36.6,0.9) -- (37.4,0.9);
  \draw (36.6,0.9) -- (36,1.5);
  \draw (36.6,0.1) -- (36,-0.5);
  \draw (38,1.5) -- (36,1.5)  -- (36,-0.5);
  \draw [domain=0:0.5] plot ({36-tan(72)*\x},{-0.5-\x});
  \draw [domain=0:0.5] plot ({36-tan(72)*\x},{1.5+\x});
  \draw (34.461,-1) -- (33.5,0.5) -- (34.461,2);
  \draw (33.5,0.5) -- (32.5,0.5);
  \draw (34.461,2) -- (33.9,2.9);
  \draw (34.461,-1) -- (34.461,-2.5) -- (38,-2.5) -- (38,-0.5);
  \draw [domain=0:-0.707] plot ({38-\x},{-2.5+ tan(45)*\x});
  \draw [domain=0:0.707] plot ({34.461-\x},{-2.5-tan(45)*\x});
  \draw [fill, white] (38,1.5) circle [radius=.08];
  \draw [thick] (38,1.5) circle [radius=.08];
  \draw [fill] (38,-0.5) circle [radius=.08];
  \draw [fill, white] (36,-0.5) circle [radius=.08];
  \draw [thick] (36,-0.5) circle [radius=.08];
  \draw [fill] (36,1.5) circle [radius=.08];
  \draw [fill, white] (37.4,0.1) circle [radius=.08];
  \draw [thick] (37.4,0.1) circle [radius=.08];
  \draw [fill] (37.4,0.9) circle [radius=.08];
  \draw [fill, white] (36.6,0.9) circle [radius=.08];
  \draw [thick] (36.6,0.9) circle [radius=.08];
  \draw [fill] (36.6,0.1) circle [radius=.08];
  \draw [fill, white] (34.461,2) circle [radius=.08];
  \draw [thick] (34.461,2) circle [radius=.08];
  \draw [fill, white] (34.461,-1) circle [radius=.08];
  \draw [thick] (34.461,-1) circle [radius=.08];
  \draw [fill] (33.5,0.5) circle [radius=.08];
  \draw [fill, white] (34.461,-2.5) circle [radius=.08];
  \draw [thick] (34.461,-2.5) circle [radius=.08];
  \draw [fill] (38,-2.5) circle [radius=.08];
  \node[blue] at (37,0.45) {\scalebox{0.9}{$x_{0}$}};%25
  \node[blue] at (36.23,-1.5) {\scalebox{0.9}{$\widehat{x}_1$}};
  \node[blue] at (36.23,2.2) {\scalebox{0.9}{$x_4$}};
  \node[blue] at (36.23,-3.4) {\scalebox{0.9}{$x_1$}};
  \node[blue] at (38.8,-0.2) {\scalebox{0.9}{$x_{5}$}};
  \node[blue] at (33.4,1.55) {\scalebox{0.9}{$x_{3}$}};
  \node[blue] at (33.5,-1.2) {\scalebox{0.9}{$x_{2}$}};
  \node at (38.9,2.35) {\scalebox{0.8}{$4$}};
  \node at (38.9,-3.3) {\scalebox{0.8}{$5$}};
  \node at (33.8,3.1) {\scalebox{0.8}{$3$}};
  \node at (33.56,-3.4) {\scalebox{0.8}{$1$}};
  \node at (32.25,0.5) {\scalebox{0.8}{$2$}};
  \draw [thick, decoration={brace,  mirror}, decorate] (32.5,-5.3) -- (39,-5.3);
  \node at (35.75,-6) {$=K(3,4)$};
  \end{tikzpicture}
    \caption{Five-particle amplitude $\mathcal{A}_{5}^{(1)}=K(2,3)+K(2,4)+K(3,4)$ from the momentum twistor recursion. The first term comes from tree-level factorisation, and the other terms come from the forward limit.
    } 
    \label{n5recursion}
    \end{center}
\end{figure}
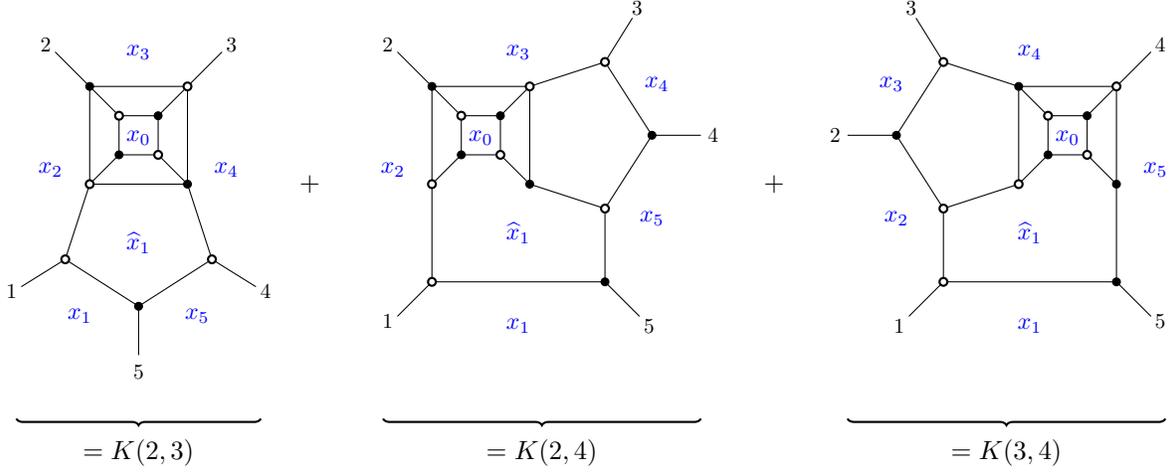

Note that the BCFW shift parameter in \eqref{momshif} can be written in terms of momentum twistors
as follows \cite{ArkaniHamed:2012nw}: 
\begin{equation}
w=\frac{\left\langle ABn1\right\rangle }{\left\langle AB1n-1\right\rangle }.\label{w}
\end{equation}
In the limit $\left\langle AB1n-1\right\rangle \rightarrow0$, $w$
blows up and the corresponding BCFW shift in momentum space \eqref{ztow}
approaches the finite value
\begin{equation}
z\rightarrow\frac{\left\langle n-1n\right\rangle }{\left\langle n-11\right\rangle }.
\label{infw}
\end{equation}
This is precisely the location of the spurious pole which arose in
our recursion, described in previous sections.
Moreover, this spurious pole cancels in the same way.
To see this more concretely, let us specialise to the case $n=5$.
In this case, there are three on-shell diagrams depicted in Figure~\ref{n5recursion}, and the one-loop MHV amplitude is given by
\[
\mathcal{A}_{5}^{(1)}=K(2,3)+K(2,4)+K(3,4)\,,
\]
where
\begin{align}
K(2,3)&=\frac{\left\langle 1234\right\rangle ^{2}}{\left\langle AB12\right\rangle \left\langle AB23\right\rangle \left\langle AB34\right\rangle \left\langle AB41\right\rangle }\,,\nonumber \\
K(2,4)&=\frac{\left(\left\langle 1234\right\rangle \left\langle AB15\right\rangle +\left\langle 1352\right\rangle \left\langle AB14\right\rangle \right)^{2}}{\left\langle AB12\right\rangle \left\langle AB23\right\rangle \left\langle AB54\right\rangle \left\langle AB51\right\rangle \left\langle AB13\right\rangle \left\langle AB14\right\rangle }\,,\label{K24}\\
K(3,4)&=\frac{\left\langle 1345\right\rangle ^{2}}{\left\langle AB13\right\rangle \left\langle AB34\right\rangle \left\langle AB45\right\rangle \left\langle AB51\right\rangle }\,,\nonumber
\end{align}
and the expression for $K(2,4)$ was obtained from \eqref{kermit} using
the Schouten identity 
\begin{equation}
0=Z_{i}\left\langle jklm\right\rangle +{\rm cyclic.}\label{schouten}
\end{equation}
Note that $K(2,3)$ comes from the tree-level factorisation in Figure~\ref{recursionfig},
while $K(2,4)$ and $K(3,4)$ come from the forward limit. 

The physical poles correspond to $\left\langle ABii+1\right\rangle $, and the
other poles are spurious. Hence, there are two spurious poles, $\left\langle AB13\right\rangle $
and $\left\langle AB14\right\rangle $. The first spurious pole $\left\langle AB13\right\rangle $ occurs
in the forward limit terms. It did not appear in the procedure of the previous sections, but it would, of course, if we evaluated the forward-limit tree amplitude via its own tree-level BCFW decomposition. 
On the other hand, $\left\langle AB14\right\rangle $ must cancel between the
forward limit and tree-level factorisation terms, as we found in section \ref{5ptmhv}. To see how this works now, first
use the Schouten identity to obtain
\[
\left\langle AB14\right\rangle \left\langle AB35\right\rangle +\left\langle AB13\right\rangle \left\langle AB54\right\rangle +\left\langle AB15\right\rangle \left\langle AB43\right\rangle =0.
\]
Taking $\left\langle AB14\right\rangle \rightarrow0$ then implies
that
\[
\left\langle AB13\right\rangle \left\langle AB54\right\rangle \rightarrow\left\langle AB15\right\rangle \left\langle AB34\right\rangle .
\]
Plugging this into the denominator of \eqref{K24}, we find that
\[
K(2,4)\rightarrow-K(2,3)
\]
as $\left\langle AB14\right\rangle \rightarrow 0$, so the spurious poles indeed cancel out. Moroever, when $\left\langle AB14\right\rangle \rightarrow0$, the edge variable $w$ given by \eqref{w} with
$n=5$ blows up. According to \eqref{infw}, the spurious pole in momentum space is then located at $z=\frac{\left\langle 45\right\rangle }{\left\langle 4 1\right\rangle }$, as we found in \ref{5ptmhv}. 

In summary, we have verified that the one-loop recursion described in previous sections is equivalent
to the one first proposed for planar $\mathcal{N}=4$ SYM by matching the BCFW shifts
and spurious poles.

\subsection{Recursion with unshifted loop momentum}
\label{sec:unshiftedloop}
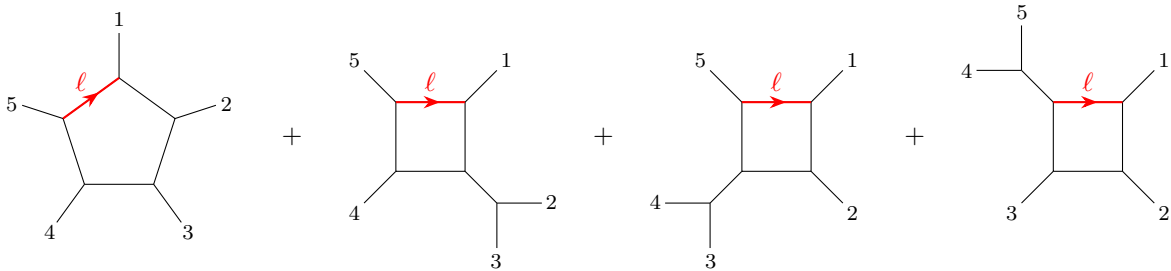
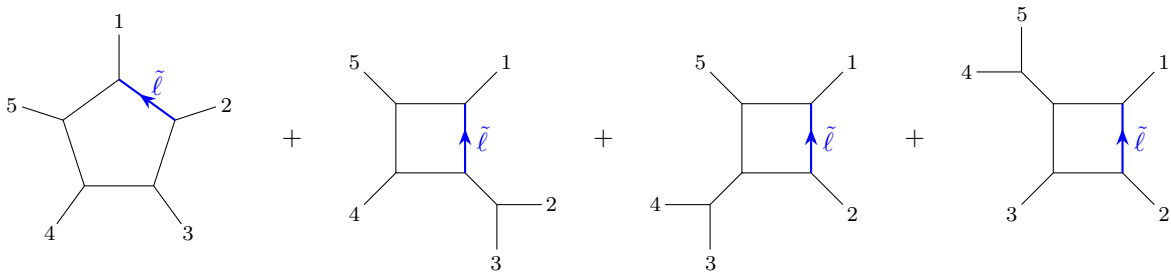
\begin{figure}[ht]
\begin{subfigure}[b]{\textwidth}
\begin{center}
    \tikzset{
  % style to add an arrow in the middle of a path, use as \draw [postaction={mid arrow }]
  mid arrow2/.style={postaction={decorate,decoration={
        markings,
        mark=at position .64 with {\arrow[#1]{Stealth[length=2.2mm]}}
      }}},
}
  \begin{tikzpicture}[scale=0.46 ]
   \draw (-1,-1.376) -- (1,-1.376) -- (1.618,0.526) -- (0,1.701) --(-1.618,0.526) -- (-1,-1.376) ;
   \draw [domain=1.6:2.8] plot ({\x},{ tan(18)*\x});
   \draw [domain=1.7:3] plot ({0},{ \x});
   \draw [domain=1.6:2.8] plot ({-\x},{ tan(18)*\x});
   \draw [domain=1:1.8] plot ({\x},{ -tan(54)*\x});
   \draw [domain=1:1.8] plot ({-\x},{- tan(54)*\x});
   \draw[red, thick,  postaction={mid arrow2}] (-1.618,0.526) --(0,1.701) ;
   \node[red] at (-1.1,1.6) {$\ell$};
   \node at (0,3.4) {\footnotesize $ 1$};
   \node at (3.1,0.95) {\footnotesize $2$};
   \node at (2,-2.75) {\footnotesize $3$};
   \node at (-2,-2.75) {\footnotesize $4$};
   \node at (-3.1,0.95) {\footnotesize $5$};
   \node at (5,0) {$+$};
   \draw (8,-1) -- (10,-1) -- (10,1) -- (8,1) -- (8,-1);
   \draw [domain=1:1.919] plot ({-\x+9},{ tan(45)*\x});
   \draw [domain=1:1.919] plot ({-\x+9},{ -tan(45)*\x});
   \draw [domain=1:1.919] plot ({\x+9},{ tan(45)*\x});
   \draw [domain=1:1.919] plot ({\x+9},{- tan(45)*\x});
   \draw (10.919,-1.919) -- (12.219,-1.919);
   \draw (10.919,-1.919) -- (10.919,-3.219);
   \draw[red, thick,  postaction={mid arrow2}] (8,1) --(10,1);
   \node[red] at (9,1.6) {$\ell$};
   \node at (11.2,2.2) {\footnotesize $ 1$};
   \node at (12.5,-1.9) {\footnotesize $2$};
   \node at (10.9,-3.6) {\footnotesize $3$};
   \node at (6.8,-2.2) {\footnotesize $4$};
   \node at (6.8,2.2) {\footnotesize $5$};
   \node at (14,0) {$+$};
   \draw (18,-1) -- (20,-1) -- (20,1) -- (18,1) -- (18,-1);
   \draw [domain=1:1.919] plot ({-\x+19},{ tan(45)*\x});
   \draw [domain=1:1.919] plot ({-\x+19},{ -tan(45)*\x});
   \draw [domain=1:1.919] plot ({\x+19},{ tan(45)*\x});
   \draw [domain=1:1.919] plot ({\x+19},{- tan(45)*\x});
   \draw (17.081,-1.919) -- (15.781,-1.919);
   \draw (17.081,-1.919) -- (17.081,-3.219);
   \draw[red, thick,  postaction={mid arrow2}] (18,1) --(20,1);
   \node[red] at (19,1.6) {$\ell$};
   \node at (21.2,2.2) {\footnotesize $ 1$};
   \node at (21.2,-2.2) {\footnotesize $2$};
   \node at (17.1,-3.6) {\footnotesize $3$};
   \node at (15.5,-1.9) {\footnotesize $4$};
   \node at (16.8,2.2) {\footnotesize $5$};
   \node at (23,0) {$+$};
   \draw (27,-1) -- (29,-1) -- (29,1) -- (27,1) -- (27,-1);
   \draw [domain=1:1.919] plot ({-\x+28},{ tan(45)*\x});
   \draw [domain=1:1.919] plot ({-\x+28},{ -tan(45)*\x});
   \draw [domain=1:1.919] plot ({\x+28},{ tan(45)*\x});
   \draw [domain=1:1.919] plot ({\x+28},{- tan(45)*\x});
   \draw (26.081,1.919) -- (24.781,1.919);
   \draw (26.081,1.919) -- (26.081,3.219);
   \draw[red, thick,  postaction={mid arrow2}] (27,1) --(29,1);
   \node[red] at (28,1.6) {$\ell$};
   \node at (30.2,2.2) {\footnotesize $ 1$};
   \node at (30.2,-2.2) {\footnotesize $2$};
   \node at (26.1,3.6) {\footnotesize $5$};
   \node at (24.5,1.9) {\footnotesize $4$};
   \node at (25.8,-2.2) {\footnotesize $3$};
  \end{tikzpicture}
  \end{center}
  \caption{Conventions for BCFW shift with \emph{shifted} loop momentum, used in previous sections.}
  \end{subfigure}\\
  
   \begin{subfigure}[b]{\textwidth}
   \begin{center}
    \tikzset{
  % style to add an arrow in the middle of a path, use as \draw [postaction={mid arrow }]
  mid arrow2/.style={postaction={decorate,decoration={
        markings,
        mark=at position .64 with {\arrow[#1]{Stealth[length=2.2mm]}}
      }}},
}
 \begin{tikzpicture}[scale=0.46 ]
   \draw (-1,-1.376) -- (1,-1.376) -- (1.618,0.526) -- (0,1.701) --(-1.618,0.526) -- (-1,-1.376) ;
   \draw [domain=1.6:2.8] plot ({\x},{ tan(18)*\x});
   \draw [domain=1.7:3] plot ({0},{ \x});
   \draw [domain=1.6:2.8] plot ({-\x},{ tan(18)*\x});
   \draw [domain=1:1.8] plot ({\x},{ -tan(54)*\x});
   \draw [domain=1:1.8] plot ({-\x},{- tan(54)*\x});
   \draw[blue, thick,  postaction={mid arrow2}] (1.618,0.526) --(0,1.701) ;
   \node[blue] at (1.1,1.6) {$\tilde\ell$};
   %\draw[fill] (-0.809,1.114) circle [radius=0.1];
   %
   \node at (0,3.4) {\footnotesize $ 1$};
   \node at (3.1,0.95) {\footnotesize $2$};
   \node at (2,-2.75) {\footnotesize $3$};
   \node at (-2,-2.75) {\footnotesize $4$};
   \node at (-3.1,0.95) {\footnotesize $5$};
   \node at (5,0) {$+$};
   \draw (8,-1) -- (10,-1) -- (10,1) -- (8,1) -- (8,-1);
   \draw [domain=1:1.919] plot ({-\x+9},{ tan(45)*\x});
   \draw [domain=1:1.919] plot ({-\x+9},{ -tan(45)*\x});
   \draw [domain=1:1.919] plot ({\x+9},{ tan(45)*\x});
   \draw [domain=1:1.919] plot ({\x+9},{- tan(45)*\x});
   \draw (10.919,-1.919) -- (12.219,-1.919);
   \draw (10.919,-1.919) -- (10.919,-3.219);
   \draw[blue, thick,  postaction={mid arrow2}] (10,-1) --(10,1);
   \node[blue] at (10.5,0) {$\tilde\ell$};
   \node at (11.2,2.2) {\footnotesize $ 1$};
   \node at (12.5,-1.9) {\footnotesize $2$};
   \node at (10.9,-3.6) {\footnotesize $3$};
   \node at (6.8,-2.2) {\footnotesize $4$};
   \node at (6.8,2.2) {\footnotesize $5$};
   \node at (14,0) {$+$};
   \draw (18,-1) -- (20,-1) -- (20,1) -- (18,1) -- (18,-1);
   \draw [domain=1:1.919] plot ({-\x+19},{ tan(45)*\x});
   \draw [domain=1:1.919] plot ({-\x+19},{ -tan(45)*\x});
   \draw [domain=1:1.919] plot ({\x+19},{ tan(45)*\x});
   \draw [domain=1:1.919] plot ({\x+19},{- tan(45)*\x});
   \draw (17.081,-1.919) -- (15.781,-1.919);
   \draw (17.081,-1.919) -- (17.081,-3.219);
   \draw[blue, thick,  postaction={mid arrow2}] (20,-1) --(20,1);
   \node[blue] at (20.5,0) {$\tilde\ell$};
   \node at (21.2,2.2) {\footnotesize $ 1$};
   \node at (21.2,-2.2) {\footnotesize $2$};
   \node at (17.1,-3.6) {\footnotesize $3$};
   \node at (15.5,-1.9) {\footnotesize $4$};
   \node at (16.8,2.2) {\footnotesize $5$};
   \node at (23,0) {$+$};
   \draw (27,-1) -- (29,-1) -- (29,1) -- (27,1) -- (27,-1);
   \draw [domain=1:1.919] plot ({-\x+28},{ tan(45)*\x});
   \draw [domain=1:1.919] plot ({-\x+28},{ -tan(45)*\x});
   \draw [domain=1:1.919] plot ({\x+28},{ tan(45)*\x});
   \draw [domain=1:1.919] plot ({\x+28},{- tan(45)*\x});
   \draw (26.081,1.919) -- (24.781,1.919);
   \draw (26.081,1.919) -- (26.081,3.219);
   \draw[blue, thick,  postaction={mid arrow2}] (29,-1) --(29,1);
   \node[blue] at (29.5,0) {$\tilde\ell$};
   \node at (30.2,2.2) {\footnotesize $ 1$};
   \node at (30.2,-2.2) {\footnotesize $2$};
   \node at (26.1,3.6) {\footnotesize $5$};
   \node at (24.5,1.9) {\footnotesize $4$};
   \node at (25.8,-2.2) {\footnotesize $3$};
  \end{tikzpicture}  
    \end{center}
      \caption{Conventions for BCFW shift with \emph{unshifted} loop momentum, used in previous literature.}
    \end{subfigure}
  \caption{Diagrams contributing to the five-point MHV amplitude, with two different conventions for the definition of the loop momentum. In both cases, we have chosen the conventions such that only one term per diagram contributes  in the forward limit.
  }
  \label{fig:placement_loop}
  \end{figure}

In this subsection, we will compare the version of the recursion described earlier in this paper with an alternative momentum space approach, in which
the loop momentum is not shifted; the most relevant references are \cite{Boels:2010nw,Boels:2011mn,Boels:2016jmi}. In this alternative approach, we only shift external legs
$1$ and $n$ as follows:
\begin{equation}
\hat{k}_{1}=k_{1}+zq,\qquad \hat{k}_{n}=k_{n}-zq,\qquad q=\lambda_{1}\tilde{\lambda}_{n}.\label{eq:momdef}
\end{equation}
We will follow the conventions in the aforementioned references and take the loop momentum, which we now denote as $\tilde{\ell}$ for clarity,
to be located between legs $2$ and $1$. In our earlier approach, we denoted as $\ell$ the loop momentum located between legs $n$ and $1$, so now that is equivalent  to $-(\tilde\ell+k_1)$, as illustrated
in Figure \ref{fig:placement_loop}; we remind the reader that the external momenta are incoming.
This BCFW deformation with unshifted loop momentum yields a recursion relation for the one-loop
integrand which is very similar to the one derived in section \eqref{contourplanar}, and
in the end the two integrands can be mapped into each other by redefining
the loop momentum as $\tilde{\ell}=-(\ell+k_{1}).$

Since the derivation of the recursion is very similar to
the one carried out in section~\ref{contourplanar}, we will not repeat the details but
rather describe it schematically and point out the main differences. We denote the deformation of the loop integrand under the shift \eqref{eq:momdef} as $\mathfrak{I}_{n}^{(1)}(z)$.
The undeformed integrand can then be obtained via the contour integral
\[
\mathfrak{I}_{n}^{(1)}=\oint_{z=0}\frac{dz}{2\pi i\,z}\, \mathfrak{I}_{n}^{(1)}(z).
\]
Compactifying the complex $z$ plane and wrapping the contour then picks up three types of poles, which correspond to a single cut, tree-level factorisation channels, and possibly a pole
at infinity. The analysis of the latter two types of poles proceeds
as before, but the analysis of the first type is slightly different,
so we will describe it in more detail. In this case, the cut occurs when
$(\tilde{\ell}+\hat{k}_{1})^{2}=0$. Writing the terms in $\mathfrak{I}_{n}^{(1)}(z)$
which contain this pole as 
\[
\frac{1}{(\tilde{\ell}+\hat{k}_{1})^{2}}\,\mathscr{I}_{n}^{(1)}(z)=\frac{1}{2\tilde{\ell}\cdot q\left(z-z_{0}\right)}\,\mathscr{I}_{n}^{(1)}(z),\qquad z_{0}=-\frac{(\tilde{\ell}+k_{1})^{2}}{2\tilde{\ell}\cdot q},
\]
we find that the corresponding residue is given by
\[
\oint_{z=z_{0}}\frac{dz}{2\pi i\,z}\,\mathfrak{I}_{n}^{(1)}(z)=\frac{1}{(\tilde{\ell}+k_{1})^{2}}\,\mathscr{I}_{n}^{(1)}(z_{0}).
\]
Note that at $z=z_{0}$ the propagator between legs $1$ and $n$
becomes on-shell, so $\mathscr{I}_{n}^{(1)}(z_{0})$ can be interpreted
as a forward limit,
\[
\mathscr{I}_{n}^{(1)}(z_{0})=\sum_{{\rm states}}\mathcal{A}_{n+2}^{(0)}(z_{0}),
\]
where the sum is over states running through the cut. Hence, we
find the following recursion for the one-loop integrand:
\begin{equation}
\mathfrak{I}_{n}^{(1)}=\sum_{I,{\rm states}}\mathcal{A}_{n_{I}}^{(0)}\,\frac{1}{K_{I}^{2}}\,\mathfrak{I}_{n-n_{I}+1}^{(1)}(z_{I})\,+\,\frac{1}{(\tilde{\ell}+k_{1})^{2}}\sum_{{\rm states}}\!\mathcal{A}_{n+2}^{(0)}(z_{0})\,+\,\mathcal{B}_{n},\label{eq:rec2}
\end{equation}
where the sum in the first term runs over tree-level factorisation
channels and the states running through them, and the last term is
a boundary term associated to a possible pole at infinity. Note that this
is essentially the same recursion as \eqref{eq:planarrecursion}. In fact, the two integrands can be mapped into each other by the change of variables $\tilde{\ell}=-\ell-k_{1}$.
Indeed, if we parametrise $\ell=\ell_{0}+\alpha q$, we find that
\[
z_{0}=\ell^{2}/2\ell\cdot q=\alpha.
\]

Let us illustrate the recursion in \eqref{eq:rec2} by computing the five-point one-loop amplitude
in $\mathcal{N}=4$ SYM. (Previous work in this approach only dealt with the four-point amplitude.) In this case, there is no boundary term and the recursion yields a forward-limit term and a tree-level factorisation term. The latter term takes the form of a one-loop four-point amplitude dressed with
a soft factor, so we will refer to it as the `soft term'. Using the
numerators in \eqref{4ptex} and \eqref{5ptex}, one finds that the forward limit can be described in terms of
a pentagon and three one-mass box integrals. The analysis is very similar
to section \ref{5ptmhv}, so we will not describe all the details but rather illustrate
how the spurious pole cancellation works. The term in
the forward limit which contains a spurious pole is given by 
\[
\mathfrak{I}_{4,5}^{\scalebox{0.8}{FL}}=\frac{N_{[4,5]123}^{{\rm box}}}{s_{4\hat{5}}\,\tilde{\ell}^{2}(\tilde{\ell}+k_{1})^{2}(\tilde{\ell}-k_{2})^{2}(\tilde{\ell}-k_{23})^{2}}\,,
\]
where $s_{4\hat{5}}=s_{45}(1-z_{0}/z_{4})$ with $z_{4}=\left\langle 45\right\rangle /\left\langle 41\right\rangle$\,,
and the soft term is given by
\[
\mathfrak{I}_{5}^{{\rm soft}}=\frac{N_{[4,5]123}^{{\rm box}}}{s_{45}\,\tilde{\ell}^{2}(\tilde{\ell}+\check{k}_{1})^{2}(\tilde{\ell}-k_{2})^{2}(\tilde{\ell}-k_{23})^{2}},\qquad \check{k}_{1}=k_{1}+z_{4}q.
\]
Noting that
\[
(\tilde{\ell}+\check{k}_{1})^{2}=(\tilde{\ell}+k_{1})^{2}(1-z_{4}/z_{0}),
\]
we find that the spurious pole cancels out as before,
\[
\mathfrak{I}_{5}^{\scalebox{0.8}{FL}}+\mathfrak{I}_{5}^{{\rm soft}}=\frac{N_{[4,5]123}^{{\rm box}}}{s_{45}\,\tilde{\ell}^{2}(\tilde{\ell}+k_{1})^{2}(\tilde{\ell}-k_{2})^{2}(\tilde{\ell}-k_{23})^{2}}.
\]
Moreover, making the change of variables $\tilde{\ell}=-\ell-k_{1}$ reproduces
the same loop integrand as section \ref{5ptmhv}.

We see that the two versions of the recursion are equivalent. In fact, using figure~\ref{recursionfig} for a comparison between the two, it turns out that the loop momentum here is $\tilde{\ell}=x_0-x_2$, whereas in the recursion defined in previous sections we took the more symmetrical choice $\ell=x_0-x_1$.

\subsection{Recursion for the integrated all-plus amplitude} \label{integralrecursion}

In this section, we will show that the recursion relation for the all-plus integrand in \eqref{allplusr} implies the recursion for the integrated one-loop amplitude described in \cite{He:2014bga}:
\begin{equation}
\mathcal{A}(1^{+},...n^{+})=\mathcal{A}^{(0)}\left(\check{n}^+,-\check{K}^-,n-1^+\right)\,\frac{1}{K^{2}}\,\mathcal{A}_{n-1}^{(1)}\left(\check{K}^+,\check{1}^+,2^+,...,n-2^+\right)\,+\,\mathcal{B}_{n},
\label{intrecur}
\end{equation}
where $\check{K}=k_{n-1}+\check{k}_{n}$ is null, $K=k_{n-1}+k_{n}$, and
$\mathcal{B}_{n}$ is the boundary term\footnote{To compare to the results in \cite{Bern:1996ja}, we must divide by $(2\pi)^{4-2\epsilon}$ since this factor was not included in our loop integration measure.}
\begin{equation}
\mathcal{B}_{n}=\frac{i \pi^{2}}{3}\frac{1}{\left\langle 12\right\rangle ...\left\langle n-11\right\rangle }\sum_{1\leq i<j\leq n-2}\left\langle ij\right\rangle \left[in\right]\left[jn\right].
\label{boundaryterm}
\end{equation}
The recursion relation in \eqref{intrecur} was deduced by BCFW shifting legs 1 and $n$ of the known one-loop all-plus amplitude, which is a rational function first conjectured in \cite{Bern:1993qk} and proven in \cite{Mahlon:1993si} using off-shell recursion \cite{Berends:1987me}. More recently, the same form of the recursion relation was used to prove a new formula for all-plus amplitudes, based on conformally invariant building blocks \cite{Henn:2019mvc}.

Comparing \eqref{intrecur} and \eqref{allplusr}, it is easy to check that the tree factorisation terms match. Therefore, the boundary term in \eqref{intrecur} can be interpreted as
the forward-limit term in \eqref{allplusr}, after loop integration. For notational brevity, we will denote here the four-dimensional part of the loop momentum as $\bar\ell$. We must have
\begin{equation}
\mathcal{B}_{n}=2 \int d^{4}\bar\ell \, d^{-2\epsilon}\mu\;
\frac{\mu^{2}}{\left\langle 12\right\rangle ...\left\langle n-1\hat{n}\right\rangle }\,
\frac{\left[\hat{1}\right|\Pi_{i=2}^{n-1}\left(D_{i}-k_{i}\,\bar L_i\right)\left|n\right]}{D_{0}... D_{n-1}},
\label{flintegral}
\end{equation}
where we plugged in \eqref{treescal} and
\[
\bar L_j=\bar\ell+\sum_{i=1}^{j}k_{j},\qquad D_{j}=\bar L_j^2-\mu^{2}.
\]
We used here a notation more convenient for separating the four- and extra-dimensional parts of the loop momentum.
It is not difficult to verify formula \eqref{flintegral} for $\mathcal{B}_n$ using direct integration \cite{Boels:2016jmi}. First note that using the dimension-shifting
formula \cite{Bern:1995db}, the loop integration can be taken to be in six dimensions,
\[
\int d^{4}\bar\ell \, d^{-2\epsilon}\mu\,\mu^{2}\,f(\bar\ell^\sigma,\mu^2)=-\frac{\epsilon}{\pi}\int d^{6-2\epsilon}\ell\, \,f(\bar\ell^\sigma,\mu^2).
\]
Moreover, we have
\[
\left\langle n-1\hat{n}\right\rangle =-\left\langle n-11\right\rangle (\alpha-z_n), \qquad [\hat1| = [1|+\alpha[n|,
\]
with
\[
z_n={\left\langle n-1n\right\rangle}/{\left\langle n-11\right\rangle}\,,
\qquad \alpha=D_{0}/\left\langle 1\right|\bar\ell\left|n\right]=\ell^2/\left\langle 1\right|\bar\ell\left|n\right].
\]
Using these relations, we find that the right-hand side of \eqref{flintegral} is
\begin{equation}
\frac{2\epsilon}{\pi\left\langle 12\right\rangle ...\left\langle n-11\right\rangle }\int\frac{d^{6-2\epsilon}\ell}{
(\alpha-z_n)D_0...D_{n-1}}\Big(
\left[1\right|\Pi_{k=2}^{n-1}(D_{i}-k_{i}\,\bar L_i)\left|n\right]+\alpha\left[n\right|\Pi_{k=2}^{n-1}(D_{i}-k_{i}\,\bar L_i)\left|n\right]\Big).
\label{2int}
\end{equation}
The first and second terms in this expression are naively linearly and quadratically divergent, respectively, but it is not difficult to show by converting to Feynman parameters that both terms are actually logarithmically divergent, i.e., $\mathcal{O}(1/\epsilon)$, and therefore give finite contributions after multiplying by the $\epsilon$ out front. In the first term, we therefore only need to keep terms with $(n-3)$ $D_{i}$'s and one $k_{i}\bar L_i$ when expanding the product in the numerator. All other terms will give vanishing contributions as $\epsilon\rightarrow0$. Integrating the first term and setting $\epsilon=0$ then gives a result which vanishes by momentum conservation. Similarly, for the second term we only need to keep $(n-4)$ $D_{i}$'s and two $k_{i}\bar L_i$'s
in the expansion of the numerator. Integrating the second term and
setting $\epsilon=0$ then gives 
\[
\frac{i \pi^2}{3}\frac{1}{\left\langle 12\right\rangle ...\left\langle n-11\right\rangle }\sum_{2\leq i<j\leq n-1}\left\langle ij\right\rangle \left[in\right]\left[jn\right],
\]
which is equivalent to \eqref{boundaryterm} due to momentum conservation. The boundary term in the all-plus recursion \eqref{intrecur} thus corresponds to 
the integrated forward-limit term \eqref{allplusr}.

%%%%%%%%%%%%%%%%%%%%%%%%%%%%%%%%
%%%%%%%%%%%%%%%%%%%%%%%%%%%%%%%%
\section{BCFW recursion for non-planar Yang-Mills and gravity} \label{nonplanar}

In this section, we discuss the one-loop BCFW recursion beyond the planar case. The residue argument is analogous to the planar story, seen in section~\ref{contourplanar}, but there are important differences. To start with, there are two ways to proceed, depending on whether we consider a single BCFW shift for the full non-planar integrand, or different BCFW shifts adapted to different (planar-like) parts of the integrand. We recall that some of the comments on boundary terms in section~\ref{boundary} apply to the non-planar case, in particular at the end of that section.

\subsection{Single BCFW shift} \label{contournonplanar} 

The first option is that of a single BCFW shift. We proceed as in the planar case, except that (i) there is no colour ordering, so there will be more tree-like poles, and (ii) the loop momentum cannot be chosen globally to lie between legs $n$ and 1 as before, so there will be more single cut contributions. We still obtain an expression of the same form,
\begin{align}
\boxed{
\;\;\mathfrak{I}_n^{(1)} \;\;=\; \sum_{I,\text{\,states}_I}\cA^{(0)}_{n_I+1}(z_I)\;\frac{1}{K_I^2}\; {\mathfrak{I}}_{n-n_I+1}^{(1)}(z_I) \;\;+
\sum_{J_\ell,\text{\,states}_{J_\ell}} \frac{1}{(\ell+K_J)^2}\;\cA^{(0,\text{reg})}_{n+2}(z_{J_\ell})
\;\;+\;\; {\mathcal B}_n \,,\;}
\label{eq:nonplanarrecursion}
\end{align}
where $(\hat \ell+\hat K_J)^2=0$ for $z=z_{J_\ell}$. This is the option taken in \cite{Boels:2010nw,Boels:2016jmi}.

We will illustrate the formula above using the BCFW shift considered in previous sections. Recall that we parametrised the loop momentum as $\ell=\ell_0+\alpha\, q$, with null momenta $\ell_0$ and $q$, such that $k_1\cdot q=k_n\cdot q=0$; for the latter conditions, we took $q=\lambda_1\tilde \lambda_n$, in the case of four-dimensional external momenta. We found it useful to formulate the BCFW residue argument by also shifting $\alpha$, as in \eqref{eq:shift}. This decomposition is not well adapted to the non-planar loop integrand, since we cannot choose the loop momentum to lie between legs $n$ and 1 beyond the planar case. We can, however, still choose the loop momentum to lie next to particle 1, which ameliorates the problem. To illustrate this, let us consider first a scalar box integrand with ordering $1324$,
\begin{equation}
\frac1{\ell^2(\ell+k_1)^2(\ell+k_{13})^2(\ell+k_{132})^2} \,.
\end{equation}
Given the shift \eqref{eq:shift}, the only pole in the residue argument is at $\hat\ell^2=0$ with $z=\alpha$, since the shift cancels in the other denominator factors, $\hat\ell+k_{\hat1...}=\ell+k_{1...}$. This is analogous to the situation we had in previous sections, in the planar case with ordering $1234$, since a single forward-limit contribution arises from this scalar box; indeed, the loop momentum still lies between legs 1 and 4. In \eqref{eq:nonplanarrecursion}, this forward-limit contribution  corresponds to the second term, with $K_J=0$.

Now, let us consider a scalar box integrand with planar ordering $1342$,
\begin{equation}
\frac1{\ell^2(\ell+k_1)^2(\ell+k_{13})^2(\ell+k_{134})^2} \,.
\end{equation}
Under the same shift as before, with $q=\lambda_1\tilde \lambda_n$, two forward-limit terms arise now: either $\hat\ell^2=0$ with $z=\alpha$, or $(\hat\ell+k_{\hat13\hat4})^2=0$ with $z=(\ell-k_2)^2/(\ell-k_2)\cdot q$. In \eqref{eq:nonplanarrecursion}, these correspond to $K_J=0$ and $K_J=k_{134}$, respectively. Hence, we get two contributions from the second term in \eqref{eq:nonplanarrecursion}. The first forward-limit contribution, from $z=\alpha$, fits nicely the decomposition $\ell=\ell_0+\alpha\,q$ of the loop momentum used in previous sections, but the second forward-limit contribution, with $z=(\ell-k_2)^2/(\ell-k_2)\cdot q$, does not. The latter case is generic in the non-planar recursion.
It is easy to see that a scalar box integrand with ordering $1423$ would give three forward-limit contributions. More generally, defining the loop momentum with $q=\lambda_1\tilde \lambda_n$ to lie next to particle 1, we get $m+1$ forward-limit contributions, where $m$ equals the number of particles between legs $n$ and 1 along the cyclic ordering.

Naturally, we could further reduce the number of forward-limit contributions by redefining the loop momentum appropriately. For instance, in the example with ordering $1423$, it would make sense to apply the shift to the loop momentum $\tilde\ell=-(\ell+k_1)$, as it would then yield a single forward-limit contribution. However, we no longer have a single BCFW shift in that case, and this reasoning leads to a decomposition of the integrand into planar-like parts, which takes us into the next subsection.

\subsection{Multiple BCFW shifts versus non-planar factorisation}

The second option is to use different BCFW shifts for planar-like sub-sectors of the loop integrand. This is particularly natural for Yang-Mills theory, or any other theory with only adjoint states. Then, we can use the Del Duca-Dixon-Maltoni colour decomposition \cite{DelDuca:1999rs} to write the colour-dressed loop integrand as
\begin{equation}
\mathfrak{I}^\text{(1),YM}_{n} = \sum_{\rho\in (S_{n}/\mathbb{Z}_n)/\mathcal{R}} c^\text{(1)}\big(\rho(1),\cdots,\rho(n)\big) \; \mathfrak{I}^\text{(1),YM}\big(\rho(1),\cdots,\rho(n)\big)  \,,
\label{eq:DDMYM}
\end{equation}
where $S_{n}/\mathbb{Z}_n$ denotes non-cyclic permutations, and $\mathcal R$ denotes reflection. The colour factors are defined as
\begin{equation}
c^\text{(1)}\big(\rho(1),\cdots,\rho(n)\big) = f^{b_0 a_{\rho(1)}b_1}f^{b_1a_{\rho(2)}b_2}\cdots f^{b_{n-1}a_{\rho(n)}b_0}\,.
\label{eq:ccoldef}
\end{equation}
The `partial integrands' obey the colour ordering $\big(\rho(1),\cdots,\rho(n)\big)$ of the external particles. It is clear, then, that we can use BCFW shifts adapted to each term and its colour ordering. In particular, $k_{\rho(1)}$ and $k_{\rho(n)}$ are shifted, $q$ is chosen accordingly, and the loop momentum lies between ${\rho(n)}$ and ${\rho(1)}$ in each term.\footnote{In our convention, $\rho$ denotes a permutation, and $\rho(i)$ is the particle in position $i$ in that permutation. Therefore, $\rho(n)$ and $\rho(1)$ are adjacent particles.}
 We are left with a recursion like \eqref{eq:planarrecursion} for each term $\mathfrak{I}^\text{(1),YM}\big(\rho(1),\cdots,\rho(n)\big)$ in \eqref{eq:DDMYM}, with a single forward-limit contribution from each term. We can also minimise the number of distinct BCFW shifts, using the cyclic symmetry of the partial amplitudes, by defining the loop momentum in each partial integrands such that $\ell$ lies just before leg 1, as in the planar case of the earlier sections. The convenient shift for each permutation $\rho$ is associated to a loop momentum $\ell=\ell_0+\alpha\,q$ with $q=\lambda_1\tilde\lambda_{\rho(i-1)}$, where $i=\rho^{-1}(1)$, i.e., $\rho(i-1)$ denotes the particle before 1 in the ordering. Therefore, we only need $n-1$ distinct shifts. Recall that, in the case of MHV integrands in $\mathcal{N}=4$ super Yang-Mills, we employed in section~\ref{sec:bdy_MHV} a representation where the (BCJ) kinematic numerators of all diagrams were invariant under a BCFW shift with $q=\lambda_1\tilde\lambda_j$ for any $j\neq 1$, if particle 1 is chosen as the reference particle in defining the numerators. This representation of the integrand is still very well adapted to the BCFW recursion in the non-planar case, along the lines discussed here.  

For generic theories, we can try to decompose the loop integrand into contributions with fixed particle ordering, and apply the same reasoning. In the case of double copy theories, such a decomposition is readily available if one-loop BCJ numerators are known \cite{Bern:2011rj}. For instance, for gravity, we can use
\begin{align}
\mathfrak{I}^\text{(1),grav}_{n} & = \sum_{\rho\in (S_{n}/\mathbb{Z}_n)/\mathcal{R}}  \tilde N^\text{(1),YM}\big(\rho(1),\cdots,\rho(n)\big)  \;\;
\mathfrak{I}^\text{(1),YM}\big(\rho(1),\cdots,\rho(n)\big)
 \,,
 \label{eq:DDMgrav}
\end{align}
where $\tilde N^\text{(1),YM}\big(\rho(1),\cdots,\rho(n)\big)$ are the BCJ numerators for $n$-gon diagrams with the prescribed particle ordering. The tilde means that the choice of states (with the same momenta) is independent between the two factors of the double copy, here in $\tilde N^\text{(1),YM}$ versus $\mathfrak{I}^\text{(1),YM}$. Importantly, when performing the sum over states on a unitarity cut, the gravity sum over states is equivalent to the independent sums over states in  $\tilde N^\text{(1),YM}$ and in $\mathfrak{I}^\text{(1),YM}$. That is, the double copy guarantees consistent gravity factorisations. The structure in \eqref{eq:DDMgrav} was first observed for one-loop MHV amplitudes in $\mathcal{N}=8$ supergravity and all-plus amplitudes in Einstein gravity \cite{Bern:1998sv}. It can also be obtained from on-shell diagrams  in $\mathcal{N}=8$ supergravity, as shown in \cite{Heslop:2016plj} for the one-loop four-point  amplitude.
The comments made for the gauge theory case regarding the minimal set of $n-1$ BCFW shifts, with $q=\lambda_1\tilde\lambda_j$, where the loop momentum lies between particles $j$ and 1 in the relevant ordering, still hold. In fact, for MHV amplitudes in $\mathcal{N}=8$ supergravity, the BCJ numerators reviewed in section~\ref{sec:bdy_MHV} are then unaffected by the shift, since they are defined with respect to particle 1 as the reference particle. Therefore, this set of BCJ numerators plays a role in  \eqref{eq:DDMgrav} that is entirely analogous to that of the colour factors in \eqref{eq:DDMYM} for gauge theory, from the point of view of the recursion.

This discussion may suggest that the non-planar case follows more or less straightforwardly from the planar case. The issue that we will illustrate here, however, is that planar BCFW factorisation does not recombine directly via the decompositions  \eqref{eq:DDMYM} and  \eqref{eq:DDMgrav} into non-planar BCFW factorisation. As an example, let us focus on ${\mathcal N}=4$ SYM at five points.
\begin{align}
\mathfrak{I}^\text{(1),YM}_5 \supset \;\; & c^\text{(1)}\big(1,2,3,4,5\big) \; \mathfrak{I}^\text{(1),YM}\big(1,2,3,4,5\big)  +
 c^\text{(1)}\big(1,2,3,5,4\big) \; \mathfrak{I}^\text{(1),YM}\big(1,2,3,5,4\big)  \,.
\end{align}
The colour factors correspond to pentagons with the prescribed ordering. Before talking about the BCFW shift, suppose that the loop momentum lies between 5 (4) and 1 in the first (second) term, and we want to look at the contributions corresponding to boxes with a massive corner $1/s_{45}$: these are
\begin{subequations}
\begin{align}
&c^\text{(1)}\big(1,2,3,4,5\big) \;
\int \frac{d^D\ell}{\ell^2} \;
\frac{ N^{\text{box}}_{[4,5]123}   }{s_{45}\, (\ell+k_1)^2  (\ell+k_{12})^2 (\ell+k_{123})^2} \\
%
%&\mathrm{and} &
&c^\text{(1)}\big(1,2,3,5,4\big) \; 
\int \frac{d^D\ell}{\ell^2} \;
\frac{ - N^{\text{box}}_{[4,5]123}   }{s_{45}\, (\ell+k_1)^2  (\ell+k_{12})^2 (\ell+k_{123})^2} 
\,,
\end{align}
\end{subequations}
where $\, N^{\text{box}}_{[5,4]123}=- N^{\text{box}}_{[4,5]123}\,$. Then the two terms combine into
\begin{equation}
c^\text{(1)}\big(1,2,3,[4,5]\big) \; 
\int \frac{d^D\ell}{\ell^2} \;
\frac{ N^{\text{box}}_{[4,5]123}   }{s_{45}\, (\ell+k_1)^2  (\ell+k_{12})^2 (\ell+k_{123})^2} 
\,,
\end{equation}
where we get the appropriate colour factor for that box with massive corner,
\begin{eqnarray}
c^\text{(1)}\big(1,2,3,4,5\big)-
c^\text{(1)}\big(1,2,3,5,4\big)
= c^\text{(1)}\big(1,2,3,[4,5]\big) =
f^{b_0 a_{1}b_1}f^{b_1a_{2}b_2}f^{b_2a_{3}b_3}f^{b_3b_4b_0} f^{b_4a_{4}a_5}  
\,.
\end{eqnarray}
Now, consider the residue argument, starting from, respectively,
\begin{equation}
\hat\ell=\ell_0 +(\alpha-z)\,q=\ell_0 +(\alpha-z)\,| 1\rangle [5| \,, \qquad \hat\ell'=\ell_0' +(\alpha'-z')\,q'=\ell_0' +(\alpha'-z')\,| 1\rangle [4| \,.
\end{equation}
Each planar ordering has a forward-limit term and a factorisation term, exactly as discussed for the planar case. The factorisation terms are
\begin{subequations}
\begin{align}
&c^\text{(1)}\big(1,2,3,4,5\big) \;
\int \frac{d^D\ell}{\ell^2} \;
\frac{\alpha}{\alpha-\frac{\langle 45 \rangle}{\langle  41 \rangle}} \;
\frac{ N^{\text{box}}_{[4,5]123}   }{s_{45}\, (\ell+k_1)^2  (\ell+k_{12})^2 (\ell+k_{123})^2} \\
&c^\text{(1)}\big(1,2,3,5,4\big) \; 
\int \frac{d^D\ell'}{\ell'^2} \;
\frac{\alpha'}{\alpha'-\frac{\langle 54 \rangle}{\langle  51 \rangle}} \;
\frac{- N^{\text{box}}_{[4,5]123}   }{s_{45}\, (\ell'+k_1)^2  (\ell'+k_{12})^2 (\ell'+k_{123})^2} 
\,.
\end{align}
\end{subequations}
Each cancels the spurious pole in the associated forward-limit term. However, it is not obvious whether they can be combined into an object whose colour factor corresponds to a box with massive 4-5 corner, which would be natural from the point of view of non-planar factorisation. This seems to be impossible. It would require that
\begin{equation}
\ell=\ell'\,,  \qquad
\frac{\alpha}{\alpha-\frac{\langle 45 \rangle}{\langle  41 \rangle}}=
\frac{\alpha'}{\alpha'-\frac{\langle 54 \rangle}{\langle  51 \rangle}}
\;\Leftrightarrow\; \alpha'=-\frac{\langle  41 \rangle}{\langle  51 \rangle} \,\alpha
 \,,
\end{equation}
and hence
\begin{equation}
\ell_0'=\ell_0+\alpha\, q- {\alpha'}\,q'=\ell_0+\alpha\,\langle 1| \left(  [5|+\frac{\langle  41 \rangle}{\langle  51 \rangle}\; [4| \right)
 \,.
\end{equation}
Since $\ell_0$ is null and $q\cdot q'=0$, the null condition on $\ell_0'$ is
$$
\alpha\,\ell_0\cdot q= \alpha' \,\ell_0\cdot q' \quad \Leftrightarrow \quad 
\frac{\ell_0\cdot q}{\ell_0\cdot q'} = - \frac{\langle  41 \rangle}{\langle  51 \rangle}
\,,
$$
which does not allow for a generic $\ell_0$. For instance, in $D=4$, with $\ell_0=|0\rangle [0|$, the requirement would be 
\begin{equation}
\frac{[05]}{[04]}=- \frac{\langle  41 \rangle}{\langle  51 \rangle}
\quad \Leftrightarrow \quad
\langle 1 | 4+5 |0]= -\langle 1 | 2+3 |0]=0  \,.
\end{equation}
If it were possible, the two terms would be combined into
\begin{equation}
c^\text{(1)}\big(1,2,3,[4,5]\big) \; 
\int \frac{d^D\ell}{\ell^2} \;
\frac{\alpha}{\alpha-\frac{\langle 45 \rangle}{\langle  41 \rangle}} \;
\frac{ N^{\text{box}}_{[4,5]123}   }{s_{45}\, (\ell+k_1)^2  (\ell+k_{12})^2 (\ell+k_{123})^2} 
\,,
\end{equation}
which is the non-planar massive box with corner 4-5. The fact that this combination seems generically not possible extends to the associated tree factorisation terms, which cancel the spurious poles. This cancellation occurs independently for each planar ordering.  

The factorisation of the planar parts is therefore not straightforwardly aligned with factorisation in the non-planar case, even in this simple MHV example. An analogous statement holds for gravity. In fact, the five-point example above for $\mathcal{N}=4$ super Yang-Mills can be translated immediately to the case of $\mathcal{N}=8$ supergravity via the substitution of $c^\text{(1)}(\cdots)$ by $\tilde N^\text{(1),\text{YM}}(\cdots)$, since the MHV kinematic numerators are unaffected by the BCFW shift.

Regarding gravity, let us make a clarification. In this subsection, in particular in \eqref{eq:DDMgrav}, we relied on the existence of BCJ numerators. This refers to BCJ numerators in the original BCJ loop-level proposal \cite{Bern:2010ue}, based on a loop integrand with quadratic (Feynman) propagators. Beyond the MHV case used here, there are known obstacles to constructing such numerators at higher multiplicity even at one loop \cite{Mafra:2014gja,Berg:2016fui} (although the difficulties may be ameliorated if the numerators are allowed to be non-local \cite{Bjerrum-Bohr:2013iza}). BCJ-type numerators in an integrand representation with `linear' propagators always exist, via a version of the forward limit from tree level, but such a representation was not the basis of our discussion. On the other hand, the formula \eqref{eq:DDMgrav} with quadratic propagators should have an extension to the framework of the generalised BCJ double copy, which does not require BCJ numerators, at the price of a less straightforward double copy prescription \cite{Bern:2017yxu}. We expect that the reasoning above for gravity can be translated into this more general framework, but this is beyond the scope of this paper. 

To conclude, let us consider the implications of this section to the following sections, where we will study worldsheet formulas that use a deformation of the `old' one-loop scattering equations in order to allow for quadratic propagators. Let us revisit the argument presented in the Introduction. There, a BCFW shift involving particles 1 and $n$ inspired an analogous deformation of the scattering equations involving those particles. In this subsection, we claimed that we need $n-1$ distinct BCFW shifts, with $q=\lambda_1\tilde\lambda_j$ and $j\neq 1$, to reproduce the non-planar loop integrand from planar-like parts. This suggests that, in the non-planar case, we will need  $n-1$ deformations of the scattering equations. That is exactly what we will describe in section~\ref{sec:wsquadnp}.

%%%%%%%%%%%%%%%%%%%%%%%%%%%%%%%%
%%%%%%%%%%%%%%%%%%%%%%%%%%%%%%%%
\section{Worldsheet formulas for quadratic propagators: planar case}
\label{sec:wsquad}

Motivated by the loop-level BCFW recursion relation and the connection it suggests between the forward-limit terms and standard Feynman-like propagators, in this section we investigate how local integrand expressions can arise from the worldsheet. Since Witten's seminal work \cite{Witten:2003nn}, several approaches --- building on \cite{Roiban:2004yf} and \cite{Cachazo:2013hca} --- have shown that amplitudes for various massless quantum field theories can be recast as worldsheet integrals with a remarkably simple structure,  revealing  relationships, such as the colour-kinematics duality, which connect a wide range of theories. 
These amplitude representations derive from a set of worldsheet models known as (ambi)twistor strings \cite{Mason:2013sva,Casali:2015vta}, and their characteristic feature is that the worldsheet integrals localise fully onto a universal set of constraints known as the scattering equations. Due to the existence of underlying worldsheet models, the tree-level results have been generalised to one and two loops \cite{Adamo:2013tsa,Adamo:2015hoa,Geyer:2015bja,Geyer:2015jch,He:2015yua, Cachazo:2015aol, Geyer:2016wjx,Geyer:2018xwu, Geyer:2019hnn, Feng:2016nrf}, but the simplest formulas\footnote{The models naturally give the higher-loop integrands as integrals over the moduli space of (marked) higher-genus surfaces. However, at least at one and two loops, these can be simplified drastically, and only receive contributions from a nodal sphere at one loop and (at least with supersymmetry) from a bi-nodal sphere at two loops. We are referring in the main text to the latter formulas, based on nodal spheres rather than higher-genus surfaces.}
result in an unorthodox integrand representation, written
 in terms of loop propagators which are `linear' in the loop momentum, as exemplified in the Introduction. It is therefore natural to ask whether it is possible to obtain standard quadratic integrands from a worldsheet formula -- in fact, this was one of the main motivations for this paper.

One way to approach this question is to map the expressions arising from the BCFW recursion into worldsheet formulas. For on-shell diagrams, this was initiated in  \cite{Farrow:2017eol} with four-point amplitudes, where the Grassmannian integral formulas arising from one-loop on-shell diagrams in $\mathcal{N}=4$ SYM and $\mathcal{N}=8$ supergravity were mapped into supersymmetric worldsheet formulas refined by MHV degree, extending the tree-level formulas in \cite{Geyer:2014fka}. Such one-loop worldsheet formulas, reviewed in Appendix \ref{wsonshell}, elegantly encode the forward-limit term in the one-loop recursion, and therefore by themselves describe only amplitudes at four points, since higher-point amplitudes require also tree-level-type factorisation terms. Extending this approach to higher points while preserving its simplicity seems challenging, since contributions from multiple on-shell diagrams have to be combined. Alternative worldsheet formulas for one-loop integrands with quadratic propagators were proposed in \cite{Gomez:2017lhy, Gomez:2017cpe, Ahmadiniaz:2018nvr, Agerskov:2019ryp} using a double-forward-limit construction, but this formalism has only been successfully applied to certain scalar expressions; we discuss its difficulties in Appendix \ref{sec:DFL}. 

In this section, we describe how to overcome the challenge of obtaining quadratic propagators by modifying the one-loop scattering equations  
in a manner inspired by our formulation of the one-loop BCFW recursion. Using these modified scattering equations, we give  a worldsheet formula for all one-loop MHV integrands in $\mathcal{N}=4$ SYM in the standard Feynman representation. We prove this formula in \cref{sec:BCFW_WS} using the one-loop BCFW recursion of \eqref{eq:MHVrecursion}.
Remarkably, it does not exhibit any of the spurious poles that we encounter at intermediate steps in the BCFW recursion.\footnote{Though the absence of other types of spurious poles which are typical of worldsheet formulations is not immediately manifest. We verify their absence in \S\ref{sec:BCFW_WS}.}  
In \cref{sec:proposal}, we present a natural idea to extend this MHV worldsheet formula to all one-loop integrands for external gluons in maximal super Yang-Mills, both in higher dimensions and in $D=4$ for different MHV sectors. However, we also discuss some caveats of this idea, and the clarification of its validity or necessary corrections is beyond the scope of this paper. We leave for section~\ref{sec:wsquadnp} the construction of the MHV formulas for non-planar $\mathcal{N}=4$ SYM and $\mathcal{N}=8$ supergravity.

%%%%%%%%%%%%%%
\subsection{Brief review of one-loop worldsheet formulas}\label{sec:review_ws}
\begin{figure}[ht]
	\centering 
	  \includegraphics[width=6cm]{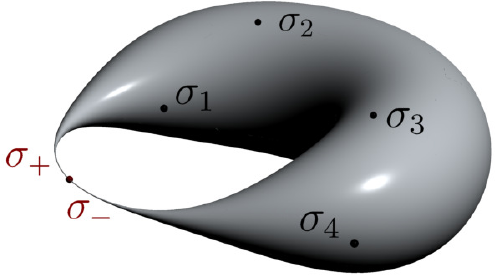}
	\caption{The nodal sphere with four punctures corresponding to external particles. The node is parametrised by two additional marked points $\sigma_+$ and $\sigma_-$.}
	\label{fig:nodal_labelled}
\end{figure}

Ambitwistor strings  \cite{Mason:2013sva, Berkovits:2013xba, Ohmori:2015sha} are the chiral worldsheet models underpinning the renowned CHY formulas  \cite{Cachazo:2013gna,Cachazo:2013hca,Cachazo:2013iea} for tree-level amplitudes in massless field theories. As such, they provide the framework for extending the scattering equations formalism to loop order,\footnote{At least when the worldsheet models satisfy certain constraints, including modular invariance at higher genus, e.g., for maximal supergravity; when they don't, one may still able to extract crucial information to reconstitute the scattering amplitudes.} and indeed correlators give the loop integrand as an integral over the moduli space of higher-genus Riemann surfaces, localised on a generalisation of the scattering equations to higher genus \cite{Adamo:2013tsa,Adamo:2015hoa,Geyer:2018xwu}. While both mathematically and conceptually appealing, the resulting higher-genus  worldsheet formulas are difficult to work with due to the dependence on Riemann theta functions, and many important features of the integrand are obscured. This suggests that a much simpler worldsheet formulation exists, which makes the rationality of the loop integrand manifest, and does not depend on higher-genus objects untypical in field theory integrands. Indeed, such a formulation was found at one loop in  \cite{Geyer:2015bja,Geyer:2015jch}, and was extended to two loops in \cite{Geyer:2016wjx,Geyer:2018xwu}; the extension to higher loops is expected, but remains conjectural. In these papers, it was shown  that  the full higher-genus expression is equivalent to a simpler formula localised on the boundary of the moduli space corresponding to a non-separating degeneration --  a nodal sphere. The equivalence can be established via a global residue theorem on the (compactified) moduli space $\widehat{\mathfrak{M}}_{g,n}$, trading the localisation on certain higher-genus scattering equations for the localisation on the appropriate boundary of the moduli space, which is the only other pole in the moduli space integrand.
The resulting worldsheet formula is based on a Riemann sphere with $g$ nodes, and closely resembles a multi-forward limit of the tree-level CHY formulas, as expected for a field theory loop integrand. 
At one loop, the loop integrand takes the form
\begin{equation}
 \mathfrak{I}^{\scalebox{0.6}{$(1)$}}_n=\frac{1}{\ell^2} \int_{\raisebox{-6pt}{\scalebox{0.7}{$\mathfrak{M}_{0,n+2}$}}}\hspace{-15pt} d\mu_{n} \;
 \cI^{\scalebox{0.6}{$(1)$}}_{\scalebox{0.5}{$1/2$}} \,\tilde{\cI}^{\scalebox{0.6}{$(1)$}}_{\scalebox{0.5}{$1/2$}}\,.
\end{equation}
Here, the measure $d\mu_n$ is defined similarly to the CHY measure at tree level: there is an integral over the $n+2$  marked points $\sigma_a$ for $a=1,2,\cdots,n,+,-$ modulo M\"{o}bius invariance, and the measure contains a set of delta-functions that fully localise the integral over the moduli space $\mathfrak{M}_{0,n+2}$,
\begin{equation}\label{eq:measure_lin}
  d\mu_{n} :=\frac{d^{n+2}\sigma}{\text{vol}\,\text{SL}(2,\C)} \,\,\, {\prod_{a=1}^{n+2}}{}' \bar\delta\left(\mathcal{E}_a\right)\,.
 \end{equation}
The marked points are the punctures $\sigma_1,\dots,\sigma_n$ corresponding to external particles, as well as $\sigma_+$ and $\sigma_-$ parametrising the node, see \cref{fig:nodal_labelled}, and the SL$(2,\mathbb{C})$ quotient allows us to fix three punctures $\sigma_{a_{1,2,3}}$ at the price of a Jacobian $\sigma_{a_1a_2}\sigma_{a_2a_3}\sigma_{a_3a_1}$.
The localisation constraints are the one-loop scattering equations  $\mathcal{E}_a =0$ on the nodal sphere, with
\begin{equation}\label{eq:SE_lin}
\mathcal{E}_i :=\sum_{j\neq i} \frac{2k_i\cdot k_j}{\sigma_{ij}} + 2\ell\cdot k_i\, \frac{\sigma_{+-}}{\sigma_{i+}\sigma_{i-}}\,,
\qquad 
\mathcal{E}_\pm :=\pm \sum_{j} \frac{2\ell\cdot k_j}{\sigma_{\pm j}} \,.
\end{equation}
Of these $n+2$ constraints, only $n-1$ are independent due to the M\"{o}bius SL$(2,\mathbb{C})$ symmetry on the sphere, as is easily verified from $\sum_a \sigma_a^m\mathcal{E}_a=0$ for $m=0,1,2$. 
This symmetry  is essential for the formalism to be well-defined,  and is guaranteed by momentum conservation,
\begin{equation}
\label{eq:momcons_v2}
\sum_{j\neq i} {2k_i\cdot k_j} =0= \sum_j 2\ell\cdot k_i  \,.
\end{equation}
The prime in the product over delta-functions in \eqref{eq:measure_lin} indicates that this symmetry should be accounted for by dropping any three of these constraints $\mathcal{E}_{b_{1,2,3}}$, and including the appropriate Jacobian $\sigma_{b_1b_2}\sigma_{b_2b_3}\sigma_{b_3b_1}$. 

Just as at tree level, this measure is universal, and all theory-dependence is carried by the `half-integrands' $\cI^{\scalebox{0.6}{$(1)$}}_{\scalebox{0.5}{$1/2$}}$ and $\tilde{\cI}^{\scalebox{0.6}{$(1)$}}_{\scalebox{0.5}{$1/2$}}$. While higher-genus worldsheet formulas have only been obtained for maximal supergravity, using colour-kinematics duality and choosing the states running in the loop allow us to extend the results on the nodal sphere to other theories as well \cite{Geyer:2015jch, He:2015yua}. Here, we give only the results for supergravity, super Yang-Mills and the scalar $n$-gon,
\begin{subequations}
\begin{align}
 \label{eq:halfintsugra}
 &\text{Supergravity:} 
 && \cI^{\scalebox{0.6}{$(1)$}}_{\scalebox{0.5}{$1/2$}}= \cI^{\scalebox{0.6}{$(1)$}}_{\scalebox{0.6}{kin}} 
 &&\tilde{\cI}^{\scalebox{0.6}{$(1)$}}_{\scalebox{0.5}{$1/2$}} =\tilde{\cI}^{\scalebox{0.6}{$(1)$}}_{\scalebox{0.6}{kin}} \\
 &
 \label{eq:halfintSYM}
 \text{Super Yang-Mills:} 
 && \cI^{\scalebox{0.6}{$(1)$}}_{\scalebox{0.5}{$1/2$}}=  \cI^{\scalebox{0.6}{$(1)$}}_{\scalebox{0.6}{kin}}
 &&\tilde{\cI}^{\scalebox{0.6}{$(1)$}}_{\scalebox{0.5}{$1/2$}} =\mathcal{C}^{\scalebox{0.6}{$(1)$}}_{\scalebox{0.6}{cyc}}(12\dots n)\\
  \label{eq:halfintngon}
 &n\text{-gon:} 
 && \cI^{\scalebox{0.6}{$(1)$}}_{\scalebox{0.5}{$1/2$}}=\frac{1}{\sigma_{+-}^2}\prod_{i=1}^n\frac{\sigma_{+-}}{\sigma_{+i}\sigma_{i-}}
 && \tilde{\cI}^{\scalebox{0.6}{$(1)$}}_{\scalebox{0.5}{$1/2$}} =\mathcal{C}^{\scalebox{0.6}{$(1)$}}_{\scalebox{0.6}{cyc}}(12\dots n)
 \,,
\end{align}
\end{subequations}
where the various colour and kinematics building blocks are defined via
\begin{align}\label{eq:half-int_lin}
 \mathcal{C}^{\scalebox{0.6}{$(1)$}}_{\scalebox{0.6}{cyc}}(12\dots n)&=\sum_{\rho\in\text{cyc}(12\dots n)} \frac{\tr(T^{a_1}T^{a_2}\dots T^{a_n})}{(+\rho_1\rho_2\ldots \rho_n-)}\,, & 
 \cI^{\scalebox{0.6}{$(1)$,MHV}}_{\scalebox{0.6}{kin}}& =\sum_{\rho\in S_n} \frac{N^{\scalebox{0.6}{$(1)$}}_{\scalebox{0.6}{$\rho$}}}{(+\rho_1\rho_2\ldots \rho_n-)}\,,
\end{align}
and $(\rho_1\rho_2\ldots \rho_n)=\sigma_{\rho_1\rho_2}\sigma_{\rho_2\rho_3}\ldots \sigma_{\rho_n\rho_1}$ are the (inverse) Parke-Taylor factors.
For super Yang-Mills, we have given here the planar colour-ordered integrand; the full colour-dressed integrand would be obtained using
\begin{equation}
 \mathcal{C}^{\scalebox{0.6}{$(1)$}}_{\scalebox{0.6}{full},n}=\sum_{\rho\in S_n} \frac{
 f^{a_0 a_{\rho{1}} b_1}  f^{b_1 a_{\rho{2}} b_2} \ldots f^{b_{n-1} a_{\rho{n}} a_0}
 }{(+\rho_1\rho_2\ldots \rho_n-)}\,.
\end{equation}
The kinematic integrand $\cI^{\scalebox{0.6}{$(1)$}}_{\scalebox{0.6}{kin}} $ can be expressed in general as a sum over Pfaffians -- a remnant of the sum over spin structures at higher genus -- which we will discuss  in  more detail in \S\ref{sec:proposal}.
However, we can also express it as in \eqref{eq:half-int_lin}; the two forms are equivalent on the support of the scattering equations. In the latter form, it precisely mirrors the structure of the full colour building block $ \mathcal{C}^{\scalebox{0.6}{$(1)$}}_{\scalebox{0.6}{full},n}$. This is a clear manifestation of the colour-kinematics duality, since both can be written as a sum over `half-ladder' Parke-Taylor factors,  dressed with BCJ numerators $N^{\scalebox{0.6}{$(1)$}}_{\scalebox{0.6}{$\rho$}}$ in the case of $\cI^{\scalebox{0.6}{$(1)$}}_{\scalebox{0.6}{kin}} $, and dressed with colour factors in the case of $ \mathcal{C}^{\scalebox{0.6}{$(1)$}}_{\scalebox{0.6}{full},n}$.
We will mostly consider the detailed form of the kinematic integrand $\cI^{\scalebox{0.6}{$(1)$}}_{\scalebox{0.6}{kin}} $ in $D=4$ for an MHV amplitude, where it simplifies significantly. The MHV numerators were derived from string theory in \cite{He:2015wgf}, and are known to all multiplicities. We have seen low-multiplicity examples in \cref{sec:BCFW_MHV}.\\

\iffalse
While we refer to \cite{Geyer:2015jch, He:2015yua} for integrands for pure Yang-Mills, pure gravity in $D=4$ and  the bi-adjoint scalar,  it is  worth highlighting a useful toy example: the (ordered) $n$-gon, with integrand
\begin{align}
 &n\text{-gon:} 
 && \cI^{\scalebox{0.6}{$(1)$}}_{\scalebox{0.5}{$1/2$}}=\frac{1}{\sigma_{+-}^2}\prod_{i=1}^n\frac{\sigma_{+-}}{\sigma_{+i}\sigma_{i-}}
 && \tilde{\cI}^{\scalebox{0.6}{$(1)$}}_{\scalebox{0.5}{$1/2$}} =\mathcal{C}^{\scalebox{0.6}{$(1)$}}_{\scalebox{0.6}{cyc}}(12\dots n)\,.
\end{align}
We will use $n$-gon expressions when calculating low-multiplicity integrands below.\fi

An important issue with these worldsheet formulas, as previously mentioned, is that they yield loop integrands in an unorthodox representation. This can be shown systematically from factorisation as done in \cite{Geyer:2015jch}, but is also intuitive from the form of the scattering equations, which determine all possible poles, and do not depend on $\ell^2$. To illustrate the point, consider a simple example: the four-particle super Yang-Mills integrand, which is a single (ordered) box, with prefactor $st A^\text{tree}(1234)$. From the worldsheet, we find
\begin{equation}\label{eq:SYM_4pt_lin}
  \mathfrak{I}^{\lin}_{4}=-\delta^8(Q)\frac{\left[12\right]\left[34\right]}{\la 12\ra \la 34\ra} \,
 \frac{1}{\ell^{2}}
 %\frac{st A^\text{tree}(1234)}{\ell^{2}}
 \; \sum_{\rho\in \text{cyc}(1234)}\frac{1}{(2\ell\cdot k_{\rho_1})\,(2\ell\cdot k_{\rho_1\rho_2}+2k_{\rho_1}\cdot k_{\rho_2})\,(-2\ell\cdot k_{\rho_4})}\,,
\end{equation}
with $k_{\rho_1\dots\rho_i} =\sum_{j=1}^i k_{\rho_j}$,
where we have included the superscript to stress that this is a `linear representation', with propagators of the form $2\ell\cdot k +k^2$. In this particular example, the relation between this linear integrand and the usual Feynman integrand is straightforward, and just relies on a partial fraction identity \cite{Geyer:2015bja}
\begin{equation}
 \frac{1}{\prod_{i}D_i}=\sum_{i=1}^n\frac{1}{D_i\prod_{j\neq i}(D_j-D_i)}
\,,
\quad\quad \text{with Feynman propagators }\;
1/D_i=1/(\ell+k_{\rho_1\dots\rho_i})^2\,,
\end{equation}
as well as shifts in the loop momentum between different terms on the right-hand side, such that $1/D_i \leadsto 1/\ell^2$. When the numerators depend on the loop momentum, the relationship is more involved, and one systematic procedure on how to obtain the linear representation from the standard Feynman description has been developed in \cite{Baadsgaard:2015voa}, which introduced the notion of `Q-cuts'.\footnote{Strictly speaking, Q-cuts are obtained after an additional step, involving the scaling of the loop momentum. They provide fully on-shell building blocks for loop integrands in the linear representation, much like the BCFW construction provides on-shell building blocks for loop integrands in the quadratic representation.} However, so far there is no systematic approach to invert this procedure, i.e., to go from linear propagators to Feynman propagators. This makes loop integration techniques difficult to apply to the integrands obtained from the worldsheet. Over the next section, we will circumvent this difficulty; instead of trying to construct Feynman propagators from linear ones, our approach will be to modify the worldsheet formulas to directly produce Feynman propagators.

%%%%%%%%%%%%%%
\subsection{Scattering equations for Feynman propagators}
\label{sec:SE}
Let us revisit the example discussed in the Introduction, and consider an $n$-gon in the linear representation, such as the box in \eqref{eq:SYM_4pt_lin} above. Heuristically, it is clear what needs to be done to turn such an expression back to Feynman propagators: we need to select a single term in the cyclic sum -- where each term corresponds to a placement of the loop momentum -- and then adjust the propagator poles accordingly. For a planar ordering, there is a very simple way to realise this, reminiscent of the BCFW shift, by simply replacing
 \begin{equation}\label{eq:subst_ell}
   2\ell\cdot k_1\mapsto (\ell+k_1)^2\qquad \text{ and }\qquad 2\ell\cdot k_n \mapsto-(\ell-k_n)^2\,.
  \end{equation}
While the above substitution looks very ad-hoc when implemented at the level of the loop integrands, it should only be taken as a motivation for how to modify the worldsheet formulas in order to obtain Feynman propagators. Since the poles of the loop integrand are determined completely by the form of the scattering equations,  \eqref{eq:subst_ell} instructs us on how to modify the constraints $\mathcal{E}_a=0$. Below, we will first discuss the form of these new $\ell^2$-deformed scattering equations, and then we will discuss in a bit more detail why this gives standard Feynman propagators. Before proceeding, however, let us notice that, in \eqref{eq:subst_ell}, we can straightforwardly introduce the usual $i\epsilon$-prescription for Feynman propagators, by taking $\ell^2+i\epsilon$ instead of $\ell^2$ on the right-hand side of the substitution. The $i\epsilon$-prescription was one of the difficulties with working with linear propagators, as discussed in \cite{Baadsgaard:2015voa}. Now, with quadratic propagators, this problem is also solved.

Motivated by the discussion above, let us consider planar integrands with ordering $(12\dots n)$. We then define new \emph{$\ell^2$-deformed scattering equations} via
\begin{equation}\label{eq:SE_main}
\boxed{\;\;
 \mathcal{E}_a^{\Def} := \mathcal{E}_a \;\Bigg|_{\substack{\,2\ell\cdot k_1 \;\mapsto\;{+ (\ell+ k_1)^2}\\ 2\ell\cdot k_n\; \mapsto\; {-(\ell- k_n)^2}}}
 \;\;}
\end{equation}
To spell this out in a bit more detail, this implies that the scattering equations for external particles $i\neq 1, n$ are unaffected, $\mathcal{E}^{\Def}_i=\mathcal{E}_i$, but the scattering equations for particles 1 and $n$ get modified to
\begin{subequations}
\begin{align}
 \mathcal{E}^{\Def}_1&=\sum_{j\neq 1} \frac{2k_1\cdot k_j}{\sigma_{1j}} + \left(\ell+ k_1\right)^2\frac{\sigma_{+-}}{\sigma_{1+}\sigma_{1-}}   \\
 \mathcal{E}^{\Def}_n&=\sum_{j\neq n} \frac{2k_n\cdot k_j}{\sigma_{nj}} - \left(\ell- k_n\right)^2\frac{\sigma_{+-}}{\sigma_{n+}\sigma_{n-}}\,. 
\end{align}
\end{subequations}
Similarly, the new scattering equations corresponding to the nodal points become
\begin{subequations}
\begin{align}
 \mathcal{E}^{\Def}_+&=+\sum_{j\neq 1,n} \frac{2\ell\cdot k_j}{\sigma_{+j}} + \frac{\left(\ell+ k_1\right)^2}{\sigma_{+1}}- \frac{\left(\ell- k_n\right)^2}{\sigma_{+n}}   \\
 \mathcal{E}^{\Def}_-&=-\sum_{j\neq 1,n} \frac{2\ell\cdot k_j}{\sigma_{-j}} - \frac{\left(\ell+ k_1\right)^2}{\sigma_{-1}}+ \frac{\left(\ell- k_n\right)^2}{\sigma_{-n}} \,.
\end{align}
\end{subequations}
Note first that these new scattering equations  still fit into a CHY-like measure, because the substitution preserves M\"{o}bius invariance: $\sum_a \sigma_a^m\mathcal{E}_a=0$ for $m=0,1,2$ still holds on momentum conservation
\begin{equation}
\sum_{j\neq i} {2k_i\cdot k_j} =0= \sum_j 2\ell\cdot k_i  \,,
\end{equation}
because all $\ell^2$ dependence cancels out. We can thus define the measure in complete analogy to the linear case \eqref{eq:measure_lin},
 \begin{equation}\label{eq:measure_quad}
  d\mu_{n}^{\Def} :=\frac{d^{n+2}\sigma}{\text{vol}\,\text{SL}(2,\C)} \,\,\, {\prod_{a=1}^{n+2}}{}' \bar\delta\left(\mathcal{E}_a^\Def\right)\,.
 \end{equation}
We briefly note here that there exists a much more compact definition of these scattering equations. Similarly to the tree-level and to the linear one-loop scattering equations, the new scattering equations encode that a certain quadratic form $\mathfrak{P}_2$ vanishes everywhere on the nodal  sphere.\footnote{For the linear one-loop scattering equations, $\mathfrak{P}_2=P^2(\sigma)-\ell^2 \omega_{+-}^2$; see \cite{Geyer:2015bja,Geyer:2015jch}.}
For the $\ell^2$-deformed scattering equations, we find \,$\mathfrak{P}_2=P^2(\sigma)-\ell^2 \omega_{+-}^2+\ell^2\omega_{+-}\omega_{1n}$\,,
with
\begin{equation}\label{eq:def_P}
 P(\sigma)=\left(\frac{\ell}{\sigma-\sigma_+}-\frac{\ell}{\sigma-\sigma_-}+ \sum_{i=1}^n\frac{k_i}{\sigma-\sigma_i}\right)d\sigma\,,\qquad\qquad
  \omega_{ab}(\sigma)=\frac{(\sigma_{a}-\sigma_b)\,d\sigma}{(\sigma-\sigma_a)(\sigma-\sigma_b)}\,.
\end{equation}
Since $\mathfrak{P}_2$ is a meromorphic quadratic form on the sphere (where a holomorphic quadratic form cannot exist), imposing $\mathfrak{P}_2=0$ is equivalent to setting the residues at its poles to zero, and in fact it is sufficient to set all but three of such residues to zero. The residues are
\begin{equation}\label{eq:SE_compact}
 \mathcal{E}_a^\Def=\mathrm{Res}_{\sigma_a}\left(P^2(\sigma)-\ell^2 \omega_{+-}^2+\ell^2\omega_{+-}\omega_{1n}\right)\,.
\end{equation}
This form of the scattering equations suggests that, just like the linear scattering equations \eqref{eq:SE_lin}, they can be obtained from the torus via a residue theorem. The difference between the two formalisms, the term $\ell^2\omega_{+-}\omega_{1n}$ in $\mathfrak{P}_2$, is exactly proportional to the `missing' torus scattering equation exchanged in the residue theorem for the localisation on the nodal sphere. We discuss this in more detail below.\\

Before discussing the full worldsheet formulas and the integrands in more detail, let us briefly outline here why this deformation leads to Feynman propagators for planar integrands. The discussion will be light on technical details, all of which can be found in \cref{sec:factorisation}.

\begin{figure}[ht]
	\centering 
	  \includegraphics[width=5cm]{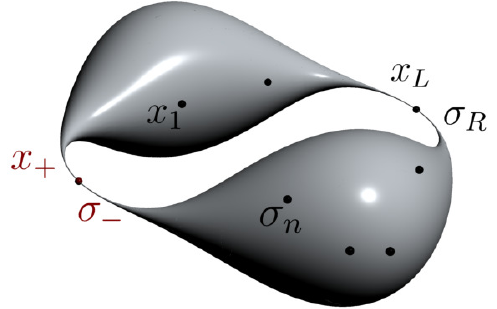}
	  \hfill
	  \raisebox{16pt}{\includegraphics[width=5cm]{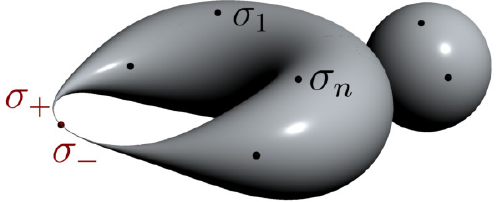}}
	  \hfill
	 \raisebox{16pt}{\includegraphics[width=5cm]{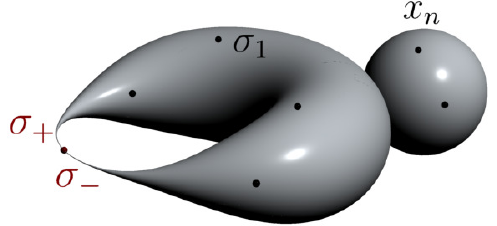}}
	\caption{All singular worldsheet configurations corresponding to physical poles. The geometry on the left, with $+$ and $1$ on one component, and $-$ and $n$ on the other, corresponds to a pole at $(\ell \pm k_L)^2=0$. The geometries in the middle and on the right correspond to tree-level poles at $k_L^2=0$, with $k_L=-\sum_{i\in L}k_i$ being the sum over all punctures on the sphere. On the right, we may exchange the roles of 1 and $n$, but the configuration with both $1$ and $n$ on the subsphere leads to unphysical loop propagators.}
	\label{fig:poles_phys}
\end{figure}

For all CHY-like worldsheet formulas, the scattering equations provide a correspondence between the boundary divisors of the moduli space and the singularity structure of the resulting formulas on momentum space. In other words, the set of possible propagator poles is fully determined by the form of the scattering equations.\footnote{Here, we assume that the only poles of the  worldsheet integrands $\cI^{\scalebox{0.6}{$(1)$}}_{\scalebox{0.5}{$1/2$}}$ are associated to the moduli of the surface.} This correspondence can be established by investigating how the scattering equations behave close to a boundary of the moduli space, where some subset $L$ of the punctures factor off on a separate sphere $\Sigma_\sL$; see \cref{fig:poles_phys}. The scattering equations descend straightforwardly to this configuration, giving two new sets of scattering equations $\mathcal{E}^{\scalebox{0.6}{$(L)$}}$ on $\Sigma_\sL$, and $\mathcal{E}^{\scalebox{0.6}{$(R)$}}$ on $\Sigma_\sR$. The pole corresponding to this configuration is determined by $ s_\sL=2\sum_{a\in R}\sigma_{a\sR}\,\mathcal{E}^{\scalebox{0.6}{$(R)$}}_a=0$, where $\sigma_\sR$ is the node connecting the two spheres.\footnote{We can recognise this as a generalisation of the momentum conservation constraints ensuring M\"{o}bius invariance.} Calculating this for all possible configurations, we find that the scattering equations \eqref{eq:SE_compact} encode standard Feynman propagators of the form $1/k_\sL^2$, $1/(\ell-k_{\scalebox{0.5}{$L$}})^2$ and $1/(\ell+k_{\scalebox{0.5}{$L$}})^2$ if $L$ is consistent with planarity, and with the loop momentum placed between particles 1 and $n$ \emph{without} massive 1-$n$ corners; see \cref{fig:poles_phys}.
\iffalse
\begin{equation*}
  s_{\scalebox{0.5}{$L$}}=
 \begin{cases}
 k_{\scalebox{0.5}{$L$}}^2   & L=L^{\scalebox{0.6}{ext}}\;\;\mathrm{or}\;\; L=\{+,- \}\cup L^{\scalebox{0.6}{ext}}\\
 %
 (\ell-k_{\scalebox{0.5}{$L$}})^2 & L=\{+\}\cup L^{\scalebox{0.6}{ext}},\;1\in L^{\scalebox{0.6}{ext}},\;n\notin L^{\scalebox{0.6}{ext}}\\ 
 %
  (\ell+k_{\scalebox{0.5}{$L$}})^2 & L=\{-\}\cup L^{\scalebox{0.6}{ext}},\;1\notin L^{\scalebox{0.6}{ext}},\;n\in L^{\scalebox{0.6}{ext}}\\ 
 %
 \mathrm{unphys} & \mathrm{else}\,. \end{cases}
 \end{equation*}\fi
They also parametrise several unphysical poles corresponding to different placements of the loop momentum, massive 1-$n$ corners\footnote{For diagrams with 1-$n$ corners, the tree-level propagators are correct, but the loop propagators are linear in $\ell$.}  and tadpoles. However, none of these unphysical poles contributes for the planar worldsheet formulas that we will construct in the next two sections.

\paragraph{Speculation on a first-principles derivation of the deformation.}  
\hypertarget{speculation}{We originally motivated the deformation of the scattering equations} \eqref{eq:SE_main} by an analogy with how the BCFW recursion reproduces Feynman propagators. However, their form \eqref{eq:SE_compact}  suggests that an alternative motivation could be given that ties in more directly with the ambitwistor string.  As briefly reviewed \hyperref[sec:review_ws]{above} and discussed in more detail in \cite{Geyer:2015jch}, the type II ambitwistor string naturally gives a one-loop integrand formulated over the torus, with a residue theorem relating this genus-one expression to a simpler formula  on the nodal sphere.  
This residue theorem trades the localisation on one of the genus-one scattering equations --- chosen to be the `modular' scattering equation
$\,\mathcal{E}_\tau:=P^2(z_0)=0\,$ fixing the modulus of the torus (with $z_0$ a reference point)  --- for the localisation on the only other pole, the moduli-space boundary corresponding to the pinched torus, i.e., the nodal sphere. On the torus, different choices of scattering equations related by adding $\mathcal{E}_i\simeq \mathcal{E}_i+\alpha_i\mathcal{E}_\tau$ are clearly equivalent. After the degeneration to the nodal sphere, however, this is not manifest any more because $\mathcal{E}_\tau=\ell^2\neq0$, and different choices of scattering equations on the torus can lead to very different integrand representations on the nodal sphere. 
The $\ell^2$-deformation in \eqref{eq:SE_main} corresponds to the choice $\alpha_1=-\alpha_n=1$ with all other $\alpha_i=0$, such that 
\begin{equation}
 \mathcal{E}^{\Def}_1= \mathcal{E}_1+\ell^2 \frac{\sigma_{+-}}{\sigma_{1+}\sigma_{1-}}\,,\qquad 
 \mathcal{E}^{\Def}_n= \mathcal{E}_n-\ell^2 \frac{\sigma_{+-}}{\sigma_{n+}\sigma_{n-}}\,.
\end{equation}
It would however be interesting to study the space of these deformations in more detail.\footnote{Clearly, the $\alpha_i$ have to satisfy $\sum_i\alpha_i=0$, which can be seen either from M\"{o}bius invariance on the nodal sphere, or translation invariance on the torus. While this seems to be the only restriction, it remains to be seen what kind of integrand representations follow from the different choices of loop scattering equations.}

One important feature of the new scattering equations \eqref{eq:SE_main} is that the deformation is adapted to planar integrands, since the loop momentum is seen to lie between particles $n$ and 1. On the other hand, the only theory for which a torus formula has been constructed directly from the ambitwistor string (as opposed to a formula that only exists on the nodal sphere, such as those for gauge theory) is type II supergravity, an inherently non-planar theory. This presents a challenge for a first-principles derivation from the ambitwistor string. Our expectation is that the torus formula for a non-planar theory can be turned into a nodal-sphere formula based not just on the scattering equations \eqref{eq:SE_main}, but on a set of distinct deformations associated to different pairs of particles (i.e., not just 1 and $n$). This would lead to nodal-sphere formulas reproducing those in section~\ref{sec:wsquadnp}, which deals with non-planar super Yang-Mills and supergravity. The decomposition of the non-planar formulas into planar-like parts along the lines discussed in that section would follow, in the ambitwistor string calculation, from a decomposition of the worldsheet correlators into parts which have the appropriate singularity structure to admit a given deformation without introducing extra poles that mess up the residue theorem from the torus to the nodal sphere. The investigation of this possibility is beyond the scope of this paper.

%%%%%%%%%%%%%%
\subsection{The MHV integrand}\label{sec:MHV}
While in principle worldsheet integrands  adapted to the new scattering equations could be constructed from first principles via BCFW recursion, this is a laborious process, and it is often difficult to recognise structures in the resulting expressions. Instead, we will  modify  the previously known worldsheet integrands to give the correct amplitudes. We will focus  now on the planar super Yang-Mills case \eqref{eq:halfintSYM}. Looking at the loop integrand $\mathfrak{I}^\lin$ in \eqref{eq:SYM_4pt_lin}, it is clear what this modification will have to achieve: it has to ensure that the loop momentum lies between particles $1$ and $n$. The reason for this if two-fold: terms with different placement of $\ell$ give unphysical poles, as is evident from  the discussion of the scattering equations above; and even if this issue could be overcome, the restriction to `a single term per $n$-gon' is still necessary to avoid overcounting Feynman diagrams involving $n$-gons by relative factors of $n$.\footnote{We discuss an alternative proposal where this requirement fails in \cref{sec:DFL}. }

While restricting to a specific placement of the loop momentum is difficult for generic integrands, it can easily be achieved for gauge theories by selecting the term in the cyclic sum of the colour half-integrand which has the correct placement of the node:
\begin{align}
 \mathcal{C}^{\scalebox{0.6}{$(1)$}}_{\scalebox{0.6}{cyc}}(12\dots n)&=\sum_{\rho\in\text{cyc}(12\dots n)} \frac{\tr(T^{a_1}T^{a_2}\dots T^{a_n})}{(+\rho_1\rho_2\ldots \rho_n-)}\supset \frac{\tr(T^{a_1}T^{a_2}\dots T^{a_n})}{(+12\ldots n-)}\,,
\end{align}
which we have highlighted here for $\ell$ between 1 and $n$. This leads to the following worldsheet proposal for the $n$-particle MHV integrand in $\cN=4$ super Yang-Mills,
\begin{equation}\label{eq:MHV-proposal}
 \boxed{\;\;
 \mathfrak{I}^{\scalebox{0.6}{$(1)$}}_{\text{sYM-MHV}}(12\dots n)=\frac{1}{\ell^2} \int_{\raisebox{-6pt}{\scalebox{0.7}{$\mathfrak{M}_{0,n+2}$}}}\hspace{-15pt}d\mu_{n}^{\Def} \, \cI^{\scalebox{0.6}{$(1)$}}_{\scalebox{0.6}{MHV}} \,\,\mathcal{C}^{\scalebox{0.6}{$(1)$}}(12\dots n)\,.
 \;\;}
\end{equation}
The measure $d\mu_{n}^{\Def}$ is defined by localising on the $\ell^2$-deformed scattering equations \eqref{eq:measure_quad}, and the colour half-integrand now only includes the term with the node between the puncture  $\sigma_1$ and $\sigma_n$, corresponding to the correct placement of the loop momentum,
\begin{align}
 \mathcal{C}^{\scalebox{0.6}{$(1)$}}(12\dots n)&= \frac{\tr(T^{a_1}T^{a_2}\dots T^{a_n})}{(+12\ldots n-)},.
\end{align}
The kinematic half-integrand is the same as for the linear integrand representation \eqref{eq:half-int_lin},
\begin{equation}
 \cI^{\scalebox{0.6}{$(1)$}}_{\scalebox{0.6}{MHV}} =\sum_{\rho\in S_n} \frac{N^{\scalebox{0.6}{$(1)$}}_{\scalebox{0.6}{$\rho$}}}{(+\rho_1\rho_2\ldots \rho_n-)}\,,
\end{equation}
where $N^{\scalebox{0.6}{$(1)$}}_{\scalebox{0.6}{$\rho$}}:=N^{\scalebox{0.6}{$(1)$,MHV}}_{\scalebox{0.6}{$\rho$}}$ are the kinematic numerators relevant to the MHV case. All-multiplicity expressions for these `half-ladder' numerators were constructed in \cite{He:2015wgf}, and we will give explicit examples below.\footnote{The algorithm extends to the BCJ numerators for $m$-gons with $m<n$, i.e. with massive corners, which can be calculated from the `half-ladder' master numerators by using Jacobi identities, e.g. 
$  \mathfrak{n}^{\scalebox{0.6}{$(1)$}}_{\scalebox{0.6}{$1[2,3]4\ldots n$}}=
  \mathfrak{n}^{\scalebox{0.6}{$(1)$}}_{\scalebox{0.6}{$1234\ldots n$}}-
  \mathfrak{n}^{\scalebox{0.6}{$(1)$}}_{\scalebox{0.6}{$1324\ldots n$}}$, for an $(n-1)$-gon with massive corner 2-3.}
Here, we only want to highlight some of their key features. First of all, supermomentum conservation is manifest, 
\begin{equation}
\label{eq:MHVnumsec7}
 N^{\scalebox{0.6}{$(1)$}}_{\scalebox{0.6}{$\rho$}}=\delta^8(Q)\;\left(\prod_{i=2}^n\frac{1}{\la 1i\ra^2}\right)\,\mathfrak{n}^{\scalebox{0.6}{$(1)$}}_{\scalebox{0.6}{$\rho$}}\,.
\end{equation}
Moreover, diagrams with particle 1 in a massive corner always have vanishing numerators due to 
\begin{equation}\label{eq:jacobi_1}
 \mathfrak{n}^{\scalebox{0.6}{$(1)$}}_{\scalebox{0.6}{$\rho_i\dots \rho_n 1\rho_2\dots\rho_{i-1}$}}=
 \mathfrak{n}^{\scalebox{0.6}{$(1)$}}_{\scalebox{0.6}{$1\rho_2\dots\rho_n$}}\,,
 \end{equation}
and thus 1 is always directly attached to the loop. In particular, this implies that there are no massive corners involving both particles 1 and $n$. This is crucial, since the scattering equations would produce unphysical poles for such corners. And lastly, another important property of these numerators is that all bubbles and triangles vanish, $\mathfrak{n}_{1A_2A_3}=0=\mathfrak{n}_{1A_2}$ for any massive corners $A_2$ and $A_3$.

\subsubsection{Toy models}
\paragraph{Toy model I: $n$-gon.} As a warm-up, it will be helpful to first formulate and calculate amplitudes in the simplest example, the $n$-gon.  In analogy with the super Yang-Mills case above, we propose the following worldsheet formula for the $n$-gon, with $n\geq 4$:
\begin{equation}\label{eq:n-gon}
 \boxed{\;\;
 \mathfrak{I}^{\scalebox{0.6}{$(1)$}}_{n\text{-gon}}(12\dots n)=\frac{1}{\ell^2} \int_{\raisebox{-6pt}{\scalebox{0.7}{$\mathfrak{M}_{0,n+2}$}}}\hspace{-15pt}d\mu_{n}^{\Def} \; \left( \frac1{\sigma_{+-}^2}\, \prod_{j=1}^n \frac{\sigma_{-+}}{\sigma_{+j}\,\sigma_{j-}}\right)\; \,\frac{1}{(+12\dots n-)}\,.
 \;\;}
\end{equation}
For four and five particles, this can be evaluated  by adapting the results of \cite{Geyer:2015jch}, giving indeed the box and the pentagon with ordering $(12\ldots n)$,
\begin{equation*}
  \mathfrak{I}^{\scalebox{0.6}{$(1)$}}_{\text{box}}=\frac{1}{\ell^{2}\,\left(\ell+k_1\right)^2\left(\ell+k_{12}\right)^2\left(\ell-k_4\right)^2}\,,\quad \quad \mathfrak{I}^{\scalebox{0.6}{$(1)$}}_{\text{pent}}=\frac{1}{\ell^{2}\,\left(\ell+k_1\right)^2\left(\ell+k_{12}\right)^2\left(\ell-k_{45}\right)^2\left(\ell-k_5\right)^2}\,.
\end{equation*}
We have also verified \eqref{eq:n-gon} numerically for $n=4,5$.
Using the BCFW recursion, we will prove in \cref{app:BCFW} that these formulas extend to all orders in $n$, and that the above worldsheet formula indeed gives
\begin{equation}
 \mathfrak{I}^{\scalebox{0.6}{$(1)$}}_{n\text{-gon}}(12\dots n)=\frac{1}{\ell^{2}\,\left(\ell+k_1\right)^2\left(\ell+k_{12}\right)^2\ldots\left(\ell-k_n\right)^2}\,.
\end{equation} 
Note that for $n=2,3$ tadpole-like configurations contribute, and so \eqref{eq:n-gon} ceases to be valid.

\paragraph{Toy model II: $(n-1)$-gon with a massive corner.}
The next step up from the $n$-gon is an $(n-1)$-gon with a single massive corner, which we label by $[i,i+1]$. For $n\geq 5$ and $i\neq n$, this is described by the following worldsheet formula,
\begin{equation}\label{eq:n-1-gon}
 \mathfrak{I}^{\scalebox{0.6}{$(1)$}}_{(n-1)\text{-gon}|[i,i+1]}
 =\frac{1}{\ell^2} \int_{\raisebox{-6pt}{\scalebox{0.7}{$\mathfrak{M}_{0,n+2}$}}}\hspace{-15pt}d\mu_{n}^{\Def} \; \left( \frac1{\sigma_{+-}^2} \,
 \frac{\sigma_{-+}}{\sigma_{+\,i\!+\!1}\,\sigma_{i\!+\!1\, i}\,\sigma_{i-}}
 \, 
 \prod_{j\neq i,i+1} \frac{\sigma_{-+}}{\sigma_{+j}\,\sigma_{j-}} 
 \right)\; \,\frac{1}{(+12\dots n-)}\,.
\end{equation}
The intuition behind this fomula is that it yields the $(n-1)$-gon (multiplied by the appropriate propagator $1/s_{i\,i\!+\!+1}$) in the limit when the punctures $\sigma_i$ and $\sigma_{i+1}$ coalesce.\footnote{This can be made explicit by studying the factorisation channels, as in \cref{app:BCFW}.} Evaluated for $n=5$, it indeed reproduces the massive box 
\begin{equation*}
  \mathfrak{I}^{\scalebox{0.6}{$(1)$}}_{\text{box}|[4,5]}=\frac{1}{s_{45}\,\ell^{2}\,\left(\ell+k_1\right)^2\left(\ell+k_{12}\right)^2\left(\ell-k_5\right)^2}\,,
\end{equation*}
which can again be verified numerically, or proven by factorisation.

\subsubsection{Four and five particles}\label{sec:4pt_WS}
Equipped with these results,  let us now calculate the four- and five-particle integrands that arise from the worldsheet representation. These will form the seed amplitudes for proving our proposal \eqref{eq:MHV-proposal} via BCFW recursion in the next section.
\paragraph{Four-particle integrand.} For four particles, all numerators coincide,
\begin{equation}\label{eq:4pt-nums}
 N^{\scalebox{0.6}{$(1)$}}_{\scalebox{0.6}{$\rho$}} =-\delta^8(Q)\frac{\left[12\right]\left[34\right]}{\la 12\ra \la 34\ra}\,,\qquad\qquad \text{for all }\;\rho \in S_4\,.
\end{equation}
Note  that the numerators are permutation invariant on the support of momentum conservation, and are proportional to the tree-level amplitude. %,  $N^{\scalebox{0.6}{$(1)$}}_{\scalebox{0.6}{$\rho$}} =st \,A^{\scalebox{0.6}{$(0)$}}(1234)$. 
We can now use the identity
\begin{equation}\label{eq:perm_id}
 \sum_{\rho\in S_n} \frac{1}{(+\rho_1\rho_2\ldots\rho_n-)}=-\frac1{\sigma_{+-}^2}\, \prod_i \frac{\sigma_{+-}}{\sigma_{+i}\,\sigma_{i-}}
 =(-1)^{n+1}\left(\frac1{\sigma_{+-}^2}\, \prod_i \frac{\sigma_{-+}}{\sigma_{+i}\,\sigma_{i-}}\right)\,,
\end{equation}
where the factor in parenthesis on the rightmost side was used in the $n$-gon expression \eqref{eq:n-gon}.
The integrand simplifies to just the ordered box familiar from the $4$-gon discussion,
\begin{equation}
 \mathfrak{I}^{\scalebox{0.6}{$(1)$}}_{\text{sYM}}(1234)=\delta^8(Q)\frac{\left[12\right]\left[34\right]}{\la 12\ra \la 34\ra}\,
 \frac{1}{\ell^{2}}\,  \int_{\mathfrak{M}_{0,4+2}}\hspace{-20pt}  d\mu_{4}^{\Def} \; \left( \frac1{\sigma_{+-}^2}\, \prod_i \frac{\sigma_{-+}}{\sigma_{+i}\,\sigma_{i-}}\right)\; \frac1{(+1234-)}\,.
\end{equation}
Using our previous results, or alternatively again adapting the work of ref.~\cite{Geyer:2015jch}, this evaluates directly to the correct integrand with Feynman propagators as given in e.g. \eqref{4ptex},\footnote{In fact, this matches \eqref{4ptex} up to a sign, due to the factor $(-1)^{n+1}$ in \eqref{eq:perm_id}. Rather than inserting this factor explicitly in a multitude of expressions, we just accept that the normalisation convention used in this section is different.}
\begin{equation}
 \mathfrak{I}^{\scalebox{0.6}{$(1)$}}_{\text{sYM}}(1234)=\delta^8(Q)\frac{\left[12\right]\left[34\right]}{\la 12\ra \la 34\ra} \,
 \frac{1}{\ell^{2}\,\left(\ell+k_1\right)^2\left(\ell+k_{12}\right)^2\left(\ell-k_4\right)^2}\,.
\end{equation}

\paragraph{Five-particle integrand.} For five particles, the relevant numerators are
\begin{subequations}
\begin{align}
 \mathfrak{n}^{\scalebox{0.6}{$(1)$}}_{\scalebox{0.6}{12345}}\;\,&=X_{2,4}X_{2,3}X_{\ell,5}+X_{3,5}X_{2+3,4}X_{\ell,2}+X_{2,5}X_{2,3}X_{2+3,4}\\
 \mathfrak{n}^{\scalebox{0.6}{$(1)$}}_{\scalebox{0.6}{$123[4,5]$}}&=X_{2,4+5}X_{2,3}X_{4,5}\,,
\end{align}
\end{subequations}
where the $X_{ab}$ were defined in \eqref{defXX}, and all other numerators are related to these  by  \eqref{eq:jacobi_1} and relabeling of the external particles. Similarly to the four-particle case, the kinematic integrand again simplifies considerably: using the Jacobi relations and the fact that triangles and bubbles vanish, we can express every numerator as a sum of $\mathfrak{n}^{\scalebox{0.6}{$(1)$}}_{\scalebox{0.6}{12345}}$ and box terms, e.g. $\mathfrak{n}^{\scalebox{0.6}{$(1)$}}_{\scalebox{0.6}{23541}}=\mathfrak{n}^{\scalebox{0.6}{$(1)$}}_{\scalebox{0.6}{12354}}=\mathfrak{n}^{\scalebox{0.6}{$(1)$}}_{\scalebox{0.6}{12345}}- \mathfrak{n}^{\scalebox{0.6}{$(1)$}}_{\scalebox{0.6}{$123|4,5]$}}$. Moreover, all of the $[i,i+1]$ box numerators are equal (c.f. \eqref{eq:4pt-nums}), and there are no $[12]$ and $[51]$ massive boxes due to the vanishing  numerators. Collecting the terms for each independent numerator, we find
\begin{align}
 \cI^{\scalebox{0.6}{$(1)$}}_{\scalebox{0.7}{MHV}}=& N^{\scalebox{0.6}{$(1)$}}_{\scalebox{0.6}{12345}}\sum_{\rho\in S_5}\frac{1}{(+\rho -)}\;
 -N^{\scalebox{0.6}{$(1)$}}_{\scalebox{0.6}{123[45]}} \sum_{\scalebox{0.9}{$\substack{\alpha\in S_3(123)\\ \rho\in\alpha\shuffle\{5,4\}}$}}\frac{1}{(+\rho -)}
   -N^{\scalebox{0.6}{$(1)$}}_{\scalebox{0.6}{12[34]5}} \sum_{\scalebox{0.9}{$\substack{\alpha\in S_3(125)\\ \rho\in\alpha\shuffle\{4,3\}}$}}\frac{1}{(+\rho -)}
  -N^{\scalebox{0.6}{$(1)$}}_{\scalebox{0.6}{1[23]45}} \sum_{\scalebox{0.9}{$\substack{\alpha\in S_3(145)\\ \rho\in\alpha\shuffle\{3,2\}}$}}\frac{1}{(+\rho -)}\,.\nonumber
\end{align}
This already resembles  the structure of one pentagon with three boxes familiar from \eqref{5ptex}. To make this more manifest, note that the KK relations \cite{KK1989}, written in \eqref{eq:KK}, allow us to simplify the box terms,
\begin{equation}
 \sum_{\scalebox{0.9}{$\substack{\alpha\in S_3(123)\\ \rho\in\alpha\shuffle\{5,4\}}$}}\frac{1}{(+\rho -)}=\sum_{\alpha\in S_3}\frac1{(+\alpha - 45)}=-\frac1{\sigma_{+-}^2}\,\,\frac{\sigma_{-+}}{\sigma_{+5}\,\sigma_{54}\,\sigma_{4-}}\prod_{i\neq 4,5} \frac{\sigma_{-+}}{\sigma_{+i}\,\sigma_{i-}}\,.
\end{equation}
Here, the first equality follows from the KK relations for Parke-Taylor factors, and the second equality uses the  relation \eqref{eq:perm_id}.  Using the latter to relate the sum over permutations $S_5$ to the pentagon, we arrive at
\begin{align}\label{eq:5pt_WS}
  \cI^{\scalebox{0.6}{$(1)$}}_{\scalebox{0.7}{MHV}}=& N^{\scalebox{0.6}{$(1)$}}_{\scalebox{0.6}{12345}} \left( \frac1{\sigma_{+-}^2}\, \prod_{i=1}^5 \frac{\sigma_{-+}}{\sigma_{+i}\,\sigma_{i-}}\right)\;+N^{\scalebox{0.6}{$(1)$}}_{\scalebox{0.6}{123[45]}} \left(  \frac1{\sigma_{+-}^2}\,\,\frac{\sigma_{-+}}{\sigma_{+5}\,\sigma_{54}\,\sigma_{4-}}
  \prod_{i\neq 4,5} \frac{\sigma_{-+}}{\sigma_{+i}\,\sigma_{i-}}\right)\\
  &\quad +N^{\scalebox{0.6}{$(1)$}}_{\scalebox{0.6}{12[34]5}} \left( \frac1{\sigma_{+-}^2}\,\frac{\sigma_{-+}}{\sigma_{+4}\,\sigma_{43}\,\sigma_{3-}} \prod_{i\neq 3,4} \frac{\sigma_{-+}}{\sigma_{+i}\,\sigma_{i-}}\right)+N^{\scalebox{0.6}{$(1)$}}_{\scalebox{0.6}{1[23]45}} \left(  \frac1{\sigma_{+-}^2}\,\,\frac{\sigma_{-+}}{\sigma_{+3}\sigma_{32}\,\sigma_{2-}}\prod_{i\neq 2,3} \frac{\sigma_{-+}}{\sigma_{+i}\,\sigma_{i-}}\right)
  \,. \nonumber
\end{align}
The worldsheet integral over \eqref{eq:5pt_WS} can be evaluated directly, e.g. adapting the results of \cite{Geyer:2015jch} to the new scattering equations, and it agrees with the five-particle integrand  \eqref{5ptex} for super Yang-Mills.

%%%%%%%%%%%%%%
\subsection{Beyond MHV}\label{sec:proposal}
Since the kinematic integrand is not modified in the MHV case, the ambitwistor string gives a natural idea for how to extend the above formulas both beyond the MHV sector, and beyond $D=4$; just replace the MHV integrand $\cI^{\scalebox{0.6}{$(1)$}}_{\scalebox{0.6}{MHV}}$ in \eqref{eq:MHV-proposal} by the general kinematic half-integrand $\cI^{\scalebox{0.6}{$(1)$}}_{\scalebox{0.6}{kin}}$ derived via a residue theorem from the original torus expression as a sum over spin structures.
\begin{equation}\label{eq:int_gen_proposal}
 \mathfrak{I}^{\scalebox{0.6}{$(1)$}}_{\text{sYM}}=\frac{1}{\ell^2} \int_{\raisebox{-6pt}{\scalebox{0.7}{$\mathfrak{M}_{0,n+2}$}}}\hspace{-15pt}d\mu_{n}^{\Def} \; \cI^{\scalebox{0.6}{$(1)$}}_{\scalebox{0.6}{kin}} \,\mathcal{C}^{\scalebox{0.6}{$(1)$}}(12\dots n)\,.
\end{equation}
This construction looks natural from the viewpoint of the ambitwistor string. There is however a cause for concern. In the MHV case discussed above, we started from a known loop integrand representation in terms of trivalent diagrams without massive 1-$n$ corners,\footnote{That is, with no trees emanating from the loop of any type that contain both 1 and $n$. For instance, in \eqref{eq:5pt_WS}, there is no diagram with massive corner 1-5. In fact, in the representation that we have been using, from \cite{He:2015wgf}, particle 1 is always directly attached to the loop, i.e., never in a massive corner.}
which turns out to be a crucial property.
It is crucial because the formula  \eqref{eq:int_gen_proposal} will not straightforwardly yield  trivalent diagrams with such massive corners. The intuitive reason for this is the form of the Parke-Taylor factor, which always places the node associated to the loop momentum $\ell$  between $1$ and $n$; we will discuss the (factorisation) details behind this intuition in the next section.
Beyond the MHV case, it is not clear whether super Yang-Mills loop integrands admit a representation without diagrams with 1-$n$ corners.\footnote{Notice that, in the MHV formula, the MHV numerators \eqref{eq:MHVnumsec7} still possess a pole as $\langle 1 n \rangle\to0$ due to the prefactor.} The form of the BCFW recursion relation suggests that this is possible, but clearly more work is needed to make that claim. 

In the absence of a known representation of the loop integrand with that property, we could still hope to prove or disprove \eqref{eq:int_gen_proposal} using the BCFW recursion relation, as in the MHV proof to be presented in \cref{sec:BCFW_WS}. This is however more technically involved due to the supersymmetric sum over states in the forward-limit term. It may be possible to resolve this using either the nodal operator construction of \cite{Roehrig:2017gbt} or the more recent numerator constructions in \cite{Edison:2020uzf, Edison:2020ehu}, but both approaches are beyond the scope of this paper.

With the above words of caution, here we  present the details of the idea \eqref{eq:int_gen_proposal}, assuming that the crucial property holds, and highlight some of its important features. 
The object to be specified is the kinematic half-integrand $\cI^{\scalebox{0.6}{$(1)$}}_{\scalebox{0.6}{kin}}$.
Two formulas for this half-integrand were presented in \cite{Geyer:2015jch}, which are equivalent on the support of the `old' one-loop scattering equations. The new, $\ell^2$-deformed scattering equations require that one of those formulas is slightly changed. For clarity, we will describe both formulas here, the changed one being the second.
The first formula, obtained originally via the residue theorem from the torus to the nodal sphere, is
\begin{equation}\label{eq:int_sum_spin}
  \cI^{\scalebox{0.6}{$(1)$}}_{\scalebox{0.6}{kin}} = \frac{1}{\sigma_{+-}^2}\left( \pf(M_{3}) \big|_{\sqrt{q}}+
  (d-2)\pf(M_{3}) \big|_{q^0}-c_\sD\,\pf(M_{2} )\big|_{q^0}\right)\,,
\end{equation}
where  $q=e^{2i\pi\tau}$ is the modulus of the torus.\footnote{Not to be confused with the BCFW shift vector $q_\mu$. Unfortunately, both objects are conventionally called $q$, but the meaning should always be clear from context. In this section, $q$ is always the modular parameter of the torus.} The different terms originate from the localisation on $q=0$ of the torus integrand multiplied by the (non-trivial) partition functions.
The constant $c_\sD$ above relates to the number of fermions in the loop, and depends on the spacetime dimension $D$. For maximal super Yang-Mills, we have $c_{10}=8$, and $c_4=2$. The matrix $M_{\mathbf\alpha}$ is a generalisation of the CHY matrix, with $\alpha$ a remnant of the sum over spin structures on the torus. It is defined as
\begin{equation}
M_\alpha =  \begin{pmatrix}
A \;\;& -C^T \\
C  \;\;&  B
\end{pmatrix}\,,
\end{equation}
where the matrix components are given by 
\begin{align}
&A_{ij} = k_i\cdot k_j \,S_\alpha(\sigma_{ij}|q), && B_{ij} =
\epsilon_i\cdot \epsilon_j \,S_\alpha(\sigma_{ij}|q), &&C_{ij} =
\epsilon_i\cdot k_j \,S_\alpha(\sigma_{ij}|q) && \text{for} \quad
i\neq j,\label{eq:Malpha}\\
&A_{ii}\, = 0, && B_{ii}\, = 0, &&C_{ii}, = -\epsilon_i\cdot P(\sigma_i), &&
\end{align}
where $P(\sigma)$ is defined as in \eqref{eq:def_P}.
The $S_\alpha$ are the nodal sphere limits of the Szeg\H{o} kernels on the torus, and to leading order we have
\begin{align}\label{eq:szegolimitsl2c}
S_2(\sigma_{ij}|q) &= \frac{1}{2}\,\frac{1}{\sigma_{i\,j}} \left(\sqrt{\frac{\sigma_{i\,\ell^+}\,\sigma_{j\,\ell^-}}{\sigma_{j\,\ell^+}\,\sigma_{i\,\ell^-}}}+ \sqrt{\frac{\sigma_{j\,\ell^+}\,\sigma_{i\,\ell^-}}{\sigma_{i\,\ell^+}\,\sigma_{j\,\ell^-}}} \right)  \sqrt{d\sigma_ i} \sqrt{d\sigma_ j} ,
  \\ 
S_3(\sigma_{ij}|q) &= \frac{1}{\sigma_{i\,j}} \left(1 +\sqrt{q} \;\frac{(\sigma_{i\,j}\,\sigma_{\ell^+\,\ell^-})^2}{\sigma_{i\,\ell^+}\,\sigma_{i\,\ell^-}\,\sigma_{j\,\ell^+}\,\sigma_{j\,\ell^-}}\right) \sqrt{d\sigma_ i} \sqrt{d\sigma_ j} ,
\end{align}
It turns out that, just as for the `old' scattering equations, the integrand expression \eqref{eq:int_sum_spin} can be simplified, this time on the $\ell^2$-deformed scattering equations. The argument follows \cite{Geyer:2015jch} closely and only relies on standard properties of a Pfaffian, so we simply present the result,
\begin{equation}\label{eq:int_gen_v2}
  \cI^{\scalebox{0.6}{$(1)$}}_{\scalebox{0.6}{kin}} =\sum_r\pf'(M_{\text{NS}}^r)-\frac{c_d}{\sigma_{\ell^+\,\ell^-}^2}\pf(M_2)\,,
  \qquad \qquad \text{with }\;
  \pf'(M_{\text{NS}}^r):=\frac{(-1)^{a+b}}{\sigma_{ab}}\pf({M_{\text{NS}}^r}_{[ab]})\,.
\end{equation}
Here $M_{\text{NS}}^r$ is an $2(n+2)\times 2(n+2)$ matrix including two additional rows and columns for each $\sigma_+$ and $\sigma_-$, and the reduced Pfaffian is defined as usual by removing two rows and columns $a,b$, denoted by the  brackets $[ab]$, and quotienting by the appropriate symmetry factor. The matrix $M_{\text{NS}}^r$ encodes the contribution from all bosonic degrees of freedom  running in the loop ---  as indicated by the subscript NS --- and 
is defined  as in \cite{Geyer:2015jch}, but with all factors of $\ell\cdot k_1$ replaced by $\frac{1}{2}(\ell+k_1)^2$, and $\ell\cdot k_n$ replaced by $-\frac{1}{2}(\ell-k_n)^2$. This results in 
\begin{equation}\label{eq:defMNS}
 M_{\text{NS}}^r=\begin{pmatrix}A & -C^T\\ C & B \end{pmatrix}\,,
\end{equation}
with entries
\begin{equation}  \label{eq:defMNS_details}
\begingroup
\renewcommand*{\arraystretch}{1.4}
 M_{\text{NS}}^r=\left( \begin{array}{cccccc:cccc}%\vspace{5pt}
    0 & 0 & \frac{1}{2}\frac{(\ell+ k_1)^2}{\sigma_{+1}} & \frac{\ell\cdot k_i}{\sigma_{+i}} & \frac{\ell\cdot k_j}{\sigma_{+j}} & -\frac{1}{2}\frac{(\ell- k_n)^2}{\sigma_{+n}} & -\epsilon^r\cdot P(\sigma_{+}) & \frac{\ell\cdot\epsilon^r}{\sigma_{+-}} & \frac{\ell\cdot\epsilon_i}{\sigma_{+i}} & \frac{\ell\cdot\epsilon_j}{\sigma_{+j}}\\%\vspace{5pt}
    & 0 & -\frac{1}{2}\frac{(\ell+k_1)^2}{\sigma_{-1}} &  -\frac{\ell\cdot k_i}{\sigma_{-i}} &  -\frac{\ell\cdot k_j}{\sigma_{-j}} & \frac{1}{2}\frac{(\ell-k_n)^2}{\sigma_{-n}} & -\frac{\ell\cdot\epsilon^r}{\sigma_{-+}} & -\epsilon^r\cdot P(\sigma_{-}) & -\frac{\ell\cdot\epsilon_i}{\sigma_{-i}} & -\frac{\ell\cdot\epsilon_j}{\sigma_{-j}}\\%\vspace{5pt}
    & & &   & 0 & \frac{k_i\cdot k_j}{\sigma_{ij}} & \frac{k_i\cdot\epsilon^r}{\sigma_{i+}} & \frac{k_i\cdot\epsilon^r}{\sigma_{i-}} & -\epsilon_i\cdot P(\sigma_{i}) & \frac{k_i\cdot\epsilon_j}{\sigma_{ij}}\\%\vspace{5pt}
    & & &   &   & 0 & \frac{k_j\cdot\epsilon^r}{\sigma_{j+}} & \frac{k_j\cdot\epsilon^r}{\sigma_{j-}}  & \frac{k_j\cdot\epsilon_i}{\sigma_{ji}} & -\epsilon_j\cdot P(\sigma_{j})\\%\vspace{5pt}
    \hdashline%\vspace{5pt}
    & & & & & & 0 & \frac{D-2}{\sigma_{\ell^+\ell^-}} & \frac{ \epsilon^r\cdot\epsilon_i}{\sigma_{\ell^+i}} & \frac{ \epsilon^r\cdot\epsilon_j}{\sigma_{\ell^+j}}\\%\vspace{5pt}
    & & & & & &   & 0 & -\frac{\epsilon^r\cdot\epsilon_i}{\sigma_{\ell^-i}} & \frac{\epsilon^r\cdot\epsilon_j}{\sigma_{\ell^-j}}\\%\vspace{5pt}
  & &  & & & &   &   & 0 & \frac{\epsilon_i\cdot \epsilon_j}{\sigma_{ij}}\\%\vspace{5pt}
   & & & & & &   &   &   & 0\\
  \end{array} \right)\,.
  \endgroup
\end{equation}
The sum in \eqref{eq:int_gen_v2} is taken over a basis of polarisation vectors $\epsilon^r$. The reduced Pfaffian is well-defined because $M_{\text{NS}}^r$ has co-rank two on support of our new scattering equations. This can be checked easily, with its kernel is spanned by the vectors $(1,\dots,1\,|\,0,\dots,0)$
and $(\sigma_{+},\sigma_{-},\sigma_1,\dots,\sigma_n\,|\,0,\dots,0)$.

\paragraph{Evidence and caveats in the idea.} While an investigation to prove or disprove the idea \eqref{eq:int_gen_proposal} is beyond the scope of this paper, the  worldsheet formula possesses many features of a super Yang-Mills integrand. 
First and foremost, it reproduces the MHV worldsheet formula \eqref{eq:MHV-proposal}  for four and five particles, as worked out in \cite{Geyer:2015jch}. This guarantees that the proposal gives the correct  four- and five-particle loop integrands (which must be MHV or $\overline{\mathrm{MHV}}$).
For higher $n$, the formula still manifests many core features of the integrand such as linearity in the polarisation data and the correct mass dimension.

Following the discussion at the end of \S\ref{sec:SE}, we can also verify that the proposal only contains physical, Feynman-like propagator poles. This relies on the structure of both half-integrands: the colour factor  $\mathcal{C}^{\scalebox{0.6}{$(1)$}}(12\dots n)$ ensures that the loop integrand is free of  unphysical poles corresponding to `wrong' placements of the loop momentum,  1-$n$ corners and unphysical loop propagators,\footnote{We prove these statements in \cref{sec:factorisation}.} while the kinematic integrand $ \cI^{\scalebox{0.6}{$(1)$}}_{\scalebox{0.6}{kin}}$ vanishes on tadpole-like worldsheet geometries \cite{Geyer:2015jch}.

However, this is insufficient to prove that \eqref{eq:int_gen_proposal} is valid in general. Returning to the MHV case, the reduction of the $D$-dimensional formulas to the four-dimensional MHV superamplitudes relies on a particular choice for the reference spinors of the polarisation vectors. In particular, particle 1 plays a privileged role, which is the reason why the objects $X_{a,b}$ used for MHV are defined as \eqref{defXX}. If an analogous representation exists beyond MHV, i.e., one without diagrams with 1-$n$ massive corners, then analogous choices may be required when using \eqref{eq:int_sum_spin} or \eqref{eq:int_gen_v2}. These issues may be related to observations made in the context of the field theory limit of superstring amplitudes. Ref.~\cite{Mafra:2014gja} used for the construction of the loop integrand a convenient set of BRST pseudo-invariant objects $C_{1|\cdots}$ where particle 1 is a reference particle. Very recently, ref.~\cite{Casali:2020knc} discussed the challenging role of trivalent diagrams with 1-$n$ massive corners in the attempt to prove the loop-level colour-kinematics duality from monodromy relations. We hope that these connections will be clarified in the future.

\paragraph{Comment on lower supersymmetry.} In this context, we would also like to comment briefly on the additional difficulty of extending the proposal to theories with lower supersymmetry. The most obvious difference here lies in the kinematic half-integrand, so we might hope for a worldsheet formula for pure Yang-Mills using the above formula but  restricted to NS-states running in the loop,\footnote{This construction does indeed give the pure Yang-Mills integrand in the `linear' formalism, i.e., with the `old' one-loop scattering equations \cite{Geyer:2015jch}.} $\cI^{\scalebox{0.6}{$(1)$}}_{\scalebox{0.6}{kin}}=\cI^{\scalebox{0.6}{$(1)$}}_{\scalebox{0.6}{NS}}=\sum_r\pf'(M_{\text{NS}}^r)$.
However, in contrast to the maximally supersymmetric case, the half-integrand $\cI^{\scalebox{0.6}{$(1)$}}_{\scalebox{0.6}{NS}}$ does \emph{not} vanish for tadpole-like worldsheet geometries,\footnote{We give more details on these tadpole-like configurations -- known as `singular solutions' to the scattering equations -- in \cref{sec:factorisation}.} and we thus expect the worldsheet formulas to lead to unphysical poles in the loop integrand. Contributions from similar poles in the `linear' representation have been shown to loop-integrate to zero in dimensional regularisation \cite{Cachazo:2015aol}, but it is not clear whether that analysis extends to the poles coming from the $\ell^2$-deformed scattering equations. 
%\todoY{I disagree here: for us, this happens one order earlier} On the other hand, the inclusion of degenerate solutions to the scattering equations may address this issue, as discussed in \cite{Cachazo:2015aol}, providing a cancellation of the unphysical poles already at the level of the loop integrand. The loop integrand will then generically fail to be gauge invariant, and one would need to show that the non-gauge invariant pieces vanish upon loop integration.

%%%%%%%%%%%%%%
%%%%%%%%%%%%%%
\section{Proof for the MHV integrand via BCFW recursion}
\label{sec:BCFW_WS}
In this section, we prove the worldsheet formula for the planar MHV integrand satisfies the BCFW recursion \eqref{eq:MHVrecursion},
\begin{align*}
{\mathfrak{I}}^{\scalebox{0.6}{$(1)$}}_{\scalebox{0.6}{MHV}}\!\left(1,2,...,n\right)=\frac{\left\langle n-1,1\right\rangle }{\left\langle n-1,n\right\rangle \left\langle n,1\right\rangle }\;{\mathfrak{I}}^{\scalebox{0.6}{$(1)$}}_{\scalebox{0.6}{MHV}}\!\left(\check{1},2,...,\check{n-1}\right)+\frac1{\ell^2}\int d^4\eta_{0}\,\mathcal{A}^{\scalebox{0.6}{$(0)$}}_{\scalebox{0.6}{NMHV}}\!\left(\ell_{0},\hat{1},2,...,\hat{n},-\ell_{0}\right)\,,
\end{align*}
and is thus a valid representation of the integrand. As before, the deformed momenta are defined via
\begin{equation*}
 \hat k_1=k_1+z\,q\,,\qquad \hat k_n=k_n-z\,q\,,\qquad \hat \ell=\ell-z\,q\,,
\end{equation*}
with $z=z_n=k_{n-1}\cdot k_n/k_{n-1}\cdot q$ in the inverse soft term, and $z=z_0=\alpha=\ell^2/2\ell\cdot q$ in the single cut.
We will first discuss how the two types of poles -- the inverse soft term and the single cut -- emerge from the worldsheet, and show that no other poles contribute.   We will then study the large-$z$ limit of the scattering equations, and check that the boundary terms vanish, ${\mathcal B}_n=0$, completing the proof.

Since many readers may be unfamiliar with some of the background, and also because the discussion is rather technical, we give here a short summary of the main steps. The rest of this section then develops the details, and will be of interest only to the most enthusiastic readers.

\paragraph{Short summary of the proof.} 
The main idea behind the proof is  straightforward: if the worldsheet formula satisfies the BCFW recursion, it gives a valid representation of the loop integrand. Since we checked the seed amplitudes in \S\ref{sec:4pt_WS}, the main technical difficulty lies in translating the recursion onto the worldsheet, which relies crucially on the support of the scattering equations. The techniques we use are thus very similar to those developed at tree-level in ref.~\cite{Dolan:2013isa} and in the linear formalism in ref.~\cite{Geyer:2015jch}, building on older results in string theory, see e.g. \cite{Polchinski:1998rq}.

Let us now take a closer look at what `satisfying the recursion' actually entails for a worldsheet formula. Retracing the steps of \cref{sec:BCFW}, the usual  BCFW argument allows us to express the original worldsheet formula as a sum over its residues at $z\neq 0$. While it is easy to identify the relevant poles in $z$  in a Feynman diagram representation, they are less obvious on the worldsheet representation, where the only familiar-looking pole arises from $\hat\ell^2=0$ in the shift of the overall factor of $1/\ell^{2}$.
All other poles in the worldsheet formula originate from marked points coalescing as solutions to the scattering equations, which depend on the kinematics and hence on $z$. This coalescence corresponds (using reparametrisation invariance on the sphere) to a singular worldsheet geometry as graphically depicted in \cref{fig:poles_phys}. The scattering equations relate these boundaries of the moduli space to singular kinematic configurations, and thus allow us to identify the poles in $z$ whose residues contribute to the recursion. As we saw at the end of \S\ref{sec:SE}, some of the poles encoded by the scattering equations match what we expect from the recursion \eqref{eq:MHVrecursion}, but we also find other, unphysical poles that could potentially contribute.

\begin{figure}[ht]
	\centering 
	  \includegraphics[width=5cm]{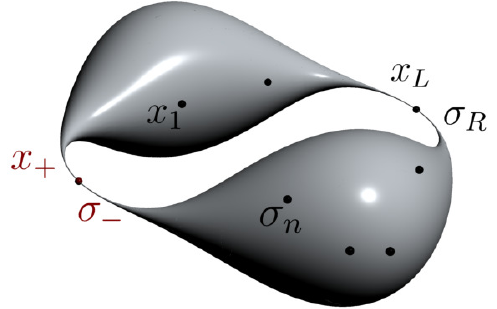}
	  \hfill
	  \raisebox{16pt}{\includegraphics[width=5cm]{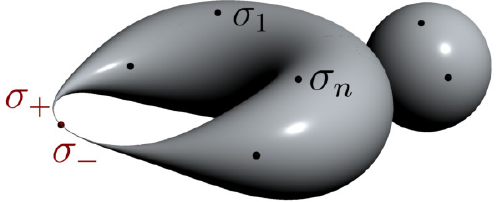}}
	  \hfill
	 \raisebox{16pt}{\includegraphics[width=5cm]{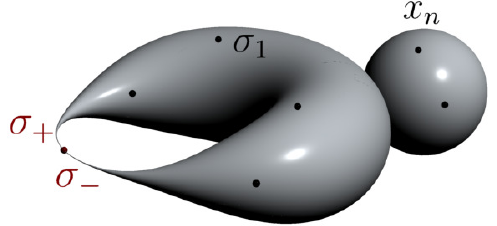}}
	\caption{All singular worldsheet configurations corresponding to physical poles. The geometry on the left gives a single cut, and the two geometries in the middle and on the right correspond to inverse soft channels. In the BCFW recursion, the middle term does not contribute since its pole is unaffected by the shift.}
	\label{fig:poles_phys_v2}
\end{figure}

The main work below is thus to verify that 
\begin{itemize}
 \item[\emph{(i)}]  the residues on the respective physical poles agree with the MHV recursion (so the single cut and the inverse soft term emerge correctly),
 \item[\emph{(ii)}] the residues on all other physical poles vanish (such as non-planar trees),  and 
 \item[\emph{(iii)}] the residues on all unphysical poles vanish.
\end{itemize}
The most straightforward of these is identifying the residue on the single cut. Due to the close similarity of the one-loop worldsheet formula with  tree level, the correct forward-limit tree amplitude just emerges directly when restricting to the cut. On the other hand, the tree factorisation term arises from the singular worldsheet  in the middle and right of \cref{fig:poles_phys_v2}, and we obtain the residues by direct calculation, using a suitable parametrisation for the worldsheet.\footnote{To be more precise, we choose a convenient parametrisation of the moduli space close to the relevant boundary divisor.} The form of the kinematic numerators then ensures that this residue agrees with the inverse soft term in the recursion. For all of point \emph{(i)},  the  form of \emph{both} half-integrands is clearly important; for example, both have to behave appropriately when the worldsheet becomes singular to give the correct residues.

\begin{figure}[ht]
	\centering 
	  \includegraphics[width=5cm]{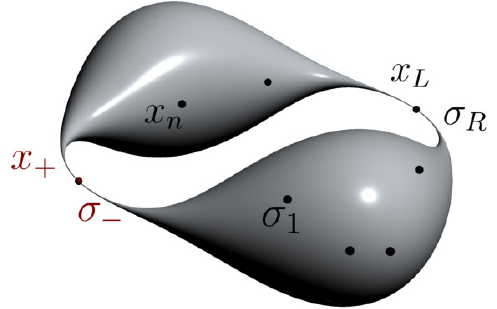}
	  \hfill
	 \raisebox{16pt}{\includegraphics[width=5cm]{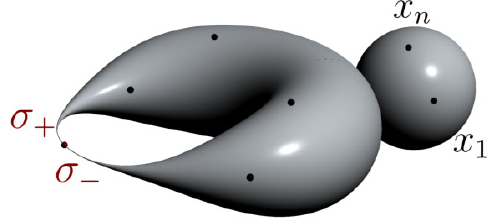}}
	\hfill
	\raisebox{18pt}{\includegraphics[width=5cm]{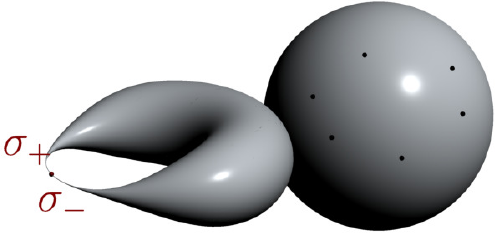}}
	\caption{The singular worldsheet configurations corresponding to the unphysical poles. The geometry on the left, with $+$ and $n$ on one component and $-$ and $1$ gives unphysical loop propagators, the middle one is the problematic 1-$n$ massive corner, and the tadpole-like geometry on the right gives an unphysical discriminant pole. }
	\label{fig:poles_unphys}
\end{figure}

Just as important as obtaining the correct residues on physical poles are \emph{(ii)} and \emph{(iii)}, that no further poles contribute to the recursion. Most concerning here is \emph{(iii)}, the possible presence of unphysical poles.
The scattering equations encode three types of unphysical poles, coming from singular worldsheet geometries  in \cref{fig:poles_unphys}:  tree-level factorisation channels involving both $1$ and $n$,  unphysical loop propagators (reminiscent of a `wrong placement' of the loop momentum), and  strange discriminant poles from tadpole-like worldsheets. For the residues on these poles to vanish, it is sufficient that the contribution from \emph{one} of the half-integrands is zero,
and indeed the two half-integrands vanish on different unphysical poles.\footnote{The language here is inaccurate: the half-integrands themselves usually do not vanish on any of these worldsheet geometries, but their behaviour causes the \emph{full} worldsheet formula -- including the measure and both half-integrands -- to become zero.} 
The colour building block $\mathcal{C}^{\scalebox{0.6}{$(1)$}}_n(12\dots n)$ vanishes on the worldsheet geometries corresponding to the unphysical loop propagators and the 1-$n$ trees, whereas the kinematic half-integrand $ \cI^{\scalebox{0.6}{$(1)$}}_{\scalebox{0.6}{MHV}} $ is zero on tadpole-like worldsheets. The two half-integrands thus jointly ensure that all residues on unphysical poles vanish.

At this point, we would like to comment briefly on the difference between these unphysical poles and the spurious poles in the BCFW recursion.   As can be seen both intuitively from the worldsheet geometries and explicitly from the form of the poles \eqref{eq:poles_unphys}, none of the unphysical poles mentioned above coincides with the spurious BCFW poles in the previous sections. The worldsheet formula is thus manifestly free of the latter. 

To finish the proof, we thus only have to verify that \emph{(ii)}, the residues of the worldsheet formula on physical poles that do not appear in the MHV recursion vanish. The poles in question are associated to tree-level propagators that are  not compatible with the planar ordering, or correspond to the sum of more than two momenta going on-shell, $k_\sL^2=0$ with $|L|>2$. 
On the worldsheet, this cancellation is again an interplay between the two half-integrands: $\mathcal{C}^{\scalebox{0.6}{$(1)$}}_n(12\dots n)$ ensures that only planar factorisation channels contribute, and $ \cI^{\scalebox{0.6}{$(1)$}}_{\scalebox{0.6}{MHV}} $ vanishes for tree-level factorisation channels  with more than two particles. 

This concludes the main part of the proof: we have checked that the worldsheet formula satisfies the BCFW recursion, up to possible boundary contributions. That these vanish can be seen heuristically from the scaling of the scattering equations, and explicitly by finding the asymptotic solutions to the scattering equations in the large-$z$ limit. The worldsheet formula thus satisfies the MHV recursion, and gives a representation of the integrand.

\subsection{The single cut}\label{sec:SC}
Below, we fill in the details for the brief summary given above. We start by mentioning two properties of the MHV numerators that are relevant to the recursion.

\paragraph{More properties of MHV numerators.} 
Throughout this section, we will need two more properties of the  MHV numerators $N^{\scalebox{0.6}{$(1)$}}_{\scalebox{0.6}{$\rho$}}$, describing how they behave under a single cut or on a factorisation channel, \cite{He:2015wgf}.
\begin{itemize}
 \item Single cut: 
 \begin{equation}
  N^{\scalebox{0.6}{$(1)$}}_{\scalebox{0.6}{$\rho$}}\Big|_{\ell\rightarrow\ell_0}=\int d^4\eta_0\;\; N^{\scalebox{0.6}{$(0)$,NMHV}}_{\scalebox{0.6}{$(\ell_0\,\rho\,-\ell_0)$}}\,.
 \end{equation}
 As the loop momentum goes on-shell, $\ell\rightarrow \ell_0$ with $\ell_0^2=0$, the numerators become tree-level NMHV numerators in the forward limit, summed over on-shell states.
\item Factorisation:
\begin{equation}\label{eq:fact_BCJ_nums}
 \mathfrak{n}^{\scalebox{0.6}{$(1)$}}_{\scalebox{0.6}{$(\rho_1[i,j]\rho_2)$}}=X_{i,j}\;\mathfrak{n}^{\scalebox{0.6}{$(1)$}}_{\scalebox{0.6}{$(\rho_1 R\,\rho_2)$}}
\end{equation}
This ensures that the numerators factorise correctly for diagrams with trees attached to an $m$-gon (with $m<n$). We used the notation $k_\sR=k_i+k_j$ for the numerator on the right. Notice that this is just a kinematic Jacobi relation between the MHV numerators of two $n$-gons (left-hand side) and an $(n-1)$-gon (right-hand side).
\end{itemize}

\paragraph{The single cut.} Consider first the single cut contribution. Since the worldsheet formula carries an explicit factor of $1/\ell^2$ (and the scattering equations don't encode a pole of the form $\ell^2$), $\mathfrak{I}^{\scalebox{0.6}{$(1)$}}_{\text{sYM}}(z)$ has a residue at $z=z_0$  given by
\begin{equation}
 \mathrm{Res}_{z_0}\,\frac{\mathfrak{I}^{\scalebox{0.6}{$(1)$}}_{\text{sYM}}(z)}{z}=\frac{1}{\ell^2}\; \left(\int_{\raisebox{-6pt}{\scalebox{0.7}{$\mathfrak{M}_{0,n+2}$}}}\hspace{-15pt}d\mu_{n}^{\Def} \;\, \cI^{\scalebox{0.6}{$(1)$}}_{\scalebox{0.6}{MHV}} \;\mathcal{C}^{\scalebox{0.6}{$(1)$}}(12\dots n)\Bigg|_{\mathrm{SC}}\right)\,,
\end{equation}
with the kinematic and colour half-integrands  as defined in the last section. We use SC to indicate that the integral is evaluated on the  single cut $\ell\rightarrow \ell_0$, $k_1\rightarrow \hat k_1=k_1+\alpha q$, $k_n\rightarrow \hat k_n=k_n-\alpha q$.  Comparing this to the BCFW recursion, we would like to show that this single cut agrees with the forward limit of the tree-level NMHV amplitude.

For the scattering equations, it is clear from the definition of the single cut that they reduce to the tree-level constraints in the forward limit,
\begin{equation}
 \mathcal{E}_a^\Def 
  \Bigg|_{\mathrm{SC}}=
  \mathcal{E}_a^{\scalebox{0.6}{$(0)$}}=\sum_{b\neq a}\frac{\hat k_a\cdot \hat k_b}{\sigma_{ab}}\,,
\end{equation}
where the sum on the right runs over   $n+2$ on-shell particles with momenta $\hat k_i=k_i$ for $i=2,\dots,n-1$, and $\hat k_1=k_1+\alpha q$, $\hat k_n=k_n-\alpha q$, as well as $\hat k_{n+1}= \ell_0$ and $\hat k_{n+2}=-\ell_0$. The measure thus reduces to the tree-level CHY measure for $n+2$ particles, again in the forward limit,
\begin{equation}
 d\mu_n^\Def 
  \Bigg|_{\mathrm{SC}}=
 d\mu_{n+2}^\CHY 
\end{equation}
The integrand is similarly straightforward: the colour factor already agrees with the forward-limit contribution, $ \mathcal{C}^{\scalebox{0.6}{$(1)$}}(12\dots n)=\mathcal{C}^{\scalebox{0.6}{$(0)$}}(+12\dots n-)$, and the kinematic half-integrand becomes
\begin{equation}
 \cI^{\scalebox{0.6}{$(1)$,MHV}}_{\scalebox{0.6}{$n$}} 
  \Bigg|_{\mathrm{SC}}=
  \cI^{\scalebox{0.6}{$(0)$,NMHV}}_{\scalebox{0.6}{$n+2$}} =\int d^4\eta_0\;\; \sum_{\rho\in S_n} \frac{N^{\scalebox{0.6}{$(0)$,NMHV}}_{\scalebox{0.6}{$(\ell_0\,\rho\,-\ell_0)$}}}{(+12\dots n-)}\,,
\end{equation}
which is the required forward limit of the NMHV integrand, summed over states running through the cut. The tree-level half-integrand appears here in the BCJ representation, with half-ladder master diagrams with endpoints $\ell_0$ and $-\ell_0$.
The single cut is thus correctly given by the forward limit of the tree-level amplitude,
\begin{equation}
 \mathrm{Res}_{\alpha}\,\frac{\mathfrak{I}^{\scalebox{0.6}{$(1)$}}_{\text{sYM}}(z)}{z}=\frac{1}{\ell^2}\int d^4\eta_0\;\mathcal{A}_{n+2}^{\scalebox{0.6}{$(0)$}}(\alpha)\,,
\end{equation}
in line with the recursion relation.

\subsection{Factorisation}\label{sec:factorisation}
 For scattering-equations-based amplitude representations, factorisation channels of the amplitude are in one-to-one correspondence with boundary divisors of the of the moduli-space  $\mathfrak{M}_{0,n}$ of $n$ marked points on the Riemann sphere  \cite{Dolan:2013isa, Geyer:2015jch}. Below, we will first show how the scattering equations provide this map, and derive the momentum-space singularity structure it entails. This fills in the  missing details for our discussion in \cref{sec:SE}. To prove that the MHV worldsheet formula factorises correctly, we will then verify that it has  
\begin{itemize}
\item[(i)] the correct 2-particle factorisation channels, including both the  poles and respective residues,
 \item[(ii)] no $k$-particle factorisation channels for $k>2$ ,
 \item[(iii)] no unphysical factorisation channels, i.e. no massive 1-$n$ corners, no non-cyclic poles and no contributions from singular solutions.
\end{itemize}

\subsubsection{Scattering equations and measure}
Except for the overall factor of $1/\ell^2$, the worldsheet integrand in \eqref{eq:MHV-proposal} carries no explicit poles in the kinematic data. This is true for all scattering equations based amplitudes formulas, and it is in this sense that the scattering equations are universal for massless theories: they dictate the possible poles. Since all poles of the worldsheet integrand are of the form $\sigma_a -\sigma_b$, corresponding to boundary divisors of the moduli space $\partial \widehat{\mathfrak{M}}_{0,n+2}$, we should study the behaviour of the worldsheet formula around these boundary divisors to identify the behaviour of the amplitude in kinematic space. As we will see, the scattering equations will provide a map from a given bounday divisor $\partial_{\sL,\sR} \widehat{\mathfrak{M}}_{0,n+2}$ to a singular kinematic configuration defined by $s_\sL=0$, where $s_\sL=s_\sL(k_i,\ell)$ depends only on the momenta of the external particles and the loop momentum.

On the sphere, the  boundary of the moduli space\footnote{We really mean the boundary of the Deligne-Mumford compactification of the moduli space.}  $\partial \widehat{\mathfrak{M}}_{0,n+2}$  is the set of separating degenerations  $\partial_{\sL,\sR} \widehat{\mathfrak{M}}_{0,n+2}$ that split the sphere $\Sigma$ into two components, $\Sigma_\sL$ and $\Sigma_\sR$, with the punctures partitioned  such that $n+2=n_\sL+n_\sR$,
\begin{equation}\label{eq:dM}
 \partial_{\sL,\sR} \widehat{\mathfrak{M}}_{0,n+2}\simeq \widehat{\mathfrak{M}}_{0,n_L+1}\times\widehat{\mathfrak{M}}_{0,n_R+1}\,.
\end{equation}
To study the behaviour of the worldsheet formula close to this boundary, we parametrise the moduli space close to  $\partial_{\sL,\sR} \widehat{\mathfrak{M}}_{0,n+2}$ by  gluing together two  Riemann spheres $\Sigma_\sL$ and $\Sigma_\sR$.  The procedure is standard, and can be found e.g. in \cite{Polchinski:1998rq}:  let us parametrise the sphere $\Sigma_\sR$ by a variable $\sigma$, and $\Sigma_\sL$ by a variable $x$. Moreover, we will  choose a point on each sphere,  $\sigma_\sR\in \Sigma_\sR$ and $x_\sL\in\Sigma_\sL$, and remove the disks $|\sigma-\sigma_\sR|<\varepsilon^{1/2}$ and $|x-x_\sL|<\varepsilon^{1/2}$. Then we can glue the two spheres into a single surface by identifying
\begin{equation}\label{eq:param_dM}
 \left(x-x_{\sL}\right)\left(\sigma-\sigma_{\sR}\right)=\varepsilon \,.
\end{equation}
In all of this, the parameter $\varepsilon$ determines the size of the disk, and thus the boundary $\partial_{\sL,\sR} \widehat{\mathfrak{M}}_{0,n+2}$ corresponds to the limiting case $\varepsilon\rightarrow 0$.\footnote{A common gauge choice is $x_\sL=0$, such that \eqref{eq:param_dM} becomes $
 \sigma=\sigma_\sR+\varepsilon \tilde x,$ with $\tilde x=x^{-1}$.}

 Let us briefly discuss how \eqref{eq:param_dM} should be understood. For any marked point $i$, the relation \eqref{eq:param_dM} provides the transition map between the $x$-parametrisation and the $\sigma$-parametrisation of the surface. If $\varepsilon\ll1$, we furthermore know that 
 \begin{align*}
  &\text{(i) \;\,either} && x_i-x_\sL \sim 1 && \Leftrightarrow  && \sigma_i-\sigma_\sR\sim\varepsilon\,, && i\in \Sigma_\sL\,,\\
&\text{(ii)\;\,or} &&x_i-x_\sL\sim \varepsilon && \Leftrightarrow  && \sigma_i-\sigma_\sR\sim1\,, && i\in \Sigma_\sR\,.
 \end{align*}
 This defines what we mean by the partition of punctures onto the subspheres $\Sigma_\sL$ and $\Sigma_\sR$. If the distance between a puncture and the node on the sphere remains of order one throughout the degeneration $\varepsilon\rightarrow 0$, e.g. $x_i-x_\sL\sim 1$, then clearly the point $i$ will end up on the sphere $\Sigma_\sL$. We can also reverse this argument; what it means for a point to lie on $\Sigma_\sL$ is that in the parametrisation of the other sphere $\Sigma_\sR$, that point must lie on the node -- which is exactly what we see from $\sigma_i-\sigma_\sR\sim\varepsilon\rightarrow0$.
 
 If the punctures $\sigma_i$ are determined by the scattering equations, we can use this parametrisation to derive  the following lemma relating $\partial_{\sL,\sR} \widehat{\mathfrak{M}}_{0,n+2}$ to kinematic constraints $s_\sL=0$.

\begin{lemma} \label{lem:CHY-fact}
If the marked points $\sigma_a$ satisfy the scattering equations $\mathcal{E}_a^\Def=0$,
then the boundary component $\sigma_a\in \partial_{\sL,\sR} \widehat{\mathfrak{M}}_{0,n}$ implies  $s_L=0$, with 
\begin{equation}
 s_{\sL}=
 \begin{cases}
 k_{\scalebox{0.5}{$L$}}^2   & L=L^{\scalebox{0.6}{ext}}\;\;\mathrm{or}\;\; L=\{+,- \}\cup L^{\scalebox{0.6}{ext}}\\
 (\ell-k_{\scalebox{0.5}{$L$}})^2 & L=\{+\}\cup L^{\scalebox{0.6}{ext}},\;1\in L^{\scalebox{0.6}{ext}},\;n\notin L^{\scalebox{0.6}{ext}}\\ 
  (\ell+k_{\scalebox{0.5}{$L$}})^2 & L=\{-\}\cup L^{\scalebox{0.6}{ext}},\;1\notin L^{\scalebox{0.6}{ext}},\;n\in L^{\scalebox{0.6}{ext}}\\ 
 \mathrm{unphys} & \mathrm{else} \end{cases}
\end{equation}
where $k_\sL=-\sum_{i\in L^{\scalebox{0.5}{ext}}}k_i$ is the sum over momenta of external particles only.\footnote{Following the same calculation as in the main text, we can easily identify the unphysical poles as
\begin{equation}\label{eq:poles_unphys}
 s_{\sL}=
 \begin{cases}
 -2\ell\cdot k_\sL +k_\sL^2 & L=\{+\}\cup L^{\scalebox{0.6}{ext}},\;1,n\in L^{\scalebox{0.6}{ext}},\;\;\mathrm{or}\;\;1,n\notin L^{\scalebox{0.6}{ext}}\\ 
 +2\ell\cdot k_\sL +k_\sL^2 & L=\{-\}\cup L,\;1,n\in L^{\scalebox{0.6}{ext}},\;\;\mathrm{or}\;\;1,n\notin L^{\scalebox{0.6}{ext}}\\ 
  -\ell^2-2\ell\cdot k_\sL+k_\sL^2& L=\{+\}\cup L^{\scalebox{0.6}{ext}},\;1\notin L^{\scalebox{0.6}{ext}},\;n\in L^{\scalebox{0.6}{ext}}\\ 
   -\ell^2+2\ell\cdot k_\sL+k_\sL^2  & L=\{-\}\cup L^{\scalebox{0.6}{ext}},\;1\in L^{\scalebox{0.6}{ext}},\;n\notin L^{\scalebox{0.6}{ext}}\\ 
  \end{cases}
\end{equation}
We see that some while some are the familar linear propagator poles, others are even more manifestly unphysical, and reminiscent of unphysical poles encoded by the two-loop scattering equations \cite{Geyer:2016wjx}. There is also another class of unphysical poles coming from so-called singular solutions with $L=\{+,-\}$, which we will discuss below.
}
\end{lemma}
\paragraph{Proof. } This can be seen as follows. Consider first the tree-level factorisation case, where we can take wlog $L=L^{\scalebox{0.6}{ext}}$ (otherwise switch the roles of $\Sigma_\sL$ and $\Sigma_\sR$). Close to the boundary of the moduli space, the scattering equations behave to leading order as
\begin{equation}\label{eq:SE_scale_epsilon}
\mathcal{E}_a^\Def=
   \begin{cases}
       \frac{x_{a\ssL}^2}{\varepsilon}\; \mathcal{E}_a^{\scalebox{0.6}{$(L)$}}\,\; & \text{if } a\in \Sigma_\sL\\
       \hspace{15pt}\mathcal{E}_a^{\scalebox{0.6}{$(R)$}}\,\; & \text{if } a\in \Sigma_\sR
   \end{cases}\,,
\end{equation}
with the natural scattering equations on each subsphere,
\begin{equation}
 \mathcal{E}_a^{\scalebox{0.6}{$(L)$}}=\mathcal{E}_a^{\scalebox{0.6}{$(0)$}}=\sum_{j\in L}\frac{2k_a\cdot k_j}{x_{aj}}+\frac{2k_a\cdot k_\sL}{x_{a\sL}}\,,\qquad  \qquad
 \mathcal{E}_a^{\scalebox{0.6}{$(R)$}}=\mathcal{E}_a^\Def\Bigg|_{\Sigma_\sR}\,.
\end{equation}
The scattering equations $\mathcal{E}_a^{\scalebox{0.6}{$(R)$}}$ are thus the $\ell^2$-deformed scattering equations for the subsphere $\Sigma_\sR$, with momentum $k_\sR=-k_\sL$ running through the nodal point $\sigma_\sR$. As an example, if $1,n \notin L$, then
\begin{subequations}
\begin{align}
 \mathcal{E}^{\scalebox{0.6}{$(R)$}}_1&=\sum_{q\in R} \frac{2k_1\cdot k_q}{\sigma_{1q}} + \frac{2k_1\cdot k_\sR}{\sigma_{1\sR}}+ \left(\ell+ k_1\right)^2\frac{\sigma_{+-}}{\sigma_{1+}\sigma_{1-}}   \\
 \mathcal{E}^{\scalebox{0.6}{$(R)$}}_n&=\sum_{q\in R} \frac{2k_n\cdot k_q}{\sigma_{nq}} + \frac{2k_n\cdot k_\sR}{\sigma_{n\sR}}- \left(\ell- k_n\right)^2\frac{\sigma_{+-}}{\sigma_{n+}\sigma_{n-}}\,\\
 %
 %\mathcal{E}^{\scalebox{0.6}{$(R)$}}_p&=\sum_{q\in R} \frac{2k_p\cdot k_q}{\sigma_{pq}}+ \frac{2k_p\cdot k_\sR}{\sigma_{p\sR}} + 2\ell\cdot k_p\frac{\sigma_{+-}}{\sigma_{p+}\sigma_{p-}}\\
 %
  \mathcal{E}^{\scalebox{0.6}{$(R)$}}_\pm&=\pm\sum_{q\in R, q\neq 1,n} \frac{2\ell\cdot k_q}{\sigma_{\pm q}}\pm \frac{2\ell\cdot k_\sR}{\sigma_{\pm \sR}} \pm \frac{\left(\ell+ k_1\right)^2}{\sigma_{\pm 1}}\mp \frac{\left(\ell- k_n\right)^2}{\sigma_{\pm n}}\,.
\end{align}
\end{subequations}
If $n\in L$, then the nodal point $\sigma_\sR$ carries the quadratic dependence in the loop momentum (playing effectively the role of $\sigma_n$),
\begin{align}
  \mathcal{E}^{\scalebox{0.6}{$(R)$}}_\pm&=\pm\sum_{q\in R, q\neq 1} \frac{2\ell\cdot k_q}{\sigma_{\pm q}}\pm \frac{\left(\ell+ k_1\right)^2}{\sigma_{\pm 1}}\mp \frac{\left(\ell- k_\sR\right)^2}{\sigma_{\pm \sR}}\,.
\end{align}
It is clear that this configuration only solves the $\ell^2$-deformed scattering equations if the kinematics are singular. To find this constraint on the kinematic data, consider the following combination of scattering equations: $s_\sL:= -\sum_{i\in L}\sigma_{i\sR}\,\mathcal{E}_i^\Def$. Clearly this must vanish since it is a linear combination of the constraint equations. On the other hand, evaluating it explicitly, we find to leading order in $\varepsilon$,
\begin{equation}\label{eq:s_L}
0=s_\sL:=-\sum_{i\in L}\sigma_{i\sR}\,\mathcal{E}_i^\Def=-\sum_{i,j\in L} x_{i\sL}\left(\frac{2k_i\cdot k_j}{ x_{ij}}+\frac{2k_i\cdot k_\sL}{ x_{i\sL}}\right)=-\sum_{i,j\in L} k_i\cdot k_j+2k_\sL^2=k_\sL^2\,.
\end{equation}
 Repeating this calculation for the various boundary components of the moduli space gives the result of \cref{lem:CHY-fact}. \hfill $\Box$\\

Above, we have discussed the tree-level case in detail since this is the only contribution to the BCFW recursion. The other physical poles correspond again to single cuts, but the BCFW shift \eqref{eq:shift} does not probe these. Note that the unphysical poles \emph{do} get deformed by the BCFW shift, so it will be important to show that they are absent from the final worldsheet formula. 

The form of the scattering equations further ensures that the measure factorises on the boundary divisor  $\partial_{\sL,\sR} \widehat{\mathfrak{M}}_{0,n}$. This has been discussed in detail in \cite{Dolan:2013isa, Geyer:2015jch}, and we refer the interested reader to these references for details of the derivation. All steps carry over straightforwardly since they only rely on the scaling \eqref{eq:SE_scale_epsilon} of the scattering equations, as well as the form $s_\sL:= -\sum_{i\in L}\sigma_{i\sR}\,\mathcal{E}_i^\Def$ of the kinematic constraint in terms of the scattering equations and marked points. For our case of interest with $L=L^{\scalebox{0.6}{ext}}$, the measure then becomes
\begin{equation}\label{eq:fact_measure}
 d\mu_n^{\Def}=\frac{\varepsilon^{2\left(n_\ssL-1\right)}}{\prod_{i\in L}x_{i\sL}^4}\frac{d\varepsilon}{\varepsilon}\,\delta\left(k_\sL^2 - \varepsilon\mathcal{F}\right)\;d\mu_{n_\ssL+1}^{\scalebox{0.6}{$(0)$}}\;d\mu_{n_\ssR+1}^{\Def}\,.
\end{equation}
Here, $d\mu_{n_\ssL+1}^{\scalebox{0.6}{$(0)$}}=d\mu_{n_\ssL+1}^\CHY$ denotes the tree-level CHY measure on $\Sigma_\sL$ and $d\mu_{n_\ssR+1}^{\Def}$ is the loop measure on $\Sigma_\sR$. The delta-function $\delta\left(k_\sL^2 - \varepsilon\mathcal{F}\right)$ ensures that $k_\sL^2$ is indeed of order $\varepsilon$,  and $\mathcal{F}$ is a function of the marked points and kinematics that will drop out of the final residue. For \emph{any} boundary component of the moduli space (so for any choice of $L$), the measure scales in $\varepsilon$ as 
\begin{equation}
 d\mu_n^{\Def}\sim \varepsilon^{2\left(n_\ssL-1\right)}\frac{d\varepsilon}{\varepsilon}\,\delta\left(s_\sL- \varepsilon\mathcal{F}\right)\,.
\end{equation}
This behaviour  of the measure ensures that if the half-integrands scale as 
\begin{itemize}
 \item $\cI^{\scalebox{0.6}{$(1)$}}_{\scalebox{0.5}{$1/2$}} \sim \varepsilon^{-m}$\, \;with $m<n_\ssL-1$:\\
 The residue at $z_\sL$ vanishes, where  $z_\sL$ is defined to probe the pole $s_\sL$, i.e. $\hat s_\sL (z_\sL)=0$.
 \begin{equation}\label{eq:int_no_res}
   \mathrm{Res}_{z_\ssL}\;\frac{\mathfrak{I}^{\scalebox{0.6}{$(1)$}}(z)}{z}= \mathrm{Res}_{z_\ssL}\;\int d\varepsilon\, \varepsilon^{\scalebox{0.6}{$M$}}\,\delta\left(\hat s_\sL- \varepsilon\mathcal{F}\right) \mathfrak{R}(z) =0\,.
 \end{equation}
 Here, $M=2(n_\sL-m-1)\geq 0$, and the (`possible residue') $\mathfrak{R}$ contains  the remaining measure factors in \eqref{eq:fact_measure}, as well as factors from the half-integrands.  It is a polynomial in $\varepsilon$, with leading term of order one, and thus the residue vanishes.
 \item $\cI^{\scalebox{0.6}{$(1)$}}_{\scalebox{0.5}{$1/2$}} \sim \varepsilon^{-\left(n_\ssL-1\right)}$:\\
 The residue at $z_\sL$ (defined as above) picks up the simple pole in $s_\sL$. 
 \begin{equation}\label{eq:int_res}
   \mathrm{Res}_{z_\ssL}\;\frac{\mathfrak{I}^{\scalebox{0.6}{$(1)$}}(z)}{z}= \mathrm{Res}_{z_\ssL}\;\int \frac{d\varepsilon}{\varepsilon}\, \,\delta\left(\hat s_\sL- \varepsilon\mathcal{F}\right) \mathfrak{R}(z)=  \frac{1}{s_\sL}\mathfrak{R} (z_\sL)\,.
 \end{equation}
 We will discuss these residues $\mathfrak{R}$ in more detail in the next section.
\end{itemize}
Note that higher-order poles are never possible for integrands formed out of Parke-Taylor factors.

\subsubsection{Wordsheet integrand}
Using the results \eqref{eq:int_no_res} and \eqref{eq:int_res}, we can now phrase more carefully what properties the worldsheet integrands must have in order to factorise correctly under BCFW.
\begin{itemize}
\item[(i)]  \emph{Correct 2-particle factorisation channels:}\\
For $L=\{i,i+1\}$ for any $i=2,\dots,n-1$, both half-integrands $\cI^{\scalebox{0.6}{$(1)$}}_{n}: =\cI^{\scalebox{0.6}{$(1)$}}_{\scalebox{0.5}{$1/2$},n}$ must behave as 
\begin{equation}\label{eq:fact_MHVbar}
 \cI^{\scalebox{0.6}{$(1)$,MHV}}_{n} = \varepsilon^{-\left(n_\ssL-1\right)}\prod_{i\in L}x_{i\sL}^2\,\;\sum_{\mathrm{states}}\cI^{\scalebox{0.6}{$(0),\overline{\mathrm{MHV}}$}}_{3} \,\cI^{\scalebox{0.6}{$(1)$,MHV}}_{n-1} \,.
\end{equation}
Here, $\cI^{\scalebox{0.6}{$(0),\overline{\mathrm{MHV}}$}}:=\cI^{\scalebox{0.6}{$(0),\overline{\mathrm{MHV}}$}}_{\scalebox{0.5}{$1/2$}}$ is the tree-level half-integrand, and the factorised form of the integrand ensures that we find the correct residue $\mathfrak{R}(z_i)=\int d\eta_0\; \mathcal{A}^{\scalebox{0.6}{$(0),\overline{\mathrm{MHV}}$}}_3\; \mathfrak{I}^{\scalebox{0.6}{$(1)$,MHV}}_{n-1}$ on the pole $s_{i,i+1}$.\footnote{This is a special case of the more general factorisation condition
\begin{equation}\label{eq:fact_int}
 \cI^{\scalebox{0.6}{$(1)$}}_{n} = \varepsilon^{-\left(n_\ssL-1\right)}\prod_{i\in L}x_{i\sL}^2\,\;\sum_{\mathrm{states}}\mathcal{I}_{n_\ssL+1}^{\scalebox{0.6}{$(0)$}} \,\mathcal{I}_{n_\ssR+1}^{\scalebox{0.6}{$(1)$}} \,.
\end{equation}
}
We also need to verify that there is no contribution from $\cI^{\scalebox{0.6}{$(0),$MHV}}_{3} $.
 \item[(ii)] \emph{No $k$-particle tree factorisation channels for $k>2$: }\\
 By property \eqref{eq:int_no_res}, for $L=L^{\scalebox{0.6}{ext}}$ at least one of the half-integrands must scale as 
 \begin{equation}\label{eq:no-res}
  \cI^{\scalebox{0.6}{$(1)$}}_{\scalebox{0.5}{$1/2$}} \sim \varepsilon^{-m}\, \qquad\;\text{with } \;m<|L|-1\,.
 \end{equation}
 Note that even if only one integrand scales like \eqref{eq:no-res}, the residue on the pole $s_\sL=k_\sL^2$ vanishes.
  \item[(iii)] \emph{No unphysical factorisation channels:}\\
  There are three possible unphysical factorisation channels:  trees not respecting the colour-ordering,  trees containing both particles $1$ and $n$, and unphysical loop contributions. In all three cases, we must verify that at least one of the half-integrands behaves as follows:
 \begin{align}
 & \text{no non-cyclic poles:} && \cI^{\scalebox{0.6}{$(1)$}}_{\scalebox{0.5}{$1/2$}} \sim \varepsilon^{-m}\, \;\text{with } \;m<k-1  && \text{for } \,L\,\text{ non-cyclic}\\
  &\text{no massive }\text{1-$n$}\text{ corners:} &&  \cI^{\scalebox{0.6}{$(1)$}}_{\scalebox{0.5}{$1/2$}} \sim \varepsilon^{-m}\, \;\text{with } \;m<n_\sL -1 &&\text{for } \,L= L^{\scalebox{0.6}{ext}}\supset\{1,n\}  \\
  &\text{no unphysical loop props: } &&  \cI^{\scalebox{0.6}{$(1)$}}_{\scalebox{0.5}{$1/2$}} \sim \varepsilon^{-m}\, \;\text{with } \;m<n_\sL-1 &&\text{for } \,L= L^{\scalebox{0.6}{unphys$(\ell)$}}
  \,,
 \end{align}
 where the unphysical loop configurations $L^{\scalebox{0.6}{unphys$(\ell)$}}$ where listed in \eqref{eq:poles_unphys}; see also \cref{fig:poles_unphys}.
 There is a fourth type of unphysical pole from tadpole-like configurations with $L=\{+,-\}$ which we will discuss separately below.
\end{itemize}

We will derive all of these features in detail below, but here let us briefly anticipate our findings, and discuss how the two half-integrands conspire to cancel all unphysical poles. As we will see, the colour building block $\mathcal{C}^{\scalebox{0.6}{$(1)$}}_n(12\dots n)$ ensures that there are no non-cyclic poles, no massive 1-$n$ corners, and no unphysical loop propagators.\footnote{This discussion will carry over to the general integrand proposal of \eqref{eq:int_gen_proposal}.} The kinematic half-integrand $ \cI^{\scalebox{0.6}{$(1)$}}_{\scalebox{0.6}{MHV}} $, on the other hand, guarantees the absence of  $k$-particle factorisation channels, and vanishes on the tadpole-like worldsheet geometries. We thus see that the form of both half-integrands is crucial for obtaining the correct behaviour.

\paragraph{Colour.} It is easy to check that the colour factor $\mathcal{C}^{\scalebox{0.6}{$(1)$}}_n(12\dots n)$ respects the ($k$-particle) factorisation condition \eqref{eq:fact_int} -- and thus in particular the 2-particle condition \eqref{eq:fact_MHVbar} -- for factorisation channels  with $L=\{i,i+1,\dots,i+|L|\}$ for some $|L|<n-i$, i.e. that are consecutive in  $(+12\ldots n-)$.
\begin{align}
 \mathcal{C}_n^{\scalebox{0.6}{$(1)$}}(12\dots n)
 &=\varepsilon^{-\left(n_\ssL-1\right)}\prod_{j\in L}x_{j\sL}^2\,\;\left(\sum_{a_\ssL, a_\ssR}\delta_{a_\ssL a_\ssR}\;\; \mathcal{C}_{n_\ssL+1}^{\scalebox{0.6}{$(0)$}}(\scalebox{0.8}{$i\dots i+|L|,L$})\;\; \mathcal{C}^{\scalebox{0.6}{$(1)$}}_{n_\ssR+1}(\scalebox{0.8}{$1\dots i-1,R,i+|L|+1\dots n$})\right)\nonumber \,.
\end{align}
Here, the sum over states is expressed as a sum over generators of the colour Lie algebra, using that\footnote{and using a U$(1)$ decoupling identity if the gauge group is SU$(N)$.}
\begin{equation}
 \tr(T^{a_1}T^{a_2}\dots T^{a_n})=\sum_{a_\ssL,a_\ssR}\delta_{a_\ssL a_\ssR} \;\tr(T^{a_i}\dots T^{a_{i+|\sL|}} T^{a_\sL})\;\tr(T^{a_1}\dots T^{a_{i-1}}T^{a_{\sR}}T^{a_{i+|\sL|+1}}\dots T^{a_n})\,.
\end{equation}
On the other hand, if $L$ is non-consecutive in $(+12\ldots n-)$, then the colour half-integrand instead behaves as 
\begin{equation}
 \mathcal{C}_n^{\scalebox{0.6}{$(1)$}}(12\ldots n)
 \sim \varepsilon^{-m}\,,\qquad\qquad \mathrm{with}\;\; m<n_\sL-1\,,
\end{equation}
and thus gives vanishing residues on the corresponding poles. But boundary divisors with $L=L^{\scalebox{0.6}{ext}}$  non-consecutive in $(+12\ldots n-)$ are in one-to one correspondence with tree-level factorisations that don't respect the colour-ordering $(12\ldots n)$ of the amplitude, and trees containing both particles $1$ and $n$. Since it is sufficient for one of the half-integrands to exhibit the behaviour  \eqref{eq:int_no_res} on an unphysical pole, we conclude that any worldsheet formula containing $\mathcal{C}^{\scalebox{0.6}{$(1)$}}_n(12\dots n)$ does not give massive  1-$n$ corners or non-colour-ordered tree-level factorisation channels.

Similarly, we can easily verify that all unphysical loop propagators correspond to boundaries with $L$ non-consecutive in $(+12\ldots n-)$. By the same argument, any worldsheet formula containing $\mathcal{C}^{\scalebox{0.6}{$(1)$}}_n(12\dots n)$ will therefore only give physical loop propagators.

\paragraph{Kinematics.} Consider first a two-particle factorisation channel, with $L=\{i,j\}$ for any $i$ and $j$.  To leading order in $\varepsilon$, the kinematic integrand $\cI^{\scalebox{0.6}{$(1)$}}_{\scalebox{0.6}{MHV}}$ becomes
\begin{align}
 \cI^{\scalebox{0.6}{$(1)$,MHV}}_{n}
 %
 %&=\sum_{m=0}^{n-2}\sum_{\substack{\rho=(\rho_1,\rho_2)\\\in S_m\times S_{n-m-2}}} \left(\frac{N^{\scalebox{0.6}{$(1)$}}_{\scalebox{0.6}{$(\rho_1\,AB\,\rho_2)$}}}{(+\rho_1\,\scalebox{0.8}{$AB$}\,\rho_2\,-)} + \frac{N^{\scalebox{0.6}{$(1)$}}_{\scalebox{0.6}{$(\rho_1\,BA\,\rho_2)$}}}{(+\rho_1\,\scalebox{0.8}{$BA$}\,\rho_2\,-)}\right)\nonumber\\
 %
 &=\varepsilon^{-\left(n_\ssL-1\right)}\prod_{l\in L}x_{l\sL}^2\,\;\left(\sum_{m=0}^{n-2}\sum_{\substack{\rho=(\rho_1,\rho_2)\\\in S_m\times S_{n-m-2}}} \frac{1}{(+\rho_1\,\scalebox{0.8}{$R$}\,\rho_2\,-)_\sigma}\left(\frac{N^{\scalebox{0.6}{$(1)$}}_{\scalebox{0.6}{$(\rho_1\,ij\,\rho_2)$}}}{(\scalebox{0.8}{$L\,ij$})_x} + \frac{N^{\scalebox{0.6}{$(1)$}}_{\scalebox{0.6}{$(\rho_1\,ji\,\rho_2)$}}}{(\scalebox{0.8}{$L\,ji$})_x}\right)\right)\nonumber\\
 &=\varepsilon^{-\left(n_\ssL-1\right)}\prod_{l\in L}x_{l\sL}^2\,\;\left(\delta^8(Q)\;\left(\prod_{a=2}^n\frac{1}{\la 1a\ra^2}\right)\;\frac{X_{ij}}{(\scalebox{0.8}{$L\,ij$})_x}\;\left(\sum_{\rho\in S_{n-1}} \frac{\mathfrak{n}^{\scalebox{0.6}{$(1)$}}_{\scalebox{0.6}{$\rho$}}}{(+\rho\,-)_{\sigma}}\right)\right)\,.
\end{align}
To improve readability, we introduced the  notation $(\rho_1\rho_2\ldots\rho_n)_\sigma=\sigma_{\rho_1\rho_2}\ldots\sigma_{\rho_n\rho_1}$ and $(\rho_1\rho_2\ldots\rho_n)_x=x_{\rho_1\rho_2}\ldots x_{\rho_n\rho_1}$ for Parke-Taylor factors on the subspheres $\Sigma_\sR$ and $\Sigma_\sL$, respectively. The first equality just relies on the factorisation properties of Parke-Taylor factors discussed above, and the second equality follows from property \eqref{eq:fact_BCJ_nums} of the BCJ numerators. At this point, we see that there is no contribution from the tree-level MHV amplitude, because for MHV the numerator vanishes,  $X_{ij}=\la 1i\ra[ij]\la j1\ra=0$ due to $[ij]=0$. The only non-vanishing two-particle factorisation channel thus comes from the 3-point $\overline{\mathrm{MHV}}$ amplitude. In that case, we can simplify the above expression further by noting that the supersymmetric delta-function factorises appropriately,
\begin{equation}
 [ij]^4\,\delta^8(Q)=\int d^4\eta_\sL\;\delta^8\big(Q^{\scalebox{0.6}{$(R)$}}\big)\;\delta^4\big([ij]\eta_\sL+[jL]\eta_i+[L i]\eta_j\big)\,,
\end{equation}
on the support of momentum conservation. This thus carries over to the numerator prefactors,
\begin{equation*}
 \delta^8(Q)\;\left(\prod_{a=2}^n\frac{1}{\la 1a\ra^2}\right) = \int d^4\eta_\sL\;\left(\delta^8\big(Q^{\scalebox{0.6}{$(R)$}}\big)\left(\frac{1}{\la 1R\ra^2}\prod_{\substack{a\in R\\a\neq 1}}\frac{1}{\la 1a\ra^2}\right)\right)\;\left(\frac{\delta^4\big([ij]\eta_\sL+[jL]\eta_i+[L i]\eta_j\big)}{[ij][jL][Li]\;X_{ij}}\right)\,.
\end{equation*}
\iffalse 
On the right we have used momentum conservation and the anti-MHV condition for the proportionalities of the $\lambda$'s,
\begin{equation}
 \lambda_i= -\frac{[L j]}{[ij]}\lambda_\sL\,,\qquad  \lambda_j= -\frac{[Li]}{[ji]}\lambda_\sL\,.
\end{equation}\fi
On the boundary $\partial_\sL\widehat{\mathfrak{M}}_{0,n+2}$, the kinematic integrand thus takes the simple form
\begin{equation}
 \cI^{\scalebox{0.6}{$(1)$,MHV}}_{n}=
 \varepsilon^{-\left(n_\ssL-1\right)}\prod_{l\in L}x_{l\sL}^2\,\;
  \int d^4\eta_\sL\;
  \frac{1}{(\scalebox{0.8}{$L\,ij$})_x}\,\mathcal{A}^{\scalebox{0.6}{$(0),\overline{\mathrm{MHV}}$}}_3\;\cI^{\scalebox{0.6}{$(1)$,MHV}}_{n-1}\,.
\end{equation}
To see that this agrees with \eqref{eq:fact_MHVbar}, note that $\mathcal{I}^{\scalebox{0.6}{$(0),\overline{\mathrm{MHV}}$}}_3=(\scalebox{0.8}{$L\,ij$})_x^{-2}\,\mathcal{A}^{\scalebox{0.6}{$(0),\overline{\mathrm{MHV}}$}}_3$ due to  $ \int d\mu_3^\CHY\; (\scalebox{0.8}{$L\,ij$})_x^{-2}=1$.

Using the factorisation of the colour factors discussed above, we can conclude that the worldsheet formula has the correct 2-particle factorisation channels, 
\begin{equation}
 \mathfrak{I}^{\scalebox{0.6}{$(1)$,MHV}}_{n}\sim\frac{1}{k_{i,i+1}^2}\int d^4\eta_\sL\;\mathcal{A}^{\scalebox{0.6}{$(0),\overline{\mathrm{MHV}}$}}_3\; \mathfrak{I}^{\scalebox{0.6}{$(1)$,MHV}}_{n-1}\,.
\end{equation}
We note that due to the special role particle 1 plays in the BCJ numerators, this only holds for $i\neq1$, the contribution from $L=\{1,2\}$ vanishes; see  \eqref{eq:jacobi_1}. In the BCFW recursion, only $k_n$ gets shifted, and thus we only pick up the expected `inverse soft' term,
\begin{equation}
 \mathrm{Res}_{z_{n}}\,\frac{\mathfrak{I}^{\scalebox{0.6}{$(1)$,MHV}}_{n}(z)}{z}=\frac{1}{k_{n-1,n}^2}\int d^4\eta_\sL\;\mathcal{A}^{\scalebox{0.6}{$(0),\overline{\mathrm{MHV}}$}}_3\; \mathfrak{I}^{\scalebox{0.6}{$(1)$,MHV}}_{n-1}(z_n)\,.
\end{equation}

\paragraph{No $k$-particle factorisation channels for $k>2$.}  At this point, we have already seen that the worldsheet formula reproduces the correct single cut and soft term, and that the unphysical poles do not contribute. We thus only have to check (ii), that there are no factorisation channels for $n_\sL>2$. In an extension of the two-particle case, we find here for the kinematic integrand
\begin{align}
 \cI^{\scalebox{0.6}{$(1)$,MHV}}_{n}
 &=\varepsilon^{-\left(n_\ssL-1\right)}\prod_{l\in L}x_{l\sL}^2\,\;\left(\sum_{m=0}^{n-k}\sum_{\substack{\rho=(\rho_1,\rho_2)\\\in S_m\times S_{n-m-k}}} \frac{1}{(+\rho_1\,\scalebox{0.8}{$R$}\,\rho_2\,-)_\sigma}
 \left(\sum_{\rho_\sL\in S_k}\frac{N^{\scalebox{0.6}{$(1)$}}_{\scalebox{0.6}{$(\rho_1\,\rho_\sL\,\rho_2)$}}}{(\scalebox{0.8}{$L\,\rho_\sL$})_x}\right)\right)\,.
\end{align}
\iffalse
Simplify:
\begin{align*}
 \sum_{\rho\in S_k}\frac{N^{\scalebox{0.6}{$(1)$}}_{\scalebox{0.6}{$(\rho_1\,\rho\,\rho_2)$}}}{(\scalebox{0.8}{$L\,\rho$})_x}
 %
 &=\sum_{\rho\in S_k}\!\frac{N^{\scalebox{0.6}{$(1)$}}_{\scalebox{0.6}{$(\rho_1\,\rho\,\rho_2)$}}}{(\scalebox{0.8}{$\rho$})_x}\left(\frac{1}{x_{\sL,\rho_{1}}}-\frac{1}{x_{\sL,\rho_{k}}}\right)%\\
 %
 =\!\sum_{\substack{i\in L\\\rho\in S_{k}/\mathbb{Z}_k}}\!\! \frac{\left(N^{\scalebox{0.6}{$(1)$}}_{\scalebox{0.6}{$(\rho_1\,i\,\rho\,\rho_2)$}}-N^{\scalebox{0.6}{$(1)$}}_{\scalebox{0.6}{$(\rho_1\,\rho\,i\,\rho_2)$}}\right)}{(\scalebox{0.8}{$i\,\rho$})_x\, x_{\sL i}} %\\
 %
 =\!\sum_{\substack{i\in L\\\rho\in S_{k-1}}}\!\frac{N^{\scalebox{0.6}{$(1)$}}_{\scalebox{0.6}{$(\rho_1\,[i,\rho]\,\rho_2)$}}}{x_{\sL i}\,(\scalebox{0.8}{$i\,\rho$})_x}\,.
\end{align*}\fi
We can simplify this expression by using the following identity:
\begin{equation}
 \sum_{\pi\in S_k}\frac{N^{\scalebox{0.6}{$(1)$}}_{\scalebox{0.6}{$(\rho_1\,\pi\,\rho_2)$}}}{(\scalebox{0.8}{$L\,\pi$})_x}
 =
 \sum_{\pi\in S_{k-1}}\frac{N^{\scalebox{0.6}{$(1)$}}_{\scalebox{0.6}{$(\rho_1\,|\,A_\pi\,|\,\rho_2)$}}}{(\scalebox{0.8}{$L\,\pi$})_x}\,,
\end{equation}
where $A_\pi=[\pi_1,[\pi_2,[\dots[\pi_{k-1},\pi_k]]]$.
This is the analogue of the well-known equivalence between an expansion into colour-ordered amplitudes using a trace-basis for the colour factors, or using a structure-constant basis (also known as a  DDM  or half-ladder basis); see e.g. \cite{Bern:2019prr} for a recent review. Using a generalisation of the numerator factorisation property, $\mathfrak{n}^{\scalebox{0.6}{$(1)$}}_{\scalebox{0.6}{$(\rho_1|A_\pi|\rho_2)$}}=X^{(A_\pi)} \mathfrak{n}^{\scalebox{0.6}{$(1)$}}_{\scalebox{0.6}{$(\rho_1\,R\,\rho_2)$}}$ where $k_\sR=\sum_{i\in L}k_i$, the MHV integrand thus factorises into contributions from each sphere. However, we can identify the worldsheet expression on $\Sigma_\sL$ as the one-minus tree amplitude,
\begin{equation}
\int_{\raisebox{-6pt}{\scalebox{0.7}{$\mathfrak{M}_{0,k+1}$}}}\hspace{-15pt}d\mu_{n}^{\CHY} \;\left( \sum_{\rho\in S_{k-1}}\frac{X^{(A_\rho)}}{(\scalebox{1}{$L\,\rho\, i$})}\right)\frac{1}{(i\ldots i\!+\!|L|\,L)}= 0\,,
\end{equation}
which vanishes on the support of the scattering equations. This was shown in \cite{Monteiro:2013rya}, and is also easily seen recursively; see \cref{app:BCFW}. Therefore, the MHV worldsheet formula contains no $k$-particle  factorisation channels for $k>2$.

\paragraph{Singular solutions.} There is a final type of unphysical pole with a slightly different status from the others, and which we therefore describe separately. This pole is a ``discriminant pole'' coming from so-called singular solutions of the scattering equations, with $L=\{+,-\}$; see refs. \cite{Cachazo:2015aol, Geyer:2015jch}.\footnote{To be precise, this discriminant pole takes the form
\begin{equation}
 s_\pm = \prod_{\text{tree sols } \sigma_i}\mathrm{Disc}(N_{\scalebox{0.7}{node}})\,,
\end{equation}
where $N_{\scalebox{0.7}{node}}(\sigma_+)$ is the numerator of the scattering equation $\mathcal{E_+}$. These poles may occur because 
 the two nodal scattering equations $\mathcal{E_+}=\mathcal{E}_{\scalebox{0.7}{node}}(\sigma_+)$ and $\mathcal{E_-}=\mathcal{E}_{\scalebox{0.7}{node}}(\sigma_-)$ have the same functional form.
} 
The degeneration of the sphere in this case resembles a tadpole, see 
 \cref{fig:poles_unphys}, and the tadpole-like structure is reflected in a different behaviour of the measure on this boundary,
\begin{equation}
 d\mu_n^{\Def}\sim d\varepsilon \,\delta\left(s_\sL- \varepsilon\mathcal{F}\right)\,.
\end{equation}
Compared to the  usual 2-particle factorisation $d\mu_n^{\Def}\sim \varepsilon\, d\varepsilon\,\delta\left(s_\sL- \varepsilon\mathcal{F}\right)$, the measure lacks a factor of $\varepsilon$, so the residue only vanishes if $\cI^{\scalebox{0.6}{$(1)$}}_{\scalebox{0.6}{MHV}}\,\mathcal{C}^{\scalebox{0.6}{$(1)$}}_n(12\dots n)\sim 1$. 

From the form of the colour integrand, it is clear that  $\mathcal{C}^{\scalebox{0.6}{$(1)$}}_n(12\dots n)\sim\varepsilon^{-1}$, and naively the kinematic half-integrand contains terms of the same order. It turns out, however, that the form of the numerators guarantees that the leading order terms cancel, so that in fact $\cI^{\scalebox{0.6}{$(1)$}}_{\scalebox{0.6}{MHV}}\sim \varepsilon$.\footnote{There is evidence from the general proposal in \S\ref{sec:proposal} that the cancellation can be extended by one order, such that $\cI^{\scalebox{0.6}{$(1)$}}_{\scalebox{0.6}{MHV}}\sim \varepsilon^2$.}  The relevant terms are
\begin{equation}
 \cI^{\scalebox{0.6}{$(1)$}}_{\scalebox{0.6}{MHV}}\Bigg|_{O(\varepsilon^m)}=\sum_{\rho\in S_n}\frac{N^{\scalebox{0.6}{$(1)$}}_{\scalebox{0.6}{$\rho$}}}{(+\rho_1\rho_2\ldots \rho_n)}\,\frac{1}{\sigma_{\rho_n+}^{m+1}}\,,\qquad\qquad \text{for }\,m=-1,0,1\,.
\end{equation}
The leading order term vanishes  by an extension of the well-known U$(1)$-decoupling identity, where we also use that numerators of massive corners involving particle 1 vanish,
\begin{equation*}
 \cI^{\scalebox{0.6}{$(1)$}}_{\scalebox{0.6}{MHV}}\Bigg|_{O(\varepsilon^{-1})}=\sum_{\rho\in S_n}\frac{N^{\scalebox{0.6}{$(1)$}}_{\scalebox{0.6}{$\rho$}}}{(+\rho_1\rho_2\ldots \rho_n)}
 =\sum_{\rho\in S_{n-1}}\frac{N^{\scalebox{0.6}{$(1)$}}_{\scalebox{0.6}{$1\rho$}}}{(+\rho_1\rho_2\ldots \rho_{n-1})}\left(\frac{\sigma_{+\rho_1}}{\sigma_{+1}\sigma_{1\rho_1}}+\ldots \frac{\sigma_{\rho_{n-1} +}}{\sigma_{\rho_{n-1} 1}\sigma_{1+}}\right)=0\,.
\end{equation*}
We can extend this argument to order $O(1)$ because the numerators for bubble diagrams also vanish,  $\mathfrak{n}_{1|A_2}=0$.
The actual calculations are rather technical, and the details can be found in \cref{app:BCFW}. Using these results, we find that the worldsheet integrand behaves as  $\cI^{\scalebox{0.6}{$(1)$}}_{\scalebox{0.6}{MHV}}\,\mathcal{C}^{\scalebox{0.6}{$(1)$}}_n(12\dots n)\sim 1$, and the residue on the discriminant pole thus vanishes.

\label{page:lower-susy} In this context, we would also like to briefly comment on possible extensions of the worldsheet formula to theories with less supersymmetry, such as pure Yang-Mills. In the `linear' framework, integrands are readily available,\footnote{by restricting the sum over spin structures to just the contribution from the NS sector.} and it is easy to verify that much of the factorisation analysis will still carry over.  The difficulty thus lies entirely in dealing with the contributions from the singular solutions: the above cancellations only hold for maximally supersymmetric theories, as evident from the reliance on the vanishing bubble- and triangle numerators. In general, it seems that non-supersymmetric theories will receive contributions from an unphysical pole. This is a familiar feature even in the linear formalism, where these poles appear when restricting to the non-singular solutions. There, however, it was proven in \cite{Cachazo:2015aol} that contributions from these discriminant poles loop-integrate to zero for many theories of physical interest, and can thus safely be discarded. Whether a similar analysis can be extended to this case is a question we leave for future research.

\subsection{Boundary terms} \label{sec:bdy}
\begin{figure}[ht]
	\centering 
	  \raisebox{7pt}{\includegraphics[width=5cm]{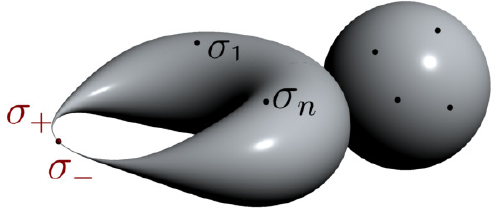}}
	  \hfill
	 \raisebox{0pt}{\includegraphics[width=5cm]{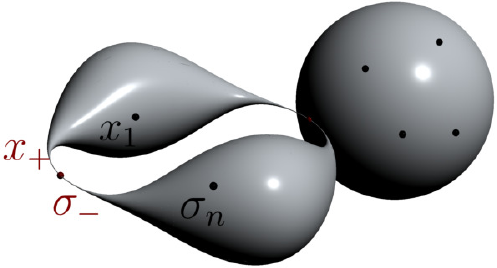}}
	\hfill
	\raisebox{7pt}{\includegraphics[width=5cm]{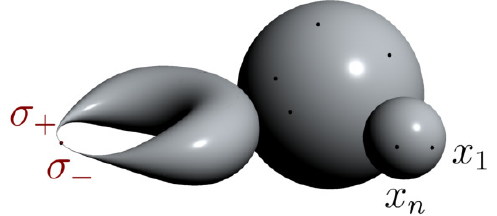}}	\caption{The singular worldsheet configurations that give a dominant balance in the limit as $z\rightarrow\infty$. For each of them, the integrand scales as $\mathfrak{I}^{(1)}_{\text{sYM}}(z) \sim z^{-1}$, and the boundary term thus vanishes.}
	\label{fig:dom-balance}
\end{figure}

To see that the boundary terms for the MHV worldsheet formula vanish, recall the argument of  \cref{sec:bdy_MHV}: since the numerators are constructed out of  supersymmetry conservation $\delta^8(Q)$ and $X$'s, they are invariant under the BCFW shift, and thus contribute $z^0$ to the limit $z\gg1$. The overall prefactor of $1/\hat \ell^2 = \alpha/((\alpha-z)\ell^2)$ scales as $z^{-1}$, and all other $z$-dependence is encoded in the scattering equations and Parke-Taylor factors. Heuristically, it is clear that in the absence of singular solutions, this can only suppress the integrand scaling by additional powers of $z$, so that the integrand scales at the worst as
\begin{equation}\label{eq:bdy_z}
 \mathfrak{I}^{\scalebox{0.6}{$(1)$}}_{\text{sYM}}(z) \sim z^{-1}\,,\qquad\qquad\text{as }z\gg1\,,
\end{equation}
and thus the boundary terms vanish
\begin{equation}
 \mathcal{B}_n =\mathrm{Res}_{\infty}\; \frac{\mathfrak{I}^{\scalebox{0.6}{$(1)$}}_{\text{sYM}}(z) }{z}=0\,.
\end{equation}
We build this heuristic argument into a full proof  in \cref{app:BCFW}, by solving the $\ell^2$-deformed scattering equations for $z\gg1$ using the `method of dominant balance' --  a standard technique for finding asymptotic solutions.\footnote{This method formalises the notion that solutions to an equation require at least two terms at leading order (such that the two can cancel against each other).} 
As depicted  in \cref{fig:dom-balance}, three types of boundary divisors give a dominant balance. In all three cases, the MHV worldsheet expression is consistent with \eqref{eq:bdy_z}, and thus the boundary terms vanish.

This concludes the proof of the one-loop MHV worldsheet formula via BCFW recursion.
 
%\subsection{Supergravity}
%\todoY{To be written}

%%%%%%%%%%%%%%%%%%%%%%%%%%%%%%%%
%%%%%%%%%%%%%%%%%%%%%%%%%%%%%%%%
\section{Worldsheet formulas for quadratic propagators: non-planar case}
\label{sec:wsquadnp}

In this section, we will very briefly present the extension of the worldsheet formalism presented above to non-planar super Yang-Mills theory and to supergravity. We will restrict ourselves to MHV super-amplitudes, since in that case we have actually proven the worldsheet formula \eqref{eq:MHV-proposal} for planar super Yang-Mills in the previous section. That formula will be our starting point. We repeat it here for convenience:
\begin{equation}\label{eq:MHV-proposal-ws_2}
% \boxed{\;\;
 %
 \mathfrak{I}^{\scalebox{0.6}{$(1)$}}_{\text{sYM-MHV}}(12\dots n)=\frac{1}{\ell^2} \int_{\raisebox{-6pt}{\scalebox{0.7}{$\mathfrak{M}_{0,n+2}$}}}\hspace{-15pt}d\mu_{n}^{\Def\text{-}\{1,n\}} \; \cI^{\scalebox{0.6}{$(1)$}}_{\scalebox{0.6}{MHV}} \,\, \frac{1}{(+12\ldots n-)}\,.
 %
 %\;\;}
\end{equation}
This is the contribution associated to the single trace $ \tr(T^{a_1}T^{a_2}\dots T^{a_n}) $ in the colour-dressed amplitude. In preparation for the following discussion, we included explicitly in the moduli-space measure the particles on which the $\ell^2$-deformation of the scattering equations is based,
 \begin{equation}\label{eq:measure_quad_ij}
  d\mu_{n}^{\Def\text{-}\{i,j\}} :=\frac{d^{n+2}\sigma}{\text{vol}\,\text{SL}(2,\C)} \,\,\, {\prod_{a=1}^{n+2}}{}' \bar\delta\left(
\mathcal{E}_a \, \big|^{\,2\ell\cdot k_i \;\mapsto\;{+ (\ell+ k_i)^2}}_{\,2\ell\cdot k_j\; \mapsto\; {-(\ell- k_j)^2}}
  \right)\,.
 \end{equation}

The reasoning in this section follows closely that in section~\ref{nonplanar}, which dealt with the BCFW recursion in the non-planar case. In particular, we will make use again of \eqref{eq:DDMYM} and  \eqref{eq:DDMgrav}. For our purposes here, we can write them as
\begin{align}
\mathfrak{I}^\text{(1)}_{n,\text{ sYM-MHV}} &= \sum_{\rho\in S_{n-1}/\mathcal{R}} c^\text{(1)}\big(1,\rho(2),\cdots,\rho(n)\big) \; \mathfrak{I}^\text{(1)}_{\text{ sYM-MHV}}\big(1,\rho(2),\cdots,\rho(n)\big)  \,,
\label{eq:DDMYM-ws}\\
\mathfrak{I}^\text{(1)}_{n,\text{ sugra-MHV}}  &= \sum_{\rho\in S_{n-1}/\mathcal{R}}  \tilde N^\text{(1)}_\text{MHV}\big(1,\rho(2),\cdots,\rho(n)\big)  \;
\mathfrak{I}^\text{(1)}_{\text{ sYM-MHV}}\big(1,\rho(2),\cdots,\rho(n)\big)
 \,.
 \label{eq:DDMgrav-ws}
\end{align}
Here, we chose to write the formula based on the permutation group $S_{n-1}$ by fixing the position of particle 1, so that the loop momentum lies between legs $\rho(n)$ and $1$; the reduction by reflections $\mathcal R$ then applies to $S_{n-1}$. The colour factors $c^\text{(1)}$ were defined in \eqref{eq:ccoldef}. The BCJ numerators $\tilde N^\text{(1)}_\text{MHV}$ were defined in \eqref{eq:MHVnumsec7}, and the tilde just means that the supermomentum conservation factors of $\delta^8(Q)$ from $\mathfrak{I}^\text{(1)}_{\text{ sYM-MHV}}$ and $\delta^8(\tilde Q)$ from $\tilde N^\text{(1)}_\text{MHV}$ combine into the appropriate factor $\delta^{16}(Q)$ for MHV supergravity amplitudes.

Based on these decompositions of the non-planar integrands into planar-like parts, it becomes clear that \eqref{eq:MHV-proposal-ws_2} can be easily extended to non-planar super Yang-Mills as
\begin{equation}\label{eq:MHV-proposal-ws-np-sym}
 \boxed{\;\;
\mathfrak{I}^\text{(1)}_{n,\text{ sYM-MHV}} = \frac{1}{\ell^2} \sum_{\rho\in S_{n-1}/\mathcal{R}} \;\int_{\raisebox{-6pt}{\scalebox{0.7}{$\mathfrak{M}_{0,n+2}$}}}\hspace{-15pt}d\mu_{n}^{\Def\text{-}\{1,\rho(n)\}} \; \cI^{\scalebox{0.6}{$(1)$}}_{\scalebox{0.6}{MHV}} \,\, 
\frac{c^\text{(1)}\big(1,\rho(2),\cdots,\rho(n)\big)}{\big(+1\,\rho(2)\ldots \rho(n)-\big)}\,.
 \;\;}
\end{equation}
For supergravity, we have
\begin{equation}\label{eq:MHV-proposal-ws-np-sugra}
 \boxed{\;\;
\mathfrak{I}^\text{(1)}_{n,\text{ sugra-MHV}} = \frac{1}{\ell^2} \sum_{\rho\in S_{n-1}/\mathcal{R}} \;\int_{\raisebox{-6pt}{\scalebox{0.7}{$\mathfrak{M}_{0,n+2}$}}}\hspace{-15pt}d\mu_{n}^{\Def\text{-}\{1,\rho(n)\}} \; \cI^{\scalebox{0.6}{$(1)$}}_{\scalebox{0.6}{MHV}} \,\, 
\frac{\tilde N^\text{(1)}_\text{MHV}\big(1,\rho(2),\cdots,\rho(n)\big)}{\big(+1\,\rho(2)\ldots \rho(n)-\big)}\,.
 \;\;}
\end{equation}
The main feature of these formulas is that we need $n-1$ sets of scattering equations, associated to the choices of $\{1,\rho(n)\}$ for the $\ell^2$-deformation of the scattering equations \eqref{eq:measure_quad_ij}. This could already be guessed from the argument presented in the \hyperref[sec:intro]{Introduction} for the analogy between the BCFW shift and the deformed scattering equations. Since, in section~\ref{nonplanar}, we also needed $n-1$ distinct BCFW shifts (with $q=\lambda_1\tilde\lambda_{\rho(n)}$) to reconstruct the non-planar integrand from planar-like parts, we would expect to need also $n-1$ distinct sets of deformed scattering equations. Notice that we could also have written
\begin{align}
\mathfrak{I}^\text{(1)}_{n,\text{ sYM-MHV}} & = \frac{1}{2\,\ell^2}\; \sum_{i=2}^n \;\int_{\raisebox{-6pt}{\scalebox{0.7}{$\mathfrak{M}_{0,n+2}$}}}\hspace{-15pt}d\mu_{n}^{\Def\text{-}\{1,i\}}
 \; \cI^{\scalebox{0.6}{$(1)$}}_{\scalebox{0.6}{MHV}}
\sum_{\rho\in S_{n-2}} 
\frac{c^\text{(1)}\big(1,\rho(2),\cdots,\rho(n-1)\,i\big)}{\big(+1\,\rho(2)\ldots \rho(n-1)\,i-\big)}\,,
\\
\mathfrak{I}^\text{(1)}_{n,\text{ sugra-MHV}} & = \frac{1}{2\,\ell^2}\; \sum_{i=2}^n \;\int_{\raisebox{-6pt}{\scalebox{0.7}{$\mathfrak{M}_{0,n+2}$}}}\hspace{-15pt}d\mu_{n}^{\Def\text{-}\{1,i\}}
\; \cI^{\scalebox{0.6}{$(1)$}}_{\scalebox{0.6}{MHV}}
\sum_{\rho\in S_{n-2}}  
\frac{\tilde N^\text{(1)}_\text{MHV}\big(1,\rho(2),\cdots,\rho(n-1)\,i\big)}{\big(+1\,\rho(2)\ldots \rho(n-1)\,i-\big)}\,,
\end{align}
where the the factor 2 takes into account the reflection reduction in previous expressions.

The use of multiple sets of scattering equations may be considered unsatisfactory. However, this seems to be the price to pay for having non-planar loop integrands with quadratic propagators. It would be interesting to know whether this feature can be derived from the ambitwistor string formalism, via the residue theorem from the torus to the nodal sphere, along the lines discussed at the end of section~\ref{sec:SE}.

In order to obtain analogous formulas beyond the MHV case (with different MHV degree, number of supersymmetries or number of spacetime dimensions) the ideas and caveats discussed in section~\ref{sec:proposal} also apply. A more detailed analysis is beyond the scope of this paper.

%%%%%%%%%%%%%%%%%%%%%%%%%%%%%%%%
%%%%%%%%%%%%%%%%%%%%%%%%%%%%%%%%
\section{Conclusion}
\label{sec:conclusion}

The study of perturbative scattering amplitudes in quantum field theory has led to powerful techniques. Two notable examples applying to tree amplitudes and loop integrands are: recursion relations, which have given rise to novel geometric descriptions; and formulations in terms of worldsheet integrals that localise onto solutions of a universal set of equations, known as the scattering equations. Whereas the former approach yields loop integrands in terms of Feynman propagators, the best developed version of the latter gives a non-standard representation in terms of propagators whose inverse is linear in the loop momentum. While it is possible to go from a (Feynman) `quadratic' representation to a `linear' one using partial fraction identities and redefinitions of the loop momentum, going in the reverse direction is highly non-trivial.

In this paper, we investigated the relation between these two representations.
The starting point was the observation that it is possible to lift certain `linear' one-loop integrands into `quadratic' ones simply by deforming two external legs. This operation turns out to be closely related to the seed of the loop-level BCFW recursion, which is a forward limit. In addition to the forward limit, the BCFW recursion also contains tree-level-type factorisation terms, which precisely cancel the spurious poles arising in the forward-limit term. These cancellations can be made manifest by shifting the loop momentum in lower-point integrands that feed into the recursion. We found that these shifts can be elegantly derived from a contour integral argument in which not only the two external legs, but also the loop momentum are shifted. Moreover, we verified that this momentum space recursion directly matches the recursion originally proposed for planar $\mathcal{N}=4$ SYM in momentum twistor space. The momentum twistor recursion has been extended to all loop orders, and has provided beautiful insights. On the other hand, the momentum space recursion is not tied to four dimensions, supersymmetry or planarity (although it is better adapted to the planar case, as we discussed). Our approach provides an alternative viewpoint on previous work in momentum space, which we demonstrated with several non-trivial examples in maximal SYM and pure YM.

Inspired by this viewpoint on the one-loop BCFW recursion, and in particular on how it yields quadratic propagators, we proposed a simple modification of the one-loop scattering equations in general dimensions that also leads to loop integrands with quadratic propagators. These can in principle be integrated using standard methods, as opposed to the integrands with `linear' propagators. Using the new scattering equations, we constructed a worldsheet formula for one-loop MHV integrands in planar $\mathcal{N}=4$ SYM, and proved it by verifying that it satisfies the BCFW recursion. Finally, we constructed MHV formulas also for non-planar $\mathcal{N}=4$ SYM and $\mathcal{N}=8$ supergravity. We expect the new formalism to hold more generally, beyond MHV and $D=4$, possibly with modifications of the moduli-space integrands that were previously known from the `old' one-loop scattering equations story. We discussed a natural idea for this extension.

\paragraph{Outlook.} A number of interesting questions remain. Firstly, it would be interesting to use the BCFW recursion to compute one-loop amplitudes in pure Yang-Mills theory beyond the all-plus helicities case. The forward limit will need to be regulated. Moreover, we will no longer be allowed to replace gluons running through the loop with complex scalars, so would have to follow the more complicated procedure analogous to appendix~\ref{allplusg}. Secondly, we would like to extend the recursion relation in the approach we took here to higher loops. We discussed two versions of the non-planar recursion. In one of the versions, we made use of the property that one-loop amplitudes can be effectively planarised, using multiple BCFW shifts, one for each planarised part. However, this version has no obvious extension to higher loops. On the other hand, the version with a single BCFW shift certainly has a natural higher-loop extension.  

On the worldsheet approach, we would like to establish whether our construction of MHV loop integrands is indeed extendible beyond MHV, beyond $D=4$, and beyond maximal supersymmetry. To be precise, we certainly expect this to be true, but the question is whether the worldsheet formulas will still have the elegant structure that we outlined. While we discussed an idea for a worldsheet formula  for gluon amplitudes in $D$-dimensional super Yang-Mills, a direct investigation is beyond the scope of this paper due to the technical difficulties in implementing the BCFW recursion when it involves sums over fermionic degrees of freedom in an RNS-like representation. This would be important to resolve, possibly  using the nodal operator constructions of \cite{Roehrig:2017gbt} or the more recent numerator constructions in \cite{Edison:2020uzf, Edison:2020ehu}. Since we verified that only physical poles contribute, these approaches could be used to identify the residues, and to check whether they agree with the forward-limit and factorisation terms of the loop-level BCFW recursion. The construction of one-loop worldsheet formulas of the new type for theories without supersymmetry presents additional challenges due to potential contributions from tadpole-like worldsheet geometries, which would require further work.

Another important question concerns the origin of the deformation of the scattering equations. In this paper, we motivated the deformation from the BCFW recursion, but also 
\hyperlink{speculation}{discussed briefly} 
an alternative motivation  that ties in more directly with the ambitwistor string. The idea is that the deformation arises as a choice in the residue theorem linking the torus and nodal sphere worldsheet formulas. For Yang-Mills, this has very little hope of serving as more than a motivation due to the lack of torus formulas. However, it would be interesting to investigate whether the torus formula for supergravity can be turned into the non-planar worldsheet expression \eqref{eq:MHV-proposal-ws-np-sugra} by decomposing the torus integrand into a DDM-like basis as in \eqref{eq:DDMgrav-ws}, and then applying a residue theorem to each `planarised sector' individually. In this manner, one could hopefully rederive our  MHV supergravity worldsheet formula from the ambitwistor string. More importantly, this derivation would probably teach us how to extend that supergravity formula beyond MHV degree, $D=4$, and maximal supersymmetry. Results for Yang-Mills would then be reachable via the colour-kinematics duality. We also hope that this approach to field theory amplitudes can be helpful in studying their connections with amplitudes in conventional (not just ambitwistor) string theory.

Finally, we would like to understand the relation between our new worldsheet formula for loop integrands with quadratic propagators and geometric formulations such as the amplituhedron for planar $\mathcal{N}=4$ SYM \cite{Arkani-Hamed:2013jha,Damgaard:2019ztj,Ferro:2020ygk}, and the associahedron and related polytopes for the bi-adjoint scalar theory \cite{Arkani-Hamed:2017mur,Salvatori:2018aha,Kalyanapuram:2020vil}. Ultimately, we hope this interplay between recursion relations, worldsheet formulas and geometric constructions will continue to advance our understanding of the S-matrix.

%%%%%%%%%%%%%%%%%%%%%%%%%%%%%%
%%%%%%%%%%%%%%%%%%%%%%%%%%%%%%

\section*{Acknowledgements}
We would like to thank Johannes Agerskov, Emil Bjerrum-Bohr, Humberto Gomez and Oliver Schlotterer for discussions. AL and RM thank the Galileo Galilei Institute for Theoretical Physics and INFN for hospitality and partial support during the workshop ``String Theory from a worldsheet perspective", and also the Munich Institute for Astro- and Particle Physics (MIAPP) of the DFG cluster of excellence ``Origin and Structure of the Universe" for hospitality and partial support during the workshop ``Precision Gravity: From the LHC to LISA", where parts of this work were done. RM thanks Chulalongkorn University for hospitality. YG is supported by the CUniverse research promotion project ``Toward World-class Fundamental Physics'' of Chulalongkorn University (grant reference CUAASC). 
AL and RM are supported by Royal Society University Research Fellowships. RSM's studentship is also funded by the Royal Society.

\newpage

\appendix

\section{The one-loop six-point MHV integrand in maximal super-Yang-Mills}
\label{app:6ptMHV} 

With the pre-factor defined in \eqref{subamp}, the six-point MHV integrand was expressed in appendix B of \cite{He:2015wgf} as
\begin{align}
%&I_{1,2,3,4,5,6}^{1-\te{loop}}(\ell)  = \frac{ \delta^8(Q) }{\prod_{j=2}^6 \langle 1 j \rangle^2 }  
&\mathcal{I}_{1,2,3,4,5,6}(\ell)=\frac{ 1 }{\ell^2 (\ell+k_1)^2} \, \Big\{ \frac{ \num_{\underline{1}|2|3|4|5|6}  }{ (\ell+k_{12})^2 (\ell+k_{123})^2 (\ell+k_{1234})^2 (\ell+k_{12345})^2} \notag \\
&\ \ \ + \frac{  \num_{\underline{1} | [2,3] |4|5|6}  }{s_{23}     (\ell+k_{123})^2 (\ell+k_{1234})^2 (\ell+k_{12345})^2} 
 + \frac{\num_{\underline{1} | 2 | [3,4] |5|6}   }{s_{34}     (\ell+k_{12})^2 (\ell+k_{1234})^2 (\ell+k_{12345})^2} \notag \\
 &\ \ \ 
  + \frac{   \num_{\underline{1} | 2 | 3| [4,5] |6}  }{s_{45}     (\ell+k_{12})^2 (\ell+k_{123})^2 (\ell+k_{12345})^2} 
   + \frac{ \num_{\underline{1} | 2 | 3 |4|  [5,6]}   }{s_{56}    (\ell+k_{12})^2 (\ell+k_{123})^2 (\ell+k_{1234})^2}  \notag \\
   %%%%  boxes
&\ \ \ + \Big( \frac{ X_{2,3} X_{2+3,4} }{s_{23}} +   \frac{ X_{4,3} X_{4+3,2}  }{s_{34}} \Big)  \frac{ X_{2+3+4,5} X_{2+3+4,6}  }{s_{234}   (\ell+k_{1234})^2 (\ell+k_{12345})^2}    +  \frac{ X_{2,3} X_{4,5}  \, X_{2+3,4+5} X_{2+3,6} }{s_{23} s_{45}   (\ell+k_{123})^2 (\ell+k_{12345})^2}   \notag \\
&\ \ \ + \Big( \frac{ X_{3,4} X_{3+4,5} }{s_{34}} +   \frac{ X_{5,4} X_{5+4,3}  }{s_{45}} \Big)  \frac{ X_{2,3+4+5} X_{2,6}  }{s_{345}   (\ell+k_{12})^2 (\ell+k_{12345})^2} +  \frac{ X_{2,3} X_{5,6} \, X_{2+3,4} X_{2+3,5+6} }{s_{23} s_{56}   (\ell+k_{123})^2 (\ell+k_{1234})^2} \notag \\
&\ \ \ + \Big( \frac{ X_{4,5} X_{4+5,6} }{s_{45}} +   \frac{ X_{6,5} X_{6+5,4}  }{s_{56}} \Big)  \frac{ X_{2,3} X_{2,4+5+6}  }{s_{456}   (\ell+k_{12})^2 (\ell+k_{123})^2}   +  \frac{ X_{3,4} X_{5,6}  \, X_{2,3+4} X_{2,5+6} }{s_{34} s_{56}   (\ell+k_{12})^2 (\ell+k_{1234})^2}
   \Big\} \ ,
\label{6ptex}
\end{align}
where
\begin{align}
\num_{\underline{1}|2|3|4|5|6} &=   X_{2,4}X_{2,3}X_{\ell-6,5}X_{\ell,6}+X_{2,5}X_{2,3}X_{2+3,4}X_{\ell,6} +X_{2,6}X_{2,3}X_{2+3,4}X_{2+3+4,5}
\notag\\
\num_{\underline{1} | [2,3] |4|5|6}  &= X_{2,3}  ( X_{2+3,5}X_{2+3,4}X_{\ell,6}+X_{2+3,6}X_{2+3,4}X_{2+3+4,5}+X_{4,6}X_{\ell,2+3}X_{2+3+4,5} )
\notag\\
\num_{\underline{1} | 2 | [3,4] |5|6}  &=
X_{3,4}  ( X_{2,5}X_{2,3+4}X_{\ell,6}+X_{2,6}X_{2,3+4}X_{2+3+4,5}+X_{3+4,6}X_{\ell,2}X_{2+3+4,5} )
\notag \\
 \num_{\underline{1} | 2 | 3| [4,5] |6} &= 
 X_{4,5} ( X_{2,4+5}X_{2,3}X_{\ell,6}+X_{2,6}X_{2,3}X_{2+3,4+5}+X_{3,6}X_{\ell,2}X_{2+3,4+5} )
\notag \\
\num_{\underline{1} | 2 | 3 |4|  [5,6]}  &=
X_{5,6} ( X_{2,4}X_{2,3}X_{\ell,5+6}+X_{2,5+6}X_{2,3}X_{2+3,4}+X_{3,5+6}X_{\ell,2}X_{2+3,4} ) \ .
\end{align}
The first term in \eqref{6ptex} corresponds to an hexagon diagram, the following four terms to pentagon diagrams with one massive corner, and the remaining terms to box diagrams with either one or two massive corners.

\section{Technical details for the all-plus recursion} \label{app:allplus}

\subsection{All-plus integrands and the forward limit}

In this section, we shed some light on how one might come to the expression on which \eqref{eq:4ptallplusFL} is based, and more generally on forward-limit terms, for the all-plus loop integrand. The vanishing of identical helicity amplitudes in a supersymmetric Yang-Mills theory implies, via the supersymmetric Ward identities, that the all-plus one-loop amplitude in pure Yang-Mills can be computed with either a gluon or a complex scalar propagating in the loop. The integrands for these may then naturally be obtained in the following way. For an $n$-point all-plus integrand, consider the set of ordered tree-level amplitudes with $n$ gluons and 2 scalars. One may deform the momenta via a parameter $\epsilon$ such that as $\epsilon \rightarrow 0$ the scalars have back-to-back momenta and $n$-point momentum conservation is acquired amongst the $n$ gluons; the forward limit is then canonically identified with the limit $\epsilon \rightarrow 0$. The expressions for the tree-level amplitudes may be obtained from the Feynman rules given in \cite{Boels:2013bi} for a set of gluons coupled with scalars, wherein one may perform the deformations mentioned above via the parameter $\epsilon$. On taking the forward limit, the tree-level amplitudes will form `regular' loop diagrams, but also external leg bubbles and tadpoles. For the four-point integrand (to which 14 tree-level diagrams contribute) the regular graphs will go as $\mathcal{O}(1)$, the external leg bubbles go as $\mathcal{O}(1)$, and the tadpoles go as $\mathcal{O}(\epsilon)$. Thus the tadpoles vanish in the forward limit, but the finite terms from the external leg bubbles must be included to ensure the correct integrand. It can be checked that the full result as $\epsilon \rightarrow 0$ is the same as the un-shifted version of \eqref{eq:4ptallplusFL} for the four-point all-plus integrand, which hence arises as a forward limit. Note that if one symmetrises in the momenta of the scalars, then the external leg bubbles will go as $\mathcal{O}(\epsilon)$, and only regular graphs will contribute.

\subsection{All-plus recursion with gluons in the loop}
\label{allplusg}

Here we discuss the method of obtaining the four-point all-plus integrand in pure Yang-Mills by having gluons run in the loop. As mentioned in section \ref{sec:allplusn=4}, for the all-plus case the residue at infinity,\footnote{The all-plus integrands, as rational functions in $z$, go as $\mathcal{O}(z^{-2})$ for large $z$; such a function cannot have a residue at infinity, as discussed in section \ref{sec:bdy_all+}.} as well as the three-point one-loop amplitudes, vanish in the recursion, and so only the forward-limit term survives,
\begin{equation*}
\mathcal{A}^{(1)}(1^+, 2^+, 3^+, 4^+) = \int \frac{d \alpha}{\alpha} d^D \ell_0 \, \delta(\ell_0^2) \; \mathcal{A}^{(0)}(\ell_0, \one^+, 2^+, 3^+, \four^+, -\ell_0).
\end{equation*}
wherein one performs the BCFW shifts
\begin{align*}
\hat{k}_1 = k_1 + \alpha q, && \hat{k}_4 = k_4 - \alpha q.
\end{align*}
From the perspective of the scattering equation formalism, this corresponds to keeping $\sigma_+$, $\sigma_-$ positioned between $\sigma_1$ and $\sigma_4$ in the Parke-Taylor factor dressing the colour. Let us write the resulting tree-level amplitude of pure Yang-Mills in terms of the shifts above as\footnote{Note that a factor of 1/2 is implicit inside of the symmetrisation, but we have cancelled it against the factor of 2 coming from the definition of the numerators; see equation \eqref{eq:allplusnums}.}
\begin{equation*}
\mathcal{A}^{(0)}(\ell_0, \one^+, 2^+, 3^+, \four^+, -\ell_0) = \mI(\ell_0, \alpha) + \mI(-\ell_0, \alpha)
\end{equation*}
symmetrised in $\ell_0$, with
\begin{align*}
\mI(\ell_0, \alpha) = \frac{1}{\prod_i \langle \eta  \hat{i} \rangle^2} \left[ \mI^{\text{box}}_{1234} + \mI^{\text{tri}}_{[12]34} + \mI^{\text{tri}}_{[23]41} + \mI^{\text{tri}}_{[34]12} + \mI^{\text{bub}}_{[12][34]} \right],
\end{align*}
where $\prod_i \langle \eta \hat{i} \rangle = \langle \eta 1 \rangle \langle \eta 2 \rangle \langle \eta 3 \rangle \langle \eta  \four \rangle$. If we denote by $D_i$ the $i$'th propagator such that
\begin{align*}
D_2 = (\ell_0 + \hat{k}_1)^2, && D_3 = (\ell_0 + \hat{k}_1 + k_2)^2, && D_4 = (\ell_0 - \hat{k}_4)^2,
\end{align*}
then the sub-integrands above are expressed as
\begingroup
\allowdisplaybreaks
\begin{align}
\mI^{\text{box}}_{1234} &= \frac{X_{\tell,\one} X_{\tell+\one,2} X_{\tell-\four,3} X_{\tell,\four}}{D_2 D_3 D_4} \label{eq:box} \\[10pt]
\mI^{\text{tri}}_{[12]34} &= \frac{1}{(2\hat{k}_1 \cdot k_2)}\frac{X_{\one,2}X_{\tell,\one+2}X_{\tell-\four,3}X_{\tell,\four}}{ D_3 D_4} \label{eq:tri1} \\[10pt]
\mI^{\text{tri}}_{[23]41} &= \frac{1}{(2k_2 \cdot k_3)} \frac{X_{2,3}X_{\tell,\one}X_{\tell+\one,2+3}X_{\tell,\four}}{D_2 D_4} \label{eq:tri2} \\[10pt]
\mI^{\text{tri}}_{[34]12} &= \frac{1}{(2k_3 \cdot \hat{k}_4)} \frac{X_{3,\four}X_{\tell \one}X_{\tell+\one,2}X_{\tell,3+\four}}{D_2 D_3} \label{eq:tri3} \\[10pt]
\mI^{\text{bub}}_{[12][34]} &= \frac{1}{(2\hat{k}_1 \cdot k_2)(2k_3 \cdot \hat{k}_4)} \frac{X_{\one,2}X_{3,\four}X_{\tell,\one+2}X_{\tell,3+\four}}{D_3}. \label{eq:bub}
\end{align}
\endgroup
This expression can be derived from the gluonic forward limit discussed in the previous subsection, using the Feynman rules for the self-dual sector of Yang-Mills in light-cone gauge, as noted in section \ref{sec:bdy_all+}, keeping the loop momentum positioned between particles 4 and 1, and performing the BCFW shifts with $z = \alpha$. The symmetrisation in $\pm\ell_0$ used above will provide convenient cancellations. We recall that the $X$ variables are defined as
\begin{equation*}
X_{A,B} := \langle \eta | K_A K_B | \eta \rangle
\end{equation*}
for $K_A$ and $K_B$ possibly off-shell, and $\eta = |\eta \rangle [\eta|$ is an auxiliary null reference vector. Note that when $\tell$ is inside a spinor bracket it is understood to be the four-dimensional part of $\tell$, i.e. $\ell_0^{(4D)}$, for which ${\ell_0^{(4D)}}^2 = \mu^2$. Alternatively, the  forward-limit-type expression above can be obtained from the $D$-dimensional worldsheet formulas presented in \cite{Geyer:2017ela}, specialised to four dimensions and all-plus.\footnote{Here, however, we will get quadratic propagators, instead of the linear propagators there in \cite{Geyer:2017ela}. Firstly, we consider here only a single Parke-Taylor term in the colour factor, not the cyclic sum as there; we take the term with the loop punctures lying between the punctures for 4 and 1. Secondly, we have here $\ell_0$ instead of $\ell$ there, but $\alpha$ appears in the BCFW shift of particles 1 and 4. Together, $\ell_0$ and $\alpha$ in the BCFW shift will give quadratic propagators in the full $\ell$.}

 Whilst presently in a form unrecognisable to the known result, the latter may be obtained through a series of manipulations involving the spinor anti-commutation relations.\footnote{Similar manipulations are performed in e.g. \cite{Brandhuber:2006bf}.} As an example, for four-dimensional massless momenta $K_A$ and $K_B$ we have
\begin{equation*}
X_{\tell, A} X_{\tell, B} = \frac{\langle \eta A \rangle \langle \eta B \rangle}{\langle A B \rangle} \left[ (2\ell_0 \cdot k_A) X_{B, \tell} - (2\ell_0 \cdot k_B) X_{A, \tell} + \mu^2 X_{A,B} \right].
\end{equation*}
With these manipulations one can derive e.g.
\begin{align}
X_{\tell, \one}X_{\tell + \one, 2} &=\frac{\langle \eta 1 \rangle \langle \eta 2 \rangle}{\langle 12 \rangle} \left[ D_2 \, X_{\one + 2, \tell} - D_3 \, X_{\one, \tell} + \mu^2 \, X_{\one,2} \right] \label{eq:x12} \\[5pt]
X_{\tell - \four, 3} X_{\tell, \four} &= \frac{\langle \eta 3 \rangle \langle \eta \four \rangle}{\langle 3 \four \rangle} \left[D_4\,X_{3+\four, \tell} - D_3 X_{\four, \tell} + \mu^2 \, X_{3,\four} \right] \label{eq:x34} ,
\end{align}
such that the integrand corresponding to the box is equivalent to
\begin{align}
\mI^{\text{box}}_{1234} = \frac{\prod_i \langle \eta \hat{i} \rangle}{\langle \one 2 \rangle \langle 3 \four \rangle} &\left[ \frac{X_{\one + 2,\tell} \, X_{3 + \four,\tell}}{D_3} - \frac{X_{\one + 2,\tell} \, X_{\four, \tell}}{D_4} + \mu^2 \frac{X_{\one + 2,\tell} \, X_{3,\four}}{D_3 D_4} \right.
\nonumber \\[5pt]
&- \frac{X_{3+\four, \tell} \, X_{\one, \tell}}{D_2} + D_3 \frac{X_{\one, \tell} \, X_{\four,\tell}}{D_2 D_4} - \mu^2 \frac{X_{\one, \tell} \, X_{3,\four}}{D_2 D_4} \label{eq:Xbox}
\\[5pt]
&+ \left. \mu^2 \frac{X_{\one,2} \, X_{3+\four,\tell}}{D_2 D_3} - \mu^2 \frac{X_{\one,2} \, X_{\four, \tell}}{D_2 D_4} + \mu^4 \frac{X_{\one,2} \, X_{3,\four}}{D_2 D_3 D_4} \right]. \nonumber
\end{align}
The last term will give the correct result, since
\begin{equation*}
\frac{1}{\prod_i \langle \eta \hat{i} \rangle^2} \frac{\prod_i \langle \eta \hat{i} \rangle}{\langle \one 2 \rangle \langle 3 \four \rangle} \mu^4 \frac{X_{\one, 2} X_{3, \four}}{D_2 D_3 D_4} = \frac{[\one 2][34]}{\langle 12 \rangle \langle 3\four \rangle} \frac{\mu^4}{D_2 D_3 D_4} = \frac{[12][34]}{\langle 12 \rangle \langle 34 \rangle} \frac{\mu^4}{D_2 D_3 D_4}.
\end{equation*}
Similar manipulations will result in two of the triangle sub-integrands being expressible as
\begin{align}
\mI^{\text{tri}}_{[12]34} &= -\frac{\prod_i \langle \eta \hat{i} \rangle}{\langle 12 \rangle \langle 3 \four \rangle} \left[ \frac{X_{\one + 2, \tell} \, X_{3+ \four, \tell}}{D_3} - \frac{X_{\one + 2, \tell} \, X_{\four, \tell}}{D_4} + \mu^2 \frac{X_{\one + 2, \tell} \, X_{3, \four}}{D_3 D_4}  \right] \label{eq:Xtri1}
\\[5pt]
\mI^{\text{tri}}_{[34]12} &= -\frac{\prod_i \langle \eta \hat{i} \rangle}{\langle 12 \rangle \langle 3 \four \rangle} \left[ \frac{X_{\one + 2, \tell} \, X_{3+ \four, \tell}}{D_3} - \frac{X_{\one, \tell} \, X_{3+ \four, \tell}}{D_2} + \mu^2 \frac{X_{\one, 2} \, X_{3+ \four,\tell}}{D_2 D_3} \right]. \label{eq:Xtri3}
\end{align}
Notice that by combining the sub-integrands at this point a number of cancellations occur. For example, the bubble terms fully cancel, and all the terms in \eqref{eq:Xtri1} and \eqref{eq:Xtri3} are cancelled by terms in \eqref{eq:Xbox}. What remains is
\begin{align}
\begin{gathered}
\mI(\ell_0, \alpha) = \frac{[12][34]}{\langle 12 \rangle \langle 34 \rangle} \frac{\mu^4}{D_2 D_3 D_4} + \frac{1}{\prod \langle \eta \hat{i} \rangle^2} \mI^{\text{tri}}_{[23]41}
\\[5pt]
+ \frac{1}{\prod_i \langle \eta \hat{i} \rangle} \frac{1}{\langle 12 \rangle \langle 3 \four \rangle} \left[ D_3 \frac{X_{\one, \tell} \, X_{\four,\tell}}{D_2 D_4} - \mu^2 \frac{X_{\one,2} \, X_{\four, \tell}}{D_2 D_4} - \mu^2 \frac{X_{\one, \tell} \, X_{3,\four}}{D_2 D_4} \right].
\end{gathered} \label{eq:l0int}
\end{align}
Now let us symmetrise\footnote{If one does not perform this symmetrisation, then the correct result will be obtained up to terms that vanish non-trivially upon loop integration.} in $\ell_0$, and pay close inspection to the second line of \eqref{eq:l0int}. Keeping in mind that $D_2 = (2\ell_0 \cdot \hat{k}_1)$ and $D_4 = -(2\ell_0 \cdot \hat{k}_4)$, the terms proportional to $\mu^2$ are anti-symmetric in $\ell_0$, and so are cancelled in the symmetrisation. What remains in the second line of \eqref{eq:l0int} is simply the first term and its symmetrisation, which can be written as
\begin{align}
\begin{gathered}
\frac{1}{\prod_i \langle \eta \hat{i} \rangle^2} \frac{D_3}{(2\hat{k}_1 \cdot k_2)^2} \frac{X_{\one, 2}X_{3, \four}X_{\one, \tell} X_{\four, \tell}}{D_2 D_4} + (\ell_0 \rightarrow - \ell_0)
= \frac{2}{\prod_i \langle \eta \hat{i} \rangle^2} \frac{X_{\one, 2}X_{3, \four}}{(2\hat{k}_1 \cdot k_2)} \frac{X_{\one, \tell} X_{\four, \tell}}{D_2 D_4} .\label{eq:symsimp1}
\end{gathered}
\end{align}
Now, let us look at the symmetrisation of the term in \eqref{eq:l0int} proportional to $\mI^{\text{tri}}_{[23]41}$. From \eqref{eq:tri2}, this results in
\begin{align}
\begin{gathered}
\frac{1}{\prod \langle \eta \hat{i} \rangle^2} \frac{X_{2,3}X_{\tell+\one,2+3}}{(2k_2 \cdot k_3)} \frac{X_{\one, \tell}X_{\four,\tell}}{D_2 D_4} + (\ell_0 \rightarrow -\ell_0)
= \frac{-2}{\prod \langle \eta \hat{i} \rangle^2} \frac{X_{2,3}X_{\one, \four}}{(2k_2 \cdot k_3)} \frac{X_{\one, \tell}X_{\four, \tell}}{D_2 D_4}. \label{eq:symsimp2}
\end{gathered}
\end{align}
Finally, by comparing \eqref{eq:symsimp1} and \eqref{eq:symsimp2} and noting that
\begin{equation*}
\frac{X_{\one, 2}X_{3, \four}}{(2\hat{k}_1 \cdot k_2)} - \frac{X_{2,3}X_{\one, \four}}{(2k_2 \cdot k_3)} = 0,
\end{equation*}
all terms in \eqref{eq:l0int} aside from the first vanish in the symmetrisation of $\ell_0$, and thus one is left with
\begin{equation*}
\mI(\ell_0, \alpha) + \mI(-\ell_0, \alpha) = \frac{[12][34]}{\langle 12 \rangle \langle 34 \rangle} \frac{\mu^4}{D_2 D_3 D_4} + (\ell_0 \rightarrow -\ell_0),
\end{equation*}
which, by taking $\ell = \ell_0 + \alpha q$ in the first term and $\ell = \ell_0 - \alpha q$ in the second term, and changing variables back to the full $\ell$, gives the one-loop amplitude to be
\begin{align}
\mathcal{A}^{(1)}(1^+, 2^+, 3^+, 4^+) = \frac{[12][34]}{\langle 12 \rangle \langle 34 \rangle} \int \frac{d^D\ell}{\ell^2} \, &\left[ \frac{\mu^4}{(\ell + k_1)^2 (\ell + k_1 + k_2)^2 (\ell - k_4)^2} \right. \nonumber \\[5pt]
+ &\hspace{5pt} \left. \frac{\mu^4}{(\ell - k_1)^2 (\ell - k_1 - k_2)^2 (\ell + k_4)^2} \right]
\end{align}
which matches directly known results for this amplitude.% This analysis has also been checked for the case of different BCFW shifts, e.g. on particles 2 and 3. Note that in this case that the initial integrand changes, since in the worldsheet formula $\sigma_+$ and $\sigma_-$ will now be positioned between $\sigma_2$ and $\sigma_3$ in the Parke-Taylor factor associated with the colour.

\subsection{Identities for traces of six slashed vectors}
\label{sec:slashedvecs}
In this appendix, we prove the two identities~\eqref{eq:parityminustrace} for traces of 6 slashed vectors, which remove the term linear in $\ell$ from the five-point all-plus YM integrand in equation~(\ref{eq:ym5ptallplus1}).

First consider the parity plus gamma matrix trace. In the first term, anticommute $\ell$ past $k_5$ using half of the Dirac algebra in the Weyl basis, that $k_5^{\dot{\alpha } \alpha } \ell_{\alpha  \dot{\beta }}^{(4D)}+\ell^{\dot{\alpha } \alpha }_{(4D)} k_{5 \alpha  \dot{\beta }}=2 \delta ^{\dot{\alpha }}{}_{\dot{\beta }} \,\ell\cdot k_5$. Continue anticommuting $\ell$ past each of $k_4,
k_3, k_2$ and $k_1$ in this way. Each anticommutation gives one term with an $\ell\cdot k_i$, and after all of the anticommutations we get $ -\langle 23\rangle [34]\langle 45\rangle [5|\ell|1\rangle [12]$ which cancels with $[12]\langle 23\rangle [34]\langle 45\rangle [5|\ell|1\rangle$. Hence,
\begin{align*}
\text{Tr}\left(\slashed{k}_1 \slashed{k}_2 \slashed{k}_3\slashed{k}_4\slashed{k}_5\slashed{\ell}\right)
&= \langle 12\rangle [23]\langle 34\rangle [45]\langle 5|\ell|1] + [12]\langle 23\rangle [34]\langle 45\rangle [5|\ell|1\rangle \\
&= (2\ell\cdot k_5) \langle 12\rangle [23] \langle 34\rangle [41]- (2\ell\cdot k_4) \langle 12\rangle [23] \langle 35\rangle [51] + (2\ell\cdot k_3) \langle 12\rangle [24] \langle 45\rangle [51]\\
&\qquad - (2\ell\cdot k_2) \langle 13\rangle [34] \langle 45\rangle [51] + (2\ell\cdot k_1) \langle 23\rangle [34] \langle 45\rangle [52].
\end{align*}
Now use five-point momentum conservation first in the spinor bracket factors and then in the  $\ell\cdot k_i$. Add and subtract $s_{45} s_{51} s_{12}$ to arrive at
\begin{equation*}
\text{Tr}\left(\slashed{k}_1 \slashed{k}_2 \slashed{k}_3\slashed{k}_4\slashed{k}_5\slashed{\ell}\right)= -(2\ell\cdot k_1) s_{34} s_{45} + \left(2 \ell\cdot k_{45}-s_{45}\right) s_{12} s_{51}+ \left(2 \ell\cdot k_{12}+s_{12}\right)s_{45} s_{51} - (2\ell\cdot k_5) s_{12} s_{23}.
\end{equation*}
Use the definition  $D_i=\left(\ell + k_{1...i} \right)^2$ to substitute in differences of the propagators to arrive at
\begin{align*}
\text{Tr}& \left(\slashed{k}_1 \slashed{k}_2 \slashed{k}_3\slashed{k}_4\slashed{k}_5\slashed{\ell}\right)=\left(D_0-D_1\right) s_{34} s_{45}-\left(D_0-D_2\right) s_{45} s_{51} +\left(D_0-D_3\right) s_{12} s_{51} -\left(D_0-D_4\right) s_{12} s_{23} \\
& =-D_1 s_{34} s_{45}+D_2 s_{45} s_{51} + D_4 s_{12} s_{23}-D_3 s_{51} s_{12} + D_0 \left(s_{34} s_{45}-s_{45} s_{51}+s_{51} s_{12}-s_{12} s_{23}\right).
\end{align*}
Finally, we apply some additional manipulations to the expression for usage in simplifying the amplitude. On support of five-point momentum conservation, we have the following non-trivial quadratic identity in Mandelstam invariants
\eq{
s_{34} s_{45}-s_{45} s_{51}+s_{51} s_{12}-s_{12} s_{23}+s_{23} s_{34} +s_{52} s_{34}-s_{53} s_{24}+s_{54} s_{23}=0.
}
Additionally, using that $\text{Tr}\left(\slashed{k}_5 \slashed{k}_2 \slashed{k}_3\slashed{k}_4\right)=s_{52} s_{34}-s_{53} s_{24}+s_{54} s_{23}$, we can write
\begin{equation*}
\text{Tr}\left(\slashed{k}_1 \slashed{k}_2 \slashed{k}_3\slashed{k}_4\slashed{k}_5\slashed{\ell}\right)=-D_0\left(s_{23} s_{34}+\text{Tr}\left(\slashed{k}_5 \slashed{k}_2 \slashed{k}_3\slashed{k}_4\right)\right)-D_1 s_{34} s_{45}+D_2 s_{45} s_{51}-D_3 s_{51} s_{12}+D_4 s_{12} s_{23}.
\end{equation*}
Now, consider the parity minus trace. The argument in this section is inspired by that of \cite{NigelGlover:2008ur}. 
\eqs{
\label{eq:parityminustrace2}
\text{Tr}\left(\gamma_5 \slashed{k}_1 \slashed{k}_2 \slashed{k}_3\slashed{k}_4\slashed{k}_5\slashed{\ell}\right) &=
\langle 12\rangle [23]\langle 34\rangle [45]\langle 5|\ell|1] - [12]\langle 23\rangle [34]\langle 45\rangle [5|\ell|1\rangle \\
&= 4 i \left(-s_{45} \varepsilon  \left(k_1 k_2 k_5 \ell\right)-s_{12} \varepsilon  \left(k_1 k_4 k_5 \ell\right)\right),
}
where we eliminated $k_3$ using five-point momentum conservation, and used the spinor bracket expressions for the $\varepsilon$ tensors, e.g. $4 i\varepsilon  \left(k_1 k_4 k_5 \ell\right) = \langle 14\rangle [45]\langle 5|\ell|1] - [14]\langle 45\rangle [5|\ell|1\rangle$.
We now introduce terms quadratic in $\ell$ which add up to zero using Cramer's rule for the linear dependence of five vectors in four dimensions, in particular
$$\ell^{(4D)} \varepsilon \left(k_1 k_2 k_4 k_5\right) + k_1 \varepsilon \left(k_2 k_4 k_5 \ell\right) + k_2 \varepsilon \left(k_4 k_5 \ell k_1\right) + k_4 \varepsilon \left(k_5 \ell k_1 k_2\right) + k_5 \varepsilon \left(\ell k_1 k_2 k_4\right) = 0.$$
Note that there are many similar ways to introduce quadratic terms in $\ell$ which add up to zero by Cramer's rule, but this is the unique way that will use the Mandelstam invariants in both terms of equation (\ref{eq:parityminustrace2}) to write an expression in terms of propagators,
\begin{align*}
\text{Tr}\left(\gamma_5 \slashed{k}_1 \slashed{k}_2 \slashed{k}_3\slashed{k}_4\slashed{k}_5\slashed{\ell}\right)&=
4 i \left(-s_{45} \varepsilon  \left(k_1 k_2 k_5 \ell\right)-s_{12} \varepsilon  \left(k_1 k_4 k_5 \ell\right)\right)+ 2\ell\cdot (\ell^{(4D)} \varepsilon \left(k_1 k_2 k_4 k_5\right) + k_1 \varepsilon \left(k_2 k_4 k_5 \ell\right)+
\\&\qquad k_2 \varepsilon \left(k_4 k_5 \ell k_1\right) + k_4 \varepsilon \left(k_5 \ell k_1 k_2\right) + k_5 \varepsilon \left(\ell k_1 k_2 k_4\right)) \\[5pt]
 &= 4 i (2 \ell^2_{(4D)} \varepsilon \left(k_1 k_2 k_3 k_4\right) + 2\ell\cdot k_1 \varepsilon  \left(k_2 k_4 k_5 \ell\right)-\left(2\ell\cdot k_2 + s_{12}\right)\varepsilon  \left(k_1 k_4 k_5 \ell\right)
 \\&\qquad -2  \left(s_{45}- 2\ell\cdot k_4\right)\varepsilon  \left(k_1 k_2 k_5 \ell\right) - 2\ell\cdot k_5  \varepsilon\left(k_1 k_2 k_4 \ell\right)).
\end{align*}
The pre-factors of each $\varepsilon$ can now be written as a difference of propagators using the definition that  $D_i=\left(k_{1 ... i}+\ell\right){}^2$,
\begin{align*}
\text{Tr}\left(\gamma_5 \slashed{k}_1 \slashed{k}_2 \slashed{k}_3\slashed{k}_4\slashed{k}_5\slashed{\ell}\right)
&=
4 i (2 \left(D_0+\mu ^2\right)\varepsilon \left(k_1 k_2 k_3 k_4\right)
+\left(D_1-D_0\right) \varepsilon  \left(k_2 k_4 k_5 \ell\right)
\\&\qquad-\left(D_2-D_1\right) \varepsilon  \left(k_1 k_4 k_5 \ell\right)
-\left(D_3-D_4\right) \varepsilon  \left(k_1 k_2 k_5 \ell\right)
+\left(D_4-D_0\right) \varepsilon  \left(k_1 k_2 k_4 \ell\right))  \\[5pt]
&= 4i (2 \left(D_0+\mu ^2\right)\varepsilon \left(k_1 k_2 k_3 k_4\right) 
-D_0 \varepsilon  \left(k_2 k_3 k_4 \ell\right)
-D_1 \varepsilon  \left(k_3 k_4 k_5 \ell\right)
\\&\qquad-D_2 \left(\varepsilon  \left(k_4 k_5 k_1 \ell\right)\right)
-D_3 \varepsilon  \left(k_5 k_1 k_2 \ell\right)
-D_4 \varepsilon  \left(k_1 k_2 k_3 \ell\right) ),
\end{align*}
where we use five-point momentum conservation to combine sums of $\varepsilon$ tensors in the second equality. This form looks symmetrical and beautiful, but we break it up slightly to reflect how it will be used in simplifying the amplitude. Combine half of the $2 D_0 \varepsilon\left(k_1 k_2 k_3 k_4\right)$ with the term involving $D_0 \varepsilon \left(\ell k_2 k_3 k_4\right)$, and use five-point momentum conservation and $\text{Tr}\left(\gamma_5 \slashed{k}_5 \slashed{k}_2 \slashed{k}_3\slashed{k}_4\right)=4 i \varepsilon \left(k_5 k_2 k_3 k_4\right)$
on the other half to arrive at the final expression:
\eqs{
\text{Tr}\left(\gamma_5 \slashed{k}_1 \slashed{k}_2 \slashed{k}_3\slashed{k}_4\slashed{k}_5\slashed{\ell}\right)
&=8 i \mu ^2 \varepsilon \left(k_1 k_2 k_3 k_4\right)-D_0 \text{Tr}\left(\gamma_5 \slashed{k}_5 \slashed{k}_2 \slashed{k}_3\slashed{k}_4\right)-4 i (D_0 \varepsilon \left(k_2 k_3 k_4\left(l+k_1\right)\right) \\
&\qquad + D_1 \varepsilon \left(k_3 k_4 k_5 \ell\right) + D_2 \varepsilon \left(k_4 k_5 k_1 \ell\right)+D_3 \varepsilon \left(k_5 k_1 k_2 \ell\right)+D_4 \varepsilon \left(k_1 k_2 k_3 \ell \right)).\\
}

\section{Worldsheet formulas from on-shell diagrams} \label{wsonshell}
Worldsheet formulas for one-loop amplitudes were deduced from on-shell diagrams in \cite{Farrow:2017eol}. They take the form of integrals over the Riemann sphere which localise onto solutions of scattering
equations refined by helicity, and therefore provide a natural loop-level generalisation of the 4d ambitwistor string formulas developed in \cite{Geyer:2014fka}. Notice that these worldsheet formulas for loop integrands are strictly four-dimensional, so that even the loop momentum lies in four dimensions.
 At four-points, the external particles are split by helicity into two sets\footnote{where $L$ contains the negative helicity particles, and $R$ the positive helicity ones.} $L=\{1,2\}$
and $R=\{3,4\}$ and we associate a puncture on the worldsheet to each particle. There are also two auxiliary punctures
$\{+,-\}$ which encode the loop momentum. Each puncture is associated with a pair of complex numbers $\sigma^\alpha$, $\alpha=1,2$, corresponding to homogeneous coordinates on $\mathbb{CP}^1$. The scattering equations
for the auxiliary punctures are given by
\begin{equation}
S_{-}=\tilde{\lambda}_{0}-\sum_{r}\frac{\tilde{\lambda}_{r}}{(-r)},\,\,\, S_{+}=\lambda_{0}-\sum_{l}\frac{\lambda_{l}}{(+l)},
\label{scatteqpm}
\end{equation}
where $(ij)=\sigma_i^\alpha \sigma_j^\beta \epsilon_{\alpha \beta}$, and the remaining scattering equations are given by
\begin{equation}
S_{l}=\hat{\tilde{\lambda}}_{l}-\sum_{r}\frac{\tilde{\lambda}_{r}}{(lr)}+\frac{\tilde{\lambda}_{0}}{(l+)},\,\,\, S_{r}=\hat{\lambda}_{r}-\sum_{l}\frac{\lambda_{l}}{(rl)}-\frac{\lambda_{0}}{(r-)},
\label{scatteringeq}
\end{equation}
with $l\in\left\{ 1,2\right\}$ and $r\in\{3,4\}$, $\hat{\lambda}_{4}=\lambda_{4}-\alpha\lambda_{1}$,
and $\left(\hat{\tilde{\lambda}}_{1},\hat{\tilde{\eta}}_{1}\right)=\left(\tilde{\lambda}_{1}+\alpha\tilde{\lambda}_{4},\tilde{\eta}_{1}+\alpha\tilde{\eta}_{4}\right)$ (the hats act trivially on the other spinors). 

The scattering amplitudes are then written as worldsheet integrals
containing delta functions which impose the refined scattering equations:
\begin{equation}
\delta(\scalebox{0.9}{SE})=\delta^{2}\left(S_{+}\right)\delta^{2}\left(S_{-}\right)\prod_{l\in L}\delta^{2|4}\left(S_{l}\right)\prod_{r\in R}\delta^{2}\left(S_{r}\right)
\label{scatteqr}
\end{equation}
as well as an integral over the loop momentum $\ell=\lambda_{0}\tilde{\lambda}_{0}-\alpha\lambda_{1}\tilde{\lambda}_{4}$, which can be written as
\[
\frac{d^{4}\ell}{\ell^{2}}=\frac{d^{2}\lambda_{0}d^{2}\tilde{\lambda}_{0}}{\mathrm{vol}\,\mathrm{GL}(1)}\frac{d\alpha}{\alpha}.
\]
For $\mathcal{N}=4$ SYM, \cite{Farrow:2017eol} found the following worldsheet formula for one-loop four-point amplitude:
\begin{equation}
\mathcal{A}_{4}^{(1)}=\int\frac{d^{4}\ell}{\ell^{2}}\frac{\prod_{a=1}^{6}d^{2}\sigma_{a}}{\mathrm{vol}\,\mathrm{GL}(2)}\left(\frac{1}{(-1)...(4+)(+-)}+(+\leftrightarrow-)\right)\delta(\scalebox{0.9}{SE})
\label{1loop4pt}
\end{equation}
where $\sigma_5 = \sigma_+$ and $\sigma_6=\sigma_-$. A similar formula was also derived for $\mathcal{N}=8$ supergravity in \cite{Farrow:2017eol} using the decorated on-shell diagrams developed in \cite{Heslop:2016plj}. 

The scattering equations in \eqref{scatteqr} have a unique solution, on the support of which the worldsheet integral in \eqref{1loop4pt} gives rise to the standard loop integrand in terms of quadratic Feynman propagators. This worldsheet representation was derived by first writing the on-shell diagram for a four-point one-loop amplitude as a Grassmannian integral, and then mapping elements of the Grassmannian into 2-point functions on Riemann sphere. Unfortunately, it is unclear how to extend this formula to higher points. At five points, there are three on-shell diagrams as described in section \ref{momtwistrecursion}, two of which encode the forward limit of a 7-point tree-level amplitude and one of which encodes a one-loop four-point amplitude dressed with a soft factor. Although it is straightforward to map each on-shell diagram into a worldsheet integral, it is unclear how to combine them into into a single worldsheet formula because the scattering equations associated with the soft term appear to be incompatible with those encoding the forward limit. There is a natural generalisation of \eqref{1loop4pt} which describes the forward limit contribution to a general $n$-point one-loop N$^k$MHV amplitude:
\begin{equation}
\mathcal{A}_{n}^{(1),\scalebox{0.8}{FL}}=\int\frac{d^{4}\ell}{\ell^{2}}\frac{1}{\mathrm{vol}\,\mathrm{GL}(2)}\left(\prod_{a=1}^{n+2}\frac{d^{2}\sigma_{a}}{(a\,a+1)}+(+\leftrightarrow -)\right)\delta(\scalebox{0.9}{SE}),
\end{equation}
where the scattering equations have the same form as \eqref{scatteqr}, except that  the left set consists of $k+2$ external particles and the right set contains the remaining external particles. Here $\sigma_{n+1} = \sigma_+$ and $\sigma_{n+2}=\sigma_-$.

We will now show how quadratic propagators arise from solving the one-loop scattering equations at four-points, given in \eqref{scatteqpm} and \eqref{scatteringeq}. After replacing the scattering equations for particles 3 and 4 to give a momentum conservation delta function, and gauge fixing the punctures for particles 1 and 2, the equations form a linear system with the following solution
\[
\sigma_{\text{sol}} = \left(
\begin{array}{cccccc}
 1 & 0 & \frac{[34] \langle 12\rangle }{\langle 1 | 3 + 0 | 4]} & -\frac{[34] \langle 12\rangle }{\langle 1 | 4 + 0 | 3]} & \frac{\langle 12\rangle }{\langle 1 0\rangle } & \frac{\langle 2 | 3 + 0 | 4] \langle 2 | 4 + 0 | 3] [2 0]}{P_{03\hat{4}}^2 \langle 12\rangle  [0 3] [0 4]} \\
 0 & 1 & \frac{[34] \langle 12\rangle }{\langle 2 | 3 + 0 | 4]} & -\frac{[34] \langle 12\rangle }{\langle 2 | \hat{4} + 0 | 3]} & \frac{\langle 12\rangle }{\langle 2 0\rangle } & \frac{\langle 1 | 3 + 0 | 4] \langle 1 | 4 + 0 | 3] [0 \hat{1}]}{P_{03\hat{4}}^2 \langle 12\rangle  [0 3] [0 4]} \\
\end{array}
\right)
\]
where the columns are labeled $1,2,3,4,+,-$, and $P_{03\hat{4}}= \lambda_0 \tilde{\lambda}_0+k_3+\hat{k}_4$. Some relevant minors of this matrix are given by
\begin{align*}
(12) &= 1 &
(23) &= -\frac{[34] \langle 12\rangle }{\langle 1 | 3 + 0 | 4]}\\
(14) &= -\frac{[34] \langle 12\rangle }{\langle 2 | \hat{4} + 0 | 3]} &
(34) &=  \frac{[34]^3 \langle 12\rangle ^3 P_{03\hat{4}}^2}{\langle 1 | 3 + 0 | 4]\langle 1 | 4 + 0 | 3] \langle 2 | 3 + 0 | 4] \langle 2 | \hat{4} + 0 | 3]}\\
(-1) &= -\frac{\langle 1 | 3 + 0 | 4]\langle 1 | 4 + 0 | 3] [0 \hat{1}]}{[0 3] [0 4] \langle 12\rangle  P_{03\hat{4}}^2} &
(4+) &= \frac{[34]^2 \langle 12\rangle ^3 \langle 0 \hat{4}\rangle }{\langle 2 | \hat{4} + 0 | 3]\langle 1 | 4 + 0 | 3] \langle 1 0\rangle  \langle 2 0\rangle }\\
(+1) &= -\frac{\langle 12\rangle }{\langle 2 0\rangle } & 
(4-) &= \frac{[34]}{[0 3]}\,.
\end{align*}
%(+-) = \frac{\text{angmomsqu}(1,\{3,0\},4) \text{angmomsqu}(1,\{4,0\},3) [0 \text{hat}(1)] \langle 2 0\rangle -\text{angmomsqu}(2,\{3,0\},4) \text{angmomsqu}(2,\{\text{hat}(4),0\},3) [2 0] \langle 1 0\rangle }{[0 3] [0 4] \langle 1 0\rangle  \langle 2 0\rangle  \text{sumpsq}(0,3,\text{hat}(4))}
The Jacobian for the system, defined by $\diracd{(2\times 6)}{\scalebox{0.9}{SE}} = J \, \diracd{(4)}{P} \diracd{(8)}{\sigma - \sigma_{\text{sol}}}$, is given by
\eq{
J = \frac{[34]^8 \langle 12\rangle ^8}{P_{03\hat{4}}^2 \langle 1 | 3 + 0 | 4] \langle 1 | 4 + 0 | 3] \langle 2 | 3 + 0 | 4] \langle 2 | 4 + 0 | 3] [3 0]^2 [4 0]^2 \langle 1 0\rangle ^2 \langle 2 0\rangle ^2}\,.
}
Based on the mapping from on-shell diagrams in \cite{Farrow:2017eol} we see that the four-point one-loop all-gluon MHV amplitude in $\cN=4$ SYM can be written supported on these equations as
\begin{equation}
\mathcal{A}^{(1)}(g^-g^-g^+g^+) = \int \frac{d^4\ell}{\ell^2}\int \frac{d^{2\times 6}\sigma}{\text{vol}\,\mathrm{GL}(2)} \delta^{2\times 6} (\scalebox{0.9}{SE}) f(\sigma)\,,
\end{equation}
where $\ell = \lambdat_0\lambda_0 - \alpha\, \lambdat_4\lambda_1$, and 
\begin{equation}
f(\sigma) = \sum_{+\to -}\frac{1}{(-1)(12)(23)(34)(4+)(+-)}=\frac{1}{(-1)(12)(23)(34)(4+)}\frac{(14)}{(+1)(4-)}\,.
\end{equation}
Combining the integrand and Jacobian on support of the solution above, we see that the worldsheet integral can be evaluated to
\begin{equation*}
\int d^{2\times 6}\sigma\, \delta^{(2\times 6)}(\scalebox{0.9}{SE})\frac{(14)}{(-1)(12)(23)(34)(4+)(+1)(4-)}
=\frac{\delta^{(4)}(P)\;\;\ang{12}^2\squ{34}^2}{(\ell + k_4)^2 (\ell + k_3 + k_4)^2(\ell + k_2 + k_3 + k_4)^2},
\end{equation*}
noting that $(\lambdat_0\lambda_0 + \hat{k}_4) = (\ell + k_4)$. This is indeed the expected four-particle integrand.

It should be straightforward to do the analogous calculation in ${\mathcal N}=8$ supergravity using the worldsheet formula from \cite{Farrow:2017eol} and the solution to the scattering equations stated in this appendix.

%%%%%%%%%%%%%%

\section{Worldsheet formulas with double-forward limit scattering equations}\label{sec:DFL}
\begin{figure}[ht]
	\centering 
	  \includegraphics[width=6cm]{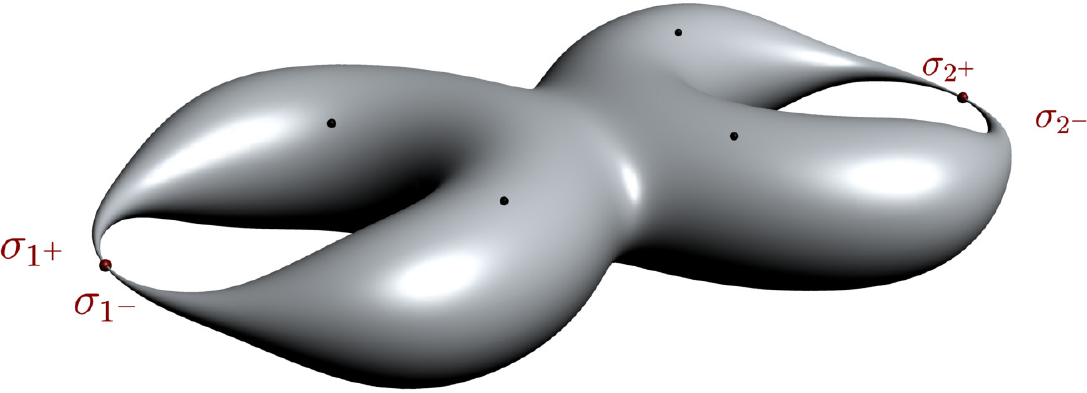}
	\caption{The bi-nodal sphere with four punctures corresponding to external particles. The loop momentum $\ell=\ell_1+\ell_2$ is parametrised by two nodes, corresponding each to a null momentum $\ell_{1,2}$.}
	\label{fig:bi-nodal_labelled}
\end{figure}

A different proposal for worldsheet formulas giving  one-loop integrands with quadratic propagators was put forward in refs. \cite{Gomez:2017lhy, Gomez:2017cpe, Ahmadiniaz:2018nvr, Agerskov:2019ryp}, relying on what we will call a `double-forward limit' (DFL) construction. The underlying idea is that every off-shell loop momentum $\ell$ can be decomposed into a sum of two null momenta: $\ell=\ell_1+\ell_2$, with $\ell_1^2=\ell_2^2=0$. These momenta can then be incorporated naturally into the scattering equation framework by introducing \emph{four} additional marked points,  $\sigma_{1^\pm}$ and $\sigma_{2^\pm}$, corresponding respectively to the momenta $\pm \ell_1$ and $\pm \ell_2$ in a DFL construction; see \cref{fig:bi-nodal_labelled}. The proposal also includes an algorithm for obtaining worldsheet integrands in this representation from known half-integrands in the linear representation, as in \eqref{eq:half-int_lin}.

In this section, we first review this proposal, and then discuss why the proposed algorithm for constructing worldsheet integrands only holds for the $n$-gon. In particular, it does {not} work for general bi-adjoint scalar integrands or Yang-Mills theory.  
In a nutshell, the problem is an `overcounting' factor for the $n$-gon in the DFL construction,  $\mathfrak{I}^{\DFL}_{\text{n-gon}}=n\,\mathfrak{I}_{\text{n-gon}}$. 
As an overall factor, this is harmless for the $n$-gon and can be absorbed by a redefinition of the integrand, but for other theories this overcounting  leads to systematic errors that cannot be easily corrected at the level of the worldsheet formulas.  This is most easily seen in a simple four-particle bi-adjoint scalar example, which we will discuss below:
\begin{equation}
 \mathfrak{I}_{\mathrm{BS}}^\DFL(1234|1243) = 3\,\mathfrak{I}_{\text{tri}}(12[34])+2\,\mathfrak{I}_{\text{bub}}([12][34])
 \;\;\;\; \neq \;\;\;\;
 \mathfrak{I}_{\mathrm{BS}}=\mathfrak{I}_{\text{tri}}(12[34])+\mathfrak{I}_{\text{bub}}([12][34])\,.\nonumber
\end{equation}
We see that the triangle has an (unphysical) numerical prefactor of 3, whereas the bubble comes with a prefactor of 2. These problems are exacerbated for Yang-Mills, where even correcting the numerical prefactors by hand fails due to problems of gauge invariance.
 We would like to stress at this point that this does \emph{not} invalidate the DFL approach in general, but instead indicates that new worldsheet integrands suitable to the DFL scattering equations will have to be constructed.

\paragraph{Review.} Let us first review the worldsheet proposal of \cite{Gomez:2017lhy, Gomez:2017cpe, Ahmadiniaz:2018nvr, Agerskov:2019ryp} in more detail. 
In contrast to the one-loop approach described in the main text, where the worldsheet is a sphere with $n + 2$ marked points (including a node $\sigma_\pm$), the DFL worldsheet is a sphere with  $n + 2 + 2$ marked points. The two nodes $\sigma_{1^\pm}$ and $\sigma_{2^\pm}$ encode pairs of  null momenta $\pm\ell_1$ and $\pm\ell_2$  in a double forward limit. Corresponding to this worldsheet, there are natural DFL scattering equations,
\begin{equation}\label{eq:SE_DFL}
 \mathcal{E}_\sA^\DFL = \sum_{\sB=1}^{n+4} \frac{2k_\sA\cdot k_\sB}{\sigma_{\sA\sB}}\,,\qquad \qquad A,B\in\{1,\dots,n,1^\pm,2^\pm\}\,,
\end{equation}
with $k_{1^\pm}=\pm\ell_1$ and $k_{2^\pm}=\pm\ell_2$. Using these scattering equations, general loop integrands with Feynman propagators can be written as \cite{Gomez:2017lhy, Gomez:2017cpe}, 
\begin{equation}
 \mathfrak{I}_n^{\DFL}=\ell^2\,\int d^D\!\left(\ell_1+\ell_2\right)\;\delta^D\left(\ell_1+\ell_2-\ell\right)\;\int_{\mathfrak{M}_{0,n+4}}  \hspace{-20pt}d\mu_{n}^\DFL\; \cI^{\DFL}_{\scalebox{0.5}{$1/2$}}\,\tilde{\cI}^{\DFL}_{\scalebox{0.5}{$1/2$}}\,.
\end{equation}
Here, the integral against the delta-function enforces $\ell=\ell_1+\ell_2$, and we have introduced an explicit factor of $\ell^2$ for later convenience. The measure $d\mu_{n}^\DFL$  is the standard CHY-measure for $n+4$ marked points $\sigma_A$, but with the loop-level scattering equations $\mathcal{E}_\sA^\DFL$ given in  \eqref{eq:SE_DFL}.  One of the crucial features of this construction, reminiscent of higher loops in the usual approach \cite{Geyer:2016wjx,Geyer:2018xwu, Geyer:2019hnn}, is that the scattering equations $\mathcal{E}_\sA^\DFL$ not only encode physical poles, but also unphysical poles  that must be absent from the final formula. Recall in this context that the set of all possible poles is determined by the form of the scattering equations; see \cref{sec:factorisation} for details. Following the same line of reasoning as in \cref{lem:CHY-fact}, the  worldsheet geometries below  correspond to physical singular kinematics,
\begin{equation}
 s_{\sL}=
 \begin{cases}
 k_{\scalebox{0.5}{$L$}}^2   & L=L^{\scalebox{0.6}{ext}}\;\;\mathrm{or}\;\; L=\{1^\pm,2^\pm \}\cup L^{\scalebox{0.6}{ext}}\\
 (\ell-k_{\scalebox{0.5}{$L$}})^2 & L=\{1^+,2^+\}\cup L^{\scalebox{0.6}{ext}}\\ 
  (\ell+k_{\scalebox{0.5}{$L$}})^2 & L=\{1^-,2^-\}\cup L^{\scalebox{0.6}{ext}}\\ 
 \mathrm{unphys} & \mathrm{else}\,. \end{cases}
\end{equation}
 for some $k_\sL=-\sum_{i\in L}k_i$. The unphysical poles, originating from worldsheet geometries where only one of the nodal points factors off, or two nodal points with misaligned loop momenta, are of the form $(\ell_1\pm k_\sL)^2$, $(\ell_2\pm k_\sL)^2$ and $(\ell_1-\ell_2 \pm k_\sL)^2$ respectively. Additionally, as occurs also for our $\ell^2$-deformed scattering equations, there are unphysical  discriminant-style poles from tadpole-like worldsheets. 
 
 Clearly the absence of these unphysical poles tightly constrains the form of the worldsheet integrands that can give physical amplitudes. 
Refs.~\cite{Gomez:2017lhy,Gomez:2017cpe, Ahmadiniaz:2018nvr, Agerskov:2019ryp} propose an algorithm for constructing these integrands for any theory with a known integrand in the standard `linear' representation, such that they never contain unphysical poles. This is done as follows:
if a half-integrand $\cI^{\scalebox{0.6}{$(1)$}}_{\scalebox{0.5}{$1/2$}}$ from the linear representation can be defined on the DFL scattering equations (we will say more about this restriction in a moment), %\footnote{This technical qualification will become clear in the case of Yang-Mills considered below.}
 the half-integrand $ \cI^{\DFL}_{\scalebox{0.5}{$1/2$}}$  is then given by
\begin{equation}
        \cI^{\DFL}_{\scalebox{0.5}{$1/2$}}=
                    \frac{(\sigma_{1^+1^-})^2}{(1^+2^+2^-1^-)}\;
           \cI^{\scalebox{0.6}{$(1)$}}_{\scalebox{0.5}{$1/2$}}(\sigma_{1^\pm})\,,
           \qquad\qquad 
           \tilde{\cI}^{\DFL}_{\scalebox{0.5}{$1/2$}}=
                    \frac{(\sigma_{2^+2^-})^2}{(1^+2^+2^-1^-)}\;
           \cI^{\scalebox{0.6}{$(1)$}}_{\scalebox{0.5}{$1/2$}}(\sigma_{2^\pm})\,.\label{eq:int_DFL}
\end{equation}
Even before discussing concrete examples, we can appreciate the beauty of this construction by seeing how it prevents all unphysical poles: the product of the two prefactors is $(\sigma_{1^+2^+}\,\sigma_{1^-2^-})^{-2}$, which ensures that the half-integrands give a vanishing contribution to all residues on unphysical poles of the form $(\ell_1\pm k_\sL)^2$, $(\ell_2\pm k_\sL)^2$ or $(\ell_1-\ell_2 \pm k_\sL)^2$. This only leaves the unphysical discriminant poles from tadpole-like worldsheets. The DFL integrand has a vanishing residue on these poles if both half-integrands remain finite on the tadpole-like worldsheets. This condition is also required for the half-integrands $\cI^{\scalebox{0.6}{$(1)$}}_{\scalebox{0.5}{$1/2$}}$ in the linear approach, and thus the construction \eqref{eq:int_DFL} guarantees that there are no unphysical poles.\footnote{The details here are easily filled in following the line of argument in \cref{sec:factorisation}; see especially the discussion around \eqref{eq:int_no_res}.}

We will now take a closer look at the proposal for the key examples discussed in \cite{Gomez:2017lhy,Gomez:2017cpe, Ahmadiniaz:2018nvr, Agerskov:2019ryp}. Using the known half-integrands reviewed in \S\ref{sec:wsquad}, the DFL worldsheet integrands  for the  $n$-gon, the bi-adjoint scalar and Yang-Mills theory are given by \eqref{eq:int_DFL} with 
\begin{align*}
 &n\text{-gon:} &&\cI^{\scalebox{0.6}{$(1)$}}_{\scalebox{0.5}{$1/2$}}=\mathcal{C}^{\scalebox{0.6}{$(1)$}}_{\scalebox{0.6}{cyc}}(\sigma_{1^\pm}|\alpha)
 %\frac1{\sigma_{1^+1^-}^2}\prod_{i=1}^n\frac{\sigma_{1^+1^-}}{\sigma_{1^+i}\sigma_{i1^-}}
 &&\tilde{\cI}^{\scalebox{0.6}{$(1)$}}_{\scalebox{0.5}{$1/2$}}=\frac1{\sigma_{2^+2^-}^2}\prod_{i=1}^n\frac{\sigma_{2^+2^-}}{\sigma_{2^+i}\sigma_{i2^-}}\,,\\
 &\text{Bi-adjoint scalar:}  &&\cI^{\scalebox{0.6}{$(1)$}}_{\scalebox{0.5}{$1/2$}} = \mathcal{C}^{\scalebox{0.6}{$(1)$}}_{\scalebox{0.6}{cyc}}(\sigma_{1^\pm}|\alpha)
 && \tilde{\cI}^{\scalebox{0.6}{$(1)$}}_{\scalebox{0.5}{$1/2$}} =\mathcal{C}^{\scalebox{0.6}{$(1)$}}_{\scalebox{0.6}{cyc}}(\sigma_{2^\pm}|\beta) \,,\\
 &\text{Yang-Mills:}  &&\cI^{\scalebox{0.6}{$(1)$}}_{\scalebox{0.5}{$1/2$}} = \mathcal{C}^{\scalebox{0.6}{$(1)$}}_{\scalebox{0.6}{cyc}}(\sigma_{1^\pm}|\alpha)
 && \tilde{\cI}^{\scalebox{0.6}{$(1)$}}_{\scalebox{0.5}{$1/2$}} =\cI^{\scalebox{0.6}{$(1)$}}_{\scalebox{0.6}{kin,pure}}(\sigma_{2^\pm})  \,,
\end{align*}
with the familiar $n$-gon and colour factors, but now with nodes $\sigma_{I^\pm}$.  To be precise, the colour factors and (pure, i.e., non-supersymmetric) kinematic  half-integrands  are given by
\begin{align}\label{eq:int_DFLCI}
 &\mathcal{C}^{\scalebox{0.6}{$(1)$}}_{\scalebox{0.6}{cyc}}(\sigma_{I^\pm}|\alpha)=\sum_{\rho\in\mathrm{cyc}(\alpha)}\frac{1}{(I^+\rho_1\rho_2\dots\rho_n I^-)}\,, 
 && \cI^{\scalebox{0.6}{$(1)$}}_{\scalebox{0.6}{kin,pure}}(\sigma_{I^\pm})=\sum_{\rho\in S_n}\frac{N_{\mathbb{A}}(\rho|\ell)}{(I^+\rho_1\rho_2\dots\rho_n I^-)}\,.
\end{align}
for any colour-ordering $\alpha $ and $I=1,2$. The kinematic half-integrand here is the one suitable for pure Yang-Mills, where $N_{\mathbb{A}}(\rho|\ell)$ are the BCJ numerators calculated from the ambitwistor string integrand in \cite{Geyer:2017ela}. As discussed in \cite{Agerskov:2019ryp}, in 4d these numerators can be interpreted as the tree-level numerators in the double-forward limit.\footnote{In the case of Yang-Mills, it becomes important that the half-integrands have to be well-defined on the DFL scattering equations. This prevents us from using the (NS version of the) half-integrand discussed in \eqref{eq:int_gen_v2}, $\cI^{\scalebox{0.6}{$(1)$}}_{\scalebox{0.6}{kin,pure}}=\sum_r \pf{}'M{\scalebox{0.6}{NS}}^r$, which is only defined on the support of the  one-loop scattering equations with $n+2$ marked points.}

\paragraph{Discussion.} As we saw above, all DFL integrands have only physical poles by construction. However, this does not guarantee that the residues at those poles give the correct on-shell amplitudes expected from locality and unitarity. To investigate this more closely, we consider first the simplest example: the $n$-gon integrand. Refs.~\cite{Geyer:2015bja, Geyer:2015jch, He:2015yua} showed that, in the linear representation,  this can be written as the worldsheet integral
\begin{equation}
 \mathfrak{I}^{\lin}_{n\text{-gon}}:=\frac{1}{\ell^2}\,\int_{\mathfrak{M}_{0,n+2}}\hspace{-20pt} d\mu_{n}^\lin\;\cI_{\scalebox{0.6}{$n$-gon}}^{\scalebox{0.5}{$1/2$}}\, \cI_{\scalebox{0.6}{$n$-gon}}^{\scalebox{0.5}{$1/2$}}
 =\frac{(-1)^n}{\ell^2}\sum_{\rho\in S_n}\prod_{i=1}^n\frac{1}{\left(2\ell\cdot k_{\rho(1\dots i)}+ k_{\rho(1\dots i)}^2\right)}\simeq \mathfrak{I}_{n\text{-gon}}\,.
\end{equation}
In the last equality, $\mathfrak{I}_{\text{n-gon}}$ is the permutation sum over the $(n-1)!$ different $n$-gons with quadratic propagators,\footnote{Here we are counting (1234) and (1432) as two separate boxes. Also note the normalisation conventions in the scattering equations imply that there is no factor of $2^n$ between the different representations.} and we used the notation $\simeq$ to stress that the equivalence only  holds up to partial fraction identities and shifts in the loop momentum. The second equality can be proven, e.g., by factorisation arguments along the lines of \S\ref{sec:factorisation}. 
On the other hand, the same line of argument for the DFL representation gives\footnote{We note that this agrees with \cite{Gomez:2017lhy}, where the right hand side was called ``symmetrised n-gon''.}
\begin{equation}
  \mathfrak{I}^{\DFL}_{n\text{-gon}}:=\ell^2\,\int_{\mathfrak{M}_{0,n+4}}\hspace{-20pt} d\mu_{n}^\DFL\;\frac{\cI_{\scalebox{0.6}{$n$-gon}}^{\scalebox{0.5}{$1/2$}}(\sigma_{1^\pm})\, \cI_{\scalebox{0.6}{$n$-gon}}^{\scalebox{0.5}{$1/2$}}(\sigma_{2^\pm})}{(\sigma_{1^+2^+}\,\sigma_{1^-2^-})^2}
  =\frac{(-1)^n}{\ell^2}\sum_{\rho\in S_n}\prod_{i=1}^n\frac{1}{\left(\ell + k_{\rho(1\dots i)}\right)^2}= n \,\mathfrak{I}_{n\text{-gon}}\,,
\end{equation}
i.e., there is an `overcounting' of the $n$-gon by a numerical factor of $n$.
In the case of the $n$-gon this overcounting is harmless, since it can be absorbed by a simple redefinition of the integrand. For other theories, however, this overcounting will lead to systematic errors that cannot be easily corrected at the level of the worldsheet formulas. To appreciate this, consider the simplest non-trivial example, the bi-adjoint scalar integrand. For simplicity, we will look at a four-particle colour-ordered integrand, with the two colour orderings given by $\alpha=(1234)$ and $\beta=(1243)$. 
The worldsheet formula in the linear representation reproduces the correct result: following, e.g., the calculation in  ref.~\cite{He:2015yua}, section III B, the  integrand is given by
\begin{equation}
\mathfrak{I}_{\mathrm{BS}}^\lin(1234|1243) \simeq \mathfrak{I}_{\text{tri}}(12[34])+\mathfrak{I}_{\text{bub}}([12][34])=\mathfrak{I}_{\mathrm{BS}}(1234|1243) \,,
\end{equation}
where $\simeq$ again indicates equality up to shifts in the loop momentum, and we have dropped the contribution from bubbles in the external legs. On the other hand, we can recycle the $n$-gon result from above for the special case of triangles and bubbles to find
\begin{equation}
\label{eq:BSovercount}
 \mathfrak{I}_{\mathrm{BS}}^\DFL(1234|1243) \simeq 3\,\mathfrak{I}_{\text{tri}}(12[34])+2\,\mathfrak{I}_{\text{bub}}([12][34])\,,
\end{equation}
from the DFL construction (again dropping bubbles in the external legs).
Since the numerical prefactors from the triangle (3) and the bubble (2) differ, the result is \emph{not} related to the bi-adjoint scalar integrand by an overall factor.
This makes it clear that even for four external particles in one of the simplest theories, this construction does not reproduce the correct integrands,
\begin{equation}
  \mathfrak{I}^{\DFL}_{\text{BS}}\neq  \, \mathfrak{I}_{\text{BS}}\,.
\end{equation}

Since the above errors are systematic, one could envision an `ad-hoc' approach to removing them \emph{after} performing the worldsheet integrals. While unsatisfying, this is at least in principle possible for the bi-adjoint scalar. For Yang-Mills theory, however, even this prescription fails, because the proposal suffers from more than a simple `overcounting' of diagrams --  the expression for the loop integrand does not obey gauge invariance (up to terms that vanish in the loop integration). 
To see this in more detail,  consider again the four-particle integrand \cite{Agerskov:2019ryp}
\begin{align}
{\mathfrak{I}}_{\text{YM}}^\DFL(1234) &=
\ell^2\,\int_{\mathfrak{M}_{0,n+4}}\hspace{-20pt} d\mu_{n}^\DFL\;
\frac{\mathcal{C}^{\scalebox{0.6}{$(1)$}}_{\scalebox{0.6}{cyc}}(\sigma_{1^\pm}|1234)\, \cI_{\scalebox{0.6}{kin,pure}}^{\scalebox{0.6}{$(1)$}}(\sigma_{2^\pm}) }{(\sigma_{1^+2^+}\,\sigma_{1^-2^-})^2}
=\, \mathfrak{I}^{\text{box}}_{\DFL}+ \mathfrak{I}^{\text{tri}}_{\DFL} +\mathfrak{I}^{\text{bub}}_{\DFL}
\ ,
\label{eq:IYM1234}
\end{align}
where the box is colour-ordered, $ \mathfrak{I}^{\text{box}}_{\DFL}= \mathfrak{I}^{\text{box}}_{\DFL}(1234)$, as are the triangle and bubble contributions,
\begin{equation*}
 \mathfrak{I}^{\text{tri}}_{\DFL}=\mathfrak{I}^{\text{tri}}_{\DFL}(1234)+\mathfrak{I}^{\text{tri}}_{\DFL}(2341)+\mathfrak{I}^{\text{tri}}_{\DFL}(3412)+\mathfrak{I}^{\text{tri}}_{\DFL}(4123)\,,\qquad\qquad 
 \mathfrak{I}^{\text{bub}}_{\DFL}=\mathfrak{I}^{\text{bub}}_{\DFL}(1234)+\mathfrak{I}^{\text{bub}}_{\DFL}(4123) \,.
\end{equation*}
Here and below, we drop contributions from bubbles in the external legs.
As shown in \cite{Agerskov:2019ryp}, we can use shifts in the loop momentum to bring the respective terms into the following form:
\begin{subequations}
 \begin{align}
   \mathfrak{I}^{\text{box}}_{\DFL}(1234)&=\frac{\mathbf{N}(1234|\ell)}{\ell^2\,(\ell+k_1)^2\,(\ell+k_{12})^2\,(\ell-k_4)^2}
  \\
   \mathfrak{I}^{\text{tri}}_{\DFL}(1234)&=\frac{\mathbf{N}([1,2]34|\ell)}{s_{12}\,\ell^2\,(\ell+k_{12})^2\,(\ell-k_4)^2}
 \\
   \mathfrak{I}^{\text{bub}}_{\DFL}(1234)&=\frac{\mathbf{N}([1,2][3,4]|\ell)}{s_{12}^2\,\ell^2\,(\ell+k_{12})^2}\,,
 \end{align}
\end{subequations}
where the numerators are now linear combinations of the  BCJ numerators $N_{\mathbb{A}}$ calculated from the ambitwistor string integrand (c.f. \eqref{eq:int_DFL}),
\begin{subequations}
 \begin{align}
 \label{eq:DFLN4}
  \mathbf{N}(1234|\ell) &= N_{\mathbb{A}}(1234|\ell)+N_{\mathbb{A}}(2341|\ell+k_1)+N_{\mathbb{A}}(3412|\ell+k_{12})+N_{\mathbb{A}}(4123|\ell-k_4)\\
   \label{eq:DFLN3}
  \mathbf{N}([12]34|\ell) & = N_{\mathbb{A}}([1,2]34|\ell)+N_{\mathbb{A}}(34[1,2]|\ell+k_{12})+N_{\mathbb{A}}(4[1,2]3|\ell-k_4)\\
     \label{eq:DFLN2}
  \mathbf{N}([1,2][3,4]|\ell) & = N_{\mathbb{A}}([1,2][3,4]|\ell)+N_{\mathbb{A}}([3,4][1,2]|\ell+k_{12})\,.
 \end{align}
\end{subequations}
We can see here the remnant of the overcounting in the $n$-gon case: four terms contribute to the box, three to the triangle and two to the bubble. While this certainly raises suspicions about the proposal, we would like to compare it directly to a known Yang-Mills integrand representation. 
A simple method to achieve this is to convert the formula into its \emph{linear} representation, using partial fractions and shifts of the loop momentum.
The reason is that terms that loop-integrate to zero in dimensional regularisation are particularly easy to identify in the linear representation: they scale as $\lambda^c$ for some integer $c$ if we rescale the loop momentum as $\ell\mapsto \lambda \ell$, or are linear combinations of terms with such a scaling. This facilitates the comparison between different integrands, accounting for equality \emph{up to} terms that integrate to zero. Expanding the DFL proposal, we find again ${\mathfrak{I}}_{\text{YM}}^\DFL(1234)  =\mathfrak{I}^{\text{box}}_{\DFL}+ \mathfrak{I}^{\text{tri}}_{\DFL} +\mathfrak{I}^{\text{bub}}_{\DFL}$ with all contributions colour-ordered as above, but now with 
\begin{equation}
  \mathfrak{I}^{\text{box}}_{\DFL}(1234)\simeq \frac{ \mathbf{N}(1234|\ell) }{\ell^2\,(2\ell\cdot k_1)\,(2\ell\cdot k_{12}+k_{12}^2)\,(-2\ell\cdot k_4)}+\mathrm{cyc}(1234)\,.
\end{equation}
Similar expressions can be derived for the triangle and the bubble in the linear representation. Here, we just write the box contribution for the sake of brevity. 

On the other hand, expressions for the Yang-Mills integrand in the linear representation were derived from the nodal sphere in \cite{Geyer:2017ela}. Similarly to the DFL proposal, the integrand takes the form ${\mathfrak{I}}_{\text{YM}}^\lin(1234)=\mathfrak{I}^{\text{box}}_{\text{YM}}+ \mathfrak{I}^{\text{tri}}_{\text{YM}} +\mathfrak{I}^{\text{bub}}_{\text{YM}}$, where each term is again a  sum over colour-ordered contributions. In contrast to the DFL proposal, however, the box term is simply given by
\begin{equation}
  \mathfrak{I}^{\text{box}}_{\text{YM}}(1234)\simeq \frac{ N_{\mathbb{A}}(1234|\ell) }{\ell^2\,(2\ell\cdot k_1)\,(2\ell\cdot k_{12}+k_{12}^2)\,(-2\ell\cdot k_4)}+\mathrm{cyc}(1234)\,.
\end{equation}
Expressions for the triangle and the bubble can be found in \cite{Geyer:2017ela}, where it also has been verified that the complete integrand agrees with the conventional BCJ representation of the integrand obtained in \cite{Bern:2013yya} up to shifts in the loop momentum and terms that integrate to zero. Comparing the two box formulas, the difference is non-zero, and can be checked not to loop-integrate to zero,
\begin{equation}
\mathfrak{I}^{\text{box}}_{\DFL}-\mathfrak{I}^{\text{box}}_{\text{YM}}
   =\frac{N_{\mathbb{A}}(2341|\ell+k_1)+N_{\mathbb{A}}(3412|\ell+k_{12})+N_{\mathbb{A}}(4123|\ell-k_4)}{\ell^2\,(2\ell\cdot k_1)\,(2\ell\cdot k_{12}+k_{12}^2)\,(-2\ell\cdot k_4)}+\mathrm{cyc}(1234) \not\simeq 0\,.
\end{equation}
We can see that there is a surplus of terms in $\mathfrak{I}^{\text{box}}_{\DFL}$, due to there being four $N_{\mathbb{A}}$'s in $\mathbf{N}(1234|\ell)$. This is analogous to the numerical factors in \eqref{eq:BSovercount} for the bi-adjoint scalar. 

%In fact, the difference admits a maximal cut (i.e., a cut of the four propagators), which cannot be corrected by triangles and bubbles. One could still argue based on this observation that this is just a difference of normalisation convention, and that this problem could potentially be fixed by considering \,$\frac1{4}\mathfrak{I}^{\text{box}}_{\DFL}-\mathfrak{I}^{\text{box}}_{\text{YM}}$\,, which would make the maximal cut correct. This is also not a solution to the problem. For an $n$-gon expression, at $n$ points, that would require that the comparison is \,$\frac1{n}\,\mathfrak{I}^{n\text{-gon}}_{\DFL}-\mathfrak{I}^{n\text{-gon}}_{\text{YM}}$\,,
%Hence, the DFL proposal \eqref{eq:int_DFL} is incorrect for both the bi-adjoint scalar and Yang-Mills theory.

At this point, we may still hope that the difference between the DFL proposal and the Yang-Mills integrand is due to numerical factors, as for the bi-adjoint scalar, and  that the correct integrand could be recovered by removing the expected factors by hand,
\begin{align}
\mathcal{P}\left({\mathfrak{I}}^\DFL(1234)\right) :=
\,\frac{1}{4} \mathfrak{I}^{\text{box}}_{\DFL}+\frac{1}{3} \mathfrak{I}^{\text{tri}}_{\DFL} +\frac{1}{2}\mathfrak{I}^{\text{bub}}_{\DFL}
\ .
\label{eq:IYM1234_v2}
\end{align}
While unsatisfying, a map $\mathcal{P}$ defined this way gives the correct bi-adjoint scalar integrand -- though of course no worldsheet formula, because $\mathcal{P}$  cannot be defined at the level of the worldsheet.\footnote{At least no  construction for $\mathcal{P}$ is known on the worldsheet. } For Yang-Mills, however, even this ad-hoc construction fails to yield the correct integrand,\footnote{In the special case of all-plus amplitudes, whose very simple numerators we described in \S\ref{sec:bdy_all+}, this prescription actually gives the correct loop integrand; the reason is that, in this case, the four terms on the right-hand side of  \eqref{eq:DFLN4} are all equal, hence the factor $1/4$ corrects that contribution, and analogously for \eqref{eq:DFLN3} and \eqref{eq:DFLN2}. Hence, the expression \eqref{eq:PIDFLIlin} is an equality. Generally, however, those terms will not be equal, and the prescription fails.}
\begin{equation}
  \mathcal{P}\left({\mathfrak{I}}_{\text{DFL}}^\lin(1234)\right)-{\mathfrak{I}}_{\text{YM}}^\lin(1234)
   \not\simeq 0\,,
   \label{eq:PIDFLIlin}
\end{equation}
as we have again checked numerically. In fact, we checked that the first term fails to obey gauge invariance (up to loop integration), whereas the second does obey it because it is the correct loop integrand.

\paragraph{} To conclude, we would like to reiterate that the above discussion only demonstrates that the specific proposal \eqref{eq:int_DFL} for worldsheet integrands does not apply generically, and this does not invalidate the broader double-forward-limit framework. It would be interesting to use the issues and problems discussed here to construct suitable integrands. This is not straightforward, however. A natural proposal would be to mimic our construction in \cref{sec:MHV} and restrict the colour half-integrand to a single cyclic ordering. This fails, because now the resulting worldsheet formula has discriminant-poles from tadpole-like worldsheets, even for maximal supersymmetry.
These poles appear because the two half-integrands in the DFL construction depend on \emph{different} nodes --- in contrast to our worldsheet formula, which only depends on a single node --- and thus the half-integrand $ \cI_{\scalebox{0.6}{kin}}^{\scalebox{0.6}{$(1)$}}$ does not vanish on the  discriminant pole contribution from the colour factor.

Valid integrands can of course be constructed (at least for the bi-adjoint scalar) by a bottom-up method, following up the $n$-gon construction by progressively more complicated $m$-gons with massive corners, and assembling them into the correct amplitudes. However, the issue of gauge invariance for Yang-Mills is challenging, and more generally it is not obvious that worldsheet formulas obtained by this method exhibit the simplicity of those associated to the linear representation.

\section{Technical aspects of the BCFW recursion on the worldsheet}
\label{app:BCFW}
In this section, we provide details for the more technical aspects of the BCFW recursion relation, such as the vanishing of the one-minus tree amplitude from the worldsheet approach, and the absence of singular solutions and boundary terms. We also give a proof of the $n$-gon and $(n-1)$-gon worldsheet formulas using factorisation.

\subsection{Vanishing of the \texorpdfstring{$X$}{X}-tree amplitude}
In this first part of the appendix, we fill in some of the details glossed over in \cref{sec:BCFW_WS}: that the MHV worldsheet formula supports no factorisation channels with $|L|>2$, or in other words that the only tree-level factorisation channels involve 
$\overline{\mathrm{MHV}}_3$ amplitudes. As discussed there, this relies on the vanishing of a tree-level contribution which we call here the $X$-amplitude (due to the structure of the numerators),
\begin{equation}\label{eq:X-ampl}
 \mathcal{X}_n=\int_{\raisebox{-6pt}{\scalebox{0.7}{$\mathfrak{M}_{0,n}$}}}\hspace{-15pt}d\mu_{n}^{\CHY} \;\left( \sum_{\rho\in S_{n-2}}\frac{X^{(A_\rho)}}{(\scalebox{1}{$1\,\rho\, n$})}\right)\frac{1}{(12\ldots n)}\,.
\end{equation}
We use the notation $A_\rho = [\rho_1,[\rho_2,[\ldots,[\rho_{n-2},n]]]$, and the numerators are given by the following combination of the $X$ variables,
\begin{equation}
 X^{(A_\rho)}=\prod_{j=2}^{n-2} X_{n+\rho_{n-2} +\ldots +\rho_{n-j},\rho_{n-j-1}}\,.
\end{equation}
It turns out that up to overall prefactors, this expression actually matches the worldsheet formula for the tree-level one-minus amplitude, which is known to vanish for $n>3$ (the case $n=3$ is $\overline{\mathrm{MHV}}_3$), but we will not rely on that argument. Instead, we will show directly via a recursion relation that these $X$-amplitudes vanish for $n>3$,  $\mathcal{X}_n=0$.

To see this, let us first consider low-particle examples. We have $\mathcal{X}_3\neq0$ for $\overline{\mathrm{MHV}}_3$, where $ \lambda_1\sim\lambda_2\sim\lambda_3$, but we have $\mathcal{X}_3=0$ for MHV${}_3$, where $\tilde \lambda_1\sim\tilde\lambda_2\sim\tilde\lambda_3$, because then $X_{23}=0$.
For $n=4$, the half-integrand carrying the $X$-dependence  vanishes on support of the scattering equations,
\begin{equation*}
 \frac{X^{([2,[3,4]])}}{(1234)}+\frac{X^{([3,[2,4]])}}{(1324)}=\frac{X_{1,2}X_{3,4} }{(1234)}+\frac{X_{1,3}X_{2,4} }{(1324)}\simeq 0\,,%\simeq \frac{1}{(234)\,\sigma_{14}}\left(\frac{x_{ni}}{x_{\sL i}}[L i][jn] -\frac{x_{nj}}{x_{\sL j}}[Lj][in]\right)
\end{equation*}
 and thus the following seed amplitudes vanish,
 \begin{equation}
 \mathcal{X}_3\Big|_{\text{MHV}}=0\,,\qquad \mathcal{X}_4=0\,.
\end{equation}
We will now establish that the worldsheet formula \eqref{eq:X-ampl} satisfies a simple tree-level BCFW recursion relation of the form
\begin{equation}\label{eq:recursion_X}
 \mathcal{X}_n= \sum_{j=2}^{n-2}\frac{1}{s_{1\dots j}}\mathcal{X}_{j+1} \;\mathcal{X}_{n-j+1}\,.
\end{equation}
This can be seen as follows. As before, we choose to shift particles 1 and $n$, so that the only relevant factorisation channels contain either 1 or $n$. Following a line of argument similar to the one in \cref{sec:factorisation}, the Parke-Taylor factor guarantees that  only poles respecting planar ordering $(12\ldots n)$ contribute. On these poles, the $X$-half-integrand factorises correctly due to 
\begin{equation}
                 X^{(A)} = X^{(A_1)}X^{(A_2)}    \qquad\qquad\text{for }\; \Big(\sum_{a\in A_1}k_a\Big)^2=0\,.
\end{equation}
Using the known factorisation formulas for the measure and Parke-Taylor factors, the worldsheet formula thus satisfies the recursion \eqref{eq:recursion_X}. But since the four-particle  amplitude vanishes, all higher-point amplitudes vanish as well. Hence, we have shown that $\mathcal{X}_n=0$ for all $n>3$.

\subsection{Singular solutions}
\label{app:sing_sol}

In this section, we give more details on why singular solutions to the scattering equations, such that $\sigma_+=\sigma_-$, do not contribute to our MHV worldsheet formula. As discussed in \cref{sec:BCFW_WS}, singular solutions are tadpole-like worldsheet geometries, with the nodal sphere factorising into a sphere with all external marked points, and a nodal sphere that carries no additional punctures besides the node. These geometries are problematic because they encode unphysical ``discriminant poles'', so it is crucial that these solutions do not contribute in worldsheet formulas. As before, we parametrise this boundary by 
\begin{equation}\label{eq:def_singular_sol}
 \sigma_-=\sigma_+ +\varepsilon\,,\qquad\qquad\varepsilon \ll 1\,.
\end{equation}
As discussed in \S\ref{sec:BCFW_WS} the measure scales as $d\mu_n^\Def \sim d\varepsilon\;\delta(s_\sL-\varepsilon\mathcal{F})$, and the colour half-integrand as  $\mathcal{C}^{\scalebox{0.6}{$(1)$}}_n(12\dots n)\sim\varepsilon^{-1}$. For the residue on the unphysical pole to vanish, we must check that the kinematic integrand behaves at most as $\cI^{\scalebox{0.6}{$(1)$}}_{\scalebox{0.6}{MHV}}\sim \varepsilon$. 
%In fact, we will see that we can do even better, and show that $\cI^{\scalebox{0.6}{$(1)$}}_{\scalebox{0.6}{MHV}}\sim \varepsilon^2$. 
On the singular solutions \eqref{eq:def_singular_sol}, the first few orders of the  kinematic half-integrand in $\varepsilon$ are given by
\begin{equation}\label{eq:sing_orders}
 \cI^{\scalebox{0.6}{$(1)$}}_{\scalebox{0.6}{MHV}}\Bigg|_{O(\varepsilon^m)}=\sum_{\rho\in S_n}\frac{N^{\scalebox{0.6}{$(1)$}}_{\scalebox{0.6}{$\rho$}}}{(+\rho_1\rho_2\ldots \rho_n)}\,\frac{1}{\sigma_{\rho_n+}^{m+1}}\,,\qquad\qquad \text{for }\,m=-1,0,1\,.
\end{equation}
We have already seen in the main text that a U$(1)$ decoupling identity in conjunction with the form of the numerators guarantees that this vanishes for $m=-1$, and below we discuss the order $m=0$.%\footnote{There is evidence from the ambitwistor string that the kinematic half-integrand also vanishes for $m=1$.}
As we will see, this term vanishes due to the absence of bubble numerators, $N_{1|A_2}=0$. We will also uncover a beautiful connection to the KK relations on the worlsheet -- a nice generalisation of the U$(1)$ decoupling identity that underpinned the $m=-1$ relation.

\paragraph{m=0:} Consider first the kinematic half-integrand at order $O(1)$. Using the same U$(1)$ decoupling identity as before, we can simplify the expression \eqref{eq:sing_orders}  to
\begin{align*}
 \cI^{\scalebox{0.6}{$(1)$}}_{\scalebox{0.6}{MHV}}\Bigg|_{O(\varepsilon^0)}
 %
 %&=\sum_{\rho\in S_n}\frac{N^{\scalebox{0.6}{$(1)$}}_{\scalebox{0.6}{$\rho$}}}{(+\rho_1\rho_2\ldots \rho_n)}\,\frac{1}{\sigma_{\rho_n+}}\\
 %
 &=\sum_{\rho\in S_{n-1}}\frac{N^{\scalebox{0.6}{$(1)$}}_{\scalebox{0.6}{$1\rho$}}}{(+\rho_1\rho_2\ldots \rho_{n-1})}\left(\left(\frac{\sigma_{+\rho_1}}{\sigma_{+1}\sigma_{1\rho_1}}+\ldots \frac{\sigma_{\rho_{n-2} \rho_{n-1}}}{\sigma_{\rho_{n-2} 1}\sigma_{1\rho_{n-1}}}\right)\frac{1}{\sigma_{\rho_{n-1}+}}+\frac{\sigma_{\rho_{n-1} +}}{\sigma_{\rho_{n-1} 1}\sigma_{1+}}\frac{1}{\sigma_{1+}}\right)\\
 &=-\frac{1}{\sigma_{1+}^2} \sum_{\rho\in S_{n-1}}\frac{N^{\scalebox{0.6}{$(1)$}}_{\scalebox{0.6}{$1\rho$}}}{(+\rho_1\rho_2\ldots \rho_{n-1})}\,.
\end{align*}
The sum over  $S_{n-1}$ runs again over permutations of the external particles excluding 1. At this point, we can simplify this further by using the KK relations on the worldsheet
\begin{equation}
\label{eq:KK}
 \frac{1}{(+\,\alpha \,n \,\beta)} = (-1)^{|\beta|}\sum_{\rho\in \alpha\shuffle\beta^T} \frac{1}{(+\,\rho\, n)} \,,
\end{equation}
which hold for any disjoint sets $\alpha$, $\beta$ such that $\alpha\cup\beta=\{2,\dots,n-1\}$. Collecting terms with the same Parke-Taylor factors gives
\begin{equation}
 \cI^{\scalebox{0.6}{$(1)$}}_{\scalebox{0.6}{MHV}}\Bigg|_{O(\varepsilon^0)}=-\frac{1}{\sigma_{1+}^2} \sum_{\rho\in S_{n-2}}\frac{\mathbf{N}^{\scalebox{0.6}{$(1)$}}_{\scalebox{0.6}{$1\rho$}}}{(+\rho_1\rho_2\ldots \rho_{n-2} \,n)}\,,
\end{equation}
where we defined new numerators $\mathbf{N}^{\scalebox{0.6}{$(1)$}}_{\scalebox{0.6}{$1\rho$}}$ as sums over the BCJ numerators,
\begin{equation}
 \mathbf{N}^{\scalebox{0.6}{$(1)$}}_{\scalebox{0.6}{$1\rho$}} = \sum_{\substack{\pi=\alpha\cup\beta\\\rho\in\alpha\shuffle\beta}}(-1)^{|\beta|}N^{\scalebox{0.6}{$(1)$}}_{\scalebox{0.6}{$1\pi$}}=N^{\scalebox{0.6}{$(1)$}}_{\scalebox{0.6}{$1\,|\,A_\rho$}}=0\,.
\end{equation}
The second identity, leading to $N^{\scalebox{0.6}{$(1)$}}_{\scalebox{0.6}{$1\,|\,A_\rho$}}$ with $A_\rho=[\rho_1,[\rho_2,[\ldots[\rho_{n-2},n]]]$, follows from  the BCJ properties of the numerators, and the last identity just uses that all bubble numerators vanish, $N^{\scalebox{0.6}{$(1)$}}_{1 |A_\rho}=0$. The kinematic integrand scales at most as $\cI^{\scalebox{0.6}{$(1)$}}_{\scalebox{0.6}{MHV}}\sim \varepsilon$ in the singular solutions, which guarantees that the residue on the pole vanishes.

\iffalse
\paragraph{m=1:} As indicated above, we can actually show more: following same steps as above, we can simplify the $O(\varepsilon)$ contribution to 
\begin{align}\label{eq:sing_order_epsilon}
 \cI^{\scalebox{0.6}{$(1)$}}_{\scalebox{0.6}{MHV}}\Bigg|_{O(\varepsilon)}=-\frac{1}{\sigma_{1+}^2} \sum_{\rho\in S_{n-1}}\frac{N^{\scalebox{0.6}{$(1)$}}_{\scalebox{0.6}{$1\rho$}}}{(+\rho_1\rho_2\ldots \rho_{n-1})}\frac{1}{\sigma_{\rho_{n-1}+}}\,.
 \end{align}
Here we used that the $O(1)$ terms vanish, and dropped a contribution of the same form. We could in principle try to simplify this further using similar method, but at this stage this approach gets rather cumbersome. 

Instead, we will use the fact that the kinematic integrand secretely carries a form degree in each of the marked points, so if all of its residues in a marked point are zero, the half-integrand itself vanishes as well. In particular,

Check e.g. residue at $(n+)^2$ (any term will have such a pole for some $i$). Residue given by
\begin{equation}
 -\frac{1}{\sigma_{1+}^2} \sum_{\rho\in S_{n-2}}\frac{N^{\scalebox{0.6}{$(1)$}}_{\scalebox{0.6}{$1\,\rho\, n$}}}{(+\rho_1\rho_2\ldots \rho_{n-2})}=-\frac{1}{\sigma_{1+}^2} \sum_{\rho\in S_{n-3}}\frac{N^{\scalebox{0.6}{$(1)$}}_{\scalebox{0.6}{$1\,|\,A_\rho\,|\, n$}}}{(+\rho_1\rho_2\ldots \rho_{n-3} \,n\!-\!1)}=0\,.
\end{equation}
First identity from following the same steps as above -- using the KK identity, collecting terms, second identity again from triangles vanishing. Similarly for all other poles, and thus constant (and equal zero) by Liouville.\fi
 
\subsection{Boundary terms}
In what follows, we make the heuristic argument given in \S\ref{sec:BCFW_WS} for the absence of boundary terms in the BCFW recursion relations mathematically precise by studying the large $z$ limit of the (shifted) scattering equations. Explicitly, we want to find asymptotic solutions to the following equations:
\begin{subequations}
 \begin{align}
  \hat{\mathcal{E}}_+
  &=-z\sum_{j\neq 1,n} \frac{2q\cdot k_j}{\sigma_{+j}}+\frac{(\ell+k_1)^2}{\sigma_{+1}}-\frac{(\ell-k_n)^2}{\sigma_{+n}}+\sum_{j\neq 1,n} \frac{2\ell\cdot k_j}{\sigma_{+j}}\\
  \hat{\mathcal{E}}_-
  &=+z\sum_{j\neq 1,n} \frac{2q\cdot k_j}{\sigma_{-j}}-\frac{(\ell+k_1)^2}{\sigma_{-1}}+\frac{(\ell-k_n)^2}{\sigma_{-n}}-\sum_{j\neq 1,n} \frac{2\ell\cdot k_j}{\sigma_{-j}}\\
  \vspace{5pt}
  \hat{\mathcal{E}}_1
  &=+z\sum_{j\neq 1,n} \frac{2q\cdot k_j}{\sigma_{1j}}+(\ell+k_1)^2 \frac{\sigma_{+-}}{\sigma_{1+}\sigma_{1-}}+\frac{2k_1\cdot k_n}{\sigma_{1n}}+\sum_{j\neq 1,n} \frac{2k_1\cdot k_j}{\sigma_{1j}}\\
  \hat{\mathcal{E}}_n
  &=-z\sum_{j\neq 1,n} \frac{2q\cdot k_j}{\sigma_{nj}}-(\ell-k_n)^2 \frac{\sigma_{+-}}{\sigma_{n+}\sigma_{n-}}+\frac{2k_1\cdot k_n}{\sigma_{n1}}+\sum_{j\neq 1,n} \frac{2k_n\cdot k_j}{\sigma_{nj}}\\
  \vspace{5pt}
  \hat{\mathcal{E}}_i
  &=+2z k_i\cdot q \left(\frac{\sigma_{1n}}{\sigma_{i1}\sigma_{in}}-\frac{\sigma_{+-}}{\sigma_{i+}\sigma_{i-}}\right)+2\ell\cdot k_i \frac{\sigma_{+-}}{\sigma_{i+}\sigma_{i-}}+\sum_{j\neq i} \frac{2k_i\cdot k_j}{\sigma_{ij}} \quad \text{for\;} i\neq1,n
  \,.
 \end{align}
\end{subequations}
These are just the $\ell^2$-deformed scattering equations under the BCFW shift. Let us first inspect the equations $\hat{\mathcal{E}}_i$ in the last line. The only possibilities for asymptotic solutions are the following three singular worldsheet configurations:
\begin{itemize}
 \item[(A)] $\sigma_1=\sigma_n$ and $\sigma_+=\sigma_-$\\
 This describes a nodal sphere split into three components as on the right of \cref{fig:dom-balance}, with one sphere $\Sigma_\sE$ containing only the marked points $1$ and $n$, one nodal sphere $\Sigma_\sN$ containing the node and no other punctures, and one `main'  sphere $\Sigma$ connecting these two on which all other marked points are located. 
 \item[(B)] $\sigma_1=\sigma_+$ and $\sigma_n=\sigma_-$\\
 Again, this describes a nodal sphere split into three components, but this time resembling the geometry in the middle of \cref{fig:dom-balance}. Here, the two components that split off carry one marked point each ($1$ and $n$ respectively), and are connected by the node.
 \item[(C)] $\sigma_1=\sigma_n=\sigma_+=\sigma_-$\\
 This is the leftmost geometry in \cref{fig:dom-balance}, and the simplest-looking, since it describes just two component spheres, $\Sigma_\sN$ carrying the node and the punctures $1$ and $n$, and $\Sigma_{\scalebox{0.6}{ext}}$ with all other marked points.
\end{itemize}
We will study these cases in turn, check if the Ans\"{a}tze above give a dominant balance (and thus asymptotic solutions), and determine the scaling of the worldsheet formula in $z$ for every asymptotic solution.

\paragraph{(A)} We can parametrise the moduli space around the boundary divisor by
\begin{subequations}\label{eq:deg_A}
\begin{align}
 \sigma_1 &= \sigma_\sE+\varepsilon_\sE\, x_1 
 & \sigma_+ = \sigma_\sN + \varepsilon_\sN \,y_+\,,\\
 \sigma_n &= \sigma_\sE+\varepsilon_\sE\, x_n 
 & \sigma_- = \sigma_\sN + \varepsilon_\sN \,y_-\,.
\end{align}
\end{subequations}
This is consistent with the description given in \S\ref{sec:BCFW_WS}: we have chosen the nodal points on $\Sigma_\sN$ and $\Sigma_\sE$ to lie at infinity,  $x_{\sE}=y_\sN=\infty$, and inverted the worldsheet variables $x$ and $y$ . This has the advantage of leading to simpler formulas, making dominant balances easier to identify. Using the SL$(2,\mathbb{C})$ gauge freedom on each sphere, we can further fix $x_n=0$ and $x_1=1$, as well as $y_-=0$, $y_+=1$ on $\Sigma_\sN$. It will be convenient to define the measure by removing the scattering equations $\hat{\mathcal{E}}_n$ and $\hat{\mathcal{E}}_-$, because this most naturally exposes the scaling. To leading order, all constraints $\hat{\mathcal{E}}_i=\mathcal{E}_i^{\scalebox{0.6}{$(0)$}}$ then reduce to the tree-level scattering equations for $i\neq 1,n$, and the other two scattering equations become
\begin{subequations}
 \begin{align}
  %z^{-1} \,\hat{\mathcal{E}}_- & = \hspace{12pt}\sum_{j\neq 1,n}\frac{2q\cdot k_j}{\sigma_{\sN j}}\\
  %
   z^{-1} \,\hat{\mathcal{E}}_+ & = \varepsilon_\sN \sum_{j\neq 1,n}\frac{2q\cdot k_j}{\sigma_{\sN j}^2}+z^{-1}\left(\frac{2\ell\cdot k_{1n}}{\sigma_{\sN\sE}}+\sum_{j\neq 1,n}\frac{2\ell\cdot k_{j}}{\sigma_{\sN j}}\right)\\
  \varepsilon_\sE\, z^{-1} \,\hat{\mathcal{E}}_1 & = \varepsilon_\sE \sum_{j\neq 1,n}\frac{2q\cdot k_j}{\sigma_{\sE j}} + z^{-1} k_1\cdot k_n \,.
 \end{align}
\end{subequations}
At this point, it is clear the our ansatz indeed gives a dominant balance, because we may solve  $\hat{\mathcal{E}}_+$ for $\varepsilon_\sN\sim z^{-1}$ and $\hat{\mathcal{E}}_1$ for $\varepsilon_\sE\sim z^{-1}$. We thus find that the measure scales as
\begin{equation}
 d\mu_n^\Def\sim z^{-3}\;\delta\left(\varepsilon_\sN -z^{-1}\mathcal{F}_\sN\right)\, \delta\left(\varepsilon_\sE -z^{-1}\mathcal{F}_\sE\right)\,d\varepsilon_\sN \, d\varepsilon_\sE\, d\hat\mu_{n-2}\,,
\end{equation}
where we collected all factors that scale of order one into  $d\hat\mu_{n-2}$, and both functions $\mathcal{F}_{\sN,\sE}$ are of order one. To determine the behaviour of the worldsheet formula, we thus only have to find the scaling of the half-integrands on \eqref{eq:deg_A}, 
\begin{align}
 &\mathcal{C}^{\scalebox{0.6}{$(1)$}}(12\dots n)\sim\varepsilon_\sN^{-1}\sim z\,, 
 &&\cI^{\scalebox{0.6}{$(1)$}}_{\scalebox{0.6}{MHV}} \sim \varepsilon_\sE^{-1}\varepsilon_\sN^{1}\sim 1\,.
\end{align}
The scaling of the colour half-integrand follows directly from the definition, and for the kinematic half-integrand we have used the results of \cref{app:sing_sol} on singular solutions, as well as the $z$-independence of the numerators.  Combining these results with the additional factor from $\hat\ell^{-2}\sim z^{-1}\ell^{-2}$, the worldsheet formula clearly receives no contributions from the geometry (A), 
\begin{equation}\label{eq:bdy_z_A}
 \mathfrak{I}^{\scalebox{0.6}{$(1)$}}_{\text{sYM-MHV}}(z)\Bigg|_{(A)} \sim z^{-3}\,,\qquad\qquad\text{as }z\gg1\,.
\end{equation}

\paragraph{(B)} Similarly to (A), we can  parametrise this ansatz by
\begin{subequations}
\begin{align}
 \sigma_1 &= \sigma_\sP+\varepsilon_\sP\, x_1 
 & \sigma_n = \sigma_\sM + \varepsilon_\sM \,y_n\,,\\
 \sigma_+ &= \sigma_\sP+\varepsilon_\sP\, x_+ 
 & \sigma_- = \sigma_\sM + \varepsilon_\sM \,y_-\,.
\end{align}
\end{subequations}
M\"{o}bius invariance again lets us set $x_1=0$ and $x_+=1$ on $\Sigma_\sP$, as well as $y_n=0$ and $y_-=1$ on $\Sigma_\sM$. All scattering equations $\hat{\mathcal{E}}_i=\mathcal{E}_i$ are again of order one (though now equal to the unshifted scattering equations, not the tree-level ones). If we define the measure by omitting $\hat{\mathcal{E}}_1$ and $\hat{\mathcal{E}}_n$ -- again, to expose the scaling --  the only other scattering equations fix the nodes, and are given to leading order by
\begin{subequations}
 \begin{align}
   \varepsilon_\sP\,z^{-1} \,\hat{\mathcal{E}}_+ & =- \varepsilon_\sP \sum_{j\neq 1,n}\frac{2q\cdot k_j}{\sigma_{\sP j}}+z^{-1}(\ell+k_1)^2\\
   \varepsilon_\sM\,z^{-1} \,\hat{\mathcal{E}}_- & =+ \varepsilon_\sM \sum_{j\neq 1,n}\frac{2q\cdot k_j}{\sigma_{\sM j}}+z^{-1}(\ell-k_n)^2\,.
 \end{align}
\end{subequations}
This is clearly a dominant balance, with  $\hat{\mathcal{E}}_+$ imposing  $\varepsilon_\sP\sim z^{-1}$ and $\hat{\mathcal{E}}_-$ giving $\varepsilon_\sM\sim z^{-1}$. On this asymptotic solution, the measure is
\begin{equation}
 d\mu_n^\Def\sim z^{-4}\;\delta\left(\varepsilon_\sP -z^{-1}\mathcal{F}_\sP\right)\, \delta\left(\varepsilon_\sM -z^{-1}\mathcal{F}_\sM\right)\,d\varepsilon_\sP \, d\varepsilon_\sM\, d\hat\mu_{n-2}\,.
\end{equation}
with  $d\hat\mu_{n-2}$ and $\mathcal{F}_{\sP,\sM}$ as before of order one. Both half-integrands exhibit the maximal scaling possible for Parke-Taylor factors,
\begin{align}
 &\mathcal{C}^{\scalebox{0.6}{$(1)$}}(12\dots n)\sim\varepsilon_\sP^{-1}\varepsilon_\sM^{-1}\sim z^2\,, 
 &&\cI^{\scalebox{0.6}{$(1)$}}_{\scalebox{0.6}{MHV}} \sim \sim\varepsilon_\sP^{-1}\varepsilon_\sM^{-1}\sim z^2\,.
\end{align}
without further cancellations. Their scaling is however fully cancelled by the measure, and the integrand scales as 
\begin{equation}\label{eq:bdy_z_B}
 \mathfrak{I}^{\scalebox{0.6}{$(1)$}}_{\text{sYM-MHV}}(z)\Bigg|_{(B)} \sim z^{-1}\,,\qquad\qquad\text{as }z\gg1\,.
\end{equation}

\paragraph{(C)} In the last case, the node and the punctures $1$ and $n$ all lie on one nodal sphere $\Sigma_\sN$, 
\begin{subequations}
\begin{align}
 \sigma_1 &= \sigma_\sN+\varepsilon\, x_1 
 & \sigma_+ = \sigma_\sN + \varepsilon \,x_+\,,\\
 \sigma_n &= \sigma_\sN+\varepsilon\, x_n 
 & \sigma_- = \sigma_\sN + \varepsilon \,x_-\,.
\end{align}
\end{subequations}
The SL$(2,\mathbb{C})$ invariance on this sphere lets us fix $x_+=0$ and $x_1=1$, so we will solve the scattering equations for $\varepsilon$, $x_-$ and $x_n$. To find the dominant balances, consider the following combinations of scattering equations,
\begin{subequations}\label{eq:case_C_dom}
\begin{align}
 \varepsilon\,\left(\hat{\mathcal{E}}_+ + \hat{\mathcal{E}}_-\right) &\sim x_{-+}\left(\frac{(\ell+k_1)^2}{x_{+1}x_{-1}}-\frac{(\ell-k_n)^2}{x_{+n}x_{-n}}\right)
\\
\varepsilon\,\left(\hat{\mathcal{E}}_+ + \hat{\mathcal{E}}_1\,\right) &\sim\frac{ (\ell+k_1)^2}{x_{-1}}+\frac{(\ell-k_n)^2}{x_{n+}}-\frac{2k_1\cdot k_n}{x_{n1}}\,. \label{eq:case_C_dom_1}
\end{align}
\end{subequations}
They are given to leading order $O(\varepsilon^{-1})$, and the combinations are chosen such that the leading order is independent of $z$. The first of these gives two fundamentally different types of solutions,
\begin{itemize}
 \item Solution 1: \;\; $x_{-+}=0$,
 \item Solution 2:\;\; $ x_{+n}x_{-n} (\ell+k_1)^2- x_{+1}x_{-1}(\ell-k_n)^2=0$ ,
\end{itemize}
where the first corresponds to a further degeneration of the nodal sphere, while the second does not. We will discuss these two in turn below.\\

\emph{Solution 1:} We can parametrise the further degeneration  by $x_-=x_++\varepsilon_\sM$, where we used M\"{o}bius invariance to set the location of the marked points representing the node to $y_+=0$  and  $y_-=1$, with $y\in \Sigma_\sM$ the parameter on the new sphere. The free variables are now $\varepsilon$, $\varepsilon_\sN$ and $x_n$, and we solve e.g. the scattering equations $ \hat{\mathcal{E}}_1$, $ \hat{\mathcal{E}}_n$ and $\hat{\mathcal{E}}_-$, and define the measure by dropping $\hat{\mathcal{E}}_+$ and $\hat{\mathcal{E}}_{i,j}$ for two external particles. To see that this gives  an asymptotic solution, note that \eqref{eq:case_C_dom_1} determines  $x_n$, and
\begin{subequations}
\begin{align}
 \varepsilon\,z^{-1} \,\hat{\mathcal{E}}_- & \sim \varepsilon \sum_{j\neq 1,n}\frac{2q\cdot k_j}{\sigma_{\sN j}}+z^{-1}\left(\frac{(\ell-k_n)^2}{x_{-n}}-\frac{(\ell+k_1)^2}{x_{-1}}\right)\\
 \varepsilon\,\left(\hat{\mathcal{E}}_1 + \hat{\mathcal{E}}_n\right) & \sim 
 \varepsilon_\sM\left(\frac{(\ell+k_1)^2}{x_{1+}^2}-\frac{(\ell-k_n)^2}{x_{n+}^2}\right)+\varepsilon\left(\sum_{j\neq 1,n}\frac{2k_{1n}\cdot k_j}{\sigma_{\sN j}}+z\,\varepsilon\, x_{n1}\sum_{j\neq 1,n}\frac{2q\cdot k_j}{\sigma_{\sN j}}\right)
\end{align}
\end{subequations}
give  $\varepsilon\sim z^{-1}$ and $\varepsilon_\sM\sim\varepsilon$, so this is indeed a dominant balance. The scattering equations thus contribute $\varepsilon^3 z^{-1} \sim z^{-4}$, which we combine with an additional $z^{-2}$ from  $d\sigma_1\,d\sigma_n\,d\sigma_-=\varepsilon^2 dx_n\,d\varepsilon\, d\varepsilon_\sM$ to find
\begin{equation}
 d\mu_n^\Def\sim z^{-6}\;\delta\left(\varepsilon -z^{-1}\mathcal{F}\right)\, \delta\left(\varepsilon_\sM -z^{-1}\mathcal{F}_\sM\right)\,d\varepsilon \, d\varepsilon_\sM\, d\hat\mu_{n-2}\,.
\end{equation}
The half-integrands contribute a further 
\begin{align}
 &\mathcal{C}^{\scalebox{0.6}{$(1)$}}(12\dots n)\sim\varepsilon^{-3}\varepsilon_\sM^{-1}\sim z^4\,, 
 &&\cI^{\scalebox{0.6}{$(1)$}}_{\scalebox{0.6}{MHV}}  \sim\varepsilon^{-2}\sim z^2\,,
\end{align}
where the improved behaviour for the  kinematic half-integrand is a consequence of the absence of bubble diagrams, generalising the argument above for the vanishing of singular solutions. The full worldsheet formula thus scales as  $ \mathfrak{I}^{\scalebox{0.6}{$(1)$}}_{\text{sYM-MHV}}(z) \sim z^{-1}$ on this solution.\\

\emph{Solution 2:} This case is actually simpler since it does not involve a further degeneration of the nodal sphere. We may simply solve \eqref{eq:case_C_dom} for $x_-$ and  $x_n$, and the only remaining scattering equation on $\Sigma_\sN$,
 \begin{align}
   \varepsilon\,z^{-1} \,\hat{\mathcal{E}}_n & =- \varepsilon \sum_{j\neq 1,n}\frac{2q\cdot k_j}{\sigma_{\sN j}}+z^{-1}\left(\frac{2k_1\cdot k_n}{x_{n1}}-(\ell-k_n)^2\frac{x_{-+}}{x_{n+}x_{n-}}\right)\,,
 \end{align}
 gives a dominant balance with $\varepsilon\sim z^{-1}$. These scattering equations again contribute $\varepsilon^3 z^{-1}$ to the measure, so that it scales again as
\begin{equation}
 d\mu_n^\Def\sim z^{-6}\;\delta\left(\varepsilon -z^{-1}\mathcal{F}\right)\, d\varepsilon\, d\hat\mu\,,
\end{equation}
where we used that $d\sigma_n d\sigma_-=\varepsilon^2 dx_n dx_-$. The integrands in this case are straightforward, since we may simply use their generic Parke-Taylor behaviour without further cancellations,
\begin{align}
 &\mathcal{C}^{\scalebox{0.6}{$(1)$}}(12\dots n)\sim\varepsilon^{-3}\sim z^3\,, 
 &&\cI^{\scalebox{0.6}{$(1)$}}_{\scalebox{0.6}{MHV}}  \sim\varepsilon_\sM^{-3}\sim z^3\,.
\end{align}
The worldsheet formula for both asymptotic solutions (C) thus scales as  
\begin{equation}\label{eq:bdy_z_C}
 \mathfrak{I}^{\scalebox{0.6}{$(1)$}}_{\text{sYM-MHV}}(z)\Bigg|_{(C)} \sim z^{-1}\,,\qquad\qquad\text{as }z\gg1\,.
\end{equation}
We have thus seen that all boundary terms vanish.

\subsection{Proof of the \texorpdfstring{$n$}{n}-gon and \texorpdfstring{$(n-1)$}{n-1}-gon formulas}
In the last part of this appendix, we give a brief proof of the $n$-gon   and $(n-1)$-gon worldsheet formulas  presented in the main text, in \eqref{eq:n-gon} and \eqref{eq:n-1-gon}, using the BCFW recursion relation. We repeat them here:
\begin{align}
 \mathfrak{I}^{\scalebox{0.6}{$(1)$}}_{n\text{-gon}}(12\dots n)&=\frac{1}{\ell^2} \int_{\raisebox{-6pt}{\scalebox{0.7}{$\mathfrak{M}_{0,n+2}$}}}\hspace{-15pt}d\mu_{n}^{\Def} \; \left( \frac1{\sigma_{+-}^2}\, \prod_{j=1}^n \frac{\sigma_{+-}}{\sigma_{+j}\,\sigma_{j-}}\right)\; \,\frac{1}{(+12\dots n-)}\\
 \mathfrak{I}^{\scalebox{0.6}{$(1)$}}_{(n-1)\text{-gon}|[i,i+1]}%(12\dots n)
 &=\frac{1}{\ell^2} \int_{\raisebox{-6pt}{\scalebox{0.7}{$\mathfrak{M}_{0,n+2}$}}}\hspace{-15pt}d\mu_{n}^{\Def} \; \left( \frac1{\sigma_{+-}^2}\, \prod_{j\neq i,i+1} \frac{\sigma_{+-}}{\sigma_{+j}\,\sigma_{j-}} \frac{\sigma_{+-}}{\sigma_{+i\!+\!1}\,\sigma_{i\!+\!1 i}\,\sigma_{i-}}\right)\; \,\frac{1}{(+12\dots n-)}\,.
\end{align}
We will be very brief, since all details follow \cref{sec:BCFW_WS} closely. In particular, we will not discuss the  general set-up and assume a working knowledge of the conditions under which half-integrands lead to a pole, c.f. \eqref{eq:int_no_res} and \eqref{eq:int_res}. This section should thus only be read \emph{after} reading \cref{sec:BCFW_WS} or similar discussions in \cite{Dolan:2013isa, Geyer:2015jch}. We will frequently use the following tree-level result.

\begin{lemma}\label{lemma:half-ladder}
Half-ladder diagrams, with ordering $(12\ldots n)$ and endpoints $1$ and $n$ are given by the worldsheet formula
\begin{equation}
 \mathcal{A}_n^{\scalebox{0.6}{half-ladder}}(12\dots n) := \int_{\raisebox{-6pt}{\scalebox{0.7}{$\mathfrak{M}_{0,n}$}}}\hspace{-15pt}d\mu_{n}^{\CHY} \; \left( \frac1{\sigma_{1n}^2}\, \prod_{j=2}^{n-1} \frac{\sigma_{1n}}{\sigma_{1j}\,\sigma_{jn}}\right)\; \,\frac{1}{(12\dots n)}=\frac{1}{k_{12}^2\,k_{123}^2\,\ldots k_{12\ldots n-2}^2}\,.
\end{equation}
\end{lemma}
A proof can be given by various methods, such as soft recursion or factorisation. Since we reviewed factorisation above, here we simply sketch the relevant steps of the latter. In contrast to loop level, the scattering equations at tree level only encode physical poles, so there is no need to verify the absence of unphysical poles. The Parke-Taylor factor clearly ensures that all poles respect the cyclic ordering $(12\ldots n)$, and the other half-integrand vanishes on any factorisation channel involving both $1$ and $n$. The residues are easily checked, and we see that the formula factorises into 
\begin{equation}
 \mathrm{Res}_{z_i}\frac{1}{z}\mathcal{A}_n^{\scalebox{0.6}{half-ladder}}(\hat 12\dots\hat  n) = \frac{1}{k_{12\ldots i}^2} \mathcal{A}_{i+1}^{\scalebox{0.6}{half-ladder}}(\hat 12\dots i,L)\;\mathcal{A}_{n-i+1}^{\scalebox{0.6}{half-ladder}}(L,i+1\dots \hat n)\,,
\end{equation}
with $z_i= k_{12\ldots i}^2 / 2q\cdot k_{2\ldots i}^2$. The three-particle seed amplitude is trivial, $\mathcal{A}_{3}^{\scalebox{0.6}{half-ladder}}=1$, which establishes the BCFW recursion for the half-ladder.

\paragraph{$n$-gon.} With the above lemma, we are now ready to prove that our worldsheet formula \eqref{eq:n-gon} satisfies the appropriate BCFW recursion for the $n$-gon, which in fact contains only the single-cut term,
\begin{equation}\label{eq:BCFW_n-gon}
 \mathfrak{I}^{\scalebox{0.6}{$(1)$}}_{n\text{-gon}}(12\dots n) = \frac{1}{\ell^2}\mathcal{A}_{n+2}^{\scalebox{0.6}{half-ladder}}(\alpha)\,,
\end{equation}
as already pointed out in the motivating example in the introduction. On the worldsheet, we thus have to check that there are no unphysical poles, no tree-level factorisation channels, and that the only contribution comes from the single cut. 

The absence of unphysical poles can be checked by the same methods used for the MHV proposal: the colour half-integrand $\mathcal{C}^{\scalebox{0.6}{$(1)$}}(12\dots n)$ gives vanishing residues on unphysical loop propagators and trees containing both $1$ and $n$, while the $n$-gon half-integrand is zero on tadpole-like geometries for $n\geq 4$. The two half-integrands thus jointly ensure that no unphysical poles contribute. The $n$-gon half-integrand further gives vanishing residues for all tree-level factorisation channels, since it does not contain poles of the form $\sigma_{ij}$  for both $i$ and $j$ external particles. The only poles in the worldsheet formula thus come from singular geometries with $L=\{+\}\cup L^{\scalebox{0.6}{ext}}$, $1\in L^{\scalebox{0.6}{ext}}$ and  $n\notin L^{\scalebox{0.6}{ext}}$, or $L=\{-\}\cup L^{\scalebox{0.6}{ext}}$, $1\notin L^{\scalebox{0.6}{ext}}$ and  $n\in L^{\scalebox{0.6}{ext}}$, which correspond to the physical loop propagators \emph{unaffected} by the BCFW shift. Hence, the only contribution to the BCFW recursion comes indeed from the single cut, and is trivially given by \eqref{eq:BCFW_n-gon} using \cref{lemma:half-ladder}.

\paragraph{$(n-1)$-gon.} The proof of the $(n-1)$-gon worldsheet formula follows the same lines as above: we would like to verify that it satisfies
\begin{equation}\label{eq:BCFW_(n-1)-gon}
 \mathfrak{I}^{\scalebox{0.6}{$(1)$}}_{n\text{-gon}}(12\dots n) = \frac{1}{\ell^2}\mathcal{A}_{n+2,[i,i+1]}^{\scalebox{0.6}{mass. h-l}}(\alpha)\,,
\end{equation}
where $\mathcal{A}_{n+2,[i,i+1]}^{\scalebox{0.6}{mass. h-l}}$ is the half-ladder diagram with $n+1$ legs, with one massive leg $[i,i+1]$ attached. If we shift legs $1$ and $n$ in the BCFW recursion, there is no additional contribution  from a factorisation term as long as the massive leg lies on $1<i<n-1$.

Then the same analysis as for the $n$-gon can be applied here: all unphysical poles are absent due to the form of the half-integrands. The only subtlety lies in the absence of 1-$n$ massive corners, but these are ruled out by definition: the $(n-1)$-gon formla only holds for  $i\neq n$. With no unphysical poles contributing, we can check the tree-level factorisation channels. The only pole (involving only external particles) present in \emph{both} half-integrands is $\sigma_{i \,i\!+\!1}$, so there is a single tree-level factorisation channel with the expected pole $k_{i\, i\!+\!1}^2$  in the massive corner. This pole, however, is not shifted under BCFW, and does not contribute to the recursion. Repeating the $n$-gon discussion, the worldsheet formula contains physical loop propagators, but again these are unaffected by the BCFW shift.

Once more, the only contribution to the recursion stems from the single cut,  which indeed agrees with the expected half-ladder diagram with a massive leg at $[i,i+1]$.

\bibliography{twistor-bib}

\providecommand{\href}[2]{#2}\begingroup\raggedright\begin{thebibliography}{100}

\bibitem{Britto:2005fq}
R.~Britto, F.~Cachazo, B.~Feng and E.~Witten, \emph{{Direct proof of tree-level
  recursion relation in Yang-Mills theory}},
  \href{http://dx.doi.org/10.1103/PhysRevLett.94.181602}{\emph{Phys.Rev.Lett.}
  {\bfseries 94} (2005) 181602},
  [\href{https://arxiv.org/abs/hep-th/0501052}{{\ttfamily hep-th/0501052}}].

\bibitem{Britto:2004ap}
R.~Britto, F.~Cachazo and B.~Feng, \emph{{New recursion relations for tree
  amplitudes of gluons}},
  \href{http://dx.doi.org/10.1016/j.nuclphysb.2005.02.030}{\emph{Nucl.Phys.}
  {\bfseries B715} (2005) 499--522},
  [\href{https://arxiv.org/abs/hep-th/0412308}{{\ttfamily hep-th/0412308}}].

\bibitem{Bedford:2005yy}
J.~Bedford, A.~Brandhuber, B.~J. Spence and G.~Travaglini, \emph{{A Recursion
  relation for gravity amplitudes}},
  \href{http://dx.doi.org/10.1016/j.nuclphysb.2005.016}{\emph{Nucl.Phys.}
  {\bfseries B721} (2005) 98--110},
  [\href{https://arxiv.org/abs/hep-th/0502146}{{\ttfamily hep-th/0502146}}].

\bibitem{Cachazo:2005ca}
F.~Cachazo and P.~Svrcek, \emph{{Tree level recursion relations in general
  relativity}},  \href{https://arxiv.org/abs/hep-th/0502160}{{\ttfamily
  hep-th/0502160}}.

\bibitem{Cachazo:2013gna}
F.~Cachazo, S.~He and E.~Y. Yuan, \emph{{Scattering equations and
  Kawai-Lewellen-Tye orthogonality}},
  \href{http://dx.doi.org/10.1103/PhysRevD.90.065001}{\emph{Phys. Rev.}
  {\bfseries D90} (2014) 065001},
  [\href{https://arxiv.org/abs/1306.6575}{{\ttfamily 1306.6575}}].

\bibitem{Cachazo:2013hca}
F.~Cachazo, S.~He and E.~Y. Yuan, \emph{{Scattering of Massless Particles in
  Arbitrary Dimensions}},
  \href{http://dx.doi.org/10.1103/PhysRevLett.113.171601}{\emph{Phys.Rev.Lett.}
  {\bfseries 113} (2014) 171601},
  [\href{https://arxiv.org/abs/1307.2199}{{\ttfamily 1307.2199}}].

\bibitem{Cachazo:2013iea}
F.~Cachazo, S.~He and E.~Y. Yuan, \emph{{Scattering of Massless Particles:
  Scalars, Gluons and Gravitons}},
  \href{http://dx.doi.org/10.1007/JHEP07(2014)033}{\emph{JHEP} {\bfseries 1407}
  (2014) 033}, [\href{https://arxiv.org/abs/1309.0885}{{\ttfamily 1309.0885}}].

\bibitem{Roiban:2004yf}
R.~Roiban, M.~Spradlin and A.~Volovich, \emph{{On the tree level S matrix of
  Yang-Mills theory}},
  \href{http://dx.doi.org/10.1103/PhysRevD.70.026009}{\emph{Phys.Rev.}
  {\bfseries D70} (2004) 026009},
  [\href{https://arxiv.org/abs/hep-th/0403190}{{\ttfamily hep-th/0403190}}].

\bibitem{Witten:2003nn}
E.~Witten, \emph{{Perturbative gauge theory as a string theory in twistor
  space}},
  \href{http://dx.doi.org/10.1007/s00220-004-1187-3}{\emph{Commun.Math.Phys.}
  {\bfseries 252} (2004) 189--258},
  [\href{https://arxiv.org/abs/hep-th/0312171}{{\ttfamily hep-th/0312171}}].

\bibitem{Mason:2013sva}
L.~Mason and D.~Skinner, \emph{{Ambitwistor strings and the scattering
  equations}}, \href{http://dx.doi.org/10.1007/JHEP07(2014)048}{\emph{JHEP}
  {\bfseries 1407} (2014) 048},
  [\href{https://arxiv.org/abs/1311.2564}{{\ttfamily 1311.2564}}].

\bibitem{Bern:2005hs}
Z.~Bern, L.~J. Dixon and D.~A. Kosower, \emph{{On-shell recurrence relations
  for one-loop QCD amplitudes}},
  \href{http://dx.doi.org/10.1103/PhysRevD.71.105013}{\emph{Phys. Rev. D}
  {\bfseries 71} (2005) 105013},
  [\href{https://arxiv.org/abs/hep-th/0501240}{{\ttfamily hep-th/0501240}}].

\bibitem{Brandhuber:2005kd}
A.~Brandhuber, B.~Spence and G.~Travaglini, \emph{{From trees to loops and
  back}}, \href{http://dx.doi.org/10.1088/1126-6708/2006/01/142}{\emph{JHEP}
  {\bfseries 01} (2006) 142},
  [\href{https://arxiv.org/abs/hep-th/0510253}{{\ttfamily hep-th/0510253}}].

\bibitem{ArkaniHamed:2008gz}
N.~Arkani-Hamed, F.~Cachazo and J.~Kaplan, \emph{{What is the Simplest Quantum
  Field Theory?}}, \href{http://dx.doi.org/10.1007/JHEP09(2010)016}{\emph{JHEP}
  {\bfseries 1009} (2010) 016},
  [\href{https://arxiv.org/abs/0808.1446}{{\ttfamily 0808.1446}}].

\bibitem{ArkaniHamed:2009dn}
N.~Arkani-Hamed, F.~Cachazo, C.~Cheung and J.~Kaplan, \emph{{A Duality For The
  S Matrix}}, \href{http://dx.doi.org/10.1007/JHEP03(2010)020}{\emph{JHEP}
  {\bfseries 1003} (2010) 020},
  [\href{https://arxiv.org/abs/0907.5418}{{\ttfamily 0907.5418}}].

\bibitem{ArkaniHamed:2010kv}
N.~Arkani-Hamed, J.~L. Bourjaily, F.~Cachazo, S.~Caron-Huot and J.~Trnka,
  \emph{{The All-Loop Integrand For Scattering Amplitudes in Planar N=4 SYM}},
  \href{http://dx.doi.org/10.1007/JHEP01(2011)041}{\emph{JHEP} {\bfseries 01}
  (2011) 041}, [\href{https://arxiv.org/abs/1008.2958}{{\ttfamily 1008.2958}}].

\bibitem{Boels:2010nw}
R.~H. Boels, \emph{{On BCFW shifts of integrands and integrals}},
  \href{http://dx.doi.org/10.1007/JHEP11(2010)113}{\emph{JHEP} {\bfseries 11}
  (2010) 113}, [\href{https://arxiv.org/abs/1008.3101}{{\ttfamily 1008.3101}}].

\bibitem{Boels:2011mn}
R.~H. Boels and R.~S. Isermann, \emph{{Yang-Mills amplitude relations at loop
  level from non-adjacent BCFW shifts}},
  \href{http://dx.doi.org/10.1007/JHEP03(2012)051}{\emph{JHEP} {\bfseries 03}
  (2012) 051}, [\href{https://arxiv.org/abs/1110.4462}{{\ttfamily 1110.4462}}].

\bibitem{Arkani-Hamed:2016byb}
N.~Arkani-Hamed, J.~L. Bourjaily, F.~Cachazo, A.~B. Goncharov, A.~Postnikov and
  J.~Trnka, \emph{{Grassmannian Geometry of Scattering Amplitudes}}.
\newblock Cambridge University Press, 4, 2016,
  \href{http://dx.doi.org/10.1017/CBO9781316091548}{10.1017/CBO9781316091548}.

\bibitem{Bianchi:2018peu}
L.~Bianchi, A.~Brandhuber, R.~Panerai and G.~Travaglini, \emph{{Form factor
  recursion relations at loop level}},
  \href{http://dx.doi.org/10.1007/JHEP02(2019)182}{\emph{JHEP} {\bfseries 02}
  (2019) 182}, [\href{https://arxiv.org/abs/1812.09001}{{\ttfamily
  1812.09001}}].

\bibitem{Edison:2019ovj}
A.~Edison, E.~Herrmann, J.~Parra-Martinez and J.~Trnka, \emph{{Gravity loop
  integrands from the ultraviolet}},
  \href{https://arxiv.org/abs/1909.02003}{{\ttfamily 1909.02003}}.

\bibitem{Adamo:2013tsa}
T.~Adamo, E.~Casali and D.~Skinner, \emph{{Ambitwistor strings and the
  scattering equations at one loop}},
  \href{http://dx.doi.org/10.1007/JHEP04(2014)104}{\emph{JHEP} {\bfseries 1404}
  (2014) 104}, [\href{https://arxiv.org/abs/1312.3828}{{\ttfamily 1312.3828}}].

\bibitem{Geyer:2015bja}
Y.~Geyer, L.~Mason, R.~Monteiro and P.~Tourkine, \emph{{Loop Integrands for
  Scattering Amplitudes from the Riemann Sphere}},
  \href{http://dx.doi.org/10.1103/PhysRevLett.115.121603}{\emph{Phys. Rev.
  Lett.} {\bfseries 115} (2015) 121603},
  [\href{https://arxiv.org/abs/1507.00321}{{\ttfamily 1507.00321}}].

\bibitem{He:2015yua}
S.~He and E.~Y. Yuan, \emph{{One-loop Scattering Equations and Amplitudes from
  Forward Limit}},
  \href{http://dx.doi.org/10.1103/PhysRevD.92.105004}{\emph{Phys. Rev.}
  {\bfseries D92} (2015) 105004},
  [\href{https://arxiv.org/abs/1508.06027}{{\ttfamily 1508.06027}}].

\bibitem{Geyer:2015jch}
Y.~Geyer, L.~Mason, R.~Monteiro and P.~Tourkine, \emph{{One-loop amplitudes on
  the Riemann sphere}},
  \href{http://dx.doi.org/10.1007/JHEP03(2016)114}{\emph{JHEP} {\bfseries 03}
  (2016) 114}, [\href{https://arxiv.org/abs/1511.06315}{{\ttfamily
  1511.06315}}].

\bibitem{Cachazo:2015aol}
F.~Cachazo, S.~He and E.~Y. Yuan, \emph{{One-Loop Corrections from Higher
  Dimensional Tree Amplitudes}},
  \href{http://dx.doi.org/10.1007/JHEP08(2016)008}{\emph{JHEP} {\bfseries 08}
  (2016) 008}, [\href{https://arxiv.org/abs/1512.05001}{{\ttfamily
  1512.05001}}].

\bibitem{Adamo:2015hoa}
T.~Adamo and E.~Casali, \emph{{Scattering equations, supergravity integrands,
  and pure spinors}},
  \href{http://dx.doi.org/10.1007/JHEP05(2015)120}{\emph{JHEP} {\bfseries 1505}
  (2015) 120}, [\href{https://arxiv.org/abs/1502.06826}{{\ttfamily
  1502.06826}}].

\bibitem{Feng:2016nrf}
B.~Feng, \emph{{CHY-construction of Planar Loop Integrands of Cubic Scalar
  Theory}}, \href{http://dx.doi.org/10.1007/JHEP05(2016)061}{\emph{JHEP}
  {\bfseries 05} (2016) 061},
  [\href{https://arxiv.org/abs/1601.05864}{{\ttfamily 1601.05864}}].

\bibitem{Geyer:2016wjx}
Y.~Geyer, L.~Mason, R.~Monteiro and P.~Tourkine, \emph{{Two-Loop Scattering
  Amplitudes from the Riemann Sphere}},
  \href{http://dx.doi.org/10.1103/PhysRevD.94.125029}{\emph{Phys. Rev.}
  {\bfseries D94} (2016) 125029},
  [\href{https://arxiv.org/abs/1607.08887}{{\ttfamily 1607.08887}}].

\bibitem{Geyer:2018xwu}
Y.~Geyer and R.~Monteiro, \emph{{Two-Loop Scattering Amplitudes from
  Ambitwistor Strings: from Genus Two to the Nodal Riemann Sphere}},
  \href{http://dx.doi.org/10.1007/JHEP11(2018)008}{\emph{JHEP} {\bfseries 11}
  (2018) 008}, [\href{https://arxiv.org/abs/1805.05344}{{\ttfamily
  1805.05344}}].

\bibitem{Geyer:2019hnn}
Y.~Geyer, R.~Monteiro and R.~Stark-Muchão, \emph{{Two-Loop Scattering
  Amplitudes: Double-Forward Limit and Colour-Kinematics Duality}},
  \href{http://dx.doi.org/10.1007/JHEP12(2019)049}{\emph{JHEP} {\bfseries 12}
  (2019) 049}, [\href{https://arxiv.org/abs/1908.05221}{{\ttfamily
  1908.05221}}].

\bibitem{Feng:2019xiq}
B.~Feng and C.~Hu, \emph{{One-loop CHY-Integrand of Bi-adjoint Scalar Theory}},
  \href{http://dx.doi.org/10.1007/JHEP02(2020)187}{\emph{JHEP} {\bfseries 02}
  (2020) 187}, [\href{https://arxiv.org/abs/1912.12960}{{\ttfamily
  1912.12960}}].

\bibitem{Wen:2020qrj}
C.~Wen and S.-Q. Zhang, \emph{{D3-Brane Loop Amplitudes from M5-Brane Tree
  Amplitudes}},  \href{https://arxiv.org/abs/2004.02735}{{\ttfamily
  2004.02735}}.

\bibitem{Edison:2020uzf}
A.~Edison, S.~He, O.~Schlotterer and F.~Teng, \emph{{One-loop Correlators and
  BCJ Numerators from Forward Limits}},
  \href{https://arxiv.org/abs/2005.03639}{{\ttfamily 2005.03639}}.

\bibitem{Spradlin:2009qr}
M.~Spradlin and A.~Volovich, \emph{{From Twistor String Theory To Recursion
  Relations}}, \href{http://dx.doi.org/10.1103/PhysRevD.80.085022}{\emph{Phys.
  Rev. D} {\bfseries 80} (2009) 085022},
  [\href{https://arxiv.org/abs/0909.0229}{{\ttfamily 0909.0229}}].

\bibitem{Dolan:2009wf}
L.~Dolan and P.~Goddard, \emph{{Gluon Tree Amplitudes in Open Twistor String
  Theory}}, \href{http://dx.doi.org/10.1088/1126-6708/2009/12/032}{\emph{JHEP}
  {\bfseries 0912} (2009) 032},
  [\href{https://arxiv.org/abs/0909.0499}{{\ttfamily 0909.0499}}].

\bibitem{Nandan:2009cc}
D.~Nandan, A.~Volovich and C.~Wen, \emph{{A Grassmannian Etude in NMHV
  Minors}}, \href{http://dx.doi.org/10.1007/JHEP07(2010)061}{\emph{JHEP}
  {\bfseries 07} (2010) 061},
  [\href{https://arxiv.org/abs/0912.3705}{{\ttfamily 0912.3705}}].

\bibitem{ArkaniHamed:2009dg}
N.~Arkani-Hamed, J.~Bourjaily, F.~Cachazo and J.~Trnka, \emph{{Unification of
  Residues and Grassmannian Dualities}},
  \href{http://dx.doi.org/10.1007/JHEP01(2011)049}{\emph{JHEP} {\bfseries 01}
  (2011) 049}, [\href{https://arxiv.org/abs/0912.4912}{{\ttfamily 0912.4912}}].

\bibitem{Bullimore:2009cb}
M.~Bullimore, L.~Mason and D.~Skinner, \emph{{Twistor-Strings, Grassmannians
  and Leading Singularities}},
  \href{http://dx.doi.org/10.1007/JHEP03(2010)070}{\emph{JHEP} {\bfseries 1003}
  (2010) 070}, [\href{https://arxiv.org/abs/0912.0539}{{\ttfamily 0912.0539}}].

\bibitem{Farrow:2017eol}
J.~A. Farrow and A.~E. Lipstein, \emph{{From 4d Ambitwistor Strings to On Shell
  Diagrams and Back}},
  \href{http://dx.doi.org/10.1007/JHEP07(2017)114}{\emph{JHEP} {\bfseries 07}
  (2017) 114}, [\href{https://arxiv.org/abs/1705.07087}{{\ttfamily
  1705.07087}}].

\bibitem{Cachazo:2012pz}
F.~Cachazo, L.~Mason and D.~Skinner, \emph{{Gravity in Twistor Space and its
  Grassmannian Formulation}},
  \href{http://dx.doi.org/10.3842/SIGMA.2014.051}{\emph{SIGMA} {\bfseries 10}
  (2014) 051}, [\href{https://arxiv.org/abs/1207.4712}{{\ttfamily 1207.4712}}].

\bibitem{Dolan:2013isa}
L.~Dolan and P.~Goddard, \emph{{Proof of the Formula of Cachazo, He and Yuan
  for Yang-Mills Tree Amplitudes in Arbitrary Dimension}},
  \href{http://dx.doi.org/10.1007/JHEP05(2014)010}{\emph{JHEP} {\bfseries 1405}
  (2014) 010}, [\href{https://arxiv.org/abs/1311.5200}{{\ttfamily 1311.5200}}].

\bibitem{Albonico:2020mge}
G.~Albonico, Y.~Geyer and L.~Mason, \emph{{Recursion and worldsheet formulae
  for 6d superamplitudes}},  \href{https://arxiv.org/abs/2001.05928}{{\ttfamily
  2001.05928}}.

\bibitem{Arkani-Hamed:2013jha}
N.~Arkani-Hamed and J.~Trnka, \emph{{The Amplituhedron}},
  \href{http://dx.doi.org/10.1007/JHEP10(2014)030}{\emph{JHEP} {\bfseries 10}
  (2014) 030}, [\href{https://arxiv.org/abs/1312.2007}{{\ttfamily 1312.2007}}].

\bibitem{Arkani-Hamed:2017mur}
N.~Arkani-Hamed, Y.~Bai, S.~He and G.~Yan, \emph{{Scattering Forms and the
  Positive Geometry of Kinematics, Color and the Worldsheet}},
  \href{http://dx.doi.org/10.1007/JHEP05(2018)096}{\emph{JHEP} {\bfseries 05}
  (2018) 096}, [\href{https://arxiv.org/abs/1711.09102}{{\ttfamily
  1711.09102}}].

\bibitem{Baadsgaard:2015twa}
C.~Baadsgaard, N.~E.~J. Bjerrum-Bohr, J.~L. Bourjaily, S.~Caron-Huot, P.~H.
  Damgaard and B.~Feng, \emph{{New Representations of the Perturbative
  S-Matrix}},
  \href{http://dx.doi.org/10.1103/PhysRevLett.116.061601}{\emph{Phys. Rev.
  Lett.} {\bfseries 116} (2016) 061601},
  [\href{https://arxiv.org/abs/1509.02169}{{\ttfamily 1509.02169}}].

\bibitem{Feynman:1963ax}
R.~P. Feynman, \emph{{Quantum theory of gravitation}}, {\emph{Acta Phys.
  Polon.} {\bfseries 24} (1963) 697--722}.

\bibitem{Feynman:1972mt}
R.~P. Feynman, \emph{{CLOSED LOOP AND TREE DIAGRAMS. (TALK)}}, .

\bibitem{Feynman:2000fh}
R.~P. Feynman, \emph{{Selected papers of Richard Feynman: With commentary}},
  \href{http://dx.doi.org/10.1142/4270}{\emph{World Sci. Ser.20th Cent. Phys.}
  {\bfseries 27} (2000) pp.1--600}.

\bibitem{CaronHuot:2010zt}
S.~Caron-Huot, \emph{{Loops and Trees}},
  \href{http://dx.doi.org/10.1007/JHEP05(2011)080}{\emph{JHEP} {\bfseries 05}
  (2011) 080}, [\href{https://arxiv.org/abs/1007.3224}{{\ttfamily 1007.3224}}].

\bibitem{Boels:2016jmi}
R.~H. Boels and H.~Luo, \emph{{On-shell recursion relations for generic
  integrands}},  \href{https://arxiv.org/abs/1610.05283}{{\ttfamily
  1610.05283}}.

\bibitem{Bern:1993qk}
Z.~Bern, G.~Chalmers, L.~J. Dixon and D.~A. Kosower, \emph{{One loop N gluon
  amplitudes with maximal helicity violation via collinear limits}},
  \href{http://dx.doi.org/10.1103/PhysRevLett.72.2134}{\emph{Phys. Rev. Lett.}
  {\bfseries 72} (1994) 2134--2137},
  [\href{https://arxiv.org/abs/hep-ph/9312333}{{\ttfamily hep-ph/9312333}}].

\bibitem{Mahlon:1993si}
G.~Mahlon, \emph{{Multi - gluon helicity amplitudes involving a quark loop}},
  \href{http://dx.doi.org/10.1103/PhysRevD.49.4438}{\emph{Phys. Rev. D}
  {\bfseries 49} (1994) 4438--4453},
  [\href{https://arxiv.org/abs/hep-ph/9312276}{{\ttfamily hep-ph/9312276}}].

\bibitem{He:2014bga}
S.~He, Y.-t. Huang and C.~Wen, \emph{{Loop Corrections to Soft Theorems in
  Gauge Theories and Gravity}},
  \href{http://dx.doi.org/10.1007/JHEP12(2014)115}{\emph{JHEP} {\bfseries 12}
  (2014) 115}, [\href{https://arxiv.org/abs/1405.1410}{{\ttfamily 1405.1410}}].

\bibitem{Bern:2008qj}
Z.~Bern, J.~Carrasco and H.~Johansson, \emph{{New Relations for Gauge-Theory
  Amplitudes}},
  \href{http://dx.doi.org/10.1103/PhysRevD.78.085011}{\emph{Phys.Rev.}
  {\bfseries D78} (2008) 085011},
  [\href{https://arxiv.org/abs/0805.3993}{{\ttfamily 0805.3993}}].

\bibitem{Bern:2010ue}
Z.~Bern, J.~J.~M. Carrasco and H.~Johansson, \emph{{Perturbative Quantum
  Gravity as a Double Copy of Gauge Theory}},
  \href{http://dx.doi.org/10.1103/PhysRevLett.105.061602}{\emph{Phys.Rev.Lett.}
  {\bfseries 105} (2010) 061602},
  [\href{https://arxiv.org/abs/1004.0476}{{\ttfamily 1004.0476}}].

\bibitem{He:2016mzd}
S.~He and O.~Schlotterer, \emph{{New Relations for Gauge-Theory and Gravity
  Amplitudes at Loop Level}},
  \href{http://dx.doi.org/10.1103/PhysRevLett.118.161601}{\emph{Phys. Rev.
  Lett.} {\bfseries 118} (2017) 161601},
  [\href{https://arxiv.org/abs/1612.00417}{{\ttfamily 1612.00417}}].

\bibitem{He:2017spx}
S.~He, O.~Schlotterer and Y.~Zhang, \emph{{New BCJ representations for one-loop
  amplitudes in gauge theories and gravity}},
  \href{http://dx.doi.org/10.1016/j.nuclphysb.2018.03.003}{\emph{Nucl. Phys.}
  {\bfseries B930} (2018) 328--383},
  [\href{https://arxiv.org/abs/1706.00640}{{\ttfamily 1706.00640}}].

\bibitem{Geyer:2017ela}
Y.~Geyer and R.~Monteiro, \emph{{Gluons and gravitons at one loop from
  ambitwistor strings}},
  \href{http://dx.doi.org/10.1007/JHEP03(2018)068}{\emph{JHEP} {\bfseries 03}
  (2018) 068}, [\href{https://arxiv.org/abs/1711.09923}{{\ttfamily
  1711.09923}}].

\bibitem{Gomez:2017lhy}
H.~Gomez, \emph{{Quadratic Feynman Loop Integrands From Massless Scattering
  Equations}}, \href{http://dx.doi.org/10.1103/PhysRevD.95.106006}{\emph{Phys.
  Rev.} {\bfseries D95} (2017) 106006},
  [\href{https://arxiv.org/abs/1703.04714}{{\ttfamily 1703.04714}}].

\bibitem{Gomez:2017cpe}
H.~Gomez, C.~Lopez-Arcos and P.~Talavera, \emph{{One-loop Parke-Taylor factors
  for quadratic propagators from massless scattering equations}},
  \href{http://dx.doi.org/10.1007/JHEP10(2017)175}{\emph{JHEP} {\bfseries 10}
  (2017) 175}, [\href{https://arxiv.org/abs/1707.08584}{{\ttfamily
  1707.08584}}].

\bibitem{Ahmadiniaz:2018nvr}
N.~Ahmadiniaz, H.~Gomez and C.~Lopez-Arcos, \emph{{Non-planar one-loop
  Parke-Taylor factors in the CHY approach for quadratic propagators}},
  \href{http://dx.doi.org/10.1007/JHEP05(2018)055}{\emph{JHEP} {\bfseries 05}
  (2018) 055}, [\href{https://arxiv.org/abs/1802.00015}{{\ttfamily
  1802.00015}}].

\bibitem{Agerskov:2019ryp}
J.~Agerskov, N.~E.~J. Bjerrum-Bohr, H.~Gomez and C.~Lopez-Arcos,
  \emph{{Yang-Mills Loop Amplitudes from Scattering Equations}},
  \href{https://arxiv.org/abs/1910.03602}{{\ttfamily 1910.03602}}.

\bibitem{ArkaniHamed:2008yf}
N.~Arkani-Hamed and J.~Kaplan, \emph{{On Tree Amplitudes in Gauge Theory and
  Gravity}}, \href{http://dx.doi.org/10.1088/1126-6708/2008/04/076}{\emph{JHEP}
  {\bfseries 04} (2008) 076},
  [\href{https://arxiv.org/abs/0801.2385}{{\ttfamily 0801.2385}}].

\bibitem{Bern:2013yya}
Z.~Bern, S.~Davies, T.~Dennen, Y.-t. Huang and J.~Nohle,
  \emph{{Color-Kinematics Duality for Pure Yang-Mills and Gravity at One and
  Two Loops}}, \href{http://dx.doi.org/10.1103/PhysRevD.92.045041}{\emph{Phys.
  Rev. D} {\bfseries 92} (2015) 045041},
  [\href{https://arxiv.org/abs/1303.6605}{{\ttfamily 1303.6605}}].

\bibitem{ArkaniHamed:2012nw}
N.~Arkani-Hamed, J.~L. Bourjaily, F.~Cachazo, A.~B. Goncharov, A.~Postnikov and
  J.~Trnka, \emph{{Grassmannian Geometry of Scattering Amplitudes}}.
\newblock Cambridge University Press, 2016,
  \href{http://dx.doi.org/10.1017/CBO9781316091548}{10.1017/CBO9781316091548}.

\bibitem{Benincasa:2015zna}
P.~Benincasa, \emph{{On-shell diagrammatics and the perturbative structure of
  planar gauge theories}},  \href{https://arxiv.org/abs/1510.03642}{{\ttfamily
  1510.03642}}.

\bibitem{Bern:2002zk}
Z.~Bern, A.~De~Freitas, L.~J. Dixon and H.~Wong, \emph{{Supersymmetric
  regularization, two loop QCD amplitudes and coupling shifts}},
  \href{http://dx.doi.org/10.1103/PhysRevD.66.085002}{\emph{Phys.\ Rev.\ D}
  {\bfseries 66} (2002) 085002},
  [\href{https://arxiv.org/abs/hep-ph/0202271}{{\ttfamily hep-ph/0202271}}].

\bibitem{Elvang:2013cua}
H.~Elvang and Y.-t. Huang, \emph{{Scattering Amplitudes}},
  \href{https://arxiv.org/abs/1308.1697}{{\ttfamily 1308.1697}}.

\bibitem{He:2015wgf}
S.~He, R.~Monteiro and O.~Schlotterer, \emph{{String-inspired BCJ numerators
  for one-loop MHV amplitudes}},
  \href{http://dx.doi.org/10.1007/JHEP01(2016)171}{\emph{JHEP} {\bfseries 01}
  (2016) 171}, [\href{https://arxiv.org/abs/1507.06288}{{\ttfamily
  1507.06288}}].

\bibitem{Mahlon:1993fe}
G.~Mahlon, \emph{{One loop multi - photon helicity amplitudes}},
  \href{http://dx.doi.org/10.1103/PhysRevD.49.2197}{\emph{Phys. Rev.}
  {\bfseries D49} (1994) 2197--2210},
  [\href{https://arxiv.org/abs/hep-ph/9311213}{{\ttfamily hep-ph/9311213}}].

\bibitem{Bern:1996ja}
Z.~Bern, L.~J. Dixon, D.~C. Dunbar and D.~A. Kosower, \emph{{One loop selfdual
  and N=4 superYang-Mills}},
  \href{http://dx.doi.org/10.1016/S0370-2693(96)01676-0}{\emph{Phys. Lett.}
  {\bfseries B394} (1997) 105--115},
  [\href{https://arxiv.org/abs/hep-th/9611127}{{\ttfamily hep-th/9611127}}].

\bibitem{Bern:1991aq}
Z.~Bern and D.~A. Kosower, \emph{{The Computation of loop amplitudes in gauge
  theories}}, \href{http://dx.doi.org/10.1016/0550-3213(92)90134-W}{\emph{Nucl.
  Phys.} {\bfseries B379} (1992) 451--561}.

\bibitem{Badger:2005zh}
S.~D. Badger, E.~W.~N. Glover, V.~V. Khoze and P.~Svrcek, \emph{{Recursion
  relations for gauge theory amplitudes with massive particles}},
  \href{http://dx.doi.org/10.1088/1126-6708/2005/07/025}{\emph{JHEP} {\bfseries
  07} (2005) 025}, [\href{https://arxiv.org/abs/hep-th/0504159}{{\ttfamily
  hep-th/0504159}}].

\bibitem{Forde:2005ue}
D.~Forde and D.~A. Kosower, \emph{{All-multiplicity amplitudes with massive
  scalars}}, \href{http://dx.doi.org/10.1103/PhysRevD.73.065007}{\emph{Phys.
  Rev.} {\bfseries D73} (2006) 065007},
  [\href{https://arxiv.org/abs/hep-th/0507292}{{\ttfamily hep-th/0507292}}].

\bibitem{Boels:2013bi}
R.~H. Boels, R.~S. Isermann, R.~Monteiro and D.~O'Connell,
  \emph{{Colour-Kinematics Duality for One-Loop Rational Amplitudes}},
  \href{http://dx.doi.org/10.1007/JHEP04(2013)107}{\emph{JHEP} {\bfseries 04}
  (2013) 107}, [\href{https://arxiv.org/abs/1301.4165}{{\ttfamily 1301.4165}}].

\bibitem{Monteiro:2011pc}
R.~Monteiro and D.~O'Connell, \emph{{The Kinematic Algebra From the Self-Dual
  Sector}}, \href{http://dx.doi.org/10.1007/JHEP07(2011)007}{\emph{JHEP}
  {\bfseries 07} (2011) 007},
  [\href{https://arxiv.org/abs/1105.2565}{{\ttfamily 1105.2565}}].

\bibitem{NigelGlover:2008ur}
E.~Nigel~Glover and C.~Williams, \emph{{One-Loop Gluonic Amplitudes from Single
  Unitarity Cuts}},
  \href{http://dx.doi.org/10.1088/1126-6708/2008/12/067}{\emph{JHEP} {\bfseries
  12} (2008) 067}, [\href{https://arxiv.org/abs/0810.2964}{{\ttfamily
  0810.2964}}].

\bibitem{Farrow:2018cqi}
J.~A. Farrow, \emph{{A Monte Carlo Approach to the 4D Scattering Equations}},
  \href{http://dx.doi.org/10.1007/JHEP08(2018)085}{\emph{JHEP} {\bfseries 08}
  (2018) 085}, [\href{https://arxiv.org/abs/1806.02732}{{\ttfamily
  1806.02732}}].

\bibitem{Hodges:2009hk}
A.~Hodges, \emph{{Eliminating spurious poles from gauge-theoretic amplitudes}},
  \href{http://dx.doi.org/10.1007/JHEP05(2013)135}{\emph{JHEP} {\bfseries 05}
  (2013) 135}, [\href{https://arxiv.org/abs/0905.1473}{{\ttfamily 0905.1473}}].

\bibitem{Lipstein:2012vs}
A.~E. Lipstein and L.~Mason, \emph{{From the holomorphic Wilson loop to `d log'
  loop-integrands for super-Yang-Mills amplitudes}},
  \href{http://dx.doi.org/10.1007/JHEP05(2013)106}{\emph{JHEP} {\bfseries 05}
  (2013) 106}, [\href{https://arxiv.org/abs/1212.6228}{{\ttfamily 1212.6228}}].

\bibitem{Berends:1987me}
F.~A. Berends and W.~Giele, \emph{{Recursive Calculations for Processes with n
  Gluons}}, \href{http://dx.doi.org/10.1016/0550-3213(88)90442-7}{\emph{Nucl.
  Phys. B} {\bfseries 306} (1988) 759--808}.

\bibitem{Henn:2019mvc}
J.~Henn, B.~Power and S.~Zoia, \emph{{Conformal Invariance of the One-Loop
  All-Plus Helicity Scattering Amplitudes}},
  \href{http://dx.doi.org/10.1007/JHEP02(2020)019}{\emph{JHEP} {\bfseries 02}
  (2020) 019}, [\href{https://arxiv.org/abs/1911.12142}{{\ttfamily
  1911.12142}}].

\bibitem{Bern:1995db}
Z.~Bern and A.~G. Morgan, \emph{{Massive loop amplitudes from unitarity}},
  \href{http://dx.doi.org/10.1016/0550-3213(96)00078-8}{\emph{Nucl. Phys.}
  {\bfseries B467} (1996) 479--509},
  [\href{https://arxiv.org/abs/hep-ph/9511336}{{\ttfamily hep-ph/9511336}}].

\bibitem{DelDuca:1999rs}
V.~Del~Duca, L.~J. Dixon and F.~Maltoni, \emph{{New color decompositions for
  gauge amplitudes at tree and loop level}},
  \href{http://dx.doi.org/10.1016/S0550-3213(99)00809-3}{\emph{Nucl.Phys.}
  {\bfseries B571} (2000) 51--70},
  [\href{https://arxiv.org/abs/hep-ph/9910563}{{\ttfamily hep-ph/9910563}}].

\bibitem{Bern:2011rj}
Z.~Bern, C.~Boucher-Veronneau and H.~Johansson, \emph{{N >= 4 Supergravity
  Amplitudes from Gauge Theory at One Loop}},
  \href{http://dx.doi.org/10.1103/PhysRevD.84.105035}{\emph{Phys. Rev. D}
  {\bfseries 84} (2011) 105035},
  [\href{https://arxiv.org/abs/1107.1935}{{\ttfamily 1107.1935}}].

\bibitem{Bern:1998sv}
Z.~Bern, L.~J. Dixon, M.~Perelstein and J.~S. Rozowsky, \emph{{Multileg one
  loop gravity amplitudes from gauge theory}},
  \href{http://dx.doi.org/10.1016/S0550-3213(99)00029-2}{\emph{Nucl. Phys.}
  {\bfseries B546} (1999) 423--479},
  [\href{https://arxiv.org/abs/hep-th/9811140}{{\ttfamily hep-th/9811140}}].

\bibitem{Heslop:2016plj}
P.~Heslop and A.~E. Lipstein, \emph{{On-shell diagrams for $ \mathcal{N} $ = 8
  supergravity amplitudes}},
  \href{http://dx.doi.org/10.1007/JHEP06(2016)069}{\emph{JHEP} {\bfseries 06}
  (2016) 069}, [\href{https://arxiv.org/abs/1604.03046}{{\ttfamily
  1604.03046}}].

\bibitem{Mafra:2014gja}
C.~R. Mafra and O.~Schlotterer, \emph{{Towards one-loop SYM amplitudes from the
  pure spinor BRST cohomology}},
  \href{http://dx.doi.org/10.1002/prop.201400076}{\emph{Fortsch. Phys.}
  {\bfseries 63} (2015) 105--131},
  [\href{https://arxiv.org/abs/1410.0668}{{\ttfamily 1410.0668}}].

\bibitem{Berg:2016fui}
M.~Berg, I.~Buchberger and O.~Schlotterer, \emph{{String-motivated one-loop
  amplitudes in gauge theories with half-maximal supersymmetry}},
  \href{http://dx.doi.org/10.1007/JHEP07(2017)138}{\emph{JHEP} {\bfseries 07}
  (2017) 138}, [\href{https://arxiv.org/abs/1611.03459}{{\ttfamily
  1611.03459}}].

\bibitem{Bjerrum-Bohr:2013iza}
N.~E.~J. Bjerrum-Bohr, T.~Dennen, R.~Monteiro and D.~O'Connell,
  \emph{{Integrand Oxidation and One-Loop Colour-Dual Numerators in N=4 Gauge
  Theory}}, \href{http://dx.doi.org/10.1007/JHEP07(2013)092}{\emph{JHEP}
  {\bfseries 07} (2013) 092},
  [\href{https://arxiv.org/abs/1303.2913}{{\ttfamily 1303.2913}}].

\bibitem{Bern:2017yxu}
Z.~Bern, J.~J. Carrasco, W.-M. Chen, H.~Johansson and R.~Roiban, \emph{{Gravity
  Amplitudes as Generalized Double Copies of Gauge-Theory Amplitudes}},
  \href{http://dx.doi.org/10.1103/PhysRevLett.118.181602}{\emph{Phys. Rev.
  Lett.} {\bfseries 118} (2017) 181602},
  [\href{https://arxiv.org/abs/1701.02519}{{\ttfamily 1701.02519}}].

\bibitem{Casali:2015vta}
E.~Casali, Y.~Geyer, L.~Mason, R.~Monteiro and K.~A. Roehrig, \emph{{New
  Ambitwistor String Theories}},
  \href{http://dx.doi.org/10.1007/JHEP11(2015)038}{\emph{JHEP} {\bfseries 11}
  (2015) 038}, [\href{https://arxiv.org/abs/1506.08771}{{\ttfamily
  1506.08771}}].

\bibitem{Geyer:2014fka}
Y.~Geyer, A.~E. Lipstein and L.~J. Mason, \emph{{Ambitwistor Strings in Four
  Dimensions}},
  \href{http://dx.doi.org/10.1103/PhysRevLett.113.081602}{\emph{Phys. Rev.
  Lett.} {\bfseries 113} (2014) 081602},
  [\href{https://arxiv.org/abs/1404.6219}{{\ttfamily 1404.6219}}].

\bibitem{Berkovits:2013xba}
N.~Berkovits, \emph{{Infinite Tension Limit of the Pure Spinor Superstring}},
  \href{http://dx.doi.org/10.1007/JHEP03(2014)017}{\emph{JHEP} {\bfseries 1403}
  (2014) 017}, [\href{https://arxiv.org/abs/1311.4156}{{\ttfamily 1311.4156}}].

\bibitem{Ohmori:2015sha}
K.~Ohmori, \emph{{Worldsheet Geometries of Ambitwistor String}},
  \href{http://dx.doi.org/10.1007/JHEP06(2015)075}{\emph{JHEP} {\bfseries 06}
  (2015) 075}, [\href{https://arxiv.org/abs/1504.02675}{{\ttfamily
  1504.02675}}].

\bibitem{Baadsgaard:2015voa}
C.~Baadsgaard, N.~E.~J. Bjerrum-Bohr, J.~L. Bourjaily and P.~H. Damgaard,
  \emph{{Integration Rules for Scattering Equations}},
  \href{http://dx.doi.org/10.1007/JHEP09(2015)129}{\emph{JHEP} {\bfseries 09}
  (2015) 129}, [\href{https://arxiv.org/abs/1506.06137}{{\ttfamily
  1506.06137}}].

\bibitem{KK1989}
R.~Kleiss and H.~Kuijf, \emph{Multigluon cross sections and 5-jet production at
  hadron colliders},
  \href{http://dx.doi.org/10.1016/0550-3213(89)90574-9}{\emph{Nuclear Physics
  B} {\bfseries 312} (1989) 616 -- 644}.

\bibitem{Roehrig:2017gbt}
K.~A. Roehrig and D.~Skinner, \emph{{A Gluing Operator for the Ambitwistor
  String}}, \href{http://dx.doi.org/10.1007/JHEP01(2018)069}{\emph{JHEP}
  {\bfseries 01} (2018) 069},
  [\href{https://arxiv.org/abs/1709.03262}{{\ttfamily 1709.03262}}].

\bibitem{Edison:2020ehu}
A.~Edison and F.~Teng, \emph{{Efficient Calculation of Crossing Symmetric BCJ
  Tree Numerators}},  \href{https://arxiv.org/abs/2005.03638}{{\ttfamily
  2005.03638}}.

\bibitem{Casali:2020knc}
E.~Casali, S.~Mizera and P.~Tourkine, \emph{{Loop amplitudes monodromy
  relations and color-kinematics duality}},
  \href{https://arxiv.org/abs/2005.05329}{{\ttfamily 2005.05329}}.

\bibitem{Polchinski:1998rq}
J.~Polchinski, \emph{{String theory. Vol. 1: An introduction to the bosonic
  string}}.
\newblock Cambridge Monographs on Mathematical Physics. Cambridge University
  Press, 12, 2007,
  \href{http://dx.doi.org/10.1017/CBO9780511816079}{10.1017/CBO9780511816079}.

\bibitem{Bern:2019prr}
Z.~Bern, J.~J. Carrasco, M.~Chiodaroli, H.~Johansson and R.~Roiban, \emph{{The
  Duality Between Color and Kinematics and its Applications}},
  \href{https://arxiv.org/abs/1909.01358}{{\ttfamily 1909.01358}}.

\bibitem{Monteiro:2013rya}
R.~Monteiro and D.~O'Connell, \emph{{The Kinematic Algebras from the Scattering
  Equations}}, \href{http://dx.doi.org/10.1007/JHEP03(2014)110}{\emph{JHEP}
  {\bfseries 03} (2014) 110},
  [\href{https://arxiv.org/abs/1311.1151}{{\ttfamily 1311.1151}}].

\bibitem{Damgaard:2019ztj}
D.~Damgaard, L.~Ferro, T.~Lukowski and M.~Parisi, \emph{{The Momentum
  Amplituhedron}}, \href{http://dx.doi.org/10.1007/JHEP08(2019)042}{\emph{JHEP}
  {\bfseries 08} (2019) 042},
  [\href{https://arxiv.org/abs/1905.04216}{{\ttfamily 1905.04216}}].

\bibitem{Ferro:2020ygk}
L.~Ferro and T.~Lukowski, \emph{{Amplituhedra, and Beyond}},
  \href{https://arxiv.org/abs/2007.04342}{{\ttfamily 2007.04342}}.

\bibitem{Salvatori:2018aha}
G.~Salvatori, \emph{{1-loop Amplitudes from the Halohedron}},
  \href{http://dx.doi.org/10.1007/JHEP12(2019)074}{\emph{JHEP} {\bfseries 12}
  (2019) 074}, [\href{https://arxiv.org/abs/1806.01842}{{\ttfamily
  1806.01842}}].

\bibitem{Kalyanapuram:2020vil}
N.~Kalyanapuram and R.~G. Jha, \emph{{Positive Geometries for all Scalar
  Theories from Twisted Intersection Theory}},
  \href{https://arxiv.org/abs/2006.15359}{{\ttfamily 2006.15359}}.

\bibitem{Brandhuber:2006bf}
A.~Brandhuber, B.~Spence and G.~Travaglini, \emph{{Amplitudes in Pure
  Yang-Mills and MHV Diagrams}},
  \href{http://dx.doi.org/10.1088/1126-6708/2007/02/088}{\emph{JHEP} {\bfseries
  02} (2007) 088}, [\href{https://arxiv.org/abs/hep-th/0612007}{{\ttfamily
  hep-th/0612007}}].

\end{thebibliography}\endgroup
\bibliographystyle{JHEP}

\end{document}